\documentclass[11pt,a4paper,english,oneside]{report} 
\usepackage{a4, amsmath, amstext, amsfonts, amsthm, amssymb, revsymb}
\usepackage[english]{babel}
\usepackage[latin1]{inputenc}
\usepackage[final]{graphicx}
\usepackage{fancyhdr}
\usepackage[Sonny]{fncychapl}
\usepackage[numbers, sort]{natbib}
\usepackage{pb-diagram,lamsarrow,pb-lams}
\usepackage[colorlinks=false, citecolor=magenta, urlcolor=blue, plainpages=false]{hyperref}
\usepackage{boxedminipage}
\newcommand{\eprint}[2][]{\href{http://arxiv.org/abs/#2}{#2}}

\usepackage{psfrag}
\newcommand{\rar}{\rotatebox[origin=c]{0}{$\circlearrowleft$}}
\newcommand{\lar}{\rotatebox[origin=c]{0}{$\circlearrowright$}}
\def\bp{\begin{picture}}
\def\ep{\end{picture}}
\def\ci{\circle}
\def\li{\line}

\ChNumVar{\Huge\bf} \ChRuleWidth{1.25pt} \ChTitleVar{\Huge\bf} \ChNameVar{\LARGE \bf}

\newtheorem{theorem}{Theorem}[section]
\newtheorem{conjecture}{Conjecture}[section]
\newtheorem{corollary}{Corollary}[section]
\newtheorem{lemma}{Lemma}[section]
\newtheorem{exam}{Example}[section]
\newtheorem{prop}{Proposition}[section]
\newtheorem{obs}{Observation}[section]
\newcommand{\ido}{\openone}
\DeclareMathOperator{\Tr}{Tr}

\DeclareMathOperator{\diag}{diag}
\newcommand{\ra}{\rangle}
\newcommand{\la}{\langle}

\begin{document}
\pagenumbering{Alph}
\begin{titlepage} 
\begin{center}
\vspace*{\stretch{1}}
{
\huge \bf
Entanglement Distillation

\vspace{2cm}
\LARGE A Discourse on Bound Entanglement in 

Quantum Information Theory}
\vspace{11cm}
\large

{\bf
Lieven Clarisse}

PhD thesis
\vspace{1cm}

University of York

Department of Mathematics

September 2006
\end{center}
\end{titlepage}


\chapter*{Abstract}
\pagenumbering{roman}
\addcontentsline{toc}{chapter}{Abstract} 
\markboth{Abstract}{Abstract}
In recent years entanglement has been recognised as a useful resource in quantum information and computation. This applies primarily to pure state entanglement which is, due to interaction with the environment, rarely available. Decoherence provides the main motivation for the study of entanglement distillation. A remarkable effect in the context of distillation is the existence of bound entangled states, states from which no pure state entanglement can be distilled. The concept of entanglement distillation also relates to a canonical way of theoretically quantifying mixed state entanglement.

This thesis is, apart from a review chapter on distillation, mainly a theoretical study of bound entanglement and the two major open problems in their classification. The first of these is the classification of PPT bound entanglement (separability problem). After having reviewed known tools we study in detail the multipartite permutation criteria, for which we present new results in their classification. We solve an open problem on the existence of certain PPT states. The Schmidt number of a quantum state is a largely unvalued concept, we analyse it in detail and introduce the Schmidt robustness. The notion of Schmidt number is exploited in the study of the second major problem, that of the existence of NPT bound entanglement. We reformulate this existence problem as particular instance of the separability problem, present a uniform treatment of distillability and numerically demonstrate that bound entanglement is primarily found in low dimensional systems. 

\newpage
\addcontentsline{toc}{chapter}{Contents}
\tableofcontents 

\listoffigures 
\addcontentsline{toc}{chapter}{List of figures}

\chapter*{Preface}
\addcontentsline{toc}{chapter}{Preface} 
\markboth{Preface}{Preface}
Quantum error correction allows quantum information to be reliably sent through noisy channels. Tremendous effort has been devoted to the study of quantum error correcting codes and capacities of quantum channels. This thesis deals with entanglement distillation, also known as counterfactual error correction \cite{Alb01}. The idea is to purify the noisy channel before trying to send any information over it, rather than the more indirect method of error correction. Personally I feel very strongly about the idea of distillation since it also relates directly to a canonical way of quantifying how much useful entanglement is present in a state. The distillable entanglement is a physical property of quantum states, bridging between a practical and theoretical study of entanglement. Bound entangled states are states from which no pure entanglement can be extracted, although entanglement is needed to construct the state. They are two main open problems in the classification of bound entangled states (the separability and the NPT bound entanglement problem) and they constituted the bulk of my research. 

\begin{figure}
\begin{boxedminipage}[t]{14.5cm}
\begin{center}
\includegraphics[width=14cm]{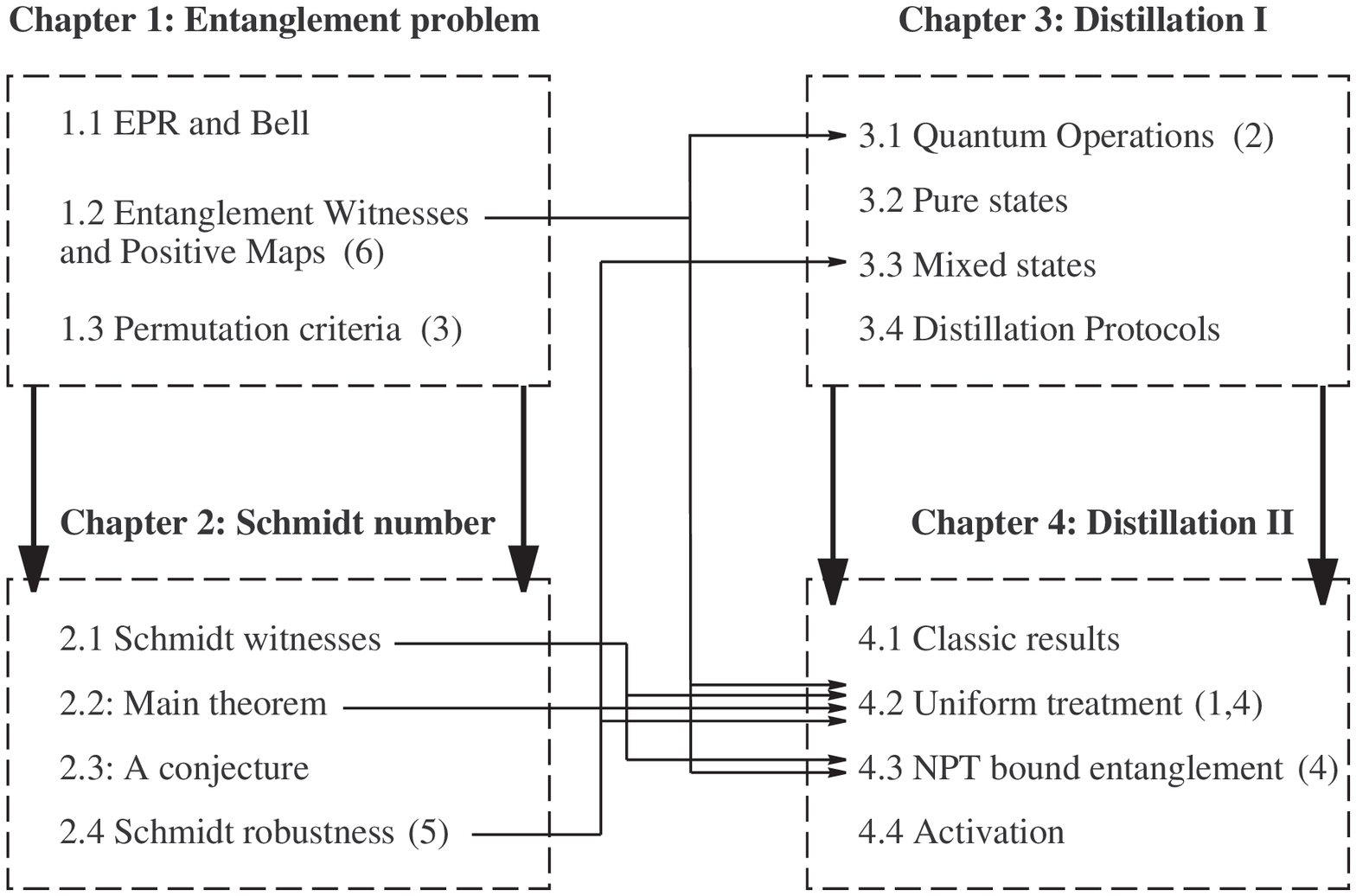}
\end{center}
\begin{enumerate}
\item L.~Clarisse. Characterization of distillability of entanglement in terms of positive maps. \emph{Physical Review A}, 71:032332, 2005, \eprint{quant-ph/0403073}.
\item L.~Clarisse, S.~Ghosh, S.~Severini, and A.~Sudbery. Entangling power of permutations. \emph{Physical Review A}, 72:012314, 2005, \eprint{quant-ph/0502040}.
\item L.~Clarisse and P.~Wocjan. On independent permutation separability criteria. \emph{Quantum Information and Computation}, 6:277--288, 2005, \eprint{quant-ph/0504160}.
\item L.~Clarisse. The distillability problem revisited. \emph{Quantum Information and Computation}, 6:539--560, 2006, \eprint{quant-ph/0510035}.
\item L.~Clarisse. On the Schmidt robustness of pure states. \emph{Journal of Physics A: Mathematical and General}, 39:4239--4249, 2006, \eprint{quant-ph/0512012}.
\item L.~Clarisse. Construction of bound entangled edge states with special ranks. 2006, \emph{To be published in Physics Letters A}, \eprint{quant-ph/0603283}.
\end{enumerate}
\end{boxedminipage}
\begin{center}
\caption[Structure of the dissertation and publications]{Structure of the dissertation and publications.}
\label{sotdap}
\end{center}
\end{figure}

During the course of my Ph.~D.\ I have published a number of papers, but in writing up, I have tried to make this thesis more than just a compilation or medley of these papers. The primary goal has been to convey the context and motivation of the different problems. Secondary, I have tried to provide a review of the state-of-the-art knowledge in entanglement distillation in the bipartite scenario. A notable gap in this presentation is a treatment of experimental entanglement distillation. The structure of the dissertation and the interrelation between the chapters, is depicted in Figure~\ref{sotdap}. The figure also contains a reference list of my papers, and their position in the thesis.

Chapter one deals with the separability problem. After having reviewed the theory of positive maps and entanglement witnesses, I present new classes of PPT entangled states with particular ranks. Next, the permutation criteria for separability are studied in detail and a new result on their classification is presented. This is the only place in the dissertation which deals with multipartite systems. 

The Schmidt number of mixed states is a generalisation of the separable versus entangled division, imposing a finer structure on the set of quantum states. This is the topic of Chapter two. The tools for the separability problem are readily generalised to deal with Schmidt numbers. I review in some detail known results. In a final section, I introduce the Schmidt robustness and derive upper and lower bounds for pure states. The motivation for the topics discussed in this chapter becomes clear in Chapter four.

In Chapter three I review the literature on bipartite entanglement distillation. After the necessary introduction of quantum operations, the main results on pure state entanglement transformation are sketched. Then I move over to mixed state entanglement. I start with a discussion on entanglement measures focusing on the distillable entanglement and end with a discussion of irreversibility in mixed state transformations. This leads to the introduction of bound entangled states. The chapter ends with a review of mixed state entanglement distillation protocols.

Chapter four is a discourse on distillability and NPT bound entanglement and describes my main results. As usual, the first sections introduce known results. In particular I review the three fundamental theorems on distillability. Next a uniform treatment of distillability theorems is presented. I illustrate the power of this reformulation by means of four applications. The most important ones are the derivation of reduction-like maps and an estimation of the volume of distillable states. In the next section the problem of the existence of NPT bound entanglement is reformulated as the problem of determining whether a particular set of states is entangled. It follows that the two major open problems on the classification of bound entangled states are really just one problem (although the problems are by no means equivalent). I end this dissertation with a short review on activation of bound entanglement.

\newpage
{\large \bf Acknowledgements}

Words fall short in gratitude to my supervisor Anthony Sudbery. He took the ungratifying task of agreeing on the supervision of an international student, with little or no references. Subsequently he arranged, together with Samuel Braunstein, financial support. I thank both of them wholeheartedly for this. From the very start Tony has been enormously encouraging and helpful, always making time for me despite a plethora of departmental duties. I have enjoyed many discussions with him; his mathematical quickness and wittiness never cease to amaze and inspire me. It is also hard not to acknowledge the mental support he offered me when I was in need for it the most. 

Many thanks go to my coauthors Sibasish Ghosh, Simone Severini, Anthony Sudbery and Pawe{\l} Wocjan; it was a true honour and pleasure working with them. During my Ph.~D.\ I enjoyed discussions with a number of researchers, including those who have helped me out clarifying technical issues on their own work. In particular, I want to thank Samuel Braunstein, Ignacio Cirac, Jens Eisert, William Hall, Otfried G\"uhne, Erik Hostens, Florian Hulpke, Maciej Lewenstein, Andreas Winter and Christof Zalka. I thank Stefan Weigert and Martin Plenio for pointing out numerous mistakes.

My girlfriend Christine has played a vital role in the completion of this Ph.~D.; Thank you Christine, for your continuous support, dedication and love. Equally appreciative I am for her commitment during the final year, which was infested by a prolonged period of illness. I would also like to express a deep gratitude to my parents and my brother and his wife who supported and encouraged me throughout. It was my parents financial support during the first year that made my move to York possible. I thank my friends in Catherine House and fellow Ph.~D.\ students for support, friendship and a wonderful time in York.

I enjoyed financial support through a W.~W.\ Smith Scholarship and EPSRC grant GR/S822176/01 for which I feel greatly indebted.

This thesis was typeset with the Miktex distribution (\url{www.miktex.de}) of {\LaTeX} and text editor Winshell (\url{www.winshell.de}). The figures were produced using GraphEq (\url{www.peda.com}), Mayura Draw (\url{www.mayura.com}) and Matlab (\url{www.mathworks.com}).

\bigskip
\noindent York \hspace{11cm} Lieven Clarisse \\
July 2006

\chapter*{Author's declaration}
\addcontentsline{toc}{chapter}{Author's declaration} 
\markboth{Author's declaration}{Author's declaration}
Most of this material has been presented before. This is true for all literature review and most of the original work. The introduction of PPT states with special ranks, Section~\ref{PPTESS} has been presented in Ref.~\cite{Clarisse06}. Theorem~\ref{racew} is one of the few results that has not been presented before. The new results on the classification of permutation criteria Sections~\ref{mtpct} and \ref{ilpct} have been obtained in collaboration with P.~Wocjan and published in Ref.~\cite{CW05}. The main idea came from myself. The results on the Schmidt robustness have been published in Ref.~\cite{Clarisse05b}. In Section~\ref{entopssec} the paper on the entangling power of permutations is briefly mentioned. This paper was written in collaboration with S. Ghosh, S. Severini and A. Sudbery, with me being the first author \cite{CGSS05}. The new results in Sections~\ref{utodt} and \ref{nptbound} have been obtained by myself and published in Ref.~\cite{Clarisse04, Clarisse05}. The Appendix contains two Matlab programs, which have not been published before. See also Figure~\ref{sotdap} (page \pageref{sotdap}) for a schematic overview of the published papers and their position in the thesis.

\chapter{Entanglement and separability} 
\pagenumbering{arabic}
\setcounter{page}{1}
\label{entsepchap}
From the 1930s, starting with the EPR thought experiment \cite{EPR35, Bohr35}, entanglement was mainly studied in the context of the completeness of quantum mechanics. It was not until the 1960s when the problem of completeness was put on solid mathematical grounds by the remarkable works of Gleason \cite{Gleason57}, Kochen and Specker \cite{KS67} and Bell \cite{Bell64, Bell66}. The work of Bell dealt directly with the EPR thought experiment and is of enormous importance as it showed that entanglement is incompatible with a certain local classical inequality which can be verified experimentally. We will discuss the EPR argument and Bell's inequality briefly in Section~\ref{eprbel}. More in-depth treatments can be found in Ref.~\cite{Redhead87, Peres93}.

In recent years, the problem of completeness seems pretty much settled and considerable effort has been put into understanding the mathematical structure of entanglement. A primary problem is, given an arbitrary quantum state, to determine whether it is entangled or not. The study of this problem came together with the realisation that Bell-like inequalities are pretty weak tests for entanglement. The breakthrough \cite{Peres96, HHH96} came when notions from convex analysis and ${\cal C}^*$-algebras were applied to the problem. In Section~\ref{mainsectionchtwo} we review these mathematical results leading to the general framework of entanglement witness and its dual formulation in terms of positive maps. We also introduce the so-called PPT states and relate them to classical problems in linear algebra. The separability problem has been shown to be NP-hard \cite{Gurvits03}, and therefore there is little hope of finding a simple analytical method which is able to distinguish all entangled states from all separable. Subsequently, interest has grown in finding good numerical methods in tackling the problem. In Section~\ref{numap} we apply tools from semidefinite programming to the separability problem. In particular we focus on the work by Doherty et~al.\ \cite{DPS03}, which besides a complete numerical solution, offers appealing analytical features. 

In Section~\ref{permcrit} we step away from the entanglement witness approach and study the permutation criteria for detecting entanglement, initiated in Ref.~\cite{Rudolph02} and further developed in Ref.~\cite{CW02,HHH02,Rudolph04, WH05b,CW05}. Although in general, these criteria cannot detect every entangled state, they are particularly appealing because of their operational simplicity and can be applied in the bipartite as well as the multipartite scenario. 

Reviews of the entanglement versus separability problem, from a variety of different viewpoints, can be found in Ref.~\cite{LBCKKSST00, Alb01, Terhal00, Terhal01, BCHHKLS01, Bruss02, EK02, EGHHKMBLS02, SSLS05}.

\section{The EPR argument and Bell's inequalities}
\label{eprbel}
The existence of entanglement \cite{Schrodinger35} follows naturally from the quantum mechanical formalism. This was first made explicit in the famous paper \cite{EPR35} by Einstein, Podolsky and Rosen (EPR), where it was used to argue that quantum mechanics as a physical theory is incomplete. Their argument runs as follows. Consider a particle with known position decaying into two equal particles. Without measurement, all we know is that the particles will drift apart with opposite momenta and that their centre of mass remains constant throughout. Assuming that both particles are well separated, there is no way a measurement on one of the particles can affect the other particle. This is the famous \emph{local realism} assumption, which dictates that well separated systems can be completely and independently described. Now, measuring the momentum or position of the first particle enables one to predict either the momentum or position of the second system, without disturbing it. In quantum mechanics position and momentum are non-commuting observables and the theory cannot predict precise values for both, hence EPR are led to conclude that quantum mechanics is incomplete. A formidable philosophical debate followed on the foundations and interpretation of quantum mechanics, starting with Bohr's reply \cite{Bohr35}. For a historical account see Ref.~\cite{Jammer74}.

In 1951 Bohm \cite{Bohm51} formulated a version of the EPR argument with an entangled spin system. Here both well separated parties ($A$ and $B$) share one half of the singlet state
$$
|\psi_-\ra=\frac{1}{\sqrt 2}(|01\ra-|10\ra).
$$
Measurement of the spin of the first half in any direction reveals the spin of the other particle in that direction. By an EPR argument the quantum mechanical description of the second half cannot be complete. It is for such a system Bell \cite{Bell64} derived his famous inequality. We will present here a proof of the more general CHSH inequality \cite{CHSH69}.

Suppose the two parties share a large number of singlet states and measure them one by one. For each measurement, observer $A$ chooses to measure the spin in direction $\vec{a}$ or $\vec{a'}$ at random (with results $a_n$ or $a'_n$). Analogously observer $B$ measures the spin of his half in direction $\vec{b}$ or $\vec{b'}$ (with results $b_n$ or $b'_n$). Let us denote with $Q(\vec{a},\vec{b})$ the average value of the product of the measurements outcomes $ab$. This quantity is a measure of the correlation between the two measurement outcomes. Now the EPR argument implies that for each singlet pair, $a_n, a'_n, b_n, b'_n$ have definite values ($\pm 1$) before a measurement has taken place. By going through all combinations, it is easy to see that 
$$
a_n(b_n+b'_n)+{a'}_n(b_n-b'_n)=\pm 2.
$$
Because the left hand side is either -2 or +2, its average over all $n$ is not larger than 2, thus we get on average
$$
| Q(\vec{a},\vec{b})+ Q(\vec{a},\vec{b'})+Q(\vec{a'},\vec{b})-Q(\vec{a'},\vec{b'}) | \leq 2.
$$
This inequality gives an upper limit to the correlation coefficients. Choosing the unit vectors $\vec{a}, \vec{b}, \vec{a'}, \vec{b'}$ in the same plane but separated by angles of $\pi/4$, one can verify that this inequality is violated quantum mechanically for the singlet state (with a value of $2\sqrt{2}$). Bell type inequalities can be derived for any number of parties, for more than two distinct measurement outcomes or for any number of alternative setups, see Ref.~\cite{Peres99}.

Violation of the Bell inequalities has been experimentally demonstrated using the polarisation of photons \cite{FC72,ADR82} and experimental results are in excellent agreement with the predictions of quantum mechanics. Apart from entanglement in polarisation, violation of Bell inequalities with photons has been demonstrated using entanglement based on position and time \cite{Franson89, TBZG98}, phase and momentum \cite{RT90} and orbital angular momentum \cite{MVWZ01}. Experiments also have been conducted without the use of photons, namely with protons \cite{LRM76} and atoms \cite{RKMSIMW01}. It should be mentioned however, that to date no experiment has been loophole free, so that technically, local realism has not been ruled out experimentally. An example of such a loophole is not having proper time-like separation of the detectors (see Ref.~\cite{CS78, FW02} for a discussion of these matters). 

\section{The main road}
\label{mainsectionchtwo}
For any entangled pure state shared by an arbitrary number of parties, it was shown that there exists a Bell inequality which is violated \cite{Gisin91, PR92}. For mixed states this does not seem the case. In this context it has been shown that although a state might not violate any Bell-like inequality, it still may be useful for teleportation \cite{Popescu94, HHH96c}. Other states only violate some inequality after a certain local generalised measurements \cite{Popescu95, Gisin96}. Yet other states only show a violation when multiple copies are measured collectively \cite{Peres96b}. It will be clear that Bell inequalities are a poor test of non-classical behaviour. It has proven convenient to characterise the set of entangled mixed states with a certain reasonable mathematical definition. The advantage of this approach is to have a workable definition of what is entangled and what not. Though only recently it has been rigorously proven that all states defined in such way behave truly non-classically (see Chapter~\ref{chapnptbound}). 

We will give the definition of entanglement for bipartite states, generalisation to multipartite states is straightforward. Let us first recall the definition of a pure state \cite{Schrodinger35, EK95}. Let $\cal H$ be a Hilbert space such that ${\cal H}= {\cal H}_A \otimes {\cal H}_B\cong {\mathbb C}^{d_1} \otimes {\mathbb C}^{d_2}$ (with integers $d_1,d_2\geq 2)$. A pure state $|\psi\ra \in {\cal H}$ shared by two parties $A$ and $B$ is called separable if and only if it can be written as 
$$
|\psi\ra=|\psi_A\ra \otimes |\psi_B\ra,
$$
with $|\psi_A\ra \in {\cal H}_A$ and $|\psi_B\ra \in {\cal H}_B$, in other words, if $|\psi\ra$ belongs to the Cartesian product ${\cal H}_A \times {\cal H}_B \subset {\cal H}_A \otimes {\cal H}_B$ \cite{Sudbery05}. Otherwise we call $|\psi\ra$ entangled.

There is a simple way of determining whether a pure state is entangled based on the Schmidt decomposition:
\begin{theorem}[Schmidt decomposition \cite{Schmidt06, EK95, NC00}]
Let $|\psi\ra \in \cal H$, then there exist orthonormal states $|a_i\ra$ and $|b_i\ra$ such that
\begin{align}
\label{sdec}
|\psi\ra=\sum_i \lambda_i |a_i\ra|b_i\ra,
\end{align}
with $\lambda_i \geq 0$ and $\sum_i \lambda^2_i=1$. The coefficients $\lambda_i$ are known as the Schmidt coefficients. The number of non-zero Schmidt coefficients is referred to as the Schmidt rank $S(\psi)$ of $|\psi\ra$. The state $|\psi\ra$ is separable if and only if it has Schmidt rank one.
\end{theorem}
\begin{proof}
Let
$$
|\psi\ra=\sum_{ij} \psi_{ij} |ij\ra,
$$
with $\Psi=\{\psi_{ij}\}$ a matrix of complex numbers. Recall that by the singular value decomposition \cite{HJ85} every matrix $\Psi$ can be written as $\Psi=UDV$, with $U$ and $V$ unitary and $D$ a diagonal matrix with nonnegative elements. So we have
$$
|\psi\ra=\sum_{ijk} U_{ji} D_{ii} V_{ik} |jk\ra.
$$
Now putting $|a_i\ra=\sum_j U_{ji} |j\ra$, $|b_i\ra=\sum_k V_{ik} |k\ra$ and $\lambda_i=d_{ii}$ we obtain the form~(\ref{sdec}). The orthonormality follows from $U$ and $V$ being unitary. To check whether a pure state is separable it is therefore sufficient to calculate the rank of $\Psi$.
\end{proof}

Moving to mixed states, let us denote with ${\cal L}({\cal H},{\cal H})={\cal L}_{\cal H}={\cal L}_{{\cal H}_A \otimes {\cal H}_B} \cong {\cal L}_{{\cal H}_A} \otimes {\cal L}_{{\cal H}_B}$ the set of linear operators on ${\cal H}$. Since we are working over a finite dimensional space ${\cal L}_{\cal H} \cong {\cal M}_{d_1d_2}$ the space of complex matrices of order $d_1d_2$. The subset of (Hermitian) positive semidefinite operators will be denoted ${\cal L}^+_{\cal H}$. The set of density operators on ${\cal H}$ constitutes then the normalised operators in ${\cal L}^+_{\cal H}$ and we will denote them with $\cal D_{\cal H}$ or simply $\cal D$. A bipartite mixed state $\rho \in \cal D$ is called separable if and only if it can be written as a convex sum of pure product states, i.e.\
$$
\rho = \sum_i p_i |\psi^A_i\ra\la \psi^A_i| \otimes |\psi^B_i\ra\la \psi^B_i|,
$$
with $p_i>0$, $\sum_i p_i = 1$, $|\psi^A_i\ra \in {\cal H}_A$ and $|\psi^B_i\ra \in {\cal H}_B$. Otherwise the state is called entangled. The set of separable states will be denoted $\cal S$ and can thus be identified with the convex hull of the normalised operators in ${\cal L}^+_{{\cal H}_A} \otimes {\cal L}^+_{{\cal H}_B} \subsetneq {\cal L}^+_{{\cal H}_A \otimes {\cal H}_B}$ . If such a decomposition exists, then from the Caratheodory theorem \cite{Horodecki97, Rockafellar70} follows that we can always replace it with a decomposition of at most $m^2$ terms where $m=d_1d_2=\dim \cal H$. This definition was first coined independently by Werner and Primas in 1983\footnote{R.~Werner, LMS Workshop on quantum information and computation (York 2005).}, but popularised in Ref.~\cite{Werner89}. The physical idea behind it is that a separable state can always be constructed by two independent parties with the aid of classical communication only (but again, a full physical justification came first later).

\subsection{Hahn-Banach theorem and entanglement witnesses}
To check whether a pure state is entangled it is sufficient to expand it in its Schmidt decomposition. For mixed states the problem is much harder as no such canonical decomposition can be obtained in a straightforward way. However, by exploiting the convexity of the set of separable states a useful characterisation can be obtained, which we will use to construct some necessary conditions for separability. 

The following theorem gives a geometrical characterization of the problem of determining whether $\rho\in D$ is contained in a certain convex subset $C\subset D$:
\begin{theorem}
\label{separation}
Let $C\subset D$ be a convex set of states in a Hilbert space ${\cal H}$, then for each $\rho\notin C$ there exists a Hermitian operator $A\in B({\cal H})$ such that 
$$
\Tr{(A\rho)}<0 \;\;\; \text{and} \;\;\; \Tr{(A\sigma)}\geq 0
$$
for all $\sigma \in C$.
\end{theorem}

This theorem is an immediate consequence of basic theorems in functional analysis \cite{Rockafellar70, Lax02}. Namely, the Hahn-Banach theorem states that a convex set and a point lying outside it can be separated by a hyperplane $W$, and the Riesz-Frechet representation theorem then characterizes such hyperplanes. The hyperplanes $W$ are commonly called witnesses \cite{Terhal01} (they witness states outside $C$).

\begin{corollary}[Horodecki et~al.\ \cite{HHH96}]
\label{corhor}
A state $\rho \in \cal D$ is separable if and only if $\Tr(\rho W)\geq 0$ for all Hermitian operators $W$ such that $\Tr([|\psi_A\ra\la \psi_A| \otimes |\psi_B\ra\la \psi_B|]W) \geq 0$, with $|\psi_A\ra \in {\cal H}_A$ and $|\psi_B\ra \in {\cal H}_B$. Such operators $W$ will be referred to as entanglement witnesses.
\end{corollary}

We have encountered an entanglement witness when we discussed the CHSH inequality. From our discussion, it is easy to see that 
\begin{align}
W_{\text{CHSH}}=2\openone -[\vec{a}\cdot\vec{\sigma}\otimes (\vec{b}+\vec{b'})\cdot\vec{\sigma}+
\vec{a'}\cdot\vec{\sigma}\otimes (\vec{b}-\vec{b'})\cdot\vec{\sigma}],
\end{align}
is an entanglement witness. Here $\vec{\sigma}=(\sigma_x,\sigma_y,\sigma_z)$ is a vector containing the Pauli spin matrices \cite{Isham95, Terhal01, HGBL05}. Note also that a general entanglement witness $W$ can be measured locally, by decomposing it as (non-convex) sum of product states. The problem of finding an optimized decomposition both in the optimal number of projectors on product vectors and in the optimal number of settings of the detectors has been partially solved in Ref.~\cite{GHBELMS02, GHBELMS02b}. 

Let us now give two examples of entanglement witnesses, which will be important for the sequel. The first one \cite{HHH96} is the so-called swap or flip operator $F$, which acts on state vectors as $F|\psi_A\ra|\psi_B\ra=|\psi_B\ra|\psi_A\ra$ or in a basis $\{|ij\ra\}$ of $\cal H$:
$$
F=\sum_{ij}|ij\ra\la ji|.
$$
It is not so hard to check that this operator is indeed positive on product states. The swap operator is an entanglement witness because $F$ has negative eigenvalues and will therefore necessarily act negative on some state (for instance for $|\psi_-\ra$). The second example is given by 
$$
W_\text{R}=\openone-dP_+
$$
with $P_+$ the projector onto the maximally entangled state $|\psi_+\ra= \frac{1}{\sqrt d}\sum_i |ii\ra$. Since the maximum overlap of a separable state with $P_+$ is $1/d$, and since $W$ has negative eigenvalues, it is an entanglement witness (for instance for $|\psi_+\ra$).

Intuitively the value $\min_W \Tr(\rho W)$ can be associated with the distance of $\rho$ to the set of separable states and in some way is related to the degree of entanglement of $\rho$. However, as such the minimum is ill-defined as the set of entanglement witnesses is unbounded. This can be overcome by imposing suitable constraints on the entanglement witnesses $W$, and it turns out that in this way a variety of geometrical intuitive entanglement measures can be obtained \cite{Brandao05, PV05}. We will come back to the topic of geometrical entanglement measures briefly in Chapter~\ref{chapsmidt}.

\subsection{Jamio{\l}kowski isomorphism and positive maps}
\label{sejaiso}
In this section we will reformulate the characterisation of separable states (Corollary~\ref{corhor}) in terms of positive maps. This can be done by making use of the Jamio{\l}\-kowski isomorphism which establishes a one-to-one correspondence between linear operators and linear maps between operator spaces. We will outline the isomorphism and illustrate it for Hermitian preserving, positive and completely positive maps. We then apply it to the examples of last section, and obtain the reduction and the partial transpose criteria, respectively. The results outlined in this section date back to the 1970-80s' \cite{Choi82} but have only recently \cite{HHH96} been used in quantum information. This section is meant as an introduction to the isomorphism and we will see more about positive maps and the context in which they were studied in Section~\ref{sdr}. 

A Hermitian map $\Lambda: {\cal L}_{{\cal H}_{A}} \rightarrow {\cal L}_{{\cal H}_{B}}$ is a \emph{linear} map that maps Hermitian operators onto Hermitian operators. A positive map $\Lambda: {\cal L}_{{\cal H}_{A}} \rightarrow {\cal L}_{{\cal H}_{B}}$ is a linear map that maps positive operators $\rho\geq 0$ onto positive operators $\Lambda(\rho)\geq 0$. A $k$-positive map is a positive map such that the induced map 
$$
\Lambda_k={\openone_k} \otimes \Lambda: {\cal L}_{{\mathbb C}^{k}} \otimes {\cal L}_{{\cal H}_{A}} \rightarrow {\cal L}_{{\mathbb C}^{k}} \otimes {\cal L}_{{\cal H}_{B}}
$$
is positive for $k\geq 1$. A complete positive map (CP map) is a $k$-positive map for all positive $k$.

In the proof of the Schmidt decomposition we found it useful to interpret the coefficients $\psi_{ij}$ in the expansion of $|\psi\ra = \sum_{ij} \psi_{ij} |ij\ra$ as the matrix $\{\psi_{ij}\}$ of an operator $\Psi$. With this notation one can equally write
\begin{align}
|\psi\ra = \sqrt{d} (\Psi \otimes \openone) |\psi_+\ra = \sqrt{d} (\openone \otimes \Psi^T) |\psi_+\ra.
\label{filtering}
\end{align}
This relation in effect establishes an explicit isomorphism between $ {\cal H}_A \otimes {\cal H}_B$ and ${\cal L}({\cal H}_A,{\cal H}_B)$ \cite{Bhatia97} and comes back frequently in quantum information (see for example \cite{VV02,Verstraete02, ZB04}). Suppose now that we have a general Hermitian operator $W \in {\cal L}_{{\cal H}_A \otimes {\cal H}_B}$, using the spectral decomposition we can write
$$
W=\sum_i w_i |\phi_i\ra\la\phi_i|,
$$
with all coefficients $w_i$ real. Using the above isomorphism we associate with every vector $|\phi_i\ra$ the matrix $\Phi_i$ and write
$$
W=d \sum_i w_i (\openone \otimes \Phi_i) P_+ (\openone \otimes \Phi^\dagger_i).
$$
Defining a map $\Lambda_W$ as acting as 
\begin{align}
\label{herfor}
\Lambda_W(A)= \sum_i w_i \Phi_i A \Phi^\dagger_i,
\end{align}
for all $A$, this can be written as
\begin{align}
\label{jaher}
W=d (\openone \otimes \Lambda_W) P_+.
\end{align}
It is clear that when a linear map $\Lambda_W$ can be written in the form of~(\ref{herfor}) it is automatically a Hermitian map. Conversely, using the above correspondence~(\ref{jaher}), it is easy to see that any Hermitian map $\Lambda_W$ will correspond to a Hermitian operator $W$ and therefore can be written in the form~(\ref{herfor}).
When $W$ is a positive operator, all the coefficients $w_i$ will be positive and $\Lambda_W$ will become a completely positive operator. Conversely, it is again easy to see that any completely positive map can be written in the form~(\ref{herfor}). These results were first derived by de Pillis \cite{Pillis67} and Choi \cite{Choi75} respectively (see also Ref.~\cite{Stinespring55, Hill73, PH81, Kraus83, HH97, Keyl02, VV02}), we summarise them in the following theorem.
\begin{theorem}
\label{choipillis}
A linear map $\Lambda: {\cal L}_{{\cal H}_{A}} \rightarrow {\cal L}_{{\cal H}_{B}}$ is Hermitian (completely positive) if and only if for all operators $A\in{\cal L}_{{\cal H}_{A}}$ it can be written as
$$
\Lambda(A)=\sum_i {\lambda_i V_i A V^\dagger_i},
$$
for some operators $V_i: {\cal L}_{{\cal H}_{A}} \rightarrow {\cal L}_{{\cal H}_{B}}$, and real (positive) numbers $\lambda_i$.
\end{theorem}

In the proof of this theorem we made use of the relation $W=d (\openone \otimes \Lambda_W) P_+$ which associates with each linear map $\Lambda_W$ an operator $W$. Note that $W$ defines uniquely the map $\Lambda_W$ since a linear map is determined by its action on the basis $\{|i\ra\la j|\}$. Indeed, it is not so hard to see that~(\ref{jaher}) can be written as
\begin{align}
\label{difforja}
\la k | \Lambda_W(|i\ra\la j|) |l\ra=\la i k |W |jl \ra. 
\end{align}
Thus in this way we have defined an isomorphism between linear maps $\Lambda: {\cal L}_{{\cal H}_{A}} \rightarrow {\cal L}_{{\cal H}_{B}}$ and operators $W \in {\cal L}_{{\cal H}_{A}} \otimes {\cal L}_{{\cal H}_{B}}$. This is the famous Jamio{\l}kowski isomorphism and can also be written as
$$
\Lambda_W(X)=\Tr_A[W(X^T \otimes \openone)].
$$

\begin{figure}
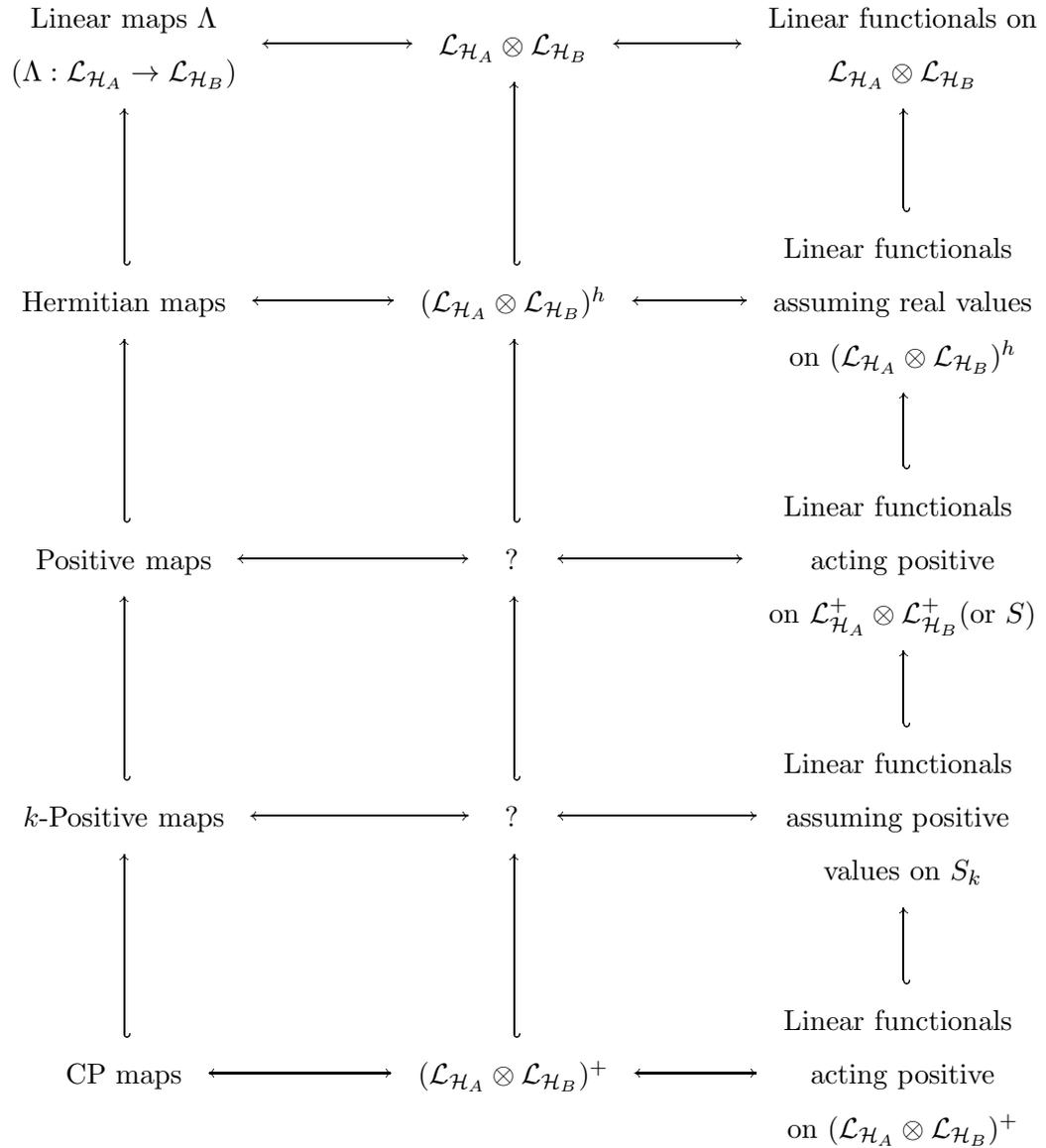

\begin{center}$
\begin{diagram}
\node{\begin{array}{c} \text{Linear maps } \Lambda \\ (\Lambda: {\cal L}_{{\cal H}_{A}} \rightarrow {\cal L}_{{\cal H}_{B}}) \end{array}} \arrow[3]{e,t,<>}{} 
\node[3]{\begin{array}{c} {\cal L}_{{\cal H}_{A}} \otimes {\cal L}_{{\cal H}_{B}} \end{array}} \arrow[3]{e,t,<>}{} 
\node[3]{\begin{array}{c} \text{Linear functionals on} \\ {\cal L}_{{\cal H}_{A}} \otimes {\cal L}_{{\cal H}_{B}} \end{array}} 
\\\\
\node{\begin{array}{c} \text{Hermitian maps} \end{array}} \arrow[3]{e,t,<>}{} \arrow[2]{n,l,L}{} 
\node[3]{\begin{array}{c} ({\cal L}_{{\cal H}_{A}} \otimes {\cal L}_{{\cal H}_{B}})^h \end{array}} \arrow[3]{e,t,<>}{} \arrow[2]{n,r,L}{} 
\node[3]{\begin{array}{c} \text{Linear functionals } \\ \text{assuming real values} \\ \text{on } ({\cal L}_{{\cal H}_{A}} \otimes {\cal L}_{{\cal H}_{B}})^h \end{array}} \arrow[2]{n,r,L}{} 
\\\\
\node{\begin{array}{c} \text{Positive maps} \end{array}} \arrow[3]{e,t,<>}{} \arrow[2]{n,l,L}{} 
\node[3]{\begin{array}{c} \text{ ? } \end{array}} \arrow[3]{e,t,<>}{} \arrow[2]{n,r,L}{} 
\node[3]{\begin{array}{c} \text{Linear functionals } \\ \text{acting positive} \\ \text{on } {\cal L}^+_{{\cal H}_{A}} \otimes {\cal L}^+_{{\cal H}_{B}} (\text{or } S) \end{array}} \arrow[2]{n,r,L}{} 
\\\\
\node{\begin{array}{c} \text{$k$-Positive maps} \end{array}} \arrow[3]{e,t,<>}{} \arrow[2]{n,l,L}{} 
\node[3]{\begin{array}{c} \text{ ? } \end{array}} \arrow[3]{e,t,<>}{} \arrow[2]{n,r,L}{} 
\node[3]{\begin{array}{c} \text{Linear functionals } \\ \text{assuming positive } \\ \text{values on } S_k \end{array}} \arrow[2]{n,r,L}{} 
\\\\
\node{\begin{array}{c} \text{CP maps} \end{array}} \arrow[3]{e,t,<>}{} \arrow[2]{n,l,L}{} 
\node[3]{\begin{array}{c} ({\cal L}_{{\cal H}_{A}} \otimes {\cal L}_{{\cal H}_{B}})^+ \end{array}} \arrow[3]{e,t,<>}{} \arrow[2]{n,r,L}{} 
\node[3]{\begin{array}{c} \text{Linear functionals } \\ \text{acting positive} \\ \text{on } ({\cal L}_{{\cal H}_{A}} \otimes {\cal L}_{{\cal H}_{B}})^+ \end{array} } \arrow[2]{n,r,L}{} 
\end{diagram}
$
\vspace{0.5cm}
\caption[The Jamio{\l}kowski isomorphism]{The Jamio{\l}kowski isomorphism and correspondences for important classes of linear maps. The correspondence for $k$-positive maps will be outlined in Chapter~\ref{chapsmidt}. Except for the addition of these maps this chart is from Ref. \cite{Choi82}.}
\label{jafigure}
\end{center}
\end{figure}

Jamio{\l}kowski \cite{Jamiolkowski72} derived this isomorphism in order to characterise positive maps. As a positive map is Hermitian we can always write it in the form of~(\ref{herfor}). However it is unclear how to distinguish between Hermitian and positive maps from this. The trick is to associate with the operator $W$ the linear functional $f_W(|\psi\ra)=\la\psi|W|\psi\ra$, this linear functional can be characterised as follows.

\begin{theorem}
\label{jamthem}
A linear map $\Lambda_W: {\cal L}_{{\cal H}_A} \rightarrow {\cal L}_{{\cal H}_B}$ is positive if and only if the corresponding functional $f_W$ on ${\cal H}_A \otimes {\cal H}_B$ is positive on product vectors.
\end{theorem}
\begin{proof}
We can prove this theorem solely by examining (\ref{difforja}) for the special case where $|i\ra=|j\ra=|x\ra$ and $|k\ra=|l\ra=|y\ra$:
$$
\la y | \Lambda_W (|x\ra\la x|) | y \ra= \la xy|W|xy\ra.
$$
If $f_W$ is positive on product vectors then $\Lambda_W$ is positive on projectors, and therefore on arbitrary positive operators also. If $\Lambda_W$ is a positive map, then from the above equation follows that $f_W$ is positive on product vectors.
\end{proof}
Thus in the language of the previous section, positive maps correspond to entanglement witnesses. Figure~\ref{jafigure} summarises the isomorphism and its relations for different classes of linear maps. With this we are in the position of stating the main result of this section, which was first published in Ref.~\cite{HHH96} (note that it can be extended to multipartite systems, for details see Ref.~\cite{HHH00}).

\begin{theorem}
\label{corhor2}
A density operator $\rho$ acting on ${\cal H}_{A}\otimes {\cal H}_{B}$ is separable if and only if
$$
({\openone} \otimes \Lambda)(\rho) \geq 0,
$$
for all positive maps $\Lambda: {\cal L}_{{\cal H}_{B}} \rightarrow {\cal L}_{{\cal H}_{A}}$.
\end{theorem}
\begin{proof}
With each linear map $\Lambda: {\cal L}_{{\cal H}_{B}} \rightarrow {\cal L}_{{\cal H}_{A}}$ there is associated an adjoint map $\Lambda^\dagger: {\cal L}_{{\cal H}_{A}} \rightarrow {\cal L}_{{\cal H}_{B}}$ defined by
$$
\Tr[A\Lambda(B)]=\Tr[\Lambda^\dagger(A)B]
$$
for all operators $A,B$. It is easy to verify that the adjoint map of a positive map is again positive. Suppose now that $\rho$ is entangled, then from Corollary~\ref{corhor} follows that there exists an entanglement witness $W$ such that $\Tr(\rho W)<0$. By virtue of the Jamio{\l}kowski isomorphism and Theorem~\ref{jamthem} this is equivalent to
$$
\Tr[(\openone \otimes \Lambda_W)(P_+)\rho]=\Tr[(\openone \otimes \Lambda^\dagger_W)(\rho)P_+]<0
$$
and since $P_+$ is positive it follows that $(\openone \otimes \Lambda^\dagger_W)(\rho)\not\geq 0$.
Conversely, let $\Lambda^\dagger_W$ be a positive map such that $(\openone \otimes \Lambda^\dagger_W)(\rho)$ has a negative eigenvalue with corresponding eigenvector $|\phi\ra=A\otimes \openone |\psi_+\ra$. With this we get
$$
\la \phi | \openone \otimes \Lambda^\dagger_W(\rho) |\phi \ra=\Tr[(A\otimes \ido) W (A^\dagger\otimes \ido) \rho]<0.
$$
Since $W$ is positive on separable states, the same applies to $(A\otimes \ido) W (A^\dagger\otimes \ido)$. This shows that $\rho$ must be entangled.
\end{proof}

The proof of Theorem~\ref{corhor2} reveals two important facts about the relation between entanglement witnesses and positive maps. Namely, given an entanglement witness $W$, the corresponding positive map $\Lambda^\dagger_W$ will always detect more states than the witness, as it detects all states detected by the class of witnesses $(A\otimes \ido) W (A^\dagger\otimes \ido)$ for all operators $A$. Furthermore, if a certain positive map $\Lambda_W$ is negative on a state $\rho$, then the proof shows us how to construct an operator $A$ such that the local transformation\footnote{See Section~\ref{qop} for more on quantum operations.}
\begin{align}
\rho \longrightarrow (A^\dagger\otimes \ido) \rho (A\otimes \ido),
\label{ewfiltering}
\end{align} 
will yield a state that can be detected by the entanglement witness $W$. 

In the previous section, we have seen two important examples of entanglement witnesses. The map associated with $W=\ido-dP_+$ is given by $\Lambda(A)=\ido \Tr(A)-A$. From this follows that all separable states $\rho$ satisfy
$$
\rho_A\otimes \ido -\rho \geq 0 \qquad \text{ and } \qquad \ido \otimes \rho_B -\rho \geq 0,
$$
while this is not necessarily the case for entangled states. It is not so hard to prove the positivity of this map directly. The associated separability criterion is called the reduction criterion and was first studied in Ref.~\cite{HH97,CAG97}.

Another entanglement witness was given by the swap operator $F$. It is not so hard to see that the associated map is given by the transposition. For a separable state $\rho = \sum_i p_i |\psi^A_i\ra\la \psi^A_i| \otimes |\psi^B_i\ra\la \psi^B_i|$ we have that $(\ido \otimes T) \rho \geq 0$, while this is not necessarily the case for entangled states. We write $(\ido \otimes T) \rho=\rho^{T_B}$ and call this operation the partial transposition. When $\rho^{T_B}\geq 0$ we say that the state has a positive partial transpose (PPT), otherwise the state has a negative partial transpose (NPT). The partial transposition criterion was first discovered by Peres \cite{Peres96} independently of the notion of positive maps. This criterion turns out to be powerful, in particular, positivity of the partial transposition is necessary and sufficient for separability when $\dim {\cal H} \leq 6$ \cite{HHH96}. It was subsequently shown also to be necessary and sufficient for states with low rank \cite{HLVC00}, in particular for pure states \cite{Terhal01}, rank two states \cite{HSTT99} and rank three states \cite{HLVC00}. In general though, it is not sufficient, and there exist PPT entangled states. We will see an example of such a PPT entangled state in the next section, where we also analyse the importance of the partial transposition in the structure of positive maps.
We will denote the set of PPT states as $\cal P$. The set $\cal P$ is, just like the set of separable states $\cal S$, compact and $\cal S \subset \cal P \subset \cal D$.

It can be shown that the reduction criterion is strictly weaker than the partial transposition criterion \cite{HH97}, except when $\dim {\cal H}_B=2$ or $\dim {\cal H}_A=2$ in which case both criteria are equivalent. This last statement follows from the identity $\ido \Tr(A)-A=(\sigma_y A \sigma_y)^T$ for two dimensional matrices $A$. The reduction and the partial transposition criterion are important in the context of distillation, they will be studied in detail in Chapter~\ref{entdis}.

\subsection{Some deeper results}
\label{sdr}
We start this section with the following classical theorem, which gives a complete characterisation of positive maps between low dimensional operators (for elementary proofs see Ref.~\cite{LCK99, EK02,VDD01b}):

\begin{theorem}[St{\o}rmer and Woronowicz \cite{Stormer63, Woronowicz76}]
Let $\Lambda: {\cal L}_{\mathbb{C}^2} \rightarrow {\cal L}_{\mathbb{C}^n}$ be a positive map, with $n\leq 3$. Then $\Lambda$ can be written as
$$
\Lambda=\Lambda_1 + \Lambda_2 \circ T,
$$
with $\Lambda_1, \Lambda_2$ completely positive maps.
\end{theorem}
We call a positive map decomposable if it can be written as a combination of a completely positive map and the combination of a completely positive map and transposition. It is easy to see that entangled states detected by a decomposable map will also be detected by the partial transposition alone. Thus from Theorem~\ref{corhor2} we obtain that for all bipartite states $\rho$ acting on $\cal H$ with $\dim {\cal H} \leq 6$, positivity of the partial transposition is a necessary and sufficient condition for separability.

The first example of an undecomposable map was given by Choi \cite{Choi75b, Choi77, Choi80}. Let $\Lambda_C: {\cal L}_{\mathbb{C}^3} \rightarrow {\cal L}_{\mathbb{C}^3}$ be the linear map defined as 
\begin{align}
\label{chmp}
\Lambda_C(A)=
\left[ \begin{array}{ccc}
 A_{11}+A_{33} & -A_{12} & -A_{13} \\
 -A_{21} & A_{22}+A_{11} & -A_{23} \\
 - A_{31} & -A_{32} & A_{33}+A_{22} 
\end{array} \right],
\end{align}
for all $3\times 3$ matrices $A=[A_{ij}]$. He originally proved that this map is positive on real positive semidefinite (and thus symmetric) matrices , but it easy to see that this implies that the map is positive on all positive semidefinite matrices \cite{Choi80}. With a counterexample, we will show that this map is not decomposable. The example is due to St{\o}rmer \cite{Stormer82} but analysed in the context of entanglement in Ref.~\cite{Alb01, HHH98b, DPS03}. Define for $0\leq \alpha \leq 5$
\begin{align}
\label{stst}
\sigma=\frac{1}{7} \left[ 2 P_+ + \alpha \sigma_+ + (5-\alpha)\sigma_- \right],
\end{align}
with 
$$
\sigma_+=\frac{1}{3}(|01\ra\la 01| +|12\ra\la 12| + |20\ra\la 20|)
$$
and $\sigma_- =F\sigma_+F$. This state has a positive partial transpose for $1\leq \alpha \leq 4$, for other $\alpha$ the state is NPT entangled. Now it is easy to verify that $(\ido \otimes \Lambda_C)(\sigma)$ is not positive for $3 \leq \alpha \leq 4$ and $1 \leq \alpha \leq 2$ since it has an eigenvalue $\lambda=(3-\alpha)/2$. This immediately shows that $\Lambda_C$ is non-decomposable and that $\sigma$ is PPT entangled in these regions. For the region $2\leq \alpha \leq 3$, the state can be shown to be separable by virtue of an explicit decomposition in terms of product states. Following Choi's example, many other non-decomposable maps have been found and studied in the literature, see for instance Ref.~\cite{Woronowicz76, TT83, Tomiyama85, Robertson85, Tang86, Osaka91, LW97, LMM03, Ha03, MM04, Hall06, CK06b}.

\begin{figure}
\begin{center}
\psfrag{r}{$\rho$}
\psfrag{t}{$\rho=(1-p)\rho_s+p\delta$}
\psfrag{e}{$\delta$}
\psfrag{s}{$\rho_s$}
\psfrag{S}{$\cal S$}
\psfrag{P}{$\cal P$}
\psfrag{D}{$\cal D$}
\includegraphics[width=10cm]{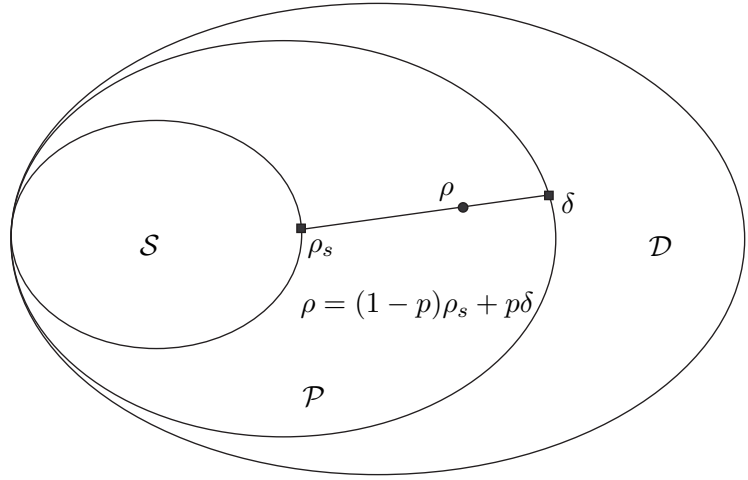}
\caption[PPT states as the sum of an edge and a separable state]{Schematic representation of the nesting $\cal S \subset \cal P \subset \cal D$ and a PPT state as the convex combination of an edge and a separable state \cite{BCHHKLS01}.}
\label{convexfig}
\end{center}
\end{figure}

The notion of decomposability can also be extended to entanglement witnesses via the Jamio{\l}kowski isomorphism. Namely it is easy to see that any decomposable map $\Lambda_W$ corresponds to a witness
$$
W=P+Q^{T_B},
$$
with $P$ and $Q$ positive semidefinite (and vice verse). We will refer to such witnesses as decomposable witnesses.
There exists \cite{LKHC00, LKCH00} a canonical representation of indecomposable entanglement witnesses, which we will derive now. Central is the idea of a so-called edge state, which is a PPT entangled state $\delta$ such that for all $\epsilon\geq0$ and all separable $|ab\rangle$,
$$
\delta-\epsilon|ab\ra\la ab|
$$
is not positive or does not have a positive partial transpose. Edge states can be thought of as lying on the boundary between PPT and NPT states, and thus in this sense, they are the most entangled PPT states. From the definition it is clear that a state $\delta$ is an edge state if and only if there is no product vector $|a b\ra$ in the range of $\delta$ such that $|a b^*\ra$ is in the range of $\delta^{T_B}$. If such a product vector exists, we could subtract it from the state and still obtain a PPT state. Repeating this process will eventually yield an edge state (see also Figure~\ref{convexfig}): 
\begin{theorem}
\label{pptdeco}
Every PPTES $\rho$ can be decomposed as
$$
\rho=(1-p)\rho_{\text{sep}}+p\delta,
$$
with $\rho_{\text{sep}}$ a separable state and $\delta$ an edge state.
\end{theorem}
The following theorem provides a canonical construction of an entanglement witness that detects a given edge state. Conversely it states that any non-decomposable entanglement witness can be written in a similar form.

\begin{theorem}
\label{edgeindec}
(i) Let $\delta$ be an edge state with $P$ and $Q$ the projector onto the kernel of $\delta$ and $\delta^{T_B}$ respectively. Let
$$
W_\delta=P+Q^{T_B},
$$
and
$$
\epsilon=\inf_{|ab\ra} \la ab| W_\delta |ab\ra,
$$
then $W_1=W_\delta -\epsilon \openone$ is an non-decomposable entanglement witness which detects $\delta$.

(ii) Any non-decomposable witness $W$ can be written as
$$
W=P+Q^{T_B} -\epsilon \openone
$$
with 
$$
0<\epsilon\leq \inf_{|ab\ra} \la ab|P+Q^{T_B} |ab\ra.
$$
Here $P,Q$ are positive semidefinite and satisfy $\Tr P\delta=\Tr Q^{T_B} \delta=0$ for some edge state $\delta$.
\end{theorem}
\begin{proof}
(i) As the range of $\delta$ contains no product vector $|a b\ra$ such that the range of $\delta^{T_B}$ contains $|a b^*\ra$, there must exist such a product vector in their respective kernels, so that $\epsilon >0$. By its very definition, $W_1$ is positive on product states and since $\Tr(W_1\delta)=-\epsilon<0$ it is an non-decomposable entanglement witness which detects $\delta$.

(ii) Let $W(\lambda)=W+\lambda \ido$ with $\lambda>0$. We denote with $\lambda_0$ the smallest $\lambda$ such that $W(\lambda_0)=0$ is decomposable. Since $W(\lambda_0)$ is decomposable we can write it as $W(\lambda_0)=P+Q^{T_B}$ and thus $W=P+Q^{T_B}-\lambda_0\ido$. Using the compactness of the set of PPT states $\cal P$ there will exist a state $\rho$ such that $\Tr(\rho W(\lambda_0))=0$. Now let $\rho=(1-p)\rho_{\text{sep}}+p\delta$, with $\delta$ an edge state, so that 
$$
 p\Tr ( W\rho_{\text{sep}})+(1-p)\Tr(W\delta)=-\lambda_0.
$$
When $p>0$ this would lead to $\Tr(W(\lambda_0)\delta)<0$, contrary to the assumption that $W(\lambda_0)$ is decomposable. So $p=0$, $\Tr(\delta W(\lambda_0))=0$ and $\Tr(\delta W)=-\lambda_0$. It follows that $\Tr P\delta=\Tr Q^{T_B} \delta=0$. The theorem follows by putting $\epsilon=\lambda_0$, where the condition on $\epsilon$ follows because $W$ is an entanglement witness.
\end{proof}
This theorem can be generalised straightforwardly to arbitrary convex sets and associated witnesses \cite{BCHHKLS01}.
In Ref.~\cite{LKHC00, LKCH00} it was shown how to optimise a non-decomposable witness $W$ (for instance constructed from an edge state). Optimising in this context means iterative subtraction of decomposable witnesses from $W$, yielding a stronger witness. We refer to the original references for details.

With an entanglement witness $W$ or alternatively the positive map $\Lambda_W$ we can associate the following bihermitian positive form:
$$
F_W(x_i,y_i)=\la xy| W | xy\ra= \la x | \Lambda_W(|y\ra\la y|)|x\ra= \sum_{ijkl} W_{ij;kl}x^*_i y^*_j x_k y_l.
$$
Suppose now that $\Lambda_W$ has real coefficients in some basis, then for real $x_i,y_i$, $F_W$ is a so-called real biquadratic form. Hilbert's 17th Problem is concerned with the question whether all real positive semidefinite forms can be written as the sum of squares (SOS) of forms (see Ref.~\cite{Reznick00} for an excellent overview on Hilbert's 17th problem). Choi~\cite{Choi75b, Choi80, Choi82, Terhal01} showed that non-decomposable maps with real coefficients allow to construct counterexamples for biquadratic forms. 
\begin{theorem}
\label{sostheo}
(i) Let $F_W$ be a real biquadratic form which cannot be written as a SOS of linear forms, then the associated map $\Lambda_W$ is non-decomposable. (ii) Conversely, if $\Lambda_W$ is a non-decomposable with real coefficients, then $F_W$ cannot be expressed as a SOS.
\end{theorem}
\begin{proof}
(i) Note that decomposable maps acting on (real) symmetric matrices act as completely positive maps, so we can write 
$$
\Lambda_W(|y\ra\la y|)= \sum_l A_l |y\ra\la y| A_l^T,
$$
and with $\la x| A_l |y\ra=a^l_{ij}$ we get
$$
F_W(x_i,y_i)=\la x | \Lambda_W(|y\ra \la y|)|x\ra=\sum_l \la x| A_l |y\ra\la y| A^T_l |x \ra= \sum_l (a^l_{ij} x_iy_j)^2.
$$
(ii) By reversing the argument of (i) it follows that if $F_W$ can be expressed as a SOS, then $\Lambda_W$ is decomposable.
\end{proof}
Earlier in this section we have seen Choi's example of an non-decomposable map. The associated real form is given by
$$
F_C=x^2_1y^2_1+x^2_2y^2_2+x^2_2y^2_2-2(x_1x_2y_1y_2+x_2x_3y_2y_3+x_3x_1y_3y_1)+x^2_1y^2_2+x^2_2y^2_3+x^2_3y^2_1.
$$
By inspection Choi showed the positivity of this form and showed that it cannot be expressed as a SOS\@. This immediately proves the non-decomposability of the associated map.

For a more abstract formulation of the separability problem, with various connections to $C^*$-algebras the reader is referred to the series of papers by Majewski et~al.\ \cite{Majewski00, MM01, Majewski04, Majewski04b}.

\subsection{PPT entangled states}
\label{PPTESS}
In this section we provide several ways of constructing PPT entangled states, with particular emphasis on the UPB construction and the range criterion. Based on this, we construct two new entangled PPT states with special ranks. For completeness, references will be listed to all other PPT entangled states known to us.

In the previous section we have seen St{\o}rmer's example \cite{Stormer82, HHH98b, DPS03} of a PPT state, constructed such that it is detected by Choi's indecomposable map. In general given an indecomposable map, one can try to guess from it some state which is PPT entangled \cite{Alb01}, see for example the PPT states constructed in Ref.~\cite{BFP04, Piani04, Breuer06}. Recently, it has been shown how to systematically construct PPT entangled states from a given indecomposable map \cite{HK03,HK05, HKP03}. Yet despite this advance, coming up with an indecomposable map is highly non-trivial and very much a matter of trial and error.

Currently, the only systematic way of constructing PPT entangled states is based on unextendible product basis (UPB) \cite{BDMSST99,DMSST99}. An UPB in a Hilbert space $\cal H = {\cal H}_A \otimes {\cal H}_B$ is a set of pure orthogonal product states which do not span\footnote{Therefore, an unextendible product basis is in fact not a basis.} $\cal H$ and such that there is no product vector orthogonal to all of them. This definition is readily extended to multipartite systems. As an example, consider the following set of vectors $S$ in ${\cal H} \cong {\mathbb{C}}^3 \otimes {\mathbb{C}}^3$:
\begin{gather*}
|\psi_0\ra=\frac{1}{\sqrt 2}|0\ra(|0\ra -|1\ra),\qquad \qquad |\psi_1\ra=\frac{1}{\sqrt 2}(|0\ra-|1\ra)|2\ra, \\
|\psi_2\ra=\frac{1}{\sqrt 2}|2\ra(|1\ra -|2\ra),\qquad \qquad |\psi_3\ra=\frac{1}{\sqrt 2}(|1\ra-|2\ra)|0\ra,\\
|\psi_4\ra=\frac{1}{3}(|0\ra+|1\ra+|2\ra)(|0\ra+|1\ra+|2\ra).
\end{gather*}
Now if this set of vectors is extendible, then there has to exist a partition of this set of vectors $S$, into two subsets $S_1$ and $S_2$, such that the local rank of the first subsystem for $S_1$ is smaller than 3, and the local rank of the second subsystem for $S_2$ is smaller than 3. In this case, one can construct a vector $|ab\ra$ such that $|a\ra$ is orthogonal to the first subsystem of $S_1$ and $|b\ra$ is orthogonal to the second subsystem of $S_2$.
But this is impossible since either $S_1$ or $S_2$ will contain at least three vectors, and it is easy to see that any three vectors, seen by either party span the full space of that party. We refer to the original papers and the review \cite{DT00} for a plethora of other examples.

From any UPB one can construct a bound entangled state as in the next theorem.
\begin{theorem}
Let the set of $n$ vectors $\{|\psi_i\ra \}$ be a UPB in $\cal H = {\cal H}_A \otimes {\cal H}_B$ , with $\dim {\cal H}=d$, then
$$
\rho=\frac{1}{d-n}(\ido - \sum_{j=1}^n |\psi_j\ra\la\psi_j|)
$$
is a PPT entangled state.
\end{theorem}
\begin{proof}
The state $\rho$ has positive partial transposition as the set $\{|\psi_j\ra\la\psi_j|^{T_B}\}$ is again a UPB.
It is easy to see that $\rho$ is entangled and in fact an edge state as, by definition, there is no product state in the space orthogonal to the UPB.
\end{proof}
In Ref.~\cite{Terhal01b} it was shown how to systematically construct entanglement witnesses detecting UPB states. The generalisation of this construction for general edge states is precisely the one we outlined in the previous section (Theorem~\ref{edgeindec}). Based on these witnesses, PPT entangled states in the neighbourhood of UPB PPT entangled states can be constructed \cite{Pittenger02, BGR04}.

As we have seen, an edge state $\rho$ contains no product vector $|a_i b_i\ra$ in its range such that $|a_i b^*_i\ra$ is in the range of $\rho^{T_B}$. This condition is clearly a sufficient for a state to be PPT entangled. We can make this criterion slightly stronger as follows \cite{Horodecki97}:
\begin{theorem}[Range criterion]
If a state $\rho$ is separable there must exist a set of product vectors $\{|a_i b_i\ra\}$ spanning the range of $\rho$ such that $\{|a_i b^*_i\ra\}$ span the range of $\rho^T_B$. 
\end{theorem}
Thus edge states violate the range criterion in the strongest way. The range criterion was originally devised by Woronowicz \cite{Woronowicz76} in the context of indecomposable maps. This criterion is remarkable simple and powerful, in particular for proving the PPT property of sparse matrices.
Examples of states violating the range criterion have been published in Ref.~\cite{Woronowicz76, Horodecki97, Choi82, BP99, HL00, FLJS06}. Various other PPT entangled states have appeared over the years \cite{DPS03, HCL01, WW01b, VW02, Ishizaka04, ABLS01, Dur01, HPHH05, HHHO03, AH04, YL04, PM06}. In most cases the states are shown to be inseparable by showing that they can help performing some non-classical tasks. In Section~\ref{actbent} we will say more about these so-called activation processes.

We conclude this section with the presentation of two new PPT states \cite{Clarisse06}. The new types are $(5,5)$ and $(6,6)$ edge states. Here in $(N,M)$, $N$ denotes the rank of the state, while $M$ denotes the rank of $\rho^{T_B}$. The question of their existence popped up in the analysis of a conjecture on 2-positive maps between $3\times 3$ matrices. We come back to this conjecture in Section~\ref{contwosdt}.

\subsubsection{A $(5,5)$ edge PPTES}
Consider the following $(5,5)$ state:
$$
\rho_{(5,5)}=\frac{1}{13}
\left[ \begin{array}{ccccccccc}
 0 & 0 & 0 & 0 & 0 & 0 & 0 & 0 & 0 \\
 0 & 2 & -1 & 0 & 0 & 0 & 0 & 0 & 1 \\
 0 & -1 & 1 & 0 & 0 & 0 & 0 & 0 & -1 \\
 0 & 0 & 0 & 3 & 0 & -1 & -1 & 0 & 0 \\
 0 & 0 & 0 & 0 & 0 & 0 & 0 & 0 & 0 \\
 0 & 0 & 0 & -1 & 0 & 1 & 1 & 0 & 0 \\
 0 & 0 & 0 & -1 & 0 & 1 & 1 & 0 & 0 \\
 0 & 0 & 0 & 0 & 0 & 0 & 0 & 2 & -2 \\
 0 & 1 & -1 & 0 & 0 & 0 & 0 & -2 & 3
\end{array} \right],
$$
and its partial transpose:
$$
\rho^{T_B}_{(5,5)}=\frac{1}{13}
\left[ \begin{array}{ccccccccc}
 0 & 0 & 0 & 0 & 0 & 0 & 0 & 0 & 0 \\
 0 & 2 & -1 & 0 & 0 & 0 & 0 & 0 & 0 \\
 0 & -1 & 1 & 0 & 0 & 0 & 0 & 1 & -1 \\
 0 & 0 & 0 & 3 & 0 & -1 & -1 & 0 & 1 \\
 0 & 0 & 0 & 0 & 0 & 0 & 0 & 0 & 0 \\
 0 & 0 & 0 & -1 & 0 & 1 & 0 & 0 & 0 \\
 0 & 0 & 0 & -1 & 0 & 0 & 1 & 0 & 0 \\
 0 & 0 & 1 & 0 & 0 & 0 & 0 & 2 & -2 \\
 0 & 0 & -1 & 1 & 0 & 0 & 0 & -2 & 3
\end{array} \right].
$$
It is not so hard to verify that both operators are positive semi-definite and have rank 5. An analytical expression of their eigenvectors and eigenvalues is however quite complex. To show that $\rho_{(5,5)}$ is an edge state we will show that it violates strongly the range criterion. For this we will use the `divide and conquer technique' from \cite{Horodecki97}.

From the matrix representation above we see that every vector in the range of $\rho_{(5,5)}$ can be written in the form
$$
V=(0,A,-E-F,C,0,D,D,E,F), \quad A,C,D,E,F \in \mathbb{C}.
$$
Now we have look at those vectors which can be written as a product
$$
V=(s,t,v)\otimes(x,y,z)=(sx,sy,sz,tx,ty,tz,vx,vy,vz).
$$
Taking these two conditions together we can therefore characterise all product vectors in the range of $\rho_{(5,5)}$. 

From the condition $sx=0$ we can distinguish the following sub cases:

1. $x=0, s\neq 0$, we have $vx=D=0=tz$ and therefore either $t=0$ or $z=0$. Without loss of generality we can also put $s=1$.

1.1. $t=0$, as $v(y+z)=-z$ we have $v=-z/(y+z)$. When $y=-z$, then $z=0=-y=x$ and we obtain the null vector. Thus the only case that remains is
$$
V=s\left(1,0,\frac{-z}{y+z}\right)\otimes(0,y,z),
$$
with $y\neq -z$.

1.2. $z=0$, from $ty=0$ follows that $t=0$ (otherwise $y=0$) thus this case is already covered in 1.1.

2. $s=0$, from $E=-F$ follows that $vy=-vz$. From $ty=0$ follows that either $t=0$ or $y=0$.

2.1 $t=0$, since $D=0=vx$ we have that $x=0$ and $v\neq 0$, and thus $z=-y$. Thus we get
$$
V=(0,0,v)\otimes(0,y,-y)
$$

2.2 $y=0$, and thus $vy=-vz=0$ so either $z=0$ or $v=0$.

2.2.1 $z=0$, and thus $vx=tz=0$. Because $x \neq 0$ we have that $v=0$:
$$
V=(0,t,0)\otimes(x,0,0)
$$

2.2.2 $v=0$ and thus $vx=tz=0$. Because $t \neq 0$ we have that $z=0$ and we arrive at 2.2.1.

We now do the same analysis for $\rho^{T_B}_{(5,5)}$. It is easy to see that every vector in its range can be written as
$$
V=(0,A,B,C,0,D,E,A+2B,-A-2B+C+D+E),
$$
with $A,B,C,D,E \in \mathbb{C}$.
We need to take the intersection of these vectors with those vectors which can be written as a product
$$
V=(s,t,v)\otimes(x,y,z)=(sx,sy,sz,tx,ty,tz,vx,vy,vz).
$$

From the condition $sx=0$ we can distinguish the following sub cases:

1. $x=0$, and since $ty=0$ we have that $y=0$ or $t=0$

1.1. $y=0$ and thus $vy=0$ and $sy=-2sz=0$. Since $z=0$ gives the null vector we get $s=0$. Now $vz=-sy-2sz+tx+tz+vx=tz$ and thus $t=v$, and thus
$$
V=(0,t,t)\otimes(0,0,z)
$$

1.2. $t=0$, and since $vz=-sy-2sz=-vy$ either $v=0$ or $z=-y$

1.2.1. $v=0$ and thus $sy=-2sz$ or since $s=0$ is trivial we get $y=-2z$ and thus
$$
V=(s,0,0)\otimes(0,-2z,z)
$$

1.2.2. $z=-y$, and thus $-vy=-sy+2sy=sy$. Since $y=-z=0$ is trivial we get $v=-s$ and
$$
V=(-s,0,s)\otimes(0,y,-y)
$$

2. $s=0$ and thus $vy=0$, we also have $ty=0$ thus either $y=0$ or $v=t=0$, but this last case gives us again the trivial vector. From $vz=-sy-2sz+tx+ty+tz+vx=tx+tz+vx$ we have $(v-t)z=(v+t)x$. What remains is
$$
V=(0,v,t)\otimes(x,0,z),
$$
with either $z=\frac{v+t}{v-t}x$ or $x=\frac{v-t}{v+t}z$ (but only one of those when $v=\pm t$).

It is now straightforward to check that when $v=(s,t,v)\otimes(x,y,z)$ belongs to the range of $\rho_{(5,5)}$ that
$v'=(s,t,v)\otimes(x^*,y^*,z^*)$ does not belong to the range of $\rho^{T_B}_{(5,5)}$. This concludes the proof that $\rho_{(5,5)}$ is a $(5,5)$ edge state.

\subsubsection{A $(6,6)$ edge PPTES}
Consider the following state
$$
\rho_{(6,6)}=\frac{1}{13}\left[ \begin{array}{ccccccccc}
 1 & 0 & 0 & 0 & 0 & 0 & 0 & 0 & -1 \\
 0 & 2 & 0 & -1 & 0 & 0 & 0 & 0 & 0 \\
 0 & 0 & 1 & 0 & 0 & 0 & 1 & 0 & 0 \\
 0 & -1 & 0 & 1 & 0 & 0 & 0 & 0 & 1 \\
 0 & 0 & 0 & 0 & 1 & 0 & 1 & 0 & 0 \\
 0 & 0 & 0 & 0 & 0 & 1 & 0 & -1 & 0 \\
 0 & 0 & 1 & 0 & 1 & 0 & 2 & 0 & 0 \\
 0 & 0 & 0 & 0 & 0 & -1 & 0 & 1 & 0 \\
 -1 & 0 & 0 & 1 & 0 & 0 & 0 & 0 & 3
\end{array} \right],
$$
with partial transpose
$$
\rho^{T_B}_{(6,6)}=\frac{1}{13}
\left[ \begin{array}{ccccccccc}
 1 & 0 & 0 & 0 & -1 & 0 & 0 & 0 & 1 \\
 0 & 2 & 0 & 0 & 0 & 0 & 0 & 0 & 0 \\
 0 & 0 & 1 & 0 & 0 & 0 & -1 & 0 & 0 \\
 0 & 0 & 0 & 1 & 0 & 0 & 0 & 1 & 0 \\
 -1 & 0 & 0 & 0 & 1 & 0 & 0 & 0 & -1 \\
 0 & 0 & 0 & 0 & 0 & 1 & 1 & 0 & 0 \\
 0 & 0 & -1 & 0 & 0 & 1 & 2 & 0 & 0 \\
 0 & 0 & 0 & 1 & 0 & 0 & 0 & 1 & 0 \\
 1 & 0 & 0 & 0 & -1 & 0 & 0 & 0 & 3
\end{array} \right].
$$
It can be checked easily that both matrices are positive definite and have rank 6. Following a similar strategy as in the previous section we now proceed to show that $\rho_{(6,6)}$ is an edge state. Every vector in the range of $\rho_{(6,6)}$ can be written in the form
\begin{align}
\label{eq1}
V=(A,B,C,D,E,F,C+E,-F,B+2D-A), 
\end{align}
with $A,B,C,D,E,F \in \mathbb{C}$, whilst every vector in the range of $\rho^{T_B}_{(6,6)}$ takes the form
\begin{align}
\label{eq2}
V'=(A,B,C,D,-A,E,E-C,D,F), 
\end{align}
with $A,B,C,D,E,F \in \mathbb{C}$. The `divide and conquer' method does not work so well here as we have no zeros to start with. Instead we will show directly that if the product vector
$$
V=(s,t,v)\otimes(x,y,z)=(sx,sy,sz,tx,ty,tz,vx,vy,vz).
$$
belongs to the range of $\rho_{(6,6)}$, then $V'=(s,t,v)\otimes(x^*,y^*,z^*)$ does not belong to the range of 
$\rho^{T_B}_{(6,6)}$. Comparing equations (\ref{eq1}) and (\ref{eq2}) with their product form we get the following set of equations:
\begin{align*}
vx&=sz+ty \\
vy&=-tz \\
vz&=sy+2tx-sx\\
ty^*&=-sx^* \\
vx^*&=tz^* - sz^* \\
vy^*&=tx^*
\end{align*}
We consider the following sub case

1. $vy\neq 0$, and it follows that all parameters are different from zero. Without loss of generality we can put $x=1$ and get
\begin{align*}
v&=sz+ty \\
vy&=-tz \\
vz&=sy+2t-s\\
ty^*&=-s \\
v&=tz^* - sz^* \\
vy^*&=t
\end{align*}
We now use $t=vy^*$ and $s=-v(y^*)^2$ to eliminate $t$ and $s$, after which we can divide every equation by $v$ resulting in
\begin{align*}
1&=-(y^*)^2z+y^*y \\
y&=-y^*z \\
z&=-(y^*)^2y+2y^*+(y^*)^2\\
1&=y^*z^* +(y^*)^2z^* \\
\end{align*}
From the first and the second equation we get $yy^*=\frac{1}{2}$, while from the second and fourth equation we get
$1=y^*(z^*-y)$ or $y=yy^*(z^*-y)=\frac{1}{2}(z^*-y)$ and thus $z=3y^*$. From the third equation we get
$$
3y^*=-(y^*)^2y+2y^*+(y^*)^2
$$
or 
$$
3=-y^*y+2+y^*
$$
and thus $y^*=\frac{3}{2}$ which contradicts $yy^*=\frac{1}{2}$. 

2. $v=y=0$, in this case it is straightforward to see that $x=y=z=0$.

3. $y=0$ and $v\neq 0$. From $sx^*=0$ and $tx^*=0$ follows that $s=t=0$ or $x=0$. In the first case we get $vx=vz=0$ or $z=x=0$ by the assumption. In the second case follows from $vz=0$ that $z=0$.

4. $v=0$ and from $tz=0$ and $tx^*=0$ follows that $t=0$ or $x=z=0$

4.1 $t=0$, from $sx^*=sz^*=0$ follows that either $s=t=v=0$ or $x=z=0=y$, but $y\neq 0$ by assumption.

4.2 $x=z=0$, from $ty=sy=0$ follows that $s=t=v=0$. 

This concludes the proof.

\subsection{A numerical approach}
\label{numap}
In this section we give an example of a numerical approach to the separability problem. Gurvits \cite{Gurvits03} reformulated the separability problem as a weak membership problem (allowing some error in the decision) and showed that even in this scenario the problem is NP-hard. Yet, as demonstrated by the partial transpose criterion, this does not exclude the existence of efficient algorithms for low dimensional systems. There have been several algorithmic proposals, which can be categorised in three classes:
\begin{description} 
\item[From the outside] Here a hierarchy of tests is devised which in every step detects entanglement of some states and provide a corresponding witness. Typically the first tests will detect highly entangled states, while further tests will detect more weakly entangled states. Ideally, the hierarchy should be complete, i.e.\ in the asymptotic regime every entangled state should be detected. Examples of such algorithms can be found in Ref.~\cite{BV04,BV04b, DPS03, DPS04, DPS01}.
\item[From the inside] In this approach, again a hierarchy of tests is devised, but this time able to deliver a certificate for separability. States detected in the first tests will be typically close to the maximally mixed state, while further tests will be able to detect states closer to the boundary of entangled states. The logic behind this is that for weakly (classical) correlated states there is more freedom in finding a convex decomposition in terms of product vectors. Examples of such algorithms can be found in Ref.~\cite{HB04, Woerdeman03}.
\item[Distance measure] The starting point here is to take an entanglement measure $E(\rho)$ which satisfies $E(\rho)=0$ for all separable states and $E(\rho)>0$ for all entangled states. Examples of such measures include the entanglement of formation \cite{BDSW96}, and geometric measures such as the best separable approximation \cite{LS97} (see Section~\ref{mesent} for an introduction to entanglement measures). The algorithm typically works then by calculating $E(\rho)$ to a certain accuracy. For examples see Ref.~\cite{LS97, EHGC04}.
\end{description}
An exhaustive review on algorithms for the separability problem, with particular emphasis on comparing the different complexities of the algorithms, can be found in Ref.~\cite{Ioannou06}.

The most successful algorithms as of today use of optimisation theory, and in particular semidefinite programming. Semidefinite programs (SDPs) are convex optimisation problems which can be written as the minimisation of a linear object function, subject to semidefinite constraints in the form of linear matrix inequalities \cite{BV04}:
\begin{align*}
\text{minimise} \qquad & c^T x, \\
\text{subject to} \qquad & F_0+\sum_i x_i F_i \geq 0,
\end{align*}
where $\{x_i\}$ is a finite set of real variables, $c$ a constant vector and $F_i$ some fixed Hermitian matrices.
Semidefinite programs have two appealing features. Firstly, efficient algorithms are available for solving SDPs in polynomial time with arbitrary accuracy. Secondly, with each (primal) SDP there is associated a dual SDP:
\begin{align*}
\text{maximise} \qquad & -\Tr[F_0 Z], \\
\text{subject to} \qquad & \Tr[F_iZ]=c_i, \quad Z\geq 0
\end{align*}
Here $Z\geq 0$ means that $Z$ is a positive semi-definite (Hermitian) operator. Now it is not so hard to see that $c^T x$ is always larger than $-\Tr[F_0 Z]$, so that solving the dual problem gives a lower bound on the primal problem. It is often the case that these values coincide, in which case the SDP's have the so-called strong duality property. Suppose now that $c=0$, then the primal problem becomes a feasibility problem. Then a positive value of $-\Tr[F_0 Z]$ shows that the primal problem is not feasible. In other words the operator $Z$ gives a certificate for the infeasibility. Needless to say that such certificates are of great value in decision problems such as the separability problem. 

The two most important algorithms that use SDP, as of today, are the algorithms of Eisert et.~al.\ \cite{EHGC04} and Doherty et.~al.\ \cite{DPS03}. The first approach is based on calculating the distance of a state to the set of separable states. In first instance this leads to a global minimisation problem with polynomial constraints. Using Lasserre's method of semi-definite relaxations, this can then be reformulated as a sequence of SDPs. The hierarchy of tests obtained is complete, as every entangled state will be found at some stage in the hierarchy. Since a global optimum will be found at some point also separability can be detected \cite{Eisert}.

Doherty's approach is based on the existence of symmetric extensions for separable states. It works from the outside and is also complete; furthermore, when a state is found to be entangled, the dual problem yields a corresponding entanglement witness. These entanglement witnesses turn out to have interesting algebraic properties, which is why we will say a bit more about (without proofs though).

Suppose a state $\rho$ on ${\cal H}_A\otimes {\cal H}_B$ is separable, then we know that it can be written as $\rho=\sum_i p_i |\phi_i\ra\la\phi_i| \otimes |\phi_i\ra\la\phi_i|$. With $\rho$, we can associate the following state:
$$
\tilde \rho=\sum_i p_i |\phi_i\ra\la\phi_i| \otimes |\psi_i\ra\la\psi_i| \otimes |\phi_i\ra\la\phi_i|
$$
on ${\cal H}_A\otimes {\cal H}_B \otimes {\cal H}_A$. This extension is naturally symmetric under interchange of the first and third party. It also has positive partial transposition (with respect to any subsystem) and by tracing out the third party, we recover the original state. An extension for an arbitrary state satisfying these three properties will be called a symmetric extension. Equally, we can consider a symmetric extension with 2 copies of the first party. Using de Quantum de Finetti theorem one can prove the following characterisation of separable states:
\begin{theorem}
A state $\rho$ on ${\cal H}_A\otimes {\cal H}_B$ is separable if and only if it has a symmetric extension to any number of copies of subsystem $A$.
\end{theorem}
Apparently this theorem has been around for some time, see Ref.~\cite{DPS03} for its history. The new result by Doherty et.~al.\ was to show that the feasibility problem of finding a symmetric extension to a fixed number of $k$ copies can be cast into a semidefinite program. In this way, one obtains a hierarchy of tests, in which each test is stronger than the other. Indeed, when an extension to $k+1$ copies exist, then naturally an extension to $k$ copies exists. The first test is the usual partial transpose criterion, while further tests go beyond that and are able to detect PPT entangled states. In practice, it turns out that the algorithm detects all known PPT entangled states in the second test of the hierarchy. The only major disadvantage is, that in practice, it is limited to density matrices acting on $\cal H$ with $\dim {\cal H} \leq 36$ because of memory constraints of existing implementations.

If the primal problem fails, the dual problem yields a corresponding entanglement witness. In Theorem~\ref{sostheo} we have seen that when a positive map $\Lambda_W$ with real coefficients is decomposable, then the associated real biquadratic form can be written as a SOS of linear forms. This result readily generalises to arbitrary positive maps. Now remarkably we have the following \cite{DPS03}:
\begin{theorem}
Let $E_W(x,y)$ be the bihermitian form associated with an entanglement witness $W$ obtained from the $(k+1)$th test of the hierarchy. Then $F(x,y)=\la x| x\ra^k E_W(x,y) $ can be written as a SOS.
\end{theorem}
With this one can prove \cite{DPS03} the following characterisation of entanglement witnesses.
\begin{corollary}
Let $W$ be an entanglement witness in the interior of the set of entanglement witnesses (that is $\Tr(\rho_{\text{sep}} W)>0$ for all separable $\rho_{\text{sep}}$). Then there is a $k>0$ such that 
$$
F(x,y)=\la x| x\ra^k E_W(x,y) 
$$ 
is a SOS.
\end{corollary}
Note that a similar characterisation can be deduced from it for positive maps. As an example consider the St{\o}rmer state $\sigma$ given by~(\ref{stst}), detected by Choi's map $\Lambda_C$ (\ref{chmp}). The second test of the hierarchy yields just the entanglement witness $W_C$ associated with $\Lambda_C$ and provides the SOS form for it:
\begin{align*}
& 4 \la xy |W_C|xy\ra \la x|x\ra= |2x_0x^*_0y_0-2x_1x^*_0y_1+x_1x^*_1y_0-x_2x^*_0y_2|^2 + 3|x_2x^*_1y_2- x_1x^*_1y_1|^2 +\\ 
& 3|y_0x_1x^*_1- x_2y_2x^*_0|^2 + 3|x_2x^*_2y_2- x_0x^*_2y_0|^2 + |2x_0x_1y^*_2-x_2x_0y^*_1-x_1x_2y^*_0|^2 + \\
& 3|x_0x_2y^*_1- x_1x_2y^*_0|^2 + |2x_0x^*_0y_2-2x_1x^*_2y_1+x_2x^*_2y_2-x_0x^*_2y_0|^2 + \\ 
& |2x_0x^*_1y_0-2x_2x^*_2y_1+x_2x^*_1y_2-x_1x^*_1y_1|^2 \geq 0.
\end{align*}

\section{Permutation criteria}
\label{permcrit}
\subsection{From cross norm over realignment to permutation criteria}
In this section we will introduce a class of separability criteria, known as the permutation criteria. We will introduce them in the way they have historically been found. Our starting point is a characterisation of separable states known as the cross norm criterion \cite{Rudolph00}. The greatest cross norm $\|\cdot\|_{\gamma}$ is defined on operators in $\rho \in{\cal L}_{{\cal H}_A} \otimes {\cal L}_{{\cal H}_B}$
as 
$$
\|\rho\|_{\gamma}= \inf_{u_i,v_i}\left[\sum_i \|u_i\|\cdot \|v_i\| \Bigg| \rho= \sum_i u_i \otimes v_i\right],
$$
where $\|A \|=\Tr[(AA^\dagger)^{1/2}]$ is the trace norm (sum of the singular values). With this Rudolph \cite{Rudolph00} proved the following theorem.
\begin{theorem}
Let $\rho$ be a density operator on ${\cal H}_A \otimes {\cal H}_B$. Then $\rho$ is separable if $\|\rho\|_{\gamma}=1$ and entangled if $\|\rho\|_{\gamma}>1$. 
\end{theorem}
The proof is technical and we omit it here. Note that the difference $\|\rho\|_{\gamma}-1$ satisfies all the usual basic requirements for an entanglement measure \cite{Rudolph00b}. Unfortunately the greatest cross norm is in practice, with the exception of highly symmetrical states, very hard to calculate.

One can equally well define other cross norms, by using a different norm in the definition. Let us denote by $\|\cdot\|_{g}$ the cross norm induced by the Hilbert-Schmidt norm $\|A \|_2=[\Tr(AA^\dagger)]^{1/2}$. From $\|\rho\|_{g}\leq\|\rho\|_{\gamma}$\footnote{This is because $\|\cdot\|_{\gamma}$ majorizes any unitarily invariant norm (hence the name greatest cross norm).}  follows that a state $\rho$ with $\|\rho\|_{g}>1$ is entangled. It turns out that $\|\rho\|_{g}$ is easily computable, the construction is the following (\cite{Rudolph02, Rudolph04}, see also \cite{CDKL01}).

In Section~\ref{sejaiso} we have seen an isomorphism between vectors $|\psi\ra \in{\cal H}_A \otimes {\cal H}_B$ and operators $\Psi \in {\cal L}({\cal H}_A,{\cal H}_B)$ defined by
$$
|\psi\ra = \sqrt{d} (\Psi \otimes \openone) |\psi_+\ra = \sqrt{d} (\openone \otimes \Psi^T) |\psi_+\ra.
$$
The space ${\cal L}({\cal H}_A,{\cal H}_B)$ is a Hilbert space, when furnished with the Hilbert-Schmidt inner product $\Tr(\Psi^\dagger \Phi)$ for $\Psi,\Phi \in {\cal L}({\cal H}_A,{\cal H}_B)$. It follows that the isomorphism is a Hilbert-space isomorphism, i.e. isometric as $\la \psi |\phi\ra= \Tr(\Psi^\dagger \Phi)$.
Let ${\cal H}={\cal H}_A \otimes {\cal H}_B$ and $\rho \in {\cal L}({\cal H})$. We can decompose $\rho$ arbitrarily as 
$$
\rho=\sum_i u_i \otimes v_i
$$
with $u_i \in {\cal L}({\cal H}_A)$ and $v_i \in {\cal L}({\cal H}_B)$. Now we apply the above isomorphism to both $u_i$ and $v_i$ and obtain
$$
|\rho\ra=\sum_i |u_i\ra \otimes |v_i\ra,
$$
with $|\rho\ra \in {\cal H}\otimes{\cal H}$. Applying the isomorphism on $|\rho\ra$ we obtain
$$
R(\rho)=\sum_i |u_i\ra\la v_i|.
$$
The operator $R(\rho) \in {\cal L}({\cal H})$ is independent of the chosen decomposition of $\rho$ and in general $\rho\neq R(\rho)$. This is because we can interpret $|\rho\ra=|\rho\rangle_{1234} \in {\cal H}\otimes{\cal H}$ as an element of $(\mathbb{C}^d)^{\otimes 4}={\cal H}_1 \otimes {\cal H}_2 \otimes {\cal H}_3 \otimes {\cal H}_4$, then with $\rho$ acting on ${\cal H}_1 \otimes {\cal H}_3$ we have
$$
|\rho\rangle=\sum_{i,j=1}^{d}(\rho_{13} \otimes \ido_{24})|\psi _{+}\rangle _{13|24}, 
$$
as opposed to
$$
|\rho\rangle=\sum_{i,j=1}^{d}(R(\rho_{12}) \otimes \ido_{34})|\psi _{+}\rangle _{12|34}.
$$
Here $|\psi _{+}\rangle _{13|24}$ denotes the maximally entangled state with respect to the $13|24$ division (and analogue for $|\Psi _{+}\rangle _{12|34}$). The next lemma is a well known identity from the theory of operator norms (see e.g.\ \cite{Schatten70})

\begin{lemma}
For any operator $R \in {\cal L}({\cal H})$ we have
$$
\|R \|=\inf_{u_i,v_i}\left[\sum_i \sqrt{|\la u_i|u_i \ra|} \cdot \sqrt{|\la v_i|v_i \ra|} \Bigg| R= \sum_i |u_i\ra\la v_i|\right].
$$
\end{lemma}

With this lemma, we can prove that $\|R(\rho)\|=\|\rho\|_g$ and from that we obtain an easily computable necessary criterion for separability:

\begin{theorem}[Realignment criterion]
For any state $\rho \in {\cal L}({\cal H})$ the identity 
$$
\|R(\rho)\|=\|\rho\|_g
$$
holds. If $\|R(\rho)\|>1$, then $\rho$ is entangled.
\end{theorem}
\begin{proof}
Let $\rho=\sum_i u_i\otimes v_i$, then $\sqrt{|\la u_i|u_i \ra|}= \sqrt{\Tr(u^\dagger_i u_i)}=\|u_i \|_2$ (and similar for $v_i$). The theorem then follows from the previous lemma together with the definition of $\|\cdot\|_g$ and inequality $\|\rho\|_g\leq \|\rho\|_\gamma\leq 1$ for separable states $\rho$.
\end{proof}

From the discussion above it follows that $R(\rho)$ is nothing more then a rearrangement of matrix entries of $\rho$. For this reason, it is often referred to as the realignment criterion \cite{CW02, Rudolph02b, OH85}. Some analytical properties of the realignment criterion have been studied in Ref. \cite{Rudolph02c}. It turns out that the realignment criterion is independent of the partial transposition and hence can detect PPT entangled states. From our discussion it is easy to see that the realignment criterion corresponds to the operation $R$ which acts on basis states as
$$
|\psi_i\ra\la \psi_j|\otimes |\psi_k\ra\la \psi_l| \stackrel{R}{\rightarrow} |\psi_i\ra\la \psi_k^*|\otimes
 |\psi_j^*\ra\la \psi_l|.
$$
Due to linearity, this operation is well defined for arbitrary quantum states. The partial transposition is also based on reordering of matrix entries. In the same notation, we denote it as the operation $T$ which acts on the second subsystem as
$$
|\psi_i\ra\la \psi_j|\otimes |\psi_k\ra\la \psi_l| \stackrel{T}{\rightarrow} |\psi_i\ra\la \psi_j|\otimes
 |\psi_l^*\ra\la \psi_k^*|.
$$
A state $\rho$ is entangled if the trace norm $\|T(\rho)\|>1$. Note that the usual partial transpose criterion
says that a state is entangled if $T(\rho)$ has some negative eigenvalues, but as $T(\rho)$ is Hermitian this is equivalent to saying that $\|T(\rho)\|>1$. This was first observed in Ref.~\cite{VW01b}, where it was shown that the quantity $\|T(\rho)\|$ gives rise to a good entanglement measure (see also \cite{CW02b}).

This striking similarity between the two criteria led to the consideration of general permutations of multipartite density matrices \cite{HHH02}. Consider an $r$-party state belonging to a Hilbert space $\cal H$, whose subsystems have the same dimension $d$. A general state $\rho\in \cal H$ can be expanded as
$$
\rho=\sum_{i_1,i_2, \ldots, i_{2r}} \rho_{i_1 i_2, i_3 i_4, \ldots, i_{2r-1} i_{2r}} | i_1 i_3\ldots i_{2r-1}\ra \la i_2 i_4 \ldots i_{2r}|,
$$
where all indices run from $1$ to $d$. Let $S_{2r}$ denote the symmetric group with $(2r)!$ elements, that is, the group of the permutations of the set $\{1,2,\ldots, 2r\}$. We define for each permutation $\sigma\in S_{2r}$ a corresponding map $\Lambda_{\sigma}$ on operators acting on ${\cal H}$ as
\begin{equation}
\label{perdeff}
\big[\Lambda_\sigma(\rho)\big]_{i_1 i_2, \ldots, i_{2r-1} i_{2r}} =
\rho_{i_{\sigma(1)} i_{\sigma(2)}, \ldots, i_{\sigma(2r-1)} i_{\sigma(2r)}}\,.
\end{equation}
We will represent permutations as $[\sigma(1)\, \sigma(2)\, \ldots\, \sigma (2r)]$ or in disjoint cycles. With this notation the partial transpose criterion corresponds to the permutation $[1\,2\,4\,3]=(3,\,4)$ while the realignment criterion corresponds to the permutation $[1\,3\,2\,4]=(2,\,3)$. 

\begin{theorem}[Permutation criteria for separability \cite{HHH02}]
Let $\rho$ be a $r$-party state acting on a Hilbert space $\rho$ and $\sigma\in S_{2r}$ an arbitrary permutation.
If 
$$
\|\Lambda_\sigma(\rho)\|>1,
$$
with $\|\cdot\|$ the trace norm, then $\rho$ is entangled (that is, not fully separable).
\end{theorem}
\begin{proof}
Let $\rho$ be a fully separable state on $\cal H$, that is
$$
\rho=\sum_i p_i |\psi^i_1\ra \la \psi^i_2| \otimes |\psi^i_3\ra \la \psi^i_4| \otimes \ldots |\psi^i_{2r-1}\ra \la \psi^i_{2r}|=\sum_i p_i \rho_i,
$$
with $\psi^i_{2k-1}=\psi^i_{2k}$ and $\rho_i$ the corresponding pure states. Then $\Lambda_\sigma(\rho)$ corresponds to
$$
\Lambda_\sigma(\rho)=\sum_i p_i |\psi^{i\#}_{\sigma(1)}\ra \la \psi^{i\#}_{\sigma(2)}| \otimes |\psi^{i\#}_{\sigma(3)}\ra \la \psi^{i\#}_{\sigma(4)}| \otimes \ldots |\psi^{i\#}_{\sigma(2r-1)}\ra \la \psi^{i\#}_{\sigma(2r)}|=\sum_i p_i \Lambda_\sigma(\rho_i),
$$
with $\psi^{i\#}_k=\psi^{i*}_k$ if $\sigma(i)$ and $i$ have different parity and $\psi^{i\#}_k=\psi^{i}_k$ otherwise.
We need to show that $\|\Lambda_\sigma(\rho)\|\leq 1$, but because of $\|A+B\|\leq \|A\|+\|B\|$ it is sufficient to show that $\|\Lambda_\sigma(\rho_i)\| \leq 1$. 
But this follows easily from $\|A\otimes B\|\leq \| A\| \|B\|$ and $\|\, |\phi\ra \la\psi|\,\|=1$ for normalised vectors $|\phi\ra$ and $|\psi\ra$.
\end{proof}

As a corollary we will show here how to construct an entanglement witness for a $r$-party state $\rho$ given that $\|\Lambda_\sigma(\rho)\|>1$ for some permutation $\sigma$. For this we generalise the bipartite construction of Chen and Wu in Ref.~\cite{CW03}, see also Ref.~\cite{Horodecki03, GMTA06, YL04}. Here and elsewhere permutations are evaluated from \emph{left} to \emph{right}.

\begin{theorem}
\label{racew}
Let $\|\Lambda_\sigma(\rho)\|>1$ for some permutation $\sigma$, and define
$$
\tilde \sigma = \tau \sigma^{-1} \tau
$$
with $\tau$ the global transposition and $\sigma^{-1}$ the inverse permutation of $\sigma$. Then 
$$
W=\ido - \Lambda_{\tilde\sigma}(V U^\dagger)
$$
is an entanglement witness with $\Tr(\rho W)=1-\|\Lambda_\sigma(\rho)\|<0$. Here $U$ and $V$ are the unitary matrices corresponding to the singular value decomposition of $\Lambda_{\sigma}(\rho)=UDV^\dagger$.
\end{theorem}
\begin{proof}
It is straightforward to check that $\Tr(\rho W)=1-\|\Lambda_\sigma(\rho)\|$ from the identity
$$
\Tr[A\Lambda_\sigma(B)]=\Tr[(\Lambda_{\sigma^{-1}}(A^T))^T B]=\Tr[\Lambda_{\tilde \sigma}(A)B],
$$
for arbitrary operators $A$ and $B$. This identity can be verified by explicitly writing the trace as a sum over matrix indices.

Now all that is left is to show that $W$ is positive on separable states. We will use of the following variational characterisation of the trace norm (see for instance \cite{HJ91}, p195):
\begin{align}
\label{varcarnor}
\|A\|= \max \{|\Tr(X A Y)|\, | XX^\dagger=YY^\dagger=\ido \},
\end{align}
for arbitrary $A$. Let $\rho_{\text{sep}}$ be an arbitrarily separable state, then
$$ 
\|\Lambda_{\sigma}(\rho_{\text{sep}})\| = \Tr(D') \leq 1,
$$
with $\Lambda_{\sigma}(\rho_{\text{sep}})=U'D'{V'}^\dagger$ the singular value decomposition of $\Lambda_{\sigma}(\rho_{\text{sep}})$. From (\ref{varcarnor}) follows that $\Tr[ X \Lambda_{\sigma} (\rho_{\text{sep}}) Y ]$ is maximal when $X=U'$ and $Y={V'}^\dagger$ and thus
\begin{align*}
\Tr(W \rho_{\text{sep}}) & =1-\Tr(\Lambda_{\tilde\sigma}(V U^\dagger)\rho_{\text{sep}}) \\
&=1- \Tr(U^\dagger \Lambda_\sigma(\rho_{\text{sep}}) V) \\
&\geq 1 - \Tr(U' \Lambda_\sigma(\rho_{\text{sep}}) {V'}^\dagger) \\
&=1- \| \Lambda_\sigma(\rho_{\text{sep}})\| \geq 0.
\end{align*}
\end{proof}

For an $r$-party state, there are $(2r)!$ permutation criteria, but not all of them are inequivalent. For instance for bipartite systems it is easy to see that partial transposition with respect to the first system is equivalent to the partial transposition with respect to the second. For bipartite systems all permutation criteria turn out to be equivalent to either the partial transposition or the realignment criterion \cite{HHH02}. For more than two parties the classification of inequivalent permutation criteria is an open problem. First steps toward such a classification have been made in Ref.~\cite{Fan02, HHH02, WH05b}. For three parties it was implied that there are $9$ inequivalent criteria \cite{Fan02, WH05b}, and for four parties at most $34$ inequivalent criteria \cite{WH05b}.

In the next section we define exactly what we mean by independent criteria and review the work of Wocjan and Horodecki \cite{WH05b} on so-called combinatorially independent permutation criteria and introduce their graphical notation. Then we point out that both papers \cite{Fan02, WH05b} overlooked an obvious fact. We show that for three parties there are just $6$ independent criteria and $22$ for four parties. We prove that the class of the combinatorially independent permutation criteria contains some equivalent criteria which always occur in pairs. We completely classify these criteria and give a new upper bound of the number of independent criteria (this is Theorem~\ref{permmainthe}). It will become clear that most likely there are no more equivalences in the remaining criteria, although we were unable to provide a complete proof of this. A proof is presented for $2$, $3$ and $4$ parties and numerical evidence for up to $8$ parties is given. These results were first published in Ref.~\cite{CW05}.
 
\subsection{Definitions, graphical notation and first results}
Let $\sigma$ and $\mu$ be two permutations in $S_{2r}$. We call the corresponding entanglement criteria $\sigma$ and $\mu$ dependent if and only if
\begin{equation}
\label{eq:equalNorm}
\| \Lambda_\sigma(\rho)\| = \|\Lambda_\mu(\rho)\|,
\end{equation}
for all quantum states. Else, the permutation criteria are called independent. Observe that if two permutation criteria $\sigma$ and $\mu$ are independent by the above definition, it could still be that they detect exactly the same entangled states. This would occur if we had 
$$
\label{eq:trulyIndependent} 
\| \Lambda_\sigma(\rho)\| > 1 \quad \Leftrightarrow \quad \|\Lambda_\mu(\rho)\| > 1
$$
for all quantum states $\rho$. However, as we will see in the next section, the condition in (\ref{eq:equalNorm}) is already good enough to reduce significantly the number of permutation separability criteria. Furthermore, we prove later that the criteria which are independent are also truly independent (in the sense that no criterion is strictly stronger than any other) for up to four subsystems. 

Our necessary condition extends the necessary condition of combinatorial independence introduced in Ref.~\cite{WH05b}. The definition of combinatorial independence is similar to the definition of independence. The decisive difference is that two permutations are combinatorially dependent if and only if equality in (\ref{eq:equalNorm}) is achieved for {\em all} operators and not only quantum states. In this case, it is easy to see that the maps $\Lambda_\sigma$ and $\Lambda_\mu$ are related by a norm-preserving map $\Lambda$. That is $\Lambda_\sigma(A)=\Lambda(\Lambda_\mu(A))$ with  $\|A\|=\|\Lambda(A)\|$ for all operators $A$. Moreover, it can be shown that if $\Lambda_\sigma$ and $\Lambda_\mu$ are related via a norm-preserving map $\Lambda$, then $\Lambda=\Lambda_\nu$ must come from some permutation $\nu$. We call such a permutation a norm-preserving permutation.

 An example of such a norm-preserving permutation is the global quantum transpose (GQT), which transposes the complete system. It can be written as 
$$
\tau=(1,2)(3,4)\cdots (2r-1,2r).
$$
Another example of a norm preserving map $\Lambda$ is the unitary transformation $\Lambda(\rho)=U\rho V$, where $U$ and $V$ are unitary operators. Reordering the different parties in the density matrix representation is an example of such a unitary transformation. Consider for instance the transformation 
$$
\rho=\sum_i \rho^A_i \otimes \rho^B_i \rightarrow \rho'=\sum_i \rho^B_i \otimes \rho^A_i.
$$
 This mapping can be implemented by means of multiplication on the left and on the right by the swap operator \cite{WH05b}.

To illustrate the necessary condition of combinatorial independence let us consider again a bipartite system. The operation $R'$ induced by the permutation $(1,\,4)$ gives rise to the same criterion as $R$ induced by the permutation $(2,\,3)$. Indeed, the permutation $R'\tau$ transforms $[1\,2\,3\,4]$ into $[2\,4\,1\,3]$, which up to reordering of the parties is equivalent to $[1\,3\,2\,4]$. This is just the transformation defined by $R$.

\begin{theorem}[Combinatorially independent criteria \cite{WH05b}]
\label{thrmone}
The group $\cal T$ of norm preserving permutations can be generated as
$$
{\cal T}= \la(2k,2l),(2k-1,2l-1),\tau \ra,
$$
where $1\leq k,l\leq r$ and $\tau$ as before denotes the GQT\@. The combinatorially independent permutation criteria correspond to the right cosets $S_{2r}/{\cal T}$ of ${\cal T}$ in $S_{2r}$. The number of independent permutation criteria is therefore not larger than $ {\frac{1}{2}}\binom{2r}{r}.$ In this number, the class of trivial norm preserving criteria is also counted.
\end{theorem}

In the same paper, the authors also devised a graphical notion of the criteria, which leads to a way of selecting a simple representative for the right cosets. They decompose any permutation as a combination of 4 elementary permutations: the partial transpose, two types of realignment or reshuffling, and the identity. The corresponding graphical notations are loops from a subsystem to itself, arrows from one subsystem to another and free subsystems (no loops or arrows), as graphically depicted in Figure~\ref{tab:basicPermutations}. We call $k$ the head and $l$ the tail if there is an arrow from $k$ to $l$. If there is a loop in $m$, then $m$ is both head and tail of the loop. We call the support of an arrow or a loop the set containing its head and tail. A configuration of arrows and loops is called disjoint, if the supports of all pairs and loops are disjoint.

\begin{figure}
\begin{center}
\begin{tabular}{c|c|c}
\quad graphical representation\quad &
\quad corresponding permutation\quad &
\quad name \quad \\ \hline\hline
$k\quad \bullet \longrightarrow \bullet \quad l$ & $(2k,\, 2l-1)$ & reshuffle
$R_{kl}$ \\\hline
$k\quad\bullet\longleftarrow \bullet\quad l$ & $(2k-1,\, 2l)$ & reshuffle
$R'_{lk}$ \\\hline
 \, $k \quad\lar$ \qquad\qquad \qquad & $(2k-1,\, 2k)$ & \quad partial
transpose\quad\\\hline
$k\quad\bullet$\quad\quad\quad\quad\quad & () & identity
\end{tabular} 
\caption[Basic permutations]{Basic permutations, $k<l$ (Table from Ref. \cite{WH05b}).}
\label{tab:basicPermutations}
\end{center}
\end{figure}

\begin{theorem}[Canonical representation of combinatorially independent criteria \cite{WH05b}]
The right cosets $S_{2r}/{\cal T}$ can always be represented by a disjoint configuration of arrows, that is, they can be decomposed as a combination of the 4 elementary permutations. It is always possible to represent a random permutation $\sigma \in S_{2r}$ by a disjoint configuration by applying the following consecutive rules 
\begin{description}
\item[Rule 1: Pruning] Writing $\sigma$ as a product of disjoint cycles, we can prune all cycles of $\sigma$ such that there are no adjacent even or odd numbers.
\item[Rule 2: Chopping] We can chop up $\sigma$ into disjoint transpositions. 
\item[Rule 3: Exchanging heads] By exchanging heads we then obtain a disjoint configuration of arrows.
\item[Rule 4: GQT] By reversing the direction of all arrows and replacing loops by free subsystems and vice versa we obtain an equivalent disjoint permutation.
\end{description}
All criteria that can be represented by a disjoint configuration of arrows are combinatorially independent up to a application of Rule 4, and possible application of Rule 3.
\end{theorem}
This theorem differs slightly from the corresponding one in the original paper \cite{WH05b}. They claimed that any two permutations represented by disjoint configurations are independent, up to application of Rule 4. As the following example illustrates, this is not always the case: by exchanging heads of $[1\,3\,2\,4\,5\,7\,6\,8]$ we obtain $[1\,7\,6\,4\,5\,3\,2\,8]$, which can be represented by a different disjoint configuration. And both configurations are not related via the GQT.

\subsection{Main theorem}
\label{mtpct}
It was shown in Ref.~\cite{WH05b} that two permutations $\mu$ and $\sigma$ are combinatorially dependent if there is a norm preserving permutation such $\nu$ that $\Lambda_\sigma(A) = \Lambda_\nu(\Lambda_\mu(A))$ for all operators $A$. One could assume that therefore Theorem~\ref{thrmone} cannot be sharpened, that is, that all combinatorially independent criteria are independent. This is true when one considers the permutation acting on arbitrary operators. In quantum mechanics however, we deal with positive operators, which are Hermitian. The following theorem exploits this fact and counts the new upper bound to the number of independent criteria.

\begin{theorem}
\label{permmainthe}
(i) The permutation criteria corresponding to the permutations $\sigma$ and $\tau\sigma$ are dependent. 

(ii). Let ${\cal Z}:=\{e,\tau\}$ be the subgroup of $S_{2r}$ generated by the GQT $\tau$. Define the action of $\cal Z$ on the right cosets $S_{2r}/{\cal T}$ by multiplication from the left of the cosets, that is, $e*\sigma {\cal T} = e\sigma \cal T$ and $\tau*\sigma {\cal T} = \tau\sigma \cal T$. The new upper bound on the number of independent criteria is the number of orbits under this action. It is given by \footnote{The integer sequence generated by the number of independent criteria equals the integer sequence A045723 from Ref.~\cite{sloan}.}
\begin{equation}
\label{uupp}
\frac{1}{4}\left[\binom{2r}{r} + 2^r + \binom{r}{r/2}\,\cdot\, even(r)\right]\,,
\end{equation}
where $even(r)=1$ if $r$ is even and $0$ otherwise. This number includes the trivial criterion given by the identity permutation.
\end{theorem}
\begin{proof}
To prove (i), let us apply the permutation $\tau\sigma$ on an arbitrary quantum state $\rho$. We have
\begin{align}
\|\Lambda_{\tau \sigma}(\rho)\|&=\|\Lambda_{\sigma}(\Lambda_{\tau
}(\rho))\|\nonumber \\
&=\|\Lambda_{\sigma}(\rho^T)\| \nonumber\\
&=\|\Lambda_{\sigma}(\bar{\rho})\| \nonumber\\
&=\|\overline{\Lambda_{\sigma}(\rho)}\|\nonumber\\
&=\|\Lambda_{\sigma}(\rho)\|.
\end{align}
Here we have used that $\rho^T=\bar{\rho}$ because $\rho$ is Hermitian and $\Lambda_{\sigma}(\bar{\rho}) =\overline{\Lambda_{\sigma}(\rho)}$ because $\Lambda$ only permutes the entries of the matrix.

Observe that multiplying a coset $\sigma \cal T$ by $\tau$ from the left is the same as conjugating it by $\tau$ because $\tau$ is contained in $\cal T$, that is, we have
$$
\tau\sigma \cal T = \tau \sigma \cal T \tau.
$$
It is readily verified that conjugation of a permutation by $\tau$ corresponds to exchanging heads (always odd numbers) and tails (always even numbers). Therefore, the direction of all (true) arrows is reversed and loops and free subsystems are not affected. An example is shown in Figure~\ref{fig:actionNewRule}. Following the four rules presented in Ref.~\cite{WH05b}, we call this Rule 5. The new rule either \emph{glues} two criteria together or does not change them. More precisely, the orbits under the action of $\cal Z$ have size $1$ or $2$. The orbit of size two is generated by the involution $\tau$:
$$
\sigma\cal T \rightarrow \tau\sigma\cal T \rightarrow \tau\tau\sigma\cal T=\sigma\cal T.
$$

\begin{figure}
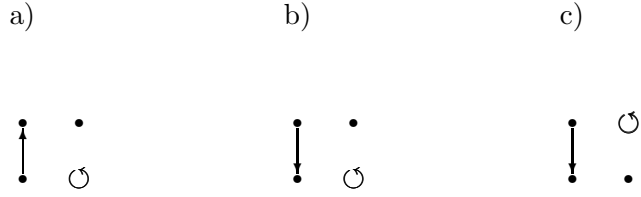

\begin{center}
\unitlength=2pt
\begin{tabular}{ccc}
a) & b) & c) \\ & & \\ & & \\
\,
\begin{minipage}{3cm}
\bp(-35,0)(-20,0)
\put(0,1){\vector(0,1){8.5}}
\put(0,10.5){\ci*{1.5}}
\put(10.5,10.5){\ci*{1.5}}
\put(0,0){\ci*{1.5}}
\put(10.5,0){\makebox(0,0){$\rar$}}
\put(10.55,0){\makebox(0,0){$\rar$}}
\put(10.5,0.05){\makebox(0,0){$\rar$}}
\ep
\end{minipage}
&
\,
\begin{minipage}{3cm}
\bp(-35,0)(-20,0)
\put(0,9.5){\vector(0,-1){8.5}}
\put(0,10.5){\ci*{1.5}}
\put(10.5,10.5){\ci*{1.5}}
\put(0,0){\ci*{1.5}}
\put(10.5,0){\makebox(0,0){$\rar$}}
\put(10.55,0){\makebox(0,0){$\rar$}}
\put(10.5,0.05){\makebox(0,0){$\rar$}}
\ep
\end{minipage}
&
\,
\begin{minipage}{3cm}
\bp(-35,0)(-20,0)
\put(0,9.5){\vector(0,-1){8.5}}
\put(0,10.5){\ci*{1.5}}
\put(10.5,10.5){\makebox(0,0){$\rar$}}
\put(10.55,10.5){\makebox(0,0){$\rar$}}
\put(10.5,10.55){\makebox(0,0){$\rar$}}
\put(0,0){\ci*{1.5}}
\put(10.5,0){\ci*{1.5}}
\ep
\end{minipage}
\end{tabular}
\caption[Equivalent permutations under the new rule]{Action of the new rule on a disjoint arrow configuration: our new rule changes a) to b). Compare this with the action of rule~4 in Ref.~\cite{WH05b} which changes a) to c). Our new rule only reverses the direction of true arrows, whereas rule~4 additionally creates loops on free subsystems and removes loops.}
\label{fig:actionNewRule}
\end{center}
\end{figure}

(ii). There are $\frac{1}{2}\binom{2r}{r}$ combinatorially independent criteria. With the new rule, there are at most
$$
\frac{1}{2}\binom{2r}{r} - \frac{1}{2}\left[\frac{1}{2}\binom{2r}{r}-K\right],
$$
independent criteria left. Here $K$ denotes the number of criteria not affected by conjugating according to the new rule. Now note that the only criteria (represented as disjoint arrow configurations) not affected by conjugation with $\tau$ are
\begin{enumerate}
\item criteria with no arrows and
\item criteria containing only arrows and having no free subsystems.
\end{enumerate}
If $r$ is odd, then situation (2) cannot occur. The number of these criteria is readily counted:
$$
\binom{r}{0}+ \binom{r}{1}+\binom{r}{2}+\cdots+ \binom{r}{\lfloor(r/2)\rfloor}=2^{r-1},
$$
where we have used an identity of binomial coefficients. So when the number of subsystems is odd, the number of criteria becomes (including the identity)
$$
\frac{1}{4}\left[\binom{2r}{r}+2^r\right].
$$

In the case $r$ is even we need to take care of situation (2). Now this number equals picking $r/2$ heads from $r$ choices, because exchanging tails does not matter. But we have to divide by two since exchanging all heads and tails does not matter either, so that the number of criteria satisfying (2) is given by
$$
\frac{1}{2}\left[\binom{r}{r/2}\right].
$$
We conclude that when $r$ is even, the number of criteria is given by (including the identity)
$$
\frac{1}{4}\left[\binom{2r}{r} + 2^r + \binom{r}{r/2} \right].
$$
To complete the proof, we have to show that the criteria in the orbits of size $2$ are combinatorially independent. Let $\sigma$ be a permutation represented by a disjoint arrow configuration. Assume that there is an arrow from subsystem $k$ to $l$ in the disjoint configuration of $\sigma$. Then there is an arrow from $l$ to $k$ in the disjoint configuration describing $\tau\sigma\tau$. Loops and free subsystems are not affected. Using these observation we see that the configuration describing $\sigma\tau\sigma\tau$ has a closed path from $k$ to $l$ and no loops. The closed path between $k$ and $l$ can be transformed into a loop on $k$ and a loop on $l$ with the help of Rule~3 (Exchanging heads) in Ref.~\cite{WH05b}. Now if we apply these arguments to all arrows of $\sigma$ we see that the permutation $\sigma\tau\sigma\tau$ is not norm-preserving. Consequently, the permutations $\sigma$ and $\tau\sigma\tau$ are combinatorially independent. This concludes the proof. 
\end{proof}

Figure~\ref{fig:actionNewRule} shows an example of how the new rule glues two combinatorially independent criteria together. In general, we have that two permutations corresponding to disjoint arrow configurations are related by the new rule if and only if they have the same (true) arrow structure (up to exchanging heads) and a complementary loop/free subsystem structure. The latter means that if the first criterion has a loop on subsystem $k$ then the subsystem $k$ is free in the second and vice versa.

\subsection{Illustrations}
\label{ilpct}
\begin{figure}
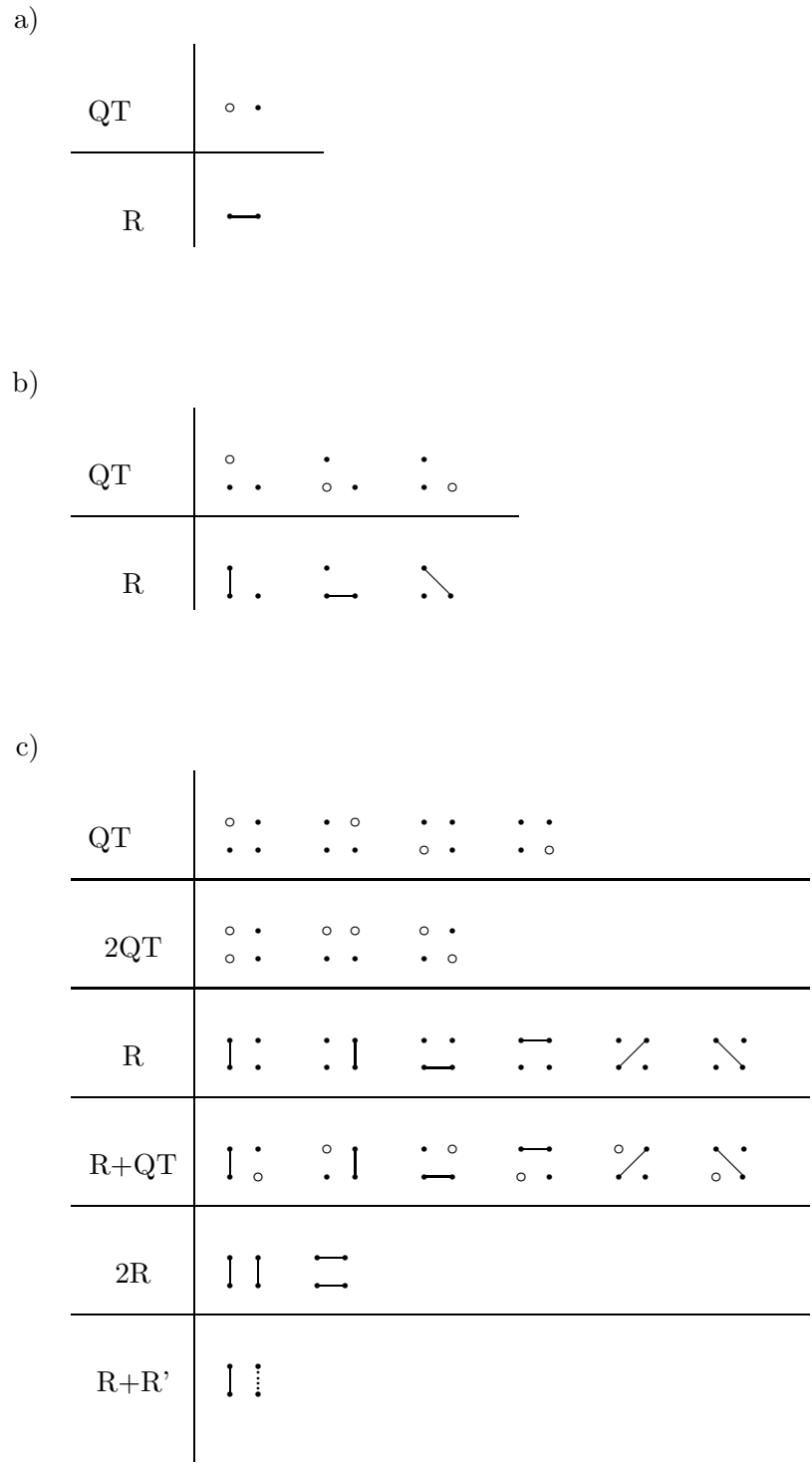

\begin{center}
\begin{tabular}{rl}
a) & \\ &
\begin{tabular}{c|l}
\unitlength=1pt
&\\[-2mm]
\begin{minipage}{1.2cm} QT \end{minipage} & \
\ $\bp(0,0)(0,-5) \put(0,0){\ci{3}}\put(10.5,0){\ci*{2}}\ep $
\quad\quad\ \
\\[2mm]
\hline
&\\[-2mm]
R & \
\ $\bp(0,0)(0,-5) \put(0,0){\li(1,0){10.1}}
\put(0,0){\ci*{2}}\put(10.5,0){\ci*{2}}\ep $
\quad\quad\ \
\end{tabular}
\\\\\\
b) & \\ &
\begin{tabular}{c|l}
\unitlength=1pt
&\\[-2mm]
\begin{minipage}{1.2cm} QT \end{minipage} & \
\ $\bp(0,0)(0,1) \put(0,10.5){\ci{3}}\put(0,0){\ci*{2}}\put(10.5,0){\ci*{2}}\ep
$
\quad\quad\ \
\ $\bp(0,0)(0,1) \put(0,10.5){\ci*{2}}\put(0,0){\ci{3}}\put(10.5,0){\ci*{2}}\ep
$
\quad\quad\ \
\ $\bp(0,0)(0,1) \put(0,10.5){\ci*{2}}\put(0,0){\ci*{2}}\put(10.5,0){\ci{3}}\ep
$
\quad\quad\ \
\\[2mm]
\hline
&\\[-2mm]
R & \
\ $\bp(0,0)(0,1) \put(0,0){\li(0,1){10.1}}
\put(0,10.5){\ci*{2}}\put(0,0){\ci*{2}}\put(10.5,0){\ci*{2}}\ep $
\quad\quad\ \
\ $\bp(0,0)(0,1) \put(0,0){\li(1,0){10.1}}
\put(0,10.5){\ci*{2}}\put(0,0){\ci*{2}}\put(10.5,0){\ci*{2}}\ep $
\quad\quad\ \
\ $\bp(0,0)(0,1) \put(0,10.5){\li(1,-1){10}}
\put(0,10.5){\ci*{2}}\put(0,0){\ci*{2}}\put(10,0){\ci*{2}}\ep $
\quad\quad\ \
\end{tabular}
\\\\\\
c) & \\ &
\begin{tabular}{c|l}
\unitlength=1pt
&\\[-2mm]
\begin{minipage}{1.2cm} QT \end{minipage} & \
\ $\bp(0,0)(0,1)
\put(0,10.5){\ci{3}}\put(10.5,10.5){\ci*{2}}\put(0,0){\ci*{2}}\put(10.5,0){\ci*{2}}\ep
$
\quad\quad\ \
\ $\bp(0,0)(0,1)
\put(0,10.5){\ci*{2}}\put(10.5,10.5){\ci{3}}\put(0,0){\ci*{2}}\put(10.5,0){\ci*{2}}\ep
$
\quad\quad\ \
\ $\bp(0,0)(0,1)
\put(0,10.5){\ci*{2}}\put(10.5,10.5){\ci*{2}}\put(0,0){\ci{3}}\put(10.5,0){\ci*{2}}\ep
$
\quad\quad\ \
\ $\bp(0,0)(0,1)
\put(0,10.5){\ci*{2}}\put(10.5,10.5){\ci*{2}}\put(0,0){\ci*{2}}\put(10.5,0){\ci{3}}\ep
$
\quad\quad\ \
\\[2mm]
\hline
&\\[-2mm]
2QT & \
\ $\bp(0,0)(0,1)
\put(0,10.5){\ci{3}}\put(10.5,10.5){\ci*{2}}\put(0,0){\ci{3}}\put(10.5,0){\ci*{2}}\ep
$
\quad\quad\ \
\ $\bp(0,0)(0,1)
\put(0,10.5){\ci{3}}\put(10.5,10.5){\ci{3}}\put(0,0){\ci*{2}}\put(10.5,0){\ci*{2}}\ep
$
\quad\quad\ \
\ $\bp(0,0)(0,1)
\put(0,10.5){\ci{3}}\put(10.5,10.5){\ci*{2}}\put(0,0){\ci*{2}}\put(10.5,0){\ci{3}}\ep
$
\quad\quad\ \
\\[2mm]
\hline
&\\[-2mm]
R & \
\ $\bp(0,0)(0,1) \put(0,0){\li(0,1){10.1}}
\put(0,10.5){\ci*{2}}\put(10.5,10.5){\ci*{2}}\put(0,0){\ci*{2}}\put(10.5,0){\ci*{2}}\ep
$
\quad\quad\ \
\ $\bp(0,0)(0,1) \put(10.5,0){\li(0,1){10.1}}
\put(0,10.5){\ci*{2}}\put(10.5,10.5){\ci*{2}}\put(0,0){\ci*{2}}\put(10.5,0){\ci*{2}}\ep
$
\quad\quad\ \
\ $\bp(0,0)(0,1) \put(0,0){\li(1,0){10.1}}
\put(0,10.5){\ci*{2}}\put(10.5,10.5){\ci*{2}}\put(0,0){\ci*{2}}\put(10.5,0){\ci*{2}}\ep
$
\quad\quad\ \
\ $\bp(0,0)(0,1) \put(0,10.5){\li(1,0){10.1}}
\put(0,10.5){\ci*{2}}\put(10.5,10.5){\ci*{2}}\put(0,0){\ci*{2}}\put(10.5,0){\ci*{2}}\ep
$
\quad\quad\ \
\ $\bp(0,0)(0,1) \put(0,0){\li(1,1){10}}
\put(0,10.5){\ci*{2}}\put(10.5,10.5){\ci*{2}}\put(0,0){\ci*{2}}\put(10,0){\ci*{2}}\ep
$
\quad\quad\ \
\ $\bp(0,0)(0,1) \put(0,10.5){\li(1,-1){10}}
\put(0,10.5){\ci*{2}}\put(10.5,10.5){\ci*{2}}\put(0,0){\ci*{2}}\put(10,0){\ci*{2}}\ep
$
\quad\quad\ \
\\[2mm]
\hline
&\\[-2mm]
R+QT & \
\ $\bp(0,0)(0,1) \put(0,0){\li(0,1){10.1}}
\put(0,10.5){\ci*{2}}\put(10.5,10.5){\ci*{2}}\put(0,0){\ci*{2}}\put(10.5,0){\ci{3}}\ep
$
\quad\quad\ \
\ $\bp(0,0)(0,1) \put(10.5,0){\li(0,1){10.1}}
\put(0,10.5){\ci{3}}\put(10.5,10.5){\ci*{2}}\put(0,0){\ci*{2}}\put(10.5,0){\ci*{2}}\ep
$
\quad\quad\ \
\ $\bp(0,0)(0,1) \put(0,0){\li(1,0){10.1}}
\put(0,10.5){\ci*{2}}\put(10.5,10.5){\ci{3}}\put(0,0){\ci*{2}}\put(10.5,0){\ci*{2}}\ep
$
\quad\quad\ \
\ $\bp(0,0)(0,1) \put(0,10.5){\li(1,0){10.1}}
\put(0,10.5){\ci*{2}}\put(10.5,10.5){\ci*{2}}\put(0,0){\ci{3}}\put(10.5,0){\ci*{2}}\ep
$
\quad\quad\ \
\ $\bp(0,0)(0,1) \put(0,0){\li(1,1){10}}
\put(0,10.5){\ci{3}}\put(10.5,10.5){\ci*{2}}\put(0,0){\ci*{2}}\put(10,0){\ci*{2}}\ep
$
\quad\quad\ \
\ $\bp(0,0)(0,1) \put(0,10.5){\li(1,-1){10}}
\put(0,10.5){\ci*{2}}\put(10.5,10.5){\ci*{2}}\put(0,0){\ci{3}}\put(10,0){\ci*{2}}\ep
$
\quad\quad\ \
\\[2mm]
\hline
&\\[-2mm]
2R & \
\ \bp(0,0)(0,1) \put(0,0){\li(0,1){10.1}}\put(10.5,0){\li(0,1){10.1}}
\put(0,10.5){\ci*{2}}\put(10.5,10.5){\ci*{2}}\put(0,0){\ci*{2}}\put(10.5,0){\ci*{2}}\ep
\quad\quad\ \
\ \bp(0,0)(0,1) \put(0,0){\li(1,0){10.1}}\put(0,10.5){\li(1,0){10.1}}
\put(0,10.5){\ci*{2}}\put(10.5,10.5){\ci*{2}}\put(0,0){\ci*{2}}\put(10.5,0){\ci*{2}}\ep
\quad\quad\ \
\\[2mm]
\hline
&\\[-2mm]
R+R' & \
\ $\bp(0,0)(0,1) \put(0,0){\li(0,1){10.1}}
\multiput(10.5,0.25)(0,2){6}{\ci*{0.5}}
\put(0,10.5){\ci*{2}}\put(10.5,10.5){\ci*{2}}\put(0,0){\ci*{2}}\put(10.5,0){\ci*{2}}\ep
$
\quad\quad\ \ \\ \\
\end{tabular}
\end{tabular}
\caption[Independent permutation criteria for $2,3$ and $4$ parties]{Independent permutation criteria for a) two, b) three, and c) four particles. The figure is adapted from Ref.~\cite{WH05b}.}
\label{fig:four}
\end{center}
\end{figure}

The different criteria resulting from our characterisation for two, three and four parties are shown graphically in Figure~\ref{fig:four}. In this figure, we have adopted the simplified graphical notation of \cite{WH05b} without arrows. Loops (QT) have been denoted by a little circle, $(2k,2l-1)$ permutations (R) as a solid line and $(2k-1,2l)$ permutations (R') as a dotted line. In order to construct this figure one first draws all disjoint configurations, and then eliminates those related by application of Rules 3, 4 and 5.

Below we show that all listed criteria are \emph{truly} independent, in the sense that no criterion detects strictly more states than any other criterion. A property we are going to use frequently is that the permutation criteria are insensitive to adding an uncorrelated ancilla. Let $\rho$ be an $r$-party state and define the $r+1$-party state $\rho'=\rho \otimes \rho_{r+1}$. Suppose $\sigma\in S_{2r}$ and define $\sigma'=\sigma \in S_{2(r+1)}$, (the same permutation, but now acting on an enlarged set of $2(r+1)$ elements). Then from $\|\Lambda_{\sigma'}(\rho') \|=\|\Lambda_\sigma(\rho) \otimes \rho_{r+1} \|= \|\Lambda_\sigma(\rho) \| \| \rho_{r+1} \|=\|\Lambda_\sigma(\rho) \| $ follows that $\|\Lambda_\sigma(\rho) \|=\|\Lambda_{\sigma'}(\rho') \|$.

\subsubsection{Two parties}
For a quantum system consisting of two parties, there are only two non trivial inequivalent permutation criteria: the partial transpose in one of the subsystems and reshuffling between the two subsystems. For the low dimensional systems ${\cal H}\cong \mathbb{C}^2 \otimes \mathbb{C}^2$ and ${\cal H}\cong \mathbb{C}^2 \otimes \mathbb{C}^3$ the positivity of the partial transpose is a necessary and sufficient condition for separability \cite{HHH96}, while this is not the case for the realignment criterion \cite{Rudolph02c, CW02}. For higher dimensional systems these criteria are truly independent, as there exist bound entangled states that are detected by the realignment criterion. As a curiosity, we have tested the realignment criterion on all known bound entangled states $\rho \in{\cal H}\cong \mathbb{C}^3 \otimes \mathbb{C}^3$ in the literature. The maximum value we have found for $\|R(\rho)\|$ is $7/6$ and is achieved for a particular chess-board state \cite{BP99}:
\begin{equation}
\rho_c=\frac{1}{12}\left[ \begin{array}{rrrrrrrrr}
 1 & 0 & 1 & 0 & 0 & 0 & \,\,\,\, 1 & 0 & 0 \\
 0 & 1 & 0 & 0 & 0 & -1 & 0 & -1 & 0 \\
 1 & 0 & 2 & 0 & -1 & 0 & 0 & 0 & 0 \\
 0 & 0 & 0 & 1 & 0 & -1 & 0 & 1 & 0 \\
 0 & 0 & -1 & 0 & 1 & 0 & 1 & 0 & 0 \\
 0 & -1 & 0 & -1 & 0 & 2 & 0 & 0 & 0 \\
 1 & 0 & 0 & 0 & 1 & 0 & 2 & 0 & 0 \\
 0 & -1 & 0 & 1 & 0 & 0 & 0 & 2 & 0 \\
 0 & 0 & 0 & 0 & 0 & 0 & 0 & 0 & 0
\end{array} \right].
\end{equation}

\subsubsection{Three parties}
For three parties we have proven that at most $6$ criteria are independent, previous work \cite{WH05b, Fan02} indicated that there were $9$. The $6$ criteria are the partial transposes (row QT) in the $3$ subsystems and the $3$ reshufflings (row R) between any of the two subsystems. On random states, it can easily be checked that the 6 criteria are indeed independent according to our definition; we now show that no criterion is strictly stronger than any other.

To show that the criteria from row QT are truly independent from one another, it is sufficient to note that there exist tripartite NPT states which are separable with regard to two splits, but not to the third one. An example of these kind of states for qubits can be found in Ref.~\cite{DC99}. Such states will obviously be detected by only one criterion from the row QT. To show the independence between the criteria in the row R, we use the property (see also \cite{Rudolph02c}) that the trace norm of the realigned density matrix remains invariant when an uncorrelated ancilla is added. Now taking a bipartite entangled state which violates the realignment criterion, and adding an uncorrelated ancilla, we find that it is only detected by only one criterion from row R (since it is separable in the other splits).

All what is left is to show that there are no dependences between the rows R and QT. So let us take two arbitrary criteria, one from row R and one from row QT. Here we can distinguish two cases, the first is where the realignment works on different subsystems than the partial transposition. As an example let us take realignment R$_{12}$ between subsystem 1 and 2, and partial transposition QT$_3$ of subsystem 3. Let $\rho$ be a bipartite entangled state detected by both the realignment and the partial transpose criterion. Then the state $\rho'=\rho\otimes \rho_3$, with $\rho_3$ an arbitrary state, will be detected by R$_{12}$, but not by QT$_3$. Conversely the state $\rho''= \rho_1 \otimes \rho$ will be detected by QT$_3$ but not by R$_{12}$. The second case, is where the realignment works on a same subsystem as the partial transposition, such as R$_{12}$ and QT$_1$. To show their independence it is sufficient to consider bipartite states that are detected by either R or QT (but not both), and adding an uncorrelated ancilla.

Note that the realignment criterion, in contrast to the partial transpose criterion can detect tripartite entangled states, that are separable under any bipartite cut. This has been demonstrated experimentally in Ref.~\cite{HHH02} with the tripartite bound entangled states from Ref.~\cite{BDMSST99}.

\subsubsection{Four parties}
For four parties, there are at 22 non trivial independent permutation criteria. As for three parties, it easy to show the independence of the criteria within a fixed row, by starting from an entangled two or three party state and adding uncorrelated ancillas. This is probably hardest to see for the row R+QT, so we illustrate this with an example. Let us take the two criteria R$_{12}$QT$_3$ and QT$_1$R$_{23}$ and let $\rho_{23}$ be a bipartite entangled state which is detected by the realignment criterion. The state $\rho_{1} \otimes \rho_{23} \otimes \rho_{4}$ (with $\rho_{1}, \rho_4$ arbitrary states) is detected by QT$_1$R$_{23}$, but not by R$_{12}$QT$_3$, as this criterion is equivalent to the criterion R'$_{12}$QT$_4$.

Let us now focus on the criteria which contain only partial transpositions. The rows QT and 2QT are easily shown to be independent by adding uncorrelated ancillas. As an example consider the criteria QT$_1$ and QT$_{12}$. Let $\rho_{12}$ be an NPT state, and add an uncorrelated ancilla $\rho_{34}$; clearly this state will be detected by QT$_{1}$ only. Conversely, constructing a state as the tensor product of an NPT entangled state $\rho_{24}$ and a separable state $\rho_{13}$ will give a state only detected by QT$_{12}$. The independence between the criteria with only loops and the criteria with at least one realignment follows easily from their independence in the bipartite case. As before this is done simply by adding suitable uncorrelated ancillas.

To show true independence within the set of realignment criteria (rows R, R+QT, 2R and R+R'), let us first consider the rows R and R+QT. Again, we consider an example, whereas the other cases can be shown similarly. Consider the criteria R$_{12}$ and QT$_4$R$_{23}$. Let $\rho$ be a bipartite entangled state which is detected by the realignment criterion and $\sigma$ a separable state. Then the state $\rho_{12} \otimes \sigma_{34}$ will be detected only by R$_{12}$, while $\sigma_{14}\otimes \rho_{23}$ only by QT$_4$R$_{23}$. As for the criteria 2R and R+R', consider the non-trivial case R$_{12}$R$_{34}$ and R$_{12}$R'$_{34}$. Their independence can be shown by applying the criteria to the 4-qubit states $\psi=(|1000\ra + |1101\ra)/\sqrt{2}$ and $\psi'= (|1000\ra + | 1110\ra)/\sqrt{2}$ (the first is detected by R$_{12}$R$_{34}$ only, while the second by R$_{12}$R'$_{34}$ only). All that is left, is to show that the criteria from row R and R+QT are independent of the criteria from rows 2R and R+R'. Without loss of generality we can take one representative from the rows 2R and R+R', say R$_{12}$R$_{34}$ (the other criteria correspond to appropriately swapping the subsystems). It is easy to find states that are detected by the criteria in rows R or R+QT but not by R$_{12}$R$_{34}$. Consider for example the criterion R$_{12}$QT$_{3}$. One can verify that it will detect the entangled state $\rho=\rho_c\otimes \openone/9$; while this state is not detected by criterion R$_{12}$R$_{34}$. Finally an example to show that the criterion R$_{12}$R$_{34}$ can detect states not detected by any criterion in the rows R or R+QT is given by $\rho=(3/4)\rho_c\otimes \rho_c+(1/4) \openone/81$.

\subsubsection{Final remarks}
In Ref.~\cite{HHH02} a powerful class of separability criteria was devised based on permutations. The class however contained many redundancies, and to give a complete characterization of the independent criteria is an open problem. In Ref.~\cite{WH05b} a graphical representation for permutations together with rules for simplifying them were introduced based on a sufficient condition for two permutations to yield dependent criteria. This equivalence meant that two combinatorially dependent criteria yield the same value of the trace norm on all operators. Combinatorially, density operators have a prominent Hermitian symmetry, that is a global quantum transposition together with a complex conjugate. We have exploited this symmetry and we have shown how this led to the dependence of particular combinatorially independent criteria. It is unlikely that there are more dependences in the criteria from Theorem~\ref{permmainthe}.

Density operators differ from arbitrary operators also because they have positive eigenvalues. But since permutations only reorder matrix entries, it is unlikely that this positiveness would lead to more criteria to be dependent. We have verified the independence of the criteria numerically on a random state (using the algorithm outlined in Ref.~\cite{ZHSL98}) for 2 up to 8 parties. In Appendix~\ref{gpmatlab} we have reproduced the matlab program which implements a random permutation. In the same way as we illustrated for three and four parties, it is easy to see that the criteria with only loops (only partial transpositions) are independent. These criteria are independent from the ones having at least one realignment since those can detect bound entangled states. To prove the independence within the class of criteria with at least one realignment one could try to generalize the arguments from the previous section.

\chapter{Schmidt number for mixed states}		
\label{chapsmidt}
In this chapter we introduce the Schmidt number for mixed states \cite{TH00}, as a generalisation of the Schmidt rank for pure states. It can be thought of as a refinement of the entangled versus separable division, where the set of mixed states is decomposed into a whole series of nested convex sets. In the first section, we study briefly the use of Schmidt witnesses to characterise these sets \cite{SBL00}. In the succeeding sections we derive the corresponding characterisation in terms of positive maps \cite{TH00, EK00} and discuss a conjecture on the decomposability of these maps in $3\otimes 3$. 

Robustness measures quantify the extent to which entangled states remain entangled under mixing. In the final part of this chapter, we generalise the notion of robustness of entanglement \cite{VT99} and introduce the Schmidt robustness and random Schmidt robustness \cite{Clarisse05b}. The latter notion is closely related to the construction of Schmidt balls around the identity. In particular we derive upper and lower bounds to the Schmidt robustness for pure states. We present two conjectures, the first one is related to the radius of inner balls around the identity in the convex set of Schmidt number $n$-states. We also conjecture a class of optimal Schmidt witnesses for pure states.

\section{Introduction and Schmidt witnesses}
\label{iswsec}
In Ref.~\cite{TH00} the notion of Schmidt rank $S(\psi)$ of pure bipartite states $|\psi\ra$ was extended to the Schmidt number $S(\rho)$ of mixed states $\rho$ as
$$
S(\rho)= \min_{\rho=\sum_i p_i |\psi_i\ra\la \psi_i|} \max_i S(\psi_i).
$$
Thus $\rho$ acting on ${\cal H}={\cal H}_A \otimes {\cal H}_B$ is said to have Schmidt number $n$ if there exists a decomposition of $\rho=\sum_i p_i |\psi_i\ra \la\psi_i |$ with all vectors $\{|\psi_i\ra\}$ having Schmidt rank at most $n$, and there exists no such decomposition with all vectors having a Schmidt rank $n-1$ or lower. This definition coincides with the Schmidt rank when $\rho$ is a pure state. 
It is clear that separable states have Schmidt number one, whereas entangled states have a Schmidt number larger than one.
Recall that we denoted the set of separable states as $\cal S$, and the complete set of states as $\cal D$. We will denote the set of density operators that have Schmidt number $n$ or less by ${\cal S}_n$. Sets of increasing Schmidt number are embedded into each other as 
$$
{\cal S}= {\cal S}_1 \subset {\cal S}_2 \subset \ldots \subset {\cal S}_d={\cal D}.
$$
The subsets ${\cal S}_i$ are all closed and convex by construction. The Schmidt rank of a pure state cannot increase under local operations and classical communication \cite{LP97}, and this holds true for the Schmidt number as well. Furthermore, the Schmidt number of a state is trivially invariant under local unitaries. This makes the Schmidt number $k$ of a state $\rho$, or equivalently the logarithmic Schmidt number ${\cal N}(\rho)=\log k$ a legitimate measure of entanglement. Most surprisingly, Terhal and Horodecki \cite{TH00} showed that ${\cal N}(\rho)$ is not an additive quantity, that is there exists states $\rho$ with Schmidt number $k$, such that
$$
{\cal N}(\rho^{\otimes n})\neq n \log k,
$$
so that in general all we can say is that
$$
{\cal N}(\rho)\leq {\cal N}(\rho^{\otimes n})\leq n{\cal N}(\rho).
$$
For multipartite systems, there is no straightforward generalisation of the Schmidt decomposition. However there have been a number of proposals, both for pure and mixed states (see e.g.\ \cite{CHS00,EB01}):

In analogy to entanglement witnesses the set of states in $S_k$ can be characterised with the aid of the so-called Schmidt witnesses \cite{SBL00,BCHHKLS01}. We call a Hermitian operator $W$ a Schmidt witness of class $k$ (for short $k$-SW) if and only if
\begin{enumerate}
\item $\Tr(W\sigma)\geq 0$ for all $\sigma \in S_{k-1}$,
\item There exists a $\rho \in S_k$ such that $\Tr(W\rho) < 0$,
\end{enumerate}
The existence of a $k$-SW for a Schmidt number $k$ state follows directly from Theorem~\ref{separation}. 
\begin{corollary}
\label{corschmidt}
A state $\rho \in \cal D$  has Schmidt number $k$ if and only if $\Tr(\rho W_{k+1})\geq 0$ for  all Schmidt witnesses $W_{k+1}$ of class $k+1$.
\end{corollary}
In a completely similar way as the discussion in Section~\ref{sdr} on PPT entangled states and their witnesses, every Schmidt number $k$ state can be decomposed as the convex sum of a Schmidt number $k-1$ state and a Schmidt number $k$ edge state (see Theorem~\ref{pptdeco}). Furthermore for every Schmidt number $k$ edge state there is a canonical witness that detects the state and conversely, every $k$-SW can be written in a canonical form with the aid of an Schmidt number $k$ edge state (see Theorem~\ref{edgeindec}). Since the theorems and proofs are almost identical we will not reproduce them here.

Consider the following example \cite{TH00} of a one-parameter set of states exhibiting all possible Schmidt numbers and a corresponding set of $k$-SWs.

\begin{exam}
Let $P_+$ be a maximally entangled state acting on a Hilbert space ${\cal H}\cong \mathbb{C}^d \otimes \mathbb{C}^d$ and the unnormalized isotropic states \cite{Rains98, HH97} defined by
\begin{align}
\label{isotropic}
\rho_\beta=\ido+\beta P_{+}, \qquad \text{with} \qquad -1\leq \beta \leq \infty.
\end{align}
The isotropic state $\rho_\beta$ has Schmidt number $n$ if and only if
$$
\frac{d((n-1)d-1)}{d-(n-1)} < \beta \leq \frac{d(nd-1)}{d-n}.
$$
The operators 
\begin{align}
\label{canwit}
W_n=\ido - d/(n-1) P_+,
\end{align}
are $n$-SW and detect in an optimal way the Schmidt number of the states $\rho_\beta$.
\label{impex}
\end{exam}

Let us for a moment assume that $W_n$ is indeed a $n$-SW\@	. Then it is not so hard to verify that when $\beta>d(nd-1)/(d-n)$, $\rho_\beta$ has Schmidt number $n+1$. The example then follows from a continuity argument, if we can show that for $\beta=d(nd-1)/(d-n)$ the state has Schmidt number $n$. This can be done by virtue of an explicit decomposition in terms of pure states \cite{TH00}. Such a decomposition can be deduced from our results in Section~\ref{smrobn}. The fact that $W_n$ is a $n$-SW follows readily from the next lemma \cite{HH97,TH00}.

\begin{lemma}
\label{fefl}
Define the fully entangled fraction of a state $\rho$ as
$$
{\cal F}(\rho)=\max_\Psi{\la\Psi|\rho|\Psi\ra},
$$
where the maximum is taken over all maximally entangled states.

(a) For a pure state $|\psi\ra$ the fully entangled fraction is given by
$$
{\cal F}(\psi)=\frac{1}{d}\Bigl[\sum_{i=1}^d \lambda_i \Bigr]^2,
$$
with $\lambda_i$ the Schmidt coefficients of $|\psi\ra$.

(b) For an arbitrary state with Schmidt number $k$ or less we have 
$$
{\cal F}(\rho)\leq \frac{k}{d},
$$
\end{lemma}
\begin{proof}
(a) See Ref.~\cite{HH97}.

(b) Suppose we have a pure state $|\psi\ra$ with Schmidt rank $k$, so the problem is to maximise
$$
\Bigl[\sum_{i=1}^k \lambda_i \Bigr]^2
$$
subject to
$$
\sum_{i=1}^k \lambda^2_i=1.
$$
We solve this using the method of Lagrange multipliers and define
$$
{\cal L}= \Bigl[\sum_{i=1}^k \lambda_i \Bigr]^2 + \mu \Bigl[\sum^k_{i=1} \lambda^2_i - 1 \Bigr].
$$
An extremum can be found by solving
$$
\frac{\partial {\cal L}}{\partial \lambda_i}=0, \qquad \qquad \frac{\partial {\cal L}}{\partial \mu}=0.
$$
The first relation leads to $\mu=-k$, and thus $\sum_i \lambda_i=k\lambda_j$, squaring this equation and carrying out summation over $j$ yields 
$$
{\cal F}(\rho)=\frac{1}{d}\Bigl[\sum_{i=1}^k \lambda_i \Bigr]^2\leq \frac{k}{d}.
$$
If the state is not pure, this inequality remains true, since a mixed state of Schmidt number $k$ can be written as a convex combination of pure states of Schmidt rank $k$.
\end{proof}

In Ref.~\cite{HBLS04} a connection was made between Schmidt witnesses and entanglement witnesses. The central idea in their construction can be deduced from the following more general lemma \cite{Clarisse05}, of which special cases have appeared in the literature over the years \cite{KLC01, DCLB99,VW02} (see also the recent \cite{PM06}). 

\begin{lemma}
\label{combinedlemma}
Let $\sigma$ be a positive operator with Schmidt number $N\geq 1$ acting on ${\cal H}_1={\cal H}_{A_1} \otimes {\cal H}_{B_1}$ and let $\eta$ be an operator acting on ${\cal H}_2={\cal H}_{A_2} \otimes {\cal H}_{B_2}$ positive on states with Schmidt number $KN$. Then the operator $\sigma \otimes \eta$ acting on ${\cal H}_{1} \otimes {\cal H}_{2}\cong {\cal H}_{A} \otimes {\cal H}_{B}$ is positive on states with Schmidt number $K$.
\end{lemma}
\begin{proof}
It is clear that it is sufficient to prove the lemma for pure states $\sigma=|\phi\ra\la\phi|$. So let 
$$
|\phi\ra=\sum_{i=1}^N \phi_i |a_i\ra_{A_1}|b_i\ra_{B_1},
$$
and take an arbitrary Schmidt rank $K$ state 
$$
|\psi\ra=\sum_{j=1}^K \psi_j |e_j\ra_{A}|f_j\ra_{B}.
$$
Then we need to prove that $\Tr(|\psi\ra\la\psi| \sigma \otimes \eta)>0$. This trace operation can be composed of tracing out the first pair, and then the second, as $\Tr(\cdot)=\Tr_2(\Tr_1(\cdot))$. Then making use of the identity $\Tr_1(C(A_1\otimes B_2))=\Tr_1(C(A_1\otimes \ido_2))B_2$ we have
$$
\Tr(|\psi\ra\la\psi| \sigma \otimes \eta)= \Tr_2( \Tr_1(|\psi\ra\la\psi| (\sigma\otimes \ido_2)) \eta)
$$
Now $\Tr_1(|\psi\ra\la\psi| (\sigma\otimes \ido_2))$ is the projector onto the pure Schmidt rank $KN$ state 
$$
|\gamma\ra_2= \la\phi|\psi\ra = \sum_{ij} \phi_i\psi_j \la a_i |e_j \ra\la b_i |f_j \ra \in {\cal H}_2
$$ 
from which the lemma follows.
\end{proof}

Let $S$ be a given $(k+1)$-SW and $P^{(k)}_+$ a maximally $k$-level entangled state, then from the above lemma follows that the operator $S'$ defined by
$$
S'=P^{(k)}_+ \otimes S, 
$$
is an entanglement witness. Thus by enlarging the Hilbert space, any SW can be mapped onto an entanglement witness; a trick which will come back in Chapter~\ref{chapnptbound}.

\section{Main theorem}
In this section we prove the main result on $k$-positive maps: the operators corresponding to $k$-positive maps via the Jamio{\l}kowski isomorphism (see Section~\ref{sejaiso}) are positive on Schmidt number $k$ states and vice verse. This is a generalisation of Theorem~\ref{jamthem}. Then we derive a characterisation of ${\cal S}_k$ in terms of positive maps, similar to Theorem~\ref{corhor2}. We start this section by giving some alternative characterisations of positive maps.

Recall that a map $\Lambda: {\cal L}_{{\cal H}_{A}} \rightarrow {\cal L}_{{\cal H}_{B}}$ is a $k$-positive map if and only if it the induced map 
$$
\Lambda_k={\openone_k} \otimes \Lambda: {\cal L}_{{\mathbb C}^{k}} \otimes {\cal L}_{{\cal H}_{A}} \rightarrow {\cal L}_{{\mathbb C}^{k}} \otimes {\cal L}_{{\cal H}_{B}}
$$
is positive for $k\geq 1$. The next lemma characterises $k$-positive maps purely on their action on pure states with Schmidt rank at most $k$ (see for instance Ref.~\cite{EK00}).
\begin{lemma}
\label{secdef}
A Hermitian map $\Lambda: {\cal L}_{{\cal H}_{A}} \rightarrow {\cal L}_{{\cal H}_{B}}$ is $k$-positive if and only if
$$
({\ido}_k \otimes \Lambda)(|\psi\ra\la\psi|)\geq 0,
$$
for all pure states $|\psi\ra\in {\mathbb C}^k \otimes {\cal H}_A$ with Schmidt rank at most $k$.
\end{lemma}
\begin{proof}
Trivial.
\end{proof}

An improvement is made by the following lemma, which states that it is sufficient to test maximally entangled vectors of rank $k$. Ref.~\cite{DSSTT00} contains a proof for $k=2$. We present here an alternative, much shorter proof for general $k$.
\begin{lemma}
\label{sufmax}
A Hermitian map $\Lambda: {\cal L}_{{\cal H}_{A}} \rightarrow {\cal L}_{{\cal H}_{B}}$ is $k$-positive if and only if 
$$
( {\ido}_k \otimes \Lambda)(|\psi\ra\la\psi|)\geq 0
$$
for all maximally entangled states $|\psi\ra\in {\mathbb C}^k \otimes {\cal H}_A$.
\end{lemma}
\begin{proof}
Let $|\psi\ra\in {\mathbb C}^k \otimes {\cal H}_A$ be an arbitrary state in Schmidt decomposition:
$$
|\psi\ra =\sum_{i=0}^{k-1}\lambda_i |e_i\ra|f_i\ra.
$$
In view of the previous lemma we need to show that
$$
D=({\ido}_k \otimes \Lambda)(|\psi\ra\la\psi|)\geq 0 \qquad \text{ if and only if } \qquad \tilde D=({\ido}_k \otimes \Lambda)(|\tilde \psi\ra\la \tilde \psi|)\geq 0,
$$
with 
$$
|\tilde \psi\ra =\sum_{i=0}^{k-1} |e_i\ra |f_i\ra.
$$
But this follows immediately from 
$$
D=(L\otimes \ido) \tilde D (L^\dagger\otimes \ido),
$$
with $L$ the diagonal matrix such that $L_{ii}=\lambda_i$.
\end{proof}

Now we are in the position to prove the following theorem by Uhlmann. The proof is from Ref.~\cite{TH00}.
\begin{theorem}
A Hermitian map $\Lambda: {\cal L}_{{\cal H}_{A}} \rightarrow {\cal L}_{{\cal H}_{B}}$ is $k$-positive if and only if
$$
\sum_{n,m=1}^k{\mu_n\mu_m \la b_n|\Lambda(|a_n\ra\la a_m|)|b_m\ra}\geq 0
$$
for all orthogonal sets of vectors $\{|a_n\ra\}_{n=1}^k$ and $\{|b_n\ra\}_{n=1}^k$ with $\sum_{n=1}^k\mu^2_n=1$ .
\end{theorem}
\begin{proof}
From the preceding lemma we have that a Hermitian map is $k$-positive if and only if 
$$
\la\phi|( {\ido}_k \otimes \Lambda)(|\psi\ra\la\psi|)|\phi\ra\geq 0,
$$
for all $\phi$ and $\psi$ with Schmidt rank $k$ and $\psi$ maximally entangled. Using the Schmidt decomposition we set $\phi=\sum_i \mu_i |c_i b_i \ra$ and $\psi=\sum_i |c_i a_i \ra$ (note that this is always possible, since $\psi$ is maximally entangled) and obtain
$$
\sum_{i,j,n,m=1}^k{\mu_n\mu_m \la c_n b_n| \ido_k\otimes \Lambda(|c_i a_i\ra\la c_j a_j|)| c_m b_m\ra}\geq 0
$$
from which the theorem follows.
\end{proof}

With this we can prove the main theorem, first published in Ref.~\cite{TH00}.
\begin{theorem}
\label{lemk1}
A linear map $\Lambda_W: {\cal L}_{{\cal H}_A} \rightarrow {\cal L}_{{\cal H}_B}$ is $k$-positive if and only if the corresponding operator $W$ acting on ${\cal H}_A \otimes {\cal H}_B$ is positive on states $\rho$ acting on ${\cal H}_A \otimes {\cal H}_B$ with Schmidt number $k$ or less (belonging to ${\cal S}_k$).
\end{theorem}
\begin{proof}
Let 
$$
W=(\ido\otimes \Lambda_W)(P_+),
$$
we have to show that $k$-positivity is equivalent to
\begin{align}
\label{shorttrace}
\Tr{W\rho}\geq 0,
\end{align}
for all $\rho\in {\cal S}_k$. It is sufficient to take $\rho$ as a pure state of Schmidt rank $k$, that is,
$$
\rho=\sum_{n,m=1}^k \mu_m \mu_n |a_m b_m\ra \la a_n b_n |.
$$
With this (\ref{shorttrace}) can be rewritten as
$$
\sum_{n,m=1}^k \mu_m \mu_n \la a_n b_n| W |a_m b_m\ra \geq 0,
$$
or
\begin{align}
\label{empold}
\sum_{i,j,n,m=1} \mu_m \mu_n \la a_n|i\ra \la a_m|j\ra \la b_n|\Lambda(|i\ra\la j|)|b_m\ra \geq 0.
\end{align}
Due to the linearity of $\Lambda$ we have that $\Lambda(|a_n\ra\la a_m|)=\sum_{ij}\la i|a_n\ra \la a_m|j\ra \Lambda(|i\ra\la j|)$, or
$$
\Lambda'(|a_n\ra\la a_m|)=\sum_{ij}\la a_n |i\ra \la a_m|j\ra \Lambda(|i\ra\la j|),
$$
with $\Lambda'=\Lambda \circ R$, where $R$ maps the vectors $|a_n\ra$ to $|a^*_n\ra$ where complex conjugation is performed with respect to the $\{|i\ra\}$ basis. Therefore (\ref{empold}) is equivalent to
$$
\sum_{n,m=1}^k{\mu_n\mu_m \la b_n|\Lambda'(|a_n\ra\la a_m|)|b_m\ra}\geq 0,
$$
and it follows that $\Lambda'$ is $k$-positive if and only if $\Tr{W\rho}\geq 0$ for all $\rho \in {\cal S}_k$.
Now the map $R$ corresponds to a unitary rotation which maps the basis $\{|a_n\ra\}$ to the basis $\{|a^*_n\ra\}$. Therefore $\Lambda$ is $k$-positive if and only if $\Lambda'$ is.
\end{proof}

Thus in the language of Schmidt witnesses: A map is $k$-positive if and only the corresponding operator is a $k+1$ Schmidt witness. As an example consider the $k+1$-SW (\ref{canwit})
$$
W_{k+1}=\ido - \frac{d}{k} P_+,
$$
the corresponding map is given by $\Lambda_k(X)=\Tr(X) \ido - kX$. The positivity properties of this map have been derived independently by Tomiyama \cite{Tomiyama85} (see also Ref.~\cite{TT83}), but from our correspondence they follow directly. Note that for $k=1$ we obtain just the reduction criterion.
\begin{theorem}
Let $\Lambda_p$ be a linear map defined by
\begin{align}
\Lambda_p(X)=\Tr(X) \ido - pX,
\end{align}
be a linear map. Then $\Lambda_p$ is $k$-positive for
\begin{align}
\frac{1}{k+1}\leq p \leq \frac{1}{k},
\end{align}
but not $k+1$-positive.
\end{theorem}

Next we give a characterisation of ${\cal S}_k$ in terms of positive maps, similar to Theorem~\ref{corhor2}.
For the proof one needs the following lemma.
\begin{lemma}
\label{lemk2}
A linear map $\Lambda: {\cal L}_{{\cal H}_{A}} \rightarrow {\cal L}_{{\cal H}_{B}}$ is $k$-positive if and only if $\Lambda^\dagger: {\cal L}_{{\cal H}_{B}} \rightarrow {\cal L}_{{\cal H}_{A}}$ is $k$-positive.
\end{lemma}
\begin{proof}
Suppose $\Lambda$ is $k$-positive, then we have
$$
\la\psi|(\ido_k\otimes\Lambda)(|\phi\ra\la\phi|)|\psi\ra\geq 0,
$$
where $|\phi\ra\in {\cal H}_{k}\otimes{\cal H}_{A}$ and $|\psi\ra \in {\cal H}_{k}\otimes {\cal H}_{B}$ are arbitrary and both have Schmidt rank $k$. Applying the definition of an adjoint map yields
$$
\la\phi|(\ido_k\otimes\Lambda^\dagger)(|\psi\ra\la\psi|)|\phi\ra\geq 0
$$
for arbitrary $|\phi\ra$ and $|\psi\ra$ of Schmidt rank $k$.
\end{proof}

\begin{corollary}
A density operator $\rho$ acting on ${\cal H}_{A}\otimes {\cal H}_{B}$ belongs to ${\cal S}_k$ if and only if
$$
({\openone} \otimes \Lambda)(\rho) \geq 0,
$$
for all $k$-positive maps $\Lambda: {\cal L}_{{\cal H}_{B}} \rightarrow {\cal L}_{{\cal H}_{A}}$.
\end{corollary}
\begin{proof}
This follows from Theorem~\ref{lemk1} and Corollary~\ref{corschmidt} in a completely analogous way as we derived Theorem~\ref{corhor2}.
\end{proof}

\section{A conjecture}
\label{contwosdt}
The detection of PPT entangled states is a difficult problem and this suggests that PPT entangled states have generically a low degree of entanglement. In particular we expect PPT entangled states to have a relative low Schmidt number. In Ref.~\cite{SBL00,EK02} evidence was given for the following conjecture.

\begin{conjecture}
All entangled PPT states on ${\cal H}= \mathbb{C}^3 \otimes \mathbb{C}^3$ have Schmidt number two.
\end{conjecture}

It is easy to see that this conjecture is equivalent to the statement that all $2$-positive maps on ${\cal L}_{\mathbb{C}^3}$ are decomposable. A first piece of evidence is that the conjecture holds for PPT entangled states of rank 4 \cite{SBL00}. 

\begin{lemma}
All entangled PPT states on ${\cal H}= \mathbb{C}^3 \otimes \mathbb{C}^3$ of rank four have Schmidt number two.
\end{lemma}
\begin{proof}
Obviously it is sufficient to show this for edge states. So let $\rho$ be a rank four edge state. We will show explicitly how to write $\rho$ as a convex sum of Schmidt rank two vectors, by iterative subtracting of such vectors (in this context, see also Ref.~\cite{LS97}).

 Since in $\mathbb{C}^3 \otimes \mathbb{C}^3$ there exists a UPB of 5 product states there must exist a product vector $|ef_1\ra$ in the kernel of $\rho$. Since $\Tr(\rho |ef_1\ra\la ef_1|)=\Tr(\rho^{T_B} |ef^*_1\ra\la ef^*_1|)=0$ and $\rho^{T_B}\geq 0$ we have that $|ef^*_1\ra$ belongs to the kernel of $\rho^{T_B}$. 

Constructing an orthogonal basis $\{|f_i\ra\}$ we have that 
$
\la f^*_i |\rho^{T_B} |ef^*_1\ra=0=\la f_1 |\rho |ef_i\ra,
$
so that for instance for $i=2$, 
$$
|\psi_1\ra = \rho |ef_2\ra= |c f_2 \ra+|d f_3\ra.
$$
Since this vector has at most Schmidt rank two and belongs to the range of $\rho$, we can subtract it and obtain a new state 
$$
\tilde \rho=\rho - \Lambda |\psi_1 \ra \la \psi_1|,
$$
with 
$$
\frac{1}{\Lambda}= \la \psi_1| \rho^{-1} |\psi_1 \ra=\la\psi_1|ef_2\ra.
$$
The state $\tilde \rho$ has rank three and it is easy to verify that $\tilde \rho |ef_1\ra=\tilde \rho |ef_2\ra=0$ but
$$
|\psi_2\ra = \tilde \rho |ef_3\ra= |\tilde c f_2 \ra+|\tilde d f_3\ra.
$$
This vector has again at most a Schmidt rank two vector, and we can again subtract it to obtain
$$
\tilde {\tilde \rho}=\tilde \rho - \tilde \Lambda |\psi_2 \ra \la \psi_2|,
$$
with 
$$
\frac{1}{\tilde \Lambda}= \la \psi_2| {\tilde \rho}^{-1} |\psi_2 \ra=\la\psi_2|ef_3\ra.
$$
The state $\tilde {\tilde \rho}$ has rank two and $\tilde {\tilde \rho} |ef_1\ra= \tilde {\tilde \rho} |ef_2\ra=\tilde {\tilde \rho} |ef_3\ra=0$.
From this follows that we can write $\rho$ as
$$
\rho=\tilde {\tilde \rho}+ \tilde \Lambda |\psi_2\ra\la\psi_2|+ \Lambda |\psi_1\ra\la\psi_1|.
$$
Since $\tilde {\tilde \rho}$ effectively acts on $\mathbb{C}^2 \otimes \mathbb{C}^3$ it has Schmidt number two and therefore $\rho$ has also Schmidt number two.
\end{proof}

In order to check the conjecture for arbitrary rank, first note that again it is sufficient to check it for edge states. Let ${\cal W}$ denote the set of all entanglement witnesses detecting these edge states. Then the conjecture is equivalent to the statement that for all $W\in \cal W$ on $\mathbb{C}^3 \otimes \mathbb{C}^3$ there exists a Schmidt rank two vector $|\psi\ra$ such that $\la \psi |W | \psi \ra<0$. Now in $\mathbb{C}^3 \otimes \mathbb{C}^3$ the sum of the ranks of $\rho$ and $\rho^{T_B}$ are not larger than 13 for edge states \cite{KCKL00,HLVC00}. For states with such a relative low rank Sanpera et~al.\ \cite{SBL00} gave an argument that such a $|\psi\ra$ always exists (for a generalisation to higher dimensions, see Ref.~\cite{HCL01}). Their argument is based on counting the degrees of freedom versus the number of constraints in constructing such a $|\psi\ra$. Their argument is especially weak for ranks $(6,6)$ and $(8,5)$ since here only a finite number of solutions can be found, for which it is by no means guaranteed they will satisfy the constraints. When the conjecture was first published only PPT edge states of rank $(4,4)$ and $(6,7)$ were known. In view of the conjecture, there was special interest in constructing PPT edge states with ranks $(6,6)$ and $(8,5)$. In a recent paper Ha and Kye \cite{HK05} managed to find edge PPTES for all ranks except $(5,5)$ and $(6,6)$. It is for this reason we have specially constructed such states in Section~\ref{PPTESS}.

For these states, we can construct witnesses in a number of different ways. First note that both states violate the realignment criterion, and thus using Theorem~\ref{racew} we obtain $W_1$. Secondly, there is the canonical construction for edge states of Theorem~\ref{edgeindec}, which gives us $W_2$. Thirdly, one can use the algorithm \cite{DPS03} from Section~\ref{numap} yielding $W_3$. Numerically we calculated all three witnesses $W_1, W_2$ and $W_3$ for $\rho_{(5,5)}$ and $\rho_{(6,6)}$. After normalisation we found that all witnesses are different and yield a different value on the states. In total this gives us six different non-decomposable witnesses. We have checked that these witnesses are negative on some Schmidt rank two state. This provides additional evidence in favour of the conjecture. In fact, we also verified this for the witnesses 
$$
\tilde W_i=W_i - (\Tr(W_i \rho) +\epsilon)\openone,
$$
 for arbitrarily small $\epsilon>0$. These witnesses $\tilde W_i$ barely detect our states and are therefore the best candidates to check the conjecture. Note, however, that this is only evidence that the states have Schmidt number two. A full proof would consist of an explicit decomposition in Schmidt rank two states. However, finding such a decomposition is in general a very hard problem and we did not attempt this.

\section{Schmidt robustness}
\label{smrobn}
In this section we generalise the notion of robustness of entanglement (see also Section~\ref{mesent}) to the robustness of Schmidt number, which we analyse for pure states. We start by recalling the definition of robustness of entanglement \cite{VT99, Steiner03, HN03}.

The $\cal K$-\emph{robustness} of a state $\rho$, $R_k(\rho)\geq 0$ is the minimal value of $R$ such that 
$$
\frac{1}{1+R}(\rho+R\rho_k)
$$
is separable, for some state $\rho_k \in \cal K$. Thus robustness measures measure how much mixing is required before a state becomes separable. With this basic definition we have
\begin{description}
\item[The robustness $R_s(\rho)$] Here $\cal K=\cal S$, the set of separable states.
\item[The random robustness $R_r(\rho)$] Here $\cal K=\{\ido\}$, the maximally mixed state)\footnote{For simplicity we have chosen not to normalise $\ido$ in the definition, in contrast to the original definition of random robustness in Ref.~\cite{VT99}.}.
\item[The generalised robustness $R_g(\rho)$] Here $\cal K=\cal D$, the set of all states.
\end{description}
From the definitions it is clear that $R_g \leq R_s \leq R_r$. Both the robustness and the generalised robustness are entanglement monotones (see Section~\ref{mesent}). It is easy to see from the definition that the random robustness is proportional to the so-called witnessed entanglement \cite{BV04b}, defined as $\max_{W\in M}[0, - \Tr (\rho W )]$ where $M$ is the set of all normalised entanglement witnesses. The following theorem gives exact values for the robustness and the random robustness for pure states.

\begin{theorem}[\cite{VT99, Steiner03, HN03}]
\label{rrth}
Let $|\psi\ra=\sum_i a_i |ii\ra$ be a pure bipartite state with ordered Schmidt coefficients $a_i\geq a_{i+1}$. 
The robustness of $\psi$ is given by 
$$
R_s(\psi)=\sum_{i\neq j}a_i a_j =(\sum_i a_i)^2-1=R_g(\psi)
$$
 and equals its generalised robustness.

A separable state that washes out the entanglement in $\psi$ most quickly is given by 
$$
\rho_s=\rho_g=\frac{1}{R_s}\sum_{i \neq j} a_i a_j |ij \ra\la ij|.
$$
The random robustness of $\psi$ is given by $R_r(\psi)=a_1a_2$.
\end{theorem}
An alternative proof of this theorem can be obtained as a corollary of our results (see Section~\ref{gsr}). In what follows, we assume that all states act on a Hilbert space ${\cal H}\cong \mathbb{C}^d \otimes \mathbb{C}^d$. Throughout we also use local operations and classical communication (LOCC, see Chapter~\ref{entdis}), as they do not increase the Schmidt number.

\subsection{Generalised Schmidt robustness}
\label{gsr}
In this section we will extend the notion of generalised robustness to generalised Schmidt robustness.
Analogously to the generalised robustness, the generalised Schmidt-$n$ robustness of a state $\rho$, $R_{gn}(\rho)$ is defined as the minimal value of $R$ such that 
$$
\frac{1}{1+R}(\rho+R\rho_{gn})
$$
has Schmidt number smaller or equal than $n$, for some $\rho_{gn}$.

Let us now analyse the generalised Schmidt robustness for pure states. 
In Theorem~\ref{rrth} we have seen that for a pure state $|\psi\ra=\sum_i a_i |ii\ra$ the state
\begin{equation}
\label{rhogcor}
\rho_g=\frac{1}{R_g}\sum_{i \neq j} a_i a_j |ij \ra\la ij|,
\end{equation}
erases most quickly the entanglement present in $|\psi\ra$. It is plausible that this same state will also erase Schmidt number in an optimal way.  We were able to prove this for the maximally entangled state. For the maximally entangled state $|\psi_+\ra=\frac{1}{\sqrt{d}} \sum_i |ii\ra$, $\rho_g$ given by \ref{rhogcor} reduces to  $\rho_g=\frac{1}{d^2-d} (\ido-Z)$, with $Z=\sum_i|ii\ra\la ii|$.

\begin{theorem}[Generalised Schmidt robustness of the maximally entangled state]
\label{gsrmes}
The state defined by 
\begin{align}
\rho(\beta)=\frac{\beta \rho_g + P_+}{1+\beta} \quad \text{where} \quad \rho_g=\frac{\ido-Z}{d^2-d}
\label{tooproof}
\end{align}
has Schmidt number $n$ for 
$$
\frac{d-n}{n} \leq \beta < \frac{d-n+1}{n-1}. 
$$
The generalised Schmidt-$n$ robustness of the maximally entangled state $P_+$ is given by 
$$
R_{gn}(P_+)=\frac{d-n}{n}.
$$
\end{theorem}
\begin{proof}
Let $S(\beta)$ be the Schmidt number of $\rho(\beta)$ and let $\beta_n=\frac{d-n}{n}$. We first show that $S(\beta_n)\leq n$.

(i) We will give an explicit decomposition of the state $\rho(\beta_n)$ in terms of Schmidt rank $n$ states. Equivalently, we show how one can construct this state locally with the aid of Schmidt rank $n$ states. In what follows we will often omit normalisation. Let us take a maximally entangled $n$-level state 
$$
|\psi_S\ra=\frac{1}{n}\sum_{i\in S}^n|ii\ra,
$$
where $S=\{i_1,\ldots,i_n \}$ is a subset of $\{1,\ldots,d\}$. We can construct such a state in $\binom{d}{n}$ possible ways, and clearly all these states have Schmidt number $n$. Now let us mix with equal weight the corresponding states $|\psi_S\ra\la\psi_S|$. Then for every $i$ and $j$ we will have
$\binom{d-1}{n-1}$ terms of the form $|ii\ra \la ii|$ and $\binom{d-2}{n-2}$ terms of the form $|ii\ra \la jj|$, $i\neq j$. Thus the resulting state will be proportional to 
$$
(d-1)Z+(n-1)\sum_{i\neq j} |ii\ra \la jj|.
$$
Therefore we have proven that the state (see Ref.~\cite{Clarisse04} for n=2)
$$
K=Z +\frac{d(n-1)}{d-n}P_+
$$
has Schmidt number at most $n$. It turns out that we can transform this state in the form (\ref{tooproof}) by applying a certain partial twirl operation. Consider the following twirl operation \cite{HH97,Rains98,VW01}
$$
\int dU (U\otimes U^*) \rho (U\otimes U^*)^\dagger,
$$
which maps any state $\rho$ into one of the form $\ido + \alpha P_+$ (an isotropic state). Here $dU$ is the uniform probability distribution on the unitary group $U(d)$. Note that the twirl can be implemented locally, both parties need to implement only one random unitary. Remarkably, it has been shown \cite{DCLB99}\footnote{In Ref.~\cite{DCLB99} the finite decomposition of the $U\otimes U$ twirl was given, but a similar decomposition can be deduced from it for the $U\otimes U^*$ twirl.} that the integral can be written as a finite sum (for qubits this was first shown in Ref.~\cite{BDSW96}). This will allow us to perform a partial twirl, just by considering a part of this finite sum. 

In the first step we apply the unitary transformation $(T\otimes T^*) K (T\otimes T^*)^\dagger$, with
$$
T=\frac{1}{\sqrt{d}} \sum_{j,k=0}^{d-1} e^{\frac{i 2\pi jk}{d}}|j\ra\la k|,
$$
which is just the quantum Fourier transform. Since $T$ is unitary, it acts as the identity on $P_+$ \cite{HH97}, while it acts on $Z$ as
$$
(T\otimes T^*) Z (T\otimes T^*)^\dagger = \frac{1}{d^2}\sum_a \sum_{j,k,s,t} e^{\frac{i 2\pi (j-t+k-s)a}{d}}|j\ra\la t| \otimes |s\ra\la k|.
$$
The terms of the form $|jj\ra\la kk|$ for $s=j$ and $t=k$ will give a contribution $P_+$, while the $|ij\ra\la ij|$ for $i\neq j$ (for $j=t$ and $k=s$) will give a contribution of $(\ido-Z)/d$, so that 
$$
K'=d (T\otimes T^*) K (T\otimes T^*)^\dagger= \ido -Z + \frac{d(d-1)n}{d-n}P_+ + L,
$$
where $L$ are terms not of the form $|ii\ra\la jj|$ or $|ij\ra\la ij|$. Now these contributions can be easily removed by repeated application of the following mixing procedure 
$$
K''=\frac{1}{2} (U\otimes U^*) K (U\otimes U^*)^\dagger + \frac{1}{2}K'.
$$
First $U$ is chosen to act as $U|k\ra=e^{i\pi\delta_{kl}}|k\ra$ for every $l=0,\ldots ,d-1$. This defines $d$ mixing procedures. Next $U$ is taken to act as $U|k\ra=e^{i\pi\delta_{kl}/2|k\ra}$ (another $d$ mixing procedures). One can readily check \cite{DCLB99} that these operations do not affect terms 
$|ii\ra\la jj|$ or $|ij\ra\la ij|$ but cancel out $L$ completely.

 Thus $S(\beta_n) \leq n$. Now for $\beta_n\leq \beta < \beta_{n-1}$, the state $\rho(\beta)$ is a convex combination of $\rho(\beta_n)$ and $\rho(\beta_{n-1})$ and therefore $S(\beta) \leq n$.

(ii) For the second part, we generalise the trick introduced in Ref.~\cite{HN03}. For any state $\sigma$, suppose $t$ is a positive number such that $P_+ +t \sigma$ has Schmidt number $n$. The operators $W_n=\ido - d/n P_+$ witness Schmidt number $n+1$ (see Example~\ref{impex}), so that we have
\begin{align*}
0 & \leq \Tr{[ (\ido-d/nP_+)(P_+ + t \sigma)]} \\
 & = 1+t-d/n\Tr{[P_+]}-d/n\Tr{[P_+ \sigma]} \\
 & \leq - \frac{d-n}{n} + t.
\end{align*}
since $\Tr{P_+ \sigma}\geq 0$. Thus $t\geq \beta_n$, and for $\sigma=\rho_g$ it follows that $S(\beta)\geq n$ for $\beta_n\leq \beta < \beta_{n-1}$.

For general $\sigma$, it follows that $R_{gn}\geq \beta_n$, but $\rho(\beta_n)=P_+ + \beta_n \rho_g$ has Schmidt number $n$, so that $R_{gn}(P_+)=\beta_n$.
\end{proof}

Note that the states (\ref{tooproof}) constitute one of the very few examples of non-trivial one parameter states for which the Schmidt number is known. To our knowledge, the only other example is that of the isotropic states, Example~\ref{impex}. This theorem allows us to present non-trivial bounds to the generalised Schmidt robustness of arbitrary pure states.\vspace{0.001cm}

\begin{corollary}[Lower and upper bounds for the generalised Schmidt robustness]
\label{corgrob}
The generalised Schmidt-$n$ robustness $R_{gn}$ of a pure state $|\psi\ra=\sum_i a_i |ii\ra$ satisfies
$$
\frac{1}{n}(\sum_i a_i)^2-1 \leq R_{gn} \leq R_g \frac{d-n}{(d-1)n},
$$
with $R_g=\sum_{i\neq j}a_i a_j =(\sum_i a_i)^2-1$ the generalised robustness of $\psi$.
\end{corollary}

\begin{proof}
In Theorem~\ref{gsrmes} we have seen that $\rho=\frac{d-n}{n} \frac{1}{d^2-d}(\ido-Z) + P_+$ has Schmidt number $n$. Performing the filtering operation $(A\otimes A) \rho (A\otimes A)^\dagger$ we hence obtain a state with Schmidt number at most $n$ (this is because local filtering cannot increase the Schmidt number of a state \cite{TH00}). If we take $A=\sum_k \sqrt{a_k} |k\ra \la k|$ we obtain that
$$
|\psi \ra\la \psi| + R_g \frac{d-n}{(d-1)n}\rho_g
$$
has Schmidt number $n$. Here $\rho_g=\frac{1}{R_g}\sum_{i \neq j} a_i a_j |ij \ra\la ij|$ as before. This gives the upper bound. The lower bound can be readily proven using exactly the same trick as in part (ii) of Theorem~\ref{gsrmes}.
\end{proof}
The lower and upper bound only coincide when $\psi$ is the maximally entangled state or when $n=1$. Note that the lower bound can be negative. The upper bound depends on the dimension of the Hilbert space in which the state is embedded, and hence will in general not match the value of the generalised Schmidt robustness.

\subsection{Random Schmidt robustness}
\label{rsrsct}
We define the random Schmidt-$n$ robustness of a state $\rho$, $R_{rn}(\rho)$ as the minimum value of $R$ such that
$$
\frac{1}{1+R}(\rho+R\ido)
$$
has Schmidt number $n$.

\subsubsection{Upper bounds}
As we have seen from Example~\ref{impex} for $\rho=P_+$ we have $R_{rn}=(d-n)/[d(nd-1)]$. For general pure states a (weak) upper bound to the random Schmidt robustness can be obtained as follows. We know that
$$
\Gamma_n=(d-n)\ido +(nd-1)dP_+
$$
has Schmidt number $n$. Local filtering $(A\otimes A) \Gamma_n (A\otimes A)^\dagger$ cannot increase the Schmidt number. So that with $A=\sum_k \sqrt{a_k} |k\ra \la k|$ we obtain
$$
\Gamma'_n=(d-n)\rho_A \otimes \rho_A +(nd-1)|\psi\ra\la\psi|,
$$
with $\rho_A$ the reduced density operator of $|\psi\ra\la\psi|$, and because $\ido-\rho_A\otimes \rho_A$ is a separable state we get as an upper bound for the random Schmidt robustness $R_{rn}(\psi) \leq (d-n)/(nd-1)$. The following theorem presents a non-trivial upper bound.

\begin{theorem}[Upper bound to the random Schmidt robustness]
The random Schmidt-$n$ robustness $R_{rn}$ of a pure state $|\psi\ra=\sum_i a_i |ii\ra$ satisfies
$$
R_{rn} \leq \frac{R_r(d-n)}{dn-1},
$$
with $R_r=a_1a_2$ the random robustness of $\psi$.
\end{theorem}
\begin{proof}
In this proof, we work again with unnormalised states. Note that, if we add two unnormalised Schmidt number $n$ states together we end up with a state of at most Schmidt number $n$. Equivalently, this can be shown by mixing the normalised states with different weights.

Analogously to the construction in Theorem~\ref{gsrmes}, let us take a maximally entangled $n$-level state 
$$
|\psi\ra_{S}=\frac{1}{n}\sum_{i\in S}^n a_i |ii\ra,
$$
where $S=\{i_1,\ldots,i_n \}$ is a subset of $\{1,\ldots,d\}$. Again, we can construct such a state in $\binom{d}{n}$ possible ways, and mixing the corresponding (unnormalised) states $|\psi_S\ra\la\psi_S|$ together, we end up with a state proportional to
$$
\sum_i a_i^2|ii\ra \la ii| +\frac{(n-1)}{d-n} |\psi\ra\la\psi|,
$$
which has at most Schmidt number $n$. Now, in Corollary~\ref{corgrob} we have seen that the state
$$
\frac{(d-1)n}{d-n} |\psi \ra\la \psi| + \sum_{i \neq j} a_i a_j |ij \ra\la ij|
$$
has Schmidt number at most $n$. Adding these two states together we find that 
$$
\frac{dn-1}{d-n} |\psi \ra\la \psi| + \sum_{i, j} a_i a_j |ij \ra\la ij|
$$
has Schmidt number no more than $n$. Mixing this state with the separable state $\sum_{i,j}(a_1a_2-a_ja_j)|ij\ra\la ij|$, we obtain finally 
$$
\frac{dn-1}{(d-n)a_1a_2} |\psi \ra\la \psi| + \ido.
$$
\end{proof}

\begin{conjecture}[Schmidt balls around the identity]
\label{mainconject}
Consider the unnormalized mixture
\begin{align}
\rho_\beta=\ido+\beta \rho, \qquad \text{with} \qquad -1\leq \beta \leq \infty,
\label{rbbb}
\end{align}
with $\rho$ an arbitrary normalized state acting on $\mathbb{C}^d \otimes \mathbb{C}^d$. The state $\rho_\beta$ has Schmidt number at most $k$ for \emph{all} $d$ and $\rho$ when
\begin{align}
\beta \leq 2(2k^2-1).
\label{vvalue}
\end{align}
\end{conjecture}

Note that it is a generalisation of the fact that $\ido +2\rho$ is separable for all normalised states $\rho$ \cite{VT99}. The conjectured value (\ref{vvalue}) can be obtained as follows. Starting from the identity, it is natural to assume that we can go fastest to a higher Schmidt number by mixing with some maximally entangled state. Taking $\rho=P^{d'}_+$ (the maximally entangled $d'$-level state) we have that $\beta\leq d'(kd'-1)/(d'-k)$, which reaches its maximum for $d'=\sqrt{k^2-1}+k$ or for integer $d'=2k$. Substituting this expression in the expression for $\beta$ gives the upper bound from the conjecture. Thus it looks like, starting from the identity, we can get most quickly to a Schmidt number $k+1$ state by mixing with the maximally entangled state in $2k \otimes 2k$. Upper bounds on the random Schmidt number two robustness are particularly useful as a distillability criterion can be deduced from it, which we will discuss in Section~\ref{utodt}.

\subsubsection{Lower bounds}
Lower bounds on the random Schmidt robustness of a pure state $\psi$ can be obtained from any Schmidt witness that detects $\psi$. Indeed, suppose $W_{n+1}$ is a normalised Schmidt number $n+1$ witness such that $\Tr(|\psi\ra\la\psi| W_{n+1})=-\alpha$, with $\alpha>0$. Then it is easy to see that 
$R_{rn}(\psi) \geq \frac{\alpha}{d^2}$. It follows that
$$
d^2R_{rn}(\psi)= -\min_{\Tr W_{n+1}=1} \Tr(|\psi\ra\la\psi| W_{n+1}).
$$
The question is now to determine this class of optimal Schmidt witnesses. We have conducted a large numerical test and have found a class which we conjecture to be optimal. The construction is the following:

Consider the Schmidt witnesses $W_{n+1}=\ido - d/n P_+$ from Example~\ref{impex}. Performing the filtering operation $(A\otimes B) W_{n+1} (A\otimes B)^\dagger$ we obtain again a Schmidt number $n+1$ witness. Let us consider the particular case where $A=B=\sum_k \sqrt{a_k} |k\ra \la k|$ diagonal and such that $\sum_i a_i^2=1$; we obtain
\begin{align}
\label{clclcl}
\tilde W_{n+1}=\frac{n\sum_{ij} a_i a_j |ij \rangle \langle ij | - \sum_{ij} a_i a_j |ii\ra\la jj|}{n(\sum_i a_i)^2-1}.
\end{align}
For a given pure state $|\psi\rangle=\sum_i b_i|ii\ra$ we therefore conjecture:
\begin{conjecture}
The random Schmidt robustness $R_{rn}$ of a pure state $|\psi\ra=\sum_i b_i |ii\ra$ is given by
\begin{align}
\label{dddd}
d^2 R_{rn}(\psi)= -\min_{\Tr \tilde W_{n+1}=1} \Tr(|\psi\ra\la\psi| \tilde W_{n+1})=-\min_{a_i} \frac{n \sum_i a_i^2 b_i^2-(\sum_i a_i b_i)^2}{n(\sum_i a_i)^2-1},
\end{align}
with $\sum_i a_i^2=1$.
\label{swcc}
\end{conjecture}
One can solve the minimisation problem with Lagrange multipliers \cite{Tony06}. The solution is given by
$$
\min_{\Tr \tilde W_{n+1}=1} \Tr(|\psi\ra\la\psi| \tilde W_{n+1})= -\lambda_{\text{min}},
$$
where $\lambda_{\text{min}}$ is the minimum eigenvalue of
$$
\left(\ido-\frac{n J }{nd-1}\right)(nD-B),
$$
with $J$ the all-ones matrix, $D$ the diagonal matrix with $D_{ii}=b^2_i $ and $B$ the matrix with coefficients $B_{ij}=b_i b_j$.

\begin{figure}
\begin{center}
\psfrag{R}{$d^2R_{r2}$}
\psfrag{a}{$a_1^2$}
\includegraphics[height=8.5cm]{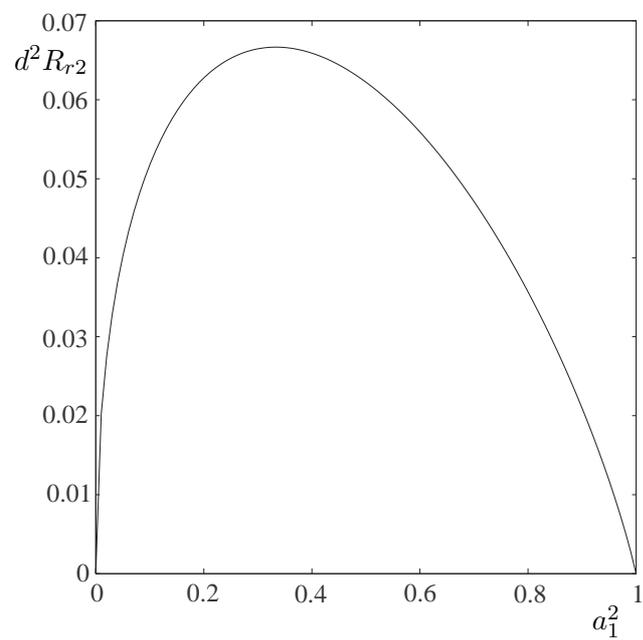}
\caption[Lower bounds on $R_{r2}(\psi)$]{Lower bounds on $R_{r2}(\psi)$ as a function of $a^2_1$, where $|\psi\ra = a_1|00\ra + a_2(|11\ra+|22\ra)$. The graph was obtained using the witnesses from Conjecture~\ref{swcc}.}
\label{figsw}
\end{center}
\end{figure}

A first step in proving this conjecture would be to show that the class of witnesses $(A\otimes B) W_{n+1} (A\otimes B)^\dagger$ with $A,B$ arbitrary matrices, is no more powerful than the class of witnesses $(A\otimes A) W_{n+1} (A\otimes A)^\dagger$ , with $A$ diagonal. We have numerically verified this for $n=2,3$ and $d=3,4$. In Figure~\ref{figsw} we have illustrated the minimization of (\ref{dddd}) for a particular set of pure states (see caption).

In conclusion, we have presented strong upper and lower bounds for the generalised and random Schmidt robustness for pure states. The problem of finding exact values is very hard, as in the end, one has to come up with an explicit convex decomposition in Schmidt rank $n$ states on the one hand, and on the other with a construction of optimal Schmidt witnesses. We hope that our results may stimulate further work, especially in proving or disproving Conjecture~\ref{mainconject}.

\chapter{Entanglement distillation: Introduction and classical results}
\label{entdis}
Distillation is the process of converting with some finite probability a large number of mixed entangled states into a smaller number of maximally entangled pure states. It relies on local collective manipulation of the systems together with classical communication between the parties (LOCC). It has become apparent that characterizing and quantifying distillation is important in understanding the nature of entanglement from a physical point of view. Considerable effort has been devoted to the characterisation of distillable states, but even for bipartite systems the matter is not settled. In the next two chapters we analyse entanglement distillation in the bipartite scenario in great detail. 

This chapter is meant as an introduction, and most of the material presented can be considered as literature review. In the first two chapters entanglement was studied as static material, however entanglement distillation involves manipulation of quantum states, so that we start this chapter with a study of quantum operations (Section~\ref{qop}). In the following section, this is then applied to the study of possible transformations between pure states both in the finite and asymptotic regime, with particular emphasis on entanglement concentration. In Section~\ref{mesent} we review the most important measures of entanglement and some of their properties. These divide natural into two classes, the geometrical entanglement measures, such as robustness measures already encountered in the previous chapter, and operational measures. An example of the latter is the entanglement of distillation, which will be the main focus of our attention. We also show that in general, entanglement is not conserved in the asymptotic regime for mixed states. We discuss two examples of such irreversibility: the maximally correlated states and the bound entangled states. We end this chapter with a sketch of the most important distillation protocols for mixed states in the asymptotic regime. 

\section{Quantum operations}
\label{qop}
\subsection{General quantum operations}
A quantum operation $\Lambda$ is a physical operation which transforms a state $\rho$ on ${\cal H}_1$ into another state $\rho'$ on ${\cal H}_2$. The aim of this section is to give a unified mathematical representation (the so-called Kraus formalism \cite{Kraus83, NCSB97, DHR01b}) of any quantum operation, independent of how this operation is to be implemented physically. Basically, there are two approaches to this \cite{NC00, Alb01}. The first is the axiomatic one: we impose on the class of maps $\Lambda: {\cal L}({\cal H}_1) \rightarrow {\cal L}({\cal H}_2)$ some physically motivated constraints and define the remaining operations as quantum operations. The second approach is constructive and the set of quantum operations is defined as the one which can be obtained by combining operations from a certain set of elementary operations. Both approaches turn out to be equivalent; we will sketch the proof of the equivalence here.

The standard quantum mechanical formalism allows us to carry out four elementary operations on quantum states: 
\begin{description}
\item[(O1)] Unitary transformations: $\rho \rightarrow U\rho U^\dagger$, with $U$ a unitary operator.
\item[(O2)] Adding an uncorrelated ancilla: $\rho \rightarrow \rho \otimes \sigma$, with $\sigma$ a density operator.
\item[(O3)] Tracing out part of the system: $\rho \rightarrow \Tr_{{\cal H}'}(\rho) $, with ${\cal H}_1={\cal H}' \otimes {\cal H}''$.
\item[(O4)] Projective measurements with possibly postselecting: $\rho \rightarrow \sum_{i=1}^k P_i\rho P_i$, with $P_i$, pairwise orthogonal projectors such that $\sum_{i=1}^n P_i=\ido$ with $k\leq n$.
\end{description}

In general a quantum operation will only be successful with a certain probability $p=\Tr(\Lambda(\rho))$, so that $\rho'=\Lambda(\rho)/p$, and $\Lambda$ does not increase the trace. It is clear that the operations (O1)-(O3) can be represented as trace preserving completely positive (CP) maps. As the composition of CP maps is CP, we see that any quantum operation composed of operations (O1)-(O3) can be represented by a trace preserving completely positive map, these operations will also be called non-measuring operations. The reverse turns out to be true: any trace preserving completely positive map can be implemented by means of the quantum operations (O1)-(O3). This result follows from the Stinespring dilation theorem \cite{Stinespring55, Alb01, Keyl02, DHR01b}. Now we invoke the Choi representation of CP maps (in this context also called the Kraus representation), which says that any CP map $\Lambda$ can be written as $\Lambda(\rho)=\sum_i L_i \rho L_i^\dagger$ for all $\rho \in {\cal L}({\cal H}_1)$ (see Theorem~\ref{choipillis}). The operators $L_i$ are commonly called the Kraus operators. For a trace preserving CP we have that $\sum_i L^\dagger_i L_i=\ido$. If we allow operations of the type (O4), we have to relax this constraint to $\sum_i L^\dagger_i L_i\leq \ido$ and renormalise the density operator after application of $\Lambda$. Another way is to keep the constraint $\sum_i L^\dagger_i L_i= \ido$ but allow postselecting on the particular outcomes of $i$. Thus we obtain (see for instance Ref.~\cite{DHR01b}):
\begin{theorem}
A quantum operation $\Lambda$ can be decomposed into operations of the form (O1)-(O4) if and only if $\Lambda$ acts as a completely positive trace non-increasing operator: $\Lambda(\rho)=\sum_i L_i \rho L^\dagger_i$, with $\sum_i L_i^\dagger L_i\leq \openone$.
\end{theorem}
From a physical point of view, the requirement that a quantum operation is CP and trace non-increasing, is obvious. Thus we could equally have started this section by demanding this, and we would also have arrived to the elementary operations (O1)-(O4). 

\subsection{Bipartite quantum operations}
Consider now a bipartite system on ${\cal H}_A\otimes {\cal H}_B$. We will review some important classes of local quantum operations where increasing degrees of communication are allowed \cite{Rains99, DHR01b, BDFMRSSW98, CDKL01}. 

$\diamondsuit $ {\bf Local operations}. This is the class of operations generated by the operators which have Kraus operators of the form $A_i\otimes \ido$ and $\ido\otimes B_i$ with $\sum_i A_i^\dagger A_i = \sum_i B_i^\dagger B_i= \openone$. In this restricted class of operations, the parties are not allowed to communicate and the operations are non-measuring. Combining them yields for an arbitrary local operation 
$\Lambda(\rho)=\sum_{ij}{(A_i\otimes B_j) \rho (A_i\otimes B_j)^\dagger}.$

$\diamondsuit $ {\bf 1-local operations}. Suppose we allow \emph{local operations with one-way communication} from $A$ to $B$. So let us suppose Alice performs a generalised measurement on her subsystem, with Kraus operators $A_i$ with $\sum_i A_i^\dagger A_i=\ido$. If the result of her operation is $i$ the operation on the state acts as (notice that this is a trace decreasing operation as it will occur only with a certain probability)
$$
\Lambda^A_i(\rho)=(A_i\otimes\ido) \rho (A^\dagger_i\otimes \ido).
$$
She will communicate Bob that she found result $i$, and depending on that outcome he will perform a certain trace preserving operation defined by $B_{ji}$ operators, with $\sum_{j}B^\dagger_{ji}B_{ji}=\ido$. Here the index $i$ denotes that the operation Bob implements depends on the result that he got from Alice. Thus the state will change as
$$
\Lambda^{AB}_i(\rho)=\sum_{j}(\ido\otimes B_{ji})(A_i\otimes\ido) \rho (A^\dagger_i\otimes \ido)(\ido\otimes B^\dagger_{ji}).
$$
Now if this same procedure is done on many particles the total ensemble will change as
$$
\Lambda^{AB}(\rho) = \sum_{ij} (A_i\otimes B_{ji}) \rho (A^\dagger_i\otimes B^{\dagger}_{ji}).
$$
Of course postselection (for certain $i$) can occur in one or more terms of this expression.

$\diamondsuit $ {\bf 2-local operations}. This class is also known as the class of local operations with two-way classical communication (LOCC). This situation is somewhat more complicated, since in this case, it is useful to do alternating measurements and communications. Following a similar strategy as for 1-local operations, we focus on one particular outcome of each measurement, and do the summation at the end. Alice starts off with her measurements, and finds $i1$ as result, therefore the relevant operator is $A_{i1}\otimes\ido$. She communicates her result to Bob and he will then perform $\ido\otimes B_{j1}(i1)$ and communicate his result to Alice who will perform $A_{i2}(i1,j1)\otimes\ido$. All these operators satisfy similar normalization properties as in the previous paragraph. In that way we get at the end of the series (for the total ensemble)
$$
\Lambda(\rho)=\sum_m(A_m\otimes B_m)\rho(A_m\otimes B_m)^\dagger,
$$
with $m=(i1,i2,\ldots,in,j1,j2,\ldots,jn)$ and
\begin{align*}
A_m &=A_{in}(i1,i2,\ldots,i(n-1),j1,j2,\ldots,j(n-1))\ldots A_{i2}(i1,j1)A_{i1}, \\
B_m &=B_{jn}(i1,i2,\ldots,in, j1,j2,\ldots,j(n-1))\ldots B_{j2}(i1,i2,j1)B_{j1}(i1).
\end{align*}
Here also, postselecting can be done for particular choices of $i1,\ldots,in,j1,\ldots,j(n-1)$.

$\diamondsuit $ {\bf Separable operations}.
Separable operations are defined as any operation which can be written as
$$
\Lambda(\rho)=\sum_i(A_i\otimes B_i)\rho(A_i\otimes B_i)^\dagger,
$$
where each $A_i$ and $B_i$ are arbitrary operations (including measurements). The normalization condition is $\sum_i (A_i\otimes B_i)^\dagger(A_i\otimes B_i)=\ido$. 
It is obvious that any LOCC is also separable, but the reverse is generally not true \cite{BDFMRSSW98}. However, we do have the following theorem \cite{CDKL01}:

\begin{theorem}
It is \emph{always} possible to simulate a separable operation using only LOCC, but with probability possibly smaller than 1. 
\end{theorem}
\begin{proof}
So, \cite{cirac} suppose we have two sets of operators $\{A_i\}$ and $\{B_i\}$ such that $\sum_i (A_i\otimes B_i)^\dagger(A_i\otimes B_i)=\ido$ and a state $\rho$. In general we will have that $\sum_i A_i^\dagger A_i\geq \ido$ and $\sum_i B_i^\dagger B_i\geq \ido$. We renormalise ${A'}_i=aA_i$ and ${B'}_i=aB_i$ so that $\sum_i {A'}_i^\dagger {A'}_i\leq \ido$ and $\sum_i {B'}_i^\dagger {B'}_i\leq \ido$. Then we let Alice and Bob perform the generalized measurements corresponding to these operators. They will obtain two measuring outcomes $k_1$ and $k_2$ and communicate their result to each other. Their strategy is to keep the state only if the outcomes of their measurement coincide, if $k_1=k_2$. Forgetting which outcome was obtained, we see that $\rho$ transforms as
\begin{align}
\rho \rightarrow N \sum_i p_i \frac{({A'}_i\otimes {B'}_i)\rho ({A'}_i\otimes {B'}_i)^\dagger}{p_i}= \sum_i(A_i\otimes B_i)\rho(A_i\otimes B_i)^\dagger,
\end{align}
with $p_i=\Tr[({A'}_i\otimes {B'}_i)\rho ({A'}_i\otimes {B'}_i)^\dagger]$ and $N$ an appropriate normalisation factor.
\end{proof}

$\diamondsuit $ {\bf PPT preserving operations}. Literally, these are operations which map PPT states to PPT states. Or with Rains' definition \cite{Rains98} operations $\Lambda$ such that the induced map 
$$
\tilde \Lambda:\rho \rightarrow \Lambda(\rho^{T_B})^{T_B}
$$
is completely positive. All separable operations are PPT preserving, as follows from \cite{Alb01}
$$
[(A\otimes B)\rho(A\otimes B)^\dagger]^{T_B}=[(A\otimes (B^\dagger)^T)\rho^{T_B}(A^\dagger\otimes B^T)].
$$
PPT preserving operations are the first class of operations which can be regarded as non-local, as for instance creation of an entangled PPT state is a PPT preserving operation. The key result on PPT preserving operations is that they can be implemented using LOCC when both parties share a specific entangled PPT state. This was first realised in the seminal paper by Cirac et.~al.\ \cite{CDKL01}. They characterised completely general quantum operations (and in particular entangling operations) on bipartite systems by means of a generalised Jamio{\l}kowski isomorphism. Because of its importance, we briefly sketch the isomorphism in the next section.

\subsection{Entangling operations}
\label{entopssec}
The Jamio{\l}kowski isomorphism (Section~\ref{sejaiso}) is an isomorphism between a linear map from an input space to an output space and an operator defined over the tensor product of these two spaces. As we have illustrated, this correspondence is useful when the map acts on one part of a bipartite system. For maps acting on compound systems it is not so easy to interpret this duality. We now introduce an extension of the Jamio{\l}kowski isomorphism with a very nice physical interpretation \cite{CDKL01}.

Suppose we have two systems $A$ and $B$, each consisting of two $d$-level subsystems $A=A_1,A_2$ and $B=B_1,B_2$ respectively. We will establish an isomorphism between a CP map $\Lambda=\Lambda_{A_1B_1}: {\cal L}({\cal H}_{A_1B_1}) \rightarrow {\cal L}( {\cal H}_{A_1B_1})$ and an operator $D=D_{A_1A_2,B_1B_2}$ acting on the total system ${\cal H}_A \otimes {\cal H}_B$. The isomorphism is given by
\begin{align}
\label{isoone}
D=\ido_{A_2B_2}\otimes \Lambda_{A_1B_1}(P_{A_1A_2}\otimes P_{B_1B_2}),
\end{align}
where $\Lambda$ is understood to act only on the systems $A_1$ and $B_1$. Here $P_{A_1A_2}$ is the projector on the maximally entangled state $|\psi\ra_{A_1A_2}=\frac{1}{\sqrt{d}} |ii\ra_{A_1A_2}$ (and analogously for $P_{B_1B_2}$).
Equivalently we can write
$$
\Lambda(\rho_{A_1B_1})=d^2\Tr_{A_2B_2}(D\rho_{A_2B_2}^T)=d^4\Tr_{A_2A_3B_2B_3}(D\rho_{A_3B_3}P_{A_2A_3}P_{B_2B_3}).
$$
From (\ref{isoone}) it follows that $D$ is the result of the action of $\Lambda$ on two systems $A_1$ and $B_1$ which are prepared in a maximally entangled state with two ancillary systems. The second form has an equally simple interpretation: if both parties share a state $D$, then they can implement the map $\Lambda$ on a certain state $\rho_{A_1B_1}$ by simultaneous projecting $\rho_{A_3B_3}$ and $D$ on the maximally entangled states of $A_2A_3$ and $B_2B_3$. From this isomorphism one can easily deduce the following correspondences:
\begin{enumerate}
\item $\Lambda$ is a separable operation if $D$ is separable, and conversely $\Lambda$ is an entangling operation if $D$ is entangled (with respect to $A$ and $B$).
\item $\Lambda$ is a PPT preserving operation if and only if $D^{T_{A}}\geq 0$. Thus any PPT preserving operation can be implemented locally with the aid of a shared PPT state.
\end{enumerate}

This isomorphism has a large number of applications (see the original reference and \cite{DC01}) which would lead us too far to go into here. A simple application is the classification of global unitaries according to their entangling power on composite systems \cite{WZ02, Zanardi00, ZZF00}. Entangling power of a unitary is defined as the average entanglement a unitary creates on product states. When $\Lambda$ is a unitary map, $D$ is a pure state and there exists a closed expression of the entangling power in terms of the entanglement of this state vector. Using this, we were able \cite{CGSS05} to show that the unitaries with highest entangling power are special kinds of permutations, with possible exception for $d=6$. The isomorphism then immediately yields a construction of generic examples of 4-qudit maximally entangled states for all dimensions except for 2 and 6.

\section{Pure states and the reversibility of entanglement}
\label{psatroe}
In this section we will study the distillation of entanglement of \emph{pure} bipartite quantum states, which is often referred to as entanglement concentration. An arbitrary pure state $\psi$ can be represented by a vector $|\psi\rangle$ in ${\cal H}_A \otimes {\cal H}_B \cong {\mathbb{C}}^d \otimes {\mathbb{C}}^d$. Without loss of generality we have taken the dimensions of the two Hilbert spaces to be equal. A way of quantifying the entanglement of $\psi$ is by looking at the local density operator $\rho_A=\Tr_B(|\psi\rangle\langle\psi|)$. For a separable state $|\psi\ra=|\psi\ra_A \otimes |\psi\ra_B$, the reduced density operator equals $\rho_A=|\psi\rangle_A\langle \psi|$, and is thus a pure state. The reduced density operator of an entangled state will have a rank larger than one. If the reduced density matrix is proportional to the identity matrix the state is maximally entangled, as is the case for the EPR singlet. For such states there is no information available locally, although the global state is known exactly \cite{Schrodinger35, Verstraete02}. A way of quantifying the purity of a local density operator $\rho_A$ is by calculating its von Neumann entropy $E(\psi)=-\Tr(\rho_A \log_2 \rho_A)$. We will call this quantity the entropy of entanglement or simply the entanglement of $\psi$. Note that the entanglement is a fully additive measure: $E(\psi_1\otimes\psi_2)=E(\psi_1) +E(\psi_2)$. Writing $\psi$ in the Schmidt basis
$$
|\psi\ra=\sum_{i=1}^d \lambda_i |ii\rangle,
$$
with $\lambda_i\geq \lambda_{i+1}\geq 0$ and $\sum_i \lambda^2_i=1$, the reduced density matrix can be written as $\rho_A=\sum_i \lambda^2_i |i\ra_A\la i|$ and the entanglement of $\psi$ as $E(\psi)=-\sum_i \lambda^2_i \log_2 \lambda^2_i$. Maximally entangled states in a $d\otimes d$ system have Schmidt coefficients $\lambda_i=1/\sqrt{d}$ and entanglement $E=\log_2 d$.

Entanglement concentration is the process of converting partially entangled states to maximally entangled states by LOCC\@. The reverse process is called entanglement dilution. We will show how in the asymptotic regime $n$ arbitrary states $\psi$ can be concentrated into $nE(\psi)$ singlets for $n\rightarrow \infty$. Conversely, $nE(\psi)$ singlets can be diluted into $n$ states $\psi$ for $n\rightarrow \infty$. From the above it follows that the entropy of entanglement quantifies entanglement of pure states in the most natural way since it is preserved in the asymptotic regime: a collection of $n$ states $\psi$ can be converted into $m_n$ states $\phi$ with LOCC and the ratio $m_n/n$ approaching $E(\psi)/E(\phi)$ for $n\rightarrow \infty$. The entanglement of a singlet is given by $E=1$ and serves thus as a proper unit of entanglement, the so-called \emph{ebit} \cite{BBPS96}. 

In the first part of this section we discuss some elementary entanglement concentration protocols, and present a dilution protocol. Next, we list some celebrated theorems of general pure state transformation in the finite and asymptotic regime. 

\subsection{Pure state distillation protocols}
\subsubsection{Local filtering -- the procrustean method}
The procrustean\footnote{Procrustes is a giant from Greek mythology. Allegedly he would capture people and tie them to an iron bed. He would then hack off their legs or stretch them to make them fit the bed.} method is conceptually maybe the simplest entanglement concentration protocol \cite{BBPS96}. It works even when both parties only share one single copy of the state. So let Alice and Bob share the state $|\psi\ra=\cos\theta |00\rangle + \sin\theta |11\ra$ with $\cos\theta \geq \sin\theta$, then we can define
\begin{align*}
A_0&=\frac{\sin\theta}{\cos\theta}|0\ra\la 0| + |1\ra\la 1|,\\
A_1&=\sqrt{\ido-A_0^\dagger A_0}.
\end{align*}
Now the protocol consists of Alice performing the measurement corresponding to these operators and communicating the result to Bob. The protocol will be successful when Alice obtains the outcome corresponding to $A_0$, in this case, occurring with probability $2 \sin^2\theta$, the output state will be a maximally entangled singlet. This protocol is optimal for single qubits and we will later see an extension of this protocol which is optimal for any pure bipartite state. Such a one-sided generalised measurement, where the state is conditionally kept will be called filtering. This is a technique which will be useful for mixed states as well.

An analogous protocol works for entanglement dilution: if Alice and Bob start out with a maximally entangled singlet and Alice applies the operation defined by 
\begin{align*}
A_0&=\cos\theta|0\ra\la 0| + \sin\theta|1\ra\la 1|,\\
A_1&=\sin\theta|0\ra\la 0| + \cos\theta|1\ra\la 1|,
\end{align*}
then they obtain the state $|\psi\ra$ with probability one.

\subsubsection{Schmidt projection}
Following \cite{BBPS96} we illustrate this method for qubits. As the method does not work for one copy we suppose that Alice and Bob share $n$ copies of the state $|\psi\ra=\cos\theta |00\ra + \sin\theta|11\ra$, so that the combined system is described by the state vector
$$
|\Psi\rangle=|\psi\rangle^{\otimes n}= \otimes_{i=1}^n [\cos\theta |00\ra_i+\sin\theta |11\ra_i].
$$
The binomial expansion of this will have $2^n$ terms of which $n+1$ will have different coefficients and the $n+1$ corresponding subspaces will be orthogonal. Let us label these subspaces by $k=0,\cdots,n$ corresponding to the power in which $\sin\theta$ appears. Now Alice can perform a von Neumann measurement corresponding to these subspaces. The probability of projecting upon the $k$-th subspace is given by 
$$
p_k=\binom{n}{k} \cos^{2(n-k)}\theta \sin^{2k}\theta.
$$
She can then communicate her result of the measurement to Bob or Bob can perform a similar measurement (which will have the same outcome). In either case they will end up with a maximally entangled state in a known $2\binom{n}{k}$-dimensional space. After distilling many such states, they can be converted to singlets without loss of entanglement.
In the asymptotic regime this protocol can be shown to be optimal: the ratio of singlets that can be distilled per input state $\psi$ approaches $E(\psi)$ for the number of input states going to infinity. To achieve this, the Schmidt projection protocol has to be applied $N$ times on batches of $n$ states, with both $N$ and $n$ going to infinity.

\subsubsection{Entanglement swapping}
The phenomenon of entanglement swapping \cite{ZZHE93, YS92} was first studied in the context of coincidence count rates of entangled pairs of photons. In the experimental setup, two different sources emit pairs of entangled photons and two halves of the pairs are being measured simultaneously. The remaining halves then become entangled. Thus entanglement swapping is the phenomenon of two particles getting entangled without direct interaction (see Ref.~\cite{PBWZ98} for an experiment demonstrating this effect). This phenomenon also occurs in a multipartite setting \cite{YS92b, BVK98}, but has to our knowledge only been demonstrated for qubits.

The idea of using entanglement swapping as an entanglement concentration procedure was first introduced in Ref.~\cite{BVK99}. Consider the following two states of photons
\begin{align*}
|\phi\ra_{12} = \cos \theta |00 \ra +\sin \theta |11 \ra, \\
|\phi\ra_{34} = \cos \theta |00 \ra +\sin \theta |11 \ra. 
\end{align*}
Suppose $|\phi\ra_{12}$ is shared by both parties Alice and Bob, and in addition Bob has the state $|\phi\ra_{34}$. When Bob measures particles 2 and 3 in the Bell basis $\{|00\ra+|11\ra, |00\ra-|11\ra, |01\ra+|10\ra, |01\ra-|10\ra\}$, the resulting state for particles 1 and 4 will be one of the four states
\begin{align*}
|\phi^\pm\ra_{14} &= \frac{1}{\sqrt{\cos^4 \theta+\sin ^4 \theta}}(\cos^2 \theta |00 \ra \pm \sin^2 \theta |11 \ra), \\
|\psi^\pm\ra_{14} &= \frac{1}{\sqrt 2}(|01 \ra \pm |10 \ra), 
\end{align*}
the first two occurring with probability $(\cos^4\theta + \sin^4\theta)/2$ and the last two with probability $\sin^2\theta\cos^2\theta$. 
In the first two cases, the entanglement is lower than in the original pair, but in the last two cases Alice and Bob end up with a maximally entangled state. In any case, however, the procedure can be repeated. Repeating the whole protocol over and over again, yields on one hand a group of maximally entangled pairs, and on the other, completely separable states. The fraction of maximally entangled pairs achievable in this way turns out to be $2 \cos^2\theta$. This number just equals the maximal probability of the initial state to be converted into a maximally entangled state. In this sense, the entanglement swapping protocol is optimal. Note that since the second pair belongs to Bob, he has complete control over it, and he could in principal use any two-qubit state. It turns out that when he uses a state which is less entangled, the protocol does not concentrate entanglement but destroys it. When the state he uses is more entangled, no entanglement is destroyed, but the concentration is less efficient. (Here the entanglement is measured as the amount of singlets distillable from the state). Extensions of this protocol have been proposed in Ref.~\cite{Hsu02, SJG00} which do not rely on such an iterative procedure since the second step of their protocol is just the procrustean method.

With this we conclude our discussion of elementary pure state entanglement concentration protocols. Various other protocols have been proposed, as for instance in Ref.~\cite{Bandyopadhyay00}. Noteworthy is the protocol outlined in Ref.~\cite{POBV02} which demonstrates that entanglement concentration is possible using the effects of quantum statistics of indistinguishable particles only.

\subsection{Finite regime}
In this section we will discuss the more general question of feasibility of arbitrary transformations between pure states. The first work on this subject was done by Lo and Popescu \cite{LP97}, in particularly they proved the following:
\begin{theorem}
\label{LPCCC}
The Schmidt number of a pure bipartite state cannot increase under local quantum operations and classical communication.
Entanglement manipulations of pure bipartite states with only one way communication are as powerful as those with two way communication, but they are more powerful than those with no communication. 
In other words, we can replace any LOCC protocol on pure states, no matter how complex, by a single local generalised measurement.
\end{theorem}

The groundbreaking work on pure state entanglement transformation was carried out by Vidal in Ref.~\cite{Vidal00}, where he introduced the notion of an entanglement monotone. His work was significantly extended by several authors in a series of papers \cite{Nielsen98, Vidal99, Hardy99, JP99, JP99b, VJN99}. Recently it became apparent that their proofs can be simplified by making use of the quantum steering theorem as recognized by Verstraete \cite{Verstraete02} to which we refer the reader for full proofs. An entanglement monotone $\mu$ is defined as a quantity associated with a quantum state which on average does not increase under local transformations. That is, if an LOCC maps a state $\rho$ with probability $p_i$ into states $\rho_i$ then $\mu(\rho)\geq \sum_i p_i \mu(\rho_i)$. The following theorem by Vidal \cite{Vidal00} is crucial:
\begin{theorem}
The restriction of an entanglement monotone $\mu$ to pure states $\psi$ is given by a unitarily invariant concave function of the local density operator $\mu(|\psi\ra)=f(\rho^B)$.
\end{theorem}

This theorem implies that any entanglement monotone acting on pure states
$$
|\psi\ra=\sum_{i=1}^d \sqrt{\lambda_i}|ii\ra
$$
 can be expressed as a concave function of its Schmidt coefficients\footnote{For convenience Schmidt coefficients in this section denote the `normal' Schmidt coefficients squared.} $\lambda_i$. An important class of entanglement monotones $E_k$, for $k=2,\ldots ,n$ is given by
$$
E_k(\psi)=\sum_{i=k}^d \lambda_i=1-\sum_{i=1}^{k-1} \lambda_i,
$$
where as usual $\lambda_i \geq \lambda_{i+1}$. The concavity of this function can be proven using the variational characterization of eigenvalues of a Hermitian matrix \cite{HJ85}.

Clearly we have that if we can convert $\phi$ to $\psi$ with certainty then we need to have $E(\psi)\leq E(\phi)$ for every entanglement monotone $E$, and in particular this has to be the case for $E=E_k$. A surprising result by Nielsen \cite{Nielsen98} is that the converse is true:
\begin{theorem}
\label{nielthe}
A necessary and sufficient condition for the transformation $\phi$ to $\psi$ to be possible with certainty is that $E_k(\psi)\leq E_k(\phi)$ for $k=2,\ldots ,n$. Alternatively, in terms of majorization, this can be expressed as $\lambda_\psi \succeq \lambda_\phi$, where $\lambda_\psi$ stands for the vector consisting of the Schmidt coefficients of $\psi$ in decreasing order.
\end{theorem}

For anyone who is familiar with unitarily invariant norms, this result will be reminiscent of Ky Fan's dominance theorem for unitarily invariant norms and the dual representation in terms of symmetric gauge functions \cite{HJ91, Bhatia97}. In fact, something much stronger can be proven \cite{Vidal99, JP99, Hardy99}:

\begin{theorem}
\label{mttt}
The probabilistic transformation from a pure bipartite state $\phi$ into the ensemble $\{p_i, \psi_i \}$ can be accomplished using LOCC if and only if $p_i\vec{\lambda}_{\psi_i} \succeq \vec{\lambda}_\phi$ or equivalently $\sum_i p_i E_k(\psi_i) \leq E_k(\phi)$ for all $k$. 
\end{theorem}

Note that from the theorem follows that the maximum probability $p$ of transforming $\phi$ into $\psi$ can be written as $p=\min_k \frac{E_k(\phi)}{E_k(\psi)}$ or $p\vec{\lambda}_{\psi} \succeq \vec{\lambda}_\phi$. The non-trivial part of these theorems is to show that there is an actual protocol achieving the transformation. The reader is referred to \cite{Verstraete02} for an elementary proof. It is there shown how the protocol achieving the transformation can be built from a few simple generalised measurements. 

Let us now return to entanglement concentration. The above theorem gives us the optimal probability of converting a pure state to a maximally entangled state of arbitrary dimension. When we allow the concentration of entanglement into different maximally entangled states (of different dimensions), we can ask how much entanglement can be distilled on average. This is answered in the following theorem \cite{JP99, Hardy99}:
\begin{theorem}
\label{findis}
The optimal entanglement concentration procedure for a single pure bipartite state $\psi$ with Schmidt coefficients $\lambda_1\geq \ldots \geq \lambda_d \geq 0$ is one that produces a maximally entangled state $\phi_j$ of $j\leq d$ levels with probability $p_j=j(\lambda_j-\lambda_{j+1})$. The corresponding optimal average distilled entanglement is 
$$
\la E\ra_{max}=\sum_{k=1}^d (\lambda_j-\lambda_{j+1})j \log_2 j.
$$
\end{theorem}

The proof of this theorem uses techniques from linear optimization, showing that there exist optimal $p_j$ compatible with Theorem~\ref{mttt}. The protocol achieving these probabilities can be obtained by going through the proof of this theorem. However, in view of Theorem~\ref{LPCCC} there exists a single local generalised measurement which does the job. The measurement operators corresponding to Alice's generalised measurement are
\begin{align}
A_j= \sum^j_{i=1}\sqrt{\frac{\lambda_j-\lambda_{j+1}}{\lambda_i}}|i\ra_A\la i|,\\
\end{align}
indeed it is easy to verify that $\sum^d_{j=1} A^\dagger_j A_j \leq \ido$ and 
\begin{align}
A_j \otimes \ido |\psi\ra = \sqrt{p_j} |\phi_j\ra.
\end{align}
This protocol, a generalised procrustean method, thus turns out to be optimal.

Another application of Theorem~\ref{mttt} is the idea of catalysis \cite{JP99b, DH02}, where a pure state is used to help converting one state into an other, such that the pure state is recovered at the end of the procedure. Mathematically this means that there exist states $\phi, \psi$ and $\chi$ such that 
$$
\lambda_\psi \not \succeq \lambda_\phi \qquad \text{ but } \qquad \lambda_{\psi\otimes \chi} \succeq \lambda_{\phi\otimes \chi}.
$$
Finally note that one can broaden the discussion about pure state transformations by including questions about the approximate transformation of one state into another, see Ref.~\cite{VJN99}.

\subsection{Asymptotic regime}
In the introduction we have already sketched the main result in the asymptotic regime, which we now state formally \cite{BBPS96, PR97, DHR01}:
\begin{theorem}
In the asymptotic regime $n$ arbitrary states $\psi$ can be concentrated into $nE(\psi)$ singlets for $n\rightarrow \infty$. Conversely, $nE(\psi)$ singlets can be diluted into $n$ states $\psi$ for $n\rightarrow \infty$. 
Thus the entropy of entanglement is conserved in the asymptotic regime and hence quantifies entanglement of pure states in a unique way.
\end{theorem}
One can prove the distillation rate using Theorem~\ref{findis} \cite{JP99}, or using the Schmidt projection distillation protocol \cite{BBPS96}. The dilution rate can be proved using the teleportation idea \cite{BBPS96}. Both the dilution and distillation rate can also be proven using Nielsen's theorem (Theorem~\ref{nielthe}). The crucial part of the latter exploits the asymptotic equipartition theorem which implies that $|\psi\ra^{\otimes n}$ can be approximated by $2^{nE(\psi)}$ terms in the Schmidt decomposition.

In Ref.~\cite{LP99} Lo and Popescu showed that pure state entanglement is a truly interconvertible, in the sense that the amount of classical communication needed in the asymptotic regime vanishes. This is trivial for the entanglement concentration, as the Schmidt projection method works without communication. The dilution step is the nontrivial part, and it can be shown that $nE(\psi)$ singlets can be diluted into $n$ pairs $\psi$ with ${\cal O}(\sqrt n )$ classical bits of communication. Recently other asymptotically optimal protocols have been proposed, some of which are universal\footnote{Sometimes also called blind protocols as neither prior knowledge of the state nor any form of quantum tomography is required.} (see for instance \cite{MH05, HM01}).

The entanglement concentration protocols we studied in the previous subsection are all of probabilistic nature, with the notable exception of Nielsen's result. As a result, the protocols reaching the entanglement yield $E(\psi)$ in the asymptotic limit are all probabilistic in a finite number of copies and only become deterministic in the asymptotic limit. The following theorem gives an upper bound to the asymptotic yield of singlets we can distill from a pure state $\psi$ if we only use protocols that are deterministic for any given number of copies \cite{Morikoshi99, MK01}.
\begin{theorem}
Let $\psi$ be a pure state with Schmidt decomposition $|\psi\ra=\sum_{i=1}^d \sqrt{\lambda_i}|ii\ra$, then the maximum number of singlets that can deterministically be concentrated is given by $E_{\text{det}}(\psi)=-\log_2 \lambda_1$. When $\psi$ is a two-qubit state this bound can be obtained by a series of two-pair manipulations.
\end{theorem}

Note that $E_{\text{det}}$ just like $E$ is a fully additive measure. The upper bound follows from Nielsen's theorem. The surprising result here is the situation for qubits where two-pair manipulations are sufficient. This is in sharp contrast with the probabilistic concentration protocols which generally need to be applied on a very large number of pairs. The values of $E_{\text{det}}$ and $E$ do not coincide, although both yields can be reached with probability one in the asymptotic limit. One can generalise this and ask for the distillable entanglement as a function of an error exponent, representing the rate of decrease in failure probability as the number of copies tends to infinity \cite{HKMMW02, Hayashi02}. 

\section{Mixed states and the irreversibility of entanglement}
\label{mesent}
As we have seen in the previous section, for pure states there are essentially a finite number of entanglement measures in the finite regime and a unique measure in the asymptotic regime. There are two obvious ways of quantifying mixed state entanglement, operational and geometrical. Operational entanglement measures are directly related to the physical applicability of entanglement. Geometrical measures can loosely be described as those quantifying the distance from a state to the set of separable states. As excellent reviews \cite{Horodecki01, PV05, Christandl06} have appeared recently in the literature, we do not attempt to give an extensive review of the subject here. In particular the geometrical entanglement measures will be treated in brief, while for the operational measures we focus mainly on the entanglement of distillation. For simplicity, entanglement measures here are defined simply as entanglement monotones and thus all presented measures vanish on separable states and are on average non-increasing under LOCC, unless otherwise stated.

\subsection{Geometric measures}
The first rigorous study of entanglement measures based on distance was carried out in Ref.~\cite{VPRK97, VP98}. Intuitively, the quantity
$$
\min_{\sigma\in {\cal S}}D(\rho||\sigma),
$$
with $D$ a suitable distance measure, can be regarded as a measure of entanglement. The most important measure of this kind arises when $D$ is the quantum relative entropy $D=S(\rho||\sigma)=\Tr(\rho\log_2\rho-\rho\log_2\sigma)$ \cite{Vedral01, SW00, Ruskai02}. The associated measure will be called the relative entropy of entanglement and denoted $E_R$. Note that for pure states, $E_R$ reduces to the entropy of entanglement $E$. Although the relative entropy of entanglement is a distance measure, it stretches much further as it has several operational interpretations (see also next section).

A related measure \cite{VP98} can be defined by using as a distance measure the Hilbert-Schmidt distance $D=\|\rho-\sigma\|_2$. Although this quantity has received considerable interest, it was shown not to be an entanglement monotone \cite{Ozawa00}. We will call this measure the best separable approximation in the Hilbert-Schmidt norm analogous to the best separable approximation $\text{BSA}(\rho)$ in the trace norm \cite{LS97}, which can be shown to be a proper measure of entanglement. It is defined as 
$$
\text{BSA}(\rho)=\min \{\|\rho-\rho_{\text{s}}\|\,|\rho\geq \rho_{\text{s}}, \rho_{\text{s}}\in {\cal S} \}.
$$
or equivalently
$$
\text{BSA}(\rho)=\min\{\lambda|\rho=(1-\lambda)\rho_{\text{s}} + \lambda\rho_{\text{e}}, \rho_{\text{s}}\in {\cal S}, \rho_{\text{e}}\in {\cal D} \setminus {\cal S}\}.
$$

\emph{The} geometric measure of entanglement (GME) is historically the first example of an entanglement measure. It was originally introduced by Shimony in Ref.~\cite{Shimony95} for pure states only as
$$
\text{GME}(\psi)=\frac{1}{2}\min \{\|\, |\psi\ra-|\phi\ra|\,\|^2, \phi \in {\cal S}\}=1-\max\{|\la\phi|\psi\ra|^2 |\,\phi \in {\cal S}\}.
$$
Thus the GME is proportional to the maximum overlap with a separable state. This measure was extensively analysed in Ref.~\cite{WG03}, including for multipartite and mixed states. The mixed state generalisation of the GME was done by the usual convex roof construction, that is as a minimisation over all convex expansions of $\rho$:
$$
\text{GME}(\rho)=\min\left\{\sum_i p_i \text{GME}(\psi_i)\Bigg| \rho=\sum_i p_i|\psi_i\ra\la\psi_i|, p_i\geq 0 \right\}.
$$

Dual to the distance measures are the robustness measures. While distance measures quantify the distance of a set to the set of separable states, robustness measures quantify to which extend entangled states remain entangled under mixing. Recall (Section~\ref{smrobn}) that the robustness $R_k$ relative to a set $\cal K$ is defined as 
$$
R_k(\rho)=\min \left\{R \Bigg| \frac{\rho+R\rho_k}{1+R}\in {\cal S}, \rho_k \in {\cal K}, R\geq 0\right\}.
$$
Taking $\cal K$ to be either $\cal S$, $\{\ido\}$ or $\cal D$ we obtain respectively the robustness $R_s$, the random robustness $R_r$ and the generalised robustness $R_g$. Note that the random robustness $R_r$ was shown not to be an entanglement monotone \cite{HN03}. However, for a given dimension, the maximum random robustness over all entangled states gives rise to a lower bound on the volume of separable states \cite{VT99}.

All the previous entanglement measures are very hard to evaluate for a general mixed state. For a NPT entangled state $\rho$, one can envisage that the sum of the negative eigenvalues, the so-called negativity ${\cal N}(\rho)$, quantifies to some extent the distance of $\rho$ to the set of PPT states \cite{ZHSL98}. And indeed, it can be shown to be an entanglement monotone. However, it is often more convenient to work with the logarithmic negativity \cite{VW01b}, defined as 
$$
E_N(\rho)=\log_2\|\rho^{T_B}\|.
$$
The reason is that $E_N$ is fully additive by construction and also has considerable operational meaning, some of which we will outline in the next section (see also \cite{APE03}).

With this, we conclude our roller coaster through the main geometric entanglement measures. Note that they also can be approached in a more unified way using either base norm Ref.~\cite{VW01, PV05} or entanglement witnesses Ref.~\cite{Brandao05}.

\begin{table}
\begin{center}
\begin{tabular}{|l|l|l|}
\hline
Measure & Definition & Key references \\\hline
Relative entropy\ & $\min\{\Tr(\rho\log_2 \frac{\rho}{\sigma}) | \sigma\in \cal S\}$ & \cite{VP98, VPRK97,HV00,PVP00,VW01}\\
BSA in trace norm & $\min \{\|\rho-\rho_{\text{s}}\|\,|\rho\geq \rho_{\text{s}}, \rho_{\text{s}}\in {\cal S} \}$ & \cite{LS97,KL00, HLVC00, LSG00,WK01} \\
BSA in HS-norm & $\min \{\|\rho-\rho_{\text{s}}\|_2 | \rho_{\text{s}}\in {\cal S} \}$ & \cite{VP98, LS00,WT98,Ozawa00, BNT02} \\
Robustness & $\min \{R\geq 0 | \rho+R\rho_s\in {\cal S}, \rho_s \in {\cal S}\}$ & \cite{VT99, HN03} \\
Gen.\ robustness & $\min \{R\geq 0 | \rho+R\rho_d\in {\cal S}, \rho_d \in {\cal D}\}$ & \cite{Steiner03, HN03, Cavalcanti06, Brandao05b} \\
Ran.\ robustness & $\min \{R\geq 0 | \rho+R\ido \in {\cal S}\}$& \cite{VT99, BV04, HN03} \\
GME (pure) & $1-\max\{|\la\phi|\psi\ra|^2 |\,\phi \in {\cal S}\}$ & \cite{Shimony95, WEGM04, Cavalcanti06, WG03,WAGM04,BEKWGHBLS03} \\
Log.\ negativity & $\log_2\|\rho^{T_B}\|$ & \cite{VW01b, Plenio05}\\
\hline
\end{tabular}
\caption{Geometrical measures of entanglement.}
\label{tblgmoe}
\end{center}
\end{table}

\subsection{Operational measures}
Having a unique measure for pure states in the asymptotic regime, it would be natural to ask if one can generalise this to mixed states. We have seen that for pure states, this measure quantifies both the amount of entanglement that can be distilled from them in the asymptotic regime and the amount of entanglement needed to construct the state in the asymptotic regime. Initially, the idea was to define two different measures for these quantities for mixed state: the distillable entanglement $E_D$ and the entanglement cost $E_C$ respectively.

A first attempt to the entanglement cost was made in Ref.~\cite{BDSW96} as the convex roof of the entanglement
$$
E_F(\rho)=\min \sum_i p_i E(|\psi_i\ra)
$$
where the minimum is taken over all convex decomposition $\rho=\sum_i p_i |\psi_i\ra\la\psi_i|$. This quantity is now known as the entanglement of formation. The evaluation of $E_F$ for a general mixed state is very hard, and analytical results are only known for high symmetric states \cite{VW01,TV00} (see also \cite{CAF05} and reference therein). The notable exception is for 2 qubit systems, where an exact analytic formula is known in terms of the concurrence \cite{Wootters97, Wootters01}.

It is however not clear that $E_F$ has the operational meaning we want it to have \cite{HHT00}. In particular $E_F(\rho)$ quantifies the entanglement needed to construct $\rho$ via \emph{a specific procedure}. Namely, given a decomposition $\rho=\sum_i p_i |\psi_i\ra\la\psi_i|$, one chooses with probability $p_i$ a particular $|\psi_i\ra$. Then, one constructs $\psi_i$ out of singlets (costing in the asymptotic regime $E(\psi_i)$ ebits) and forgets the particular $\psi_i$ chosen. In this way, we obtain one copy of $\rho$. Repeating this process over and over, then to construct $\rho$ on average we need $E_F(\rho)$ ebits. Other procedures might be more efficient, such as constructing more than one copy at the time. In particular it is easy to see that
$$
E^\infty_F(\rho)=\lim_{n\rightarrow\infty} \frac{E_F(\rho^{\otimes n})}{n}\leq E_F(\rho).
$$
This quantity $E^\infty_F(\rho)$ is known as the regularised entanglement of formation. 

In Ref.~\cite{HHT00} it was rigorously proven that $E^\infty_F(\rho)$ equals the asymptotic value of $m/n$ such that  $m$ qubits can be converted into $n$ copies of $\rho$. This asymptotic yield, we call the entanglement cost and is formally defined as (see Ref.~\cite{PV05})
$$
E_C(\rho)=\inf \bigg\{r \bigg| \lim_{n\rightarrow \infty} \big [\inf_\Lambda \|\Lambda(P^{\otimes nr}_+) - \rho^{\otimes n}\|\big] =0 \bigg\}
$$
where the second infimum is taken over all trace preserving LOCC $\Lambda$ and $P_+$ is the maximally entangled singlet state. We have used the trace norm in this definition, but some other measures work as well. Note that the equivalence $E^\infty_F(\rho)=E_C(\rho)$ is highly non-trivial as $E^\infty_F$ gives rise to a protocol with a variable number of input singlets, while $E_C$ works with a fixed number of input singlets. Now even more surprising, it is conjectured that the entanglement of formation is additive, in which case we would have $E_C=E_F$. This additivity is one of the major open problems in quantum information theory (see Ref.~\cite{Shor03, AB03, MSW02} and references therein). 
Calculating the entanglement cost is very hard problem, and its value is only known so far in cases where one can prove the additivity \cite{VDC01, Yura03, MSW02, MY03,Shimono02,HSS02b}. Both the entanglement of formation and the entanglement cost can be proven to be convex \cite{DHR01}.

We now move on to the entanglement of distillation \cite{BDSW96, Rains98, Rains99}, quantifying how much pure state entanglement we can extract from a state $\rho$. Analogously to the entanglement cost we define
\begin{align}
\label{defentdis}
E_D(\rho)=\sup \bigg\{r \bigg| \lim_{n\rightarrow \infty} \big [\inf_\Lambda \|P^{\otimes nr}_+ - \Lambda(\rho^{\otimes n})\|\big] =0 \bigg\}
\end{align}
where the second supremum is taken over all trace preserving LOCC $\Lambda$ and $P_+$ is the maximally entangled singlet state. Note that just like in the definition of $E_C$, the maximisation is over all protocols where the output has a fixed number of singlets. But similar to the entanglement cost, it has been proven rigorously \cite{Rains99} that allowing stochastic protocols does give rise to the same value for $E_D$. To elaborate a bit more on this point, suppose we have a protocol that from $k$ copies of a state $\rho$ produces $n$ singlets for half of the time and no singlets at all for the other half. However, the above definition only allows deterministic protocols, that is in our case we would like to replace this stochastic protocol by a deterministic protocol which produces exactly $n/2$ singlets. The equivalence of both definitions says that in the asymptotic regime this can always be done. The proof of the equivalence exploits the law of large numbers, as when a stochastic protocol is repeated a large number of times, it can be approximated with a deterministic protocol on a larger input.

From the above, it will be clear that $E_D\leq E_C$. The following theorem \cite{DHR01, HHH99} pushes this observation further, stating that under some reasonable conditions the entanglement cost and the entanglement of distillation are extremal measures.
\begin{theorem}
\label{mteced}
Let $E$ be a positive functional on the set of states ${\cal D}({\cal H}_A\otimes {\cal H}_B)$ satisfying for all states $\rho\in \cal D$
\begin{enumerate}
\item Normalisation: For any d-level maximally entangled state $P^d_+$, $E(P^d_+)=\log_2 d$.
\item Weak LOCC monotonicity: For any trace preserving LOCC map $\Lambda$: $E(\rho)\geq E(\Lambda(\rho))$.
\item Continuity: Let $(\rho_n)_{n\in \mathbb N}$ and $(\sigma_n)_{n\in \mathbb N}$ be any two sequences of states on $({\cal H}^A_n)_{n\in \mathbb N}$ and $({\cal H}^B_n)_{n\in \mathbb N}$ respectively. If $\|\rho_n-\sigma_n\|\rightarrow 0$ then
$$
\frac{E(\rho_n)-E(\sigma_n)}{1+\log_2 \dim {\cal H}_n}\rightarrow 0
$$ 
\end{enumerate}
Then:

(i) If $E$ is additive ($E(\rho^{\otimes n})=nE(\rho)$ for all $\rho\in \cal D$) then
$$
E_D(\rho)\leq E(\rho) \leq E_C(\rho).
$$

(ii) If the regularisation $E^\infty(\rho)=\lim_{n\rightarrow\infty} E(\rho^{\otimes n})/n$ exists for all $\rho\in \cal D$ then
$$
E_D(\rho)\leq E^\infty(\rho) \leq E_C(\rho).
$$
\end{theorem}

In some sense this theorem is a weakened version of the asymptotic reversibility for pure states. The above theorem can be used to provide both lower bounds to the entanglement cost and upper bounds to the entanglement of distillation. The conditions for the latter can be slightly relaxed \cite{HHH99}. 
\begin{theorem}
Let $E$ be a positive functional on the set of states ${\cal D}({\cal H}_A\otimes {\cal H}_B)$ satisfying for all states $\rho\in \cal D$
\begin{enumerate}
\item Weak LOCC monotonicity: For any trace preserving LOCC map $\Lambda$: $E(\rho)\geq E(\Lambda(\rho))$.
\item Partial sub-additivity: $E(\rho^{\otimes n})\leq nE(\rho)$ .
\item Continuity for the isotropic state (\ref{isotropic})
$$
\rho(\beta_d,d)=\frac{1}{d^2+\beta_d}(\ido+\beta_d P_{+}).
$$
That is, if we have a sequence of isotropic states $(\rho(\beta_d,d))_d \rightarrow P_+$ when $d\rightarrow \infty$, then we require that
$$
\lim_{d\rightarrow \infty} \frac{E(\rho(\beta_d,d))}{\log_2 d}\rightarrow 1.
$$
\end{enumerate}
Then $E_D(\rho)\leq E(\rho)$ for all $\rho$.
\end{theorem}

Thus for an entanglement measure to be an upper bound to the entanglement of distillation it is enough to be continuous in the neighborhood of the maximally entangled state. We proceed by giving some examples of entanglement measures that are strictly larger than $E_D$. The relative entropy of entanglement $E_R$ was the first example of an upper bound to $E_D$ \cite{VPRK97, VP98,HHH99, Rains98}, this follows relatively easily from the previous theorem. Another early upper bound to $E_D$ is given by the logarithmic negativity $E_N$ \cite{VW01b}, and again one easily verifies that it fulfills the conditions of the previous theorem. Note that both $E_R$ and $E_N$ do not satisfy the requirements of Theorem~\ref{mteced}. 

Rains showed in a seminal paper \cite{Rains00} that the previous two measures can be combined as
$$
E_{\text{Rains}}(\rho)=\min_{\sigma\in {\cal D}} \left(S(\rho||\sigma)+\log_2\|\sigma^{T_B}\|\right),
$$
providing an upper bound to the distillable entanglement. Note that here the minimum is taken over \emph{all} states, not just the separable ones. The Rains bound gives an upper bound to the PPT distillable entanglement, that is the entanglement that can be distilled via PPT preserving protocols. Since these protocols include LOCC operations, the Rains bound provides also an upper bound to $E_D$. Despite this, $E_{\text{Rains}}$ is probably the strongest known upper bound known on $E_D$. Note that when $\sigma=\rho$ in the definition, we simply recover $E_N$. A weaker bound can also be obtained by minimising just over the PPT states, so that the last term vanishes. The resulting quantity is known the relative entropy of entanglement with respect to PPT states \cite{Rains98,Rains00b}
$$
E^{PPT}_R(\rho)=\min_{\sigma\in {\cal P}} S(\rho||\sigma).
$$
 
Measures that satisfy the requirements of Theorem~\ref{mteced} are the regularised version $E^\infty_R$ of the entropy of entanglement $E_R$ and the regularisation of $E^{PPT}_R$, $E^{PPT\infty}_R$. There is some evidence that $E_{\text{Rains}}=E^{PPT\infty}_R$, based on the equality for Werner states, although all that is known is that for qubits $E^{PPT\infty}_R\leq E_{\text{Rains}}$ \cite{Ishizaka03,AMVW02, AEJPVM01}.

Lower bounds to the entanglement of distillation are much harder to obtain. One way of obtaining a lower bound is by providing an explicit distillation protocol (see Section~\ref{distprot}) and calculating the rate. The most important lower bound up to date is given by the hashing inequality \cite{HHH00b,DW03c,BDSW96} 
\begin{align}
 E_D(\rho) \geq E_{D_\rightarrow}(\rho)\geq \max \{0, S(\rho_B)-S(\rho), S(\rho_A)-S(\rho)\}.
\label{hashineq}
\end{align}
Here $E_{D_\rightarrow}$ is the one-way distillable entanglement, that is the entanglement of distillation where classical communication is only allowed in one direction (for instance only from Alice to Bob). Note that this relation also has important implications on the theory of quantum channel capacities. For some states the value of $E_R$ coincides with this bound, and thus allows for exact calculation of $E_D$. Examples will be given in the next section.

\begin{figure}
\begin{center}
\psfrag{f}{$\Tr(\rho F)$}
\psfrag{N}{$E_N$}
\psfrag{EF}{$E_F$}
\psfrag{RE}{$E_R$}
\psfrag{REPPT}{$E^{PPT\infty}_R$}
\includegraphics[width=14.5cm]{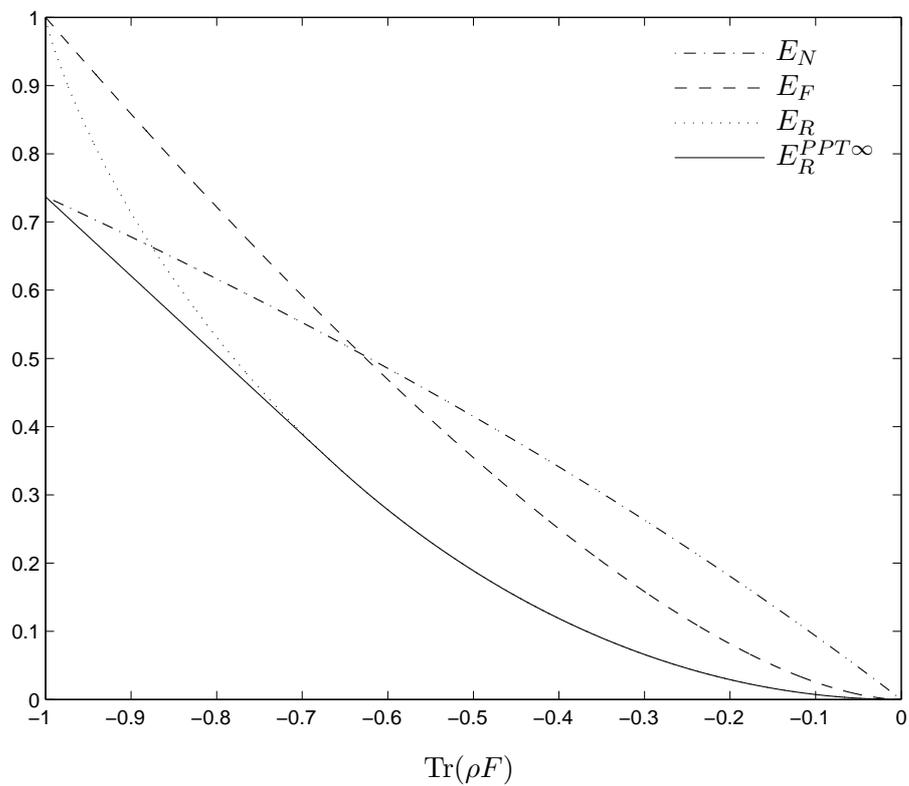}
\caption[Entanglement measures and the Werner states]{Entanglement measures $E_N$, $E_F$, $E_R=E^{PPT}_R$ and $E_{\text{Rains}}=E^{PPT\infty}_R$ as a function of the parameter $\Tr(\rho F)$ for the Werner states.}
\label{measuresd}
\end{center}
\end{figure}

From our treatment, it is easy to see that (see also \cite{Christandl06, PV05} for excellent treatments on entanglement ordering)
$$
E_F\geq E_R, E_C\geq E^\infty_R,E^{PPT}_R \geq E^{PPT\infty}_R\geq E_D
$$
and
$$
E_N,E^{PPT}_R \geq E_{\text{Rains}}\geq E_D.
$$
As an example, in Figure~\ref{measuresd} we have plotted known values of entanglement measures for the Werner states ($d=3$). The Werner states \cite{Werner89,Alb01} are states of the form 
$$
\rho = \frac{1}{d^2+\beta d}(\ido + \beta F),
$$
with $-1\leq \beta \leq 1$, which are entangled $\beta<-1/d$ or $-1\leq \Tr(\rho F)<0$. We will study this class of states extensively in the context of distillability in Section~\ref{nptbound}, see also Section~\ref{ssulo}. Because of their prominent symmetry the value of some important entanglement measures can be calculated exactly. In particular we have $E_R$=$E^{PPT}_R$ and $E_{\text{Rains}}=E^{PPT\infty}_R$ \cite{AMVW02, AEJPVM01,VW02}. Despite this, an analytical expression for both $E_C$ and $E_D$ is still out of reach. It is not even known whether the distillable entanglement for all entangled Werner states is strictly positive (see Section~\ref{nptbound})! In Section~\ref{distprot} we redraw the graph for $d=2$ for which the situation is somewhat simplified and for which good lower bounds are known on $E_D$.

\subsection{Irreversibility}
\subsubsection{Bound entanglement}
For pure states, entanglement is reversible in the sense that $E_D(\psi)=E_C(\psi)=E(\psi)$ for all pure states $\psi$. In this section we will show that this is not necessarily the case for mixed states. The first indication of this   was the discovery of bound entangled states, that is entangled states from which no entanglement can be distilled at all \cite{HHH98}. The class of bound entangled states includes the set of PPT states. We present the direct proof from Ref.~\cite{Alb01}.

\begin{theorem}
\label{pptdist}
For any PPT state $\rho$ the distillable entanglement is zero. PPT states are therefore bound entangled.
\end{theorem}
\begin{proof}
As we have seen in Section~\ref{qop} the set of PPT preserving protocols includes all LOCC transformations. Therefore LOCC cannot map a PPT state to an NPT state. Now we prove that PPT states are bounded away from the maximally entangled state. Let $\rho$ be a $d$-level PPT state. We have
\begin{align}
\Tr{\rho P_+}=\Tr{\rho^{T_B}P^{T_B}_+}=\frac{1}{d}\Tr{\rho^{T_B}F},
\end{align}
now since $\rho$ is PPT, $\tilde\rho=\rho^{T_B}$ is also a legitimate state and thus
\begin{align}
\Tr{\rho P_+}=\frac{1}{d}\Tr{\tilde\rho F}\leq \frac{1}{d},
\end{align}
where the inequality follows because the swap operator $F$ has eigenvalues $\pm 1$ only, and therefore has a mean smaller than one on any state. Thus whatever LOCC protocol we apply, PPT states can never achieve a higher overlap than $\frac{1}{d}$ with the maximally entangled state. Note that $\frac{1}{d}$ can be achieved by a product state.
\end{proof}

We come back to the question of the existence of NPT bound entanglement in Chapter~\ref{chapnptbound}. Now the above theorem would imply irreversibility if we could prove that $E_C(\rho)\geq 0$ for PPT entangled states $\rho$. However likely this might seem, it is not straightforward to prove this formally. This is because, all we know for entangled states is that $E_F(\rho)>0$, but in the asymptotic regime, it might be possible to construct PPT entangled states with an arbitrary small amount of entanglement. In Ref.~\cite{VC01} Vidal and Cirac obtained the following lower bound on $E_C$ (the proof is straightforward and we omit it here).
\begin{theorem}
Let $P$ denote the projector onto the range of a state $\rho$ on ${\cal H}_A\otimes {\cal H}_B$. If 
$$
\la ef|P^{\otimes N}|ef\ra\leq \alpha^N,
$$
for all product vectors $|ef\ra\in {\cal H}^{\otimes N}_A\otimes {\cal H}^{\otimes N}_B$, then $E_C\geq -\log_2 \alpha$.
\end{theorem}
This lower bound is relatively easy to calculate for some states \cite{VC01,VC01b}, in particular it was shown to be positive for some bound entangled state. The breakthrough however came with Ref.~\cite{YHHS05} which showed that $E_C(\rho)>0$ for all entangled states. Note that this was the one crucial piece missing in the original definition of entanglement (see Section~\ref{mainsectionchtwo}). The proof is relatively elaborate and proceeds in two steps. First a functional $G$ is derived which is positive on all entangled states and zero on all separable states. Although this quantity $G$ is by no means easy to evaluate, it serves the purpose as it is a lower bound to $E_C$.

\subsubsection{Generic irreversibility: maximally correlated states}
Irreversibility in entanglement transformation is not limited to bound entangled states. Consider for instance the class of maximally correlated states (MCS) \cite{Rains98,Rains00b} which can be written as
$$
\rho=\sum a_{ij} |ii\ra\la jj|.
$$
A necessary and sufficient condition for a state $\rho$ to be equivalent to a MCS after a possible local unitary transformation is given in Ref.~\cite{HH04}. The tag MCS reflects that when both parties perform the same measurement, they will obtain the same result. The Rains bound for this state is relatively easy to calculate and one obtains \cite{Rains98,Rains00b,WZ00,CY02b}
$$
E_{\text{Rains}}(\rho)=S(\rho_A)-S(\rho)=E_R(\rho)=E^{PPT}_R(\rho)=E^{PPT\infty}_R(\rho),
$$
and by virtue of an explicit distillation protocol it can be proven \cite{Rains00, Rains97} that this value equals the PPT distillable entanglement. Note that the relative entropy is additive when acting on maximally correlated states. In view of the hashing inequality we have:
\begin{theorem}
The distillable entanglement of a maximally correlated state $\rho$ is given by
$$
D(\rho)=S(\rho_A)-S(\rho)=E_R(\rho).
$$
\end{theorem}
Next we derive a sufficient condition for the additivity of the entanglement of formation, which allows for an exact expression of the entanglement cost $E_C$ of some MCS\@. Central is the notion of an entanglement breaking map \cite{Shor02,Ruskai03, HSR03}, this is a map ${\cal M}: {\cal L}({\cal H}_B) \rightarrow {\cal L}({\cal H}_{B'})$ such that for all $\rho \in {\cal L}({\cal H}_A) \otimes {\cal L}({\cal H}_B)$ the state $\ido_B \otimes {\cal M}(\rho)$  is separable. A map ${\cal M}$ is entanglement breaking if and only it can be written as
$$
{\cal M}(\sigma)=\sum_i \Tr(X_i\rho)\sigma_i,
$$
where $X_i\geq 0$ and $\sum_i X_i =\ido$ and $\{\sigma_i\}$ is some set of density operators. With this we have \cite{VDC01} the following theorem.
\begin{theorem}
Let $V$ be a subspace of ${\cal H}_{A_1} \otimes {\cal H}_{B_1}$. Define a map ${\cal M}: {\cal L}(V) \rightarrow {\cal L}({\cal H}_{A_1})$ by
$$
{\cal M}(\rho_V)= \Tr_{B_1} \rho_V
$$
for all $\rho_V \in {\cal L}(V)$. If $\cal M$ is entanglement breaking, then 
$$
E_f(\rho_V\otimes \sigma)=E_f(\rho_V) +E_f(\sigma)
$$
for all $\rho_V \in {\cal L}(V) \subset {\cal L}({\cal H}_{A_1} \otimes {\cal H}_{B_1})$ and all $\sigma \in {\cal L}({\cal H}_{A_2} \otimes {\cal H}_{B_2})$. In particular, for all $\rho_V \in {\cal L}(V)$ we have $E_C(\rho_V)=E_F(\rho_V)$.
\end{theorem}

Let us now apply this to the maximally correlated states \cite{HSS02b,VDC01, VWW03}. Any MCS is supported on the span of $\{|ii\ra\}$, so let $|\psi\ra=\sum_i |\psi_i\ra_{C}|ii\ra_{A_1B_1} $, then
$$
\ido_C\otimes {\cal M}(|\psi\ra\la \psi|)=\Tr_{B_1}(|\psi\ra\la \psi|)=\sum_i |\psi_i\ra\la \psi_i|_{C}\otimes |i\ra\la i|_{A_1}.
$$
In other words, the map $\cal M$ associated to $V$ is entanglement breaking, and thus $E_F(\rho)=E_C(\rho)$ for all MCS $\rho$. The entanglement of formation can be calculated exactly for qubits, and comparing the entanglement cost with the entanglement of distillation, one obtains generic irreversibility for all two-qubit MCS\@. In view of these results it would be worthwhile investigating whether a closed expression for the entanglement of formation can be obtained for MCS of arbitrary dimension. There is no reason however to suspect that irreversibility in higher dimension would not be generic.

\subsubsection{Reversibility and Distinguishability}
\label{quasipure}
The previous section raises the question if there are any reversible mixed states. Trivial examples constitute the so-called quasi-pure states \cite{EFPPW99, VWW03,SST01, HV00} of the form
$$
\rho=\sum_i p_i |ii\ra\la ii|\otimes |\psi_i\ra\la\psi_i|.
$$
A simple measurement of both parties in the $\{|i\ra \}$ basis will result in the collapse $\rho\rightarrow |ii\ra\la ii|\otimes  |\psi_i\ra\la\psi_i|$, therefore $E_C(\rho)=E_D(\rho)=\sum_i p_i E(\psi_i)$. Note that the process of identifying the tag $|ii\ra\la ii|$ is reversible in the asymptotic regime, hence the reversibility. More general quasi-pure states take the form 
$$
\rho=\sum_i p_i \rho_i \otimes |\psi_i\ra\la\psi_i|,
$$
with $\{\rho_i\}$ a set of separable and orthogonal states which are locally distinguishable. This latter condition is necessary as there are separable orthogonal states that cannot be distinguished locally \cite{BDFMRSSW98,YC03, HSSH03} with probability one. It seems to be an open question whether there are reversible truly mixed states which are not quasi-pure (see also \cite{HHH98c}).

Distinguishability has far deeper implications on distillation. We start with an example \cite{GKRSS01, CJY03}. Let $\{|B_{ij}\ra\}$ denote the set of the four Bell states $\frac{1}{2} (|00\ra\pm|11\ra)$ and $\frac{1}{2}( |01\ra\pm|10\ra)$, and define 
$$
\rho_n=\frac{1}{4} \sum_{i,j=0}^1 |B_{ij}\ra\la B_{ij}|^{\otimes n}.
$$
We focus on the case where $n$ is even. The state $\rho_2$ is separable and therefore $\rho_2^{\otimes m}$ is also separable. The relative entropy for $n=2m$ is bounded by
$$
E_r(\rho_{2m})\leq S(\rho_{2m}||\rho_2^{\otimes m})=2m-2.
$$
This last step follows by direct verification. Now the set of two copies of all four Bell states can be locally distinguished \cite{WSHV00}, so that by sacrificing two copies, one can distill $2m-2$ ebits from $\rho_{2m}$. Since the entanglement of distillation is bounded by the relative entropy we have $E_r(\rho_{2m})=E_D(\rho_{2m})=2m-2$. Other examples of using explicit LOCC distinguishing in a distillation process can be found in Ref.~\cite{GKR03}.

The link between distinguishability and distillation can be formalised as follows. First we start by recalling the Holevo bound (see for instance \cite{NC00}, Chapter 11 and 12). Suppose Alice prepares a state $\rho^x$ with probability $p_x$ and hands it over to Bob. He performs a measurement, with measurement outcome $y$. The random variable $X$ with values $x$ is characterised by the entropy $S(X)=-\sum_x p_x \log_2 p_x$ and similarly $Y$ is characterised by $S(Y)=-\sum_y p_y \log_2 p_y$. The joint entropy is then characterised by $S(X,Y)=-\sum_{x,y} p(x,y) \log_2 p(x,y)$. With these definitions the mutual information is defined by $S(X:Y)=S(X)+S(Y)-S(X,Y)$ and quantifies how much information $X$ and $Y$ have in common. The accessible information $I_{\text{acc}}$ is the maximum mutual information over all possible measurement strategies of Bob. As an example, suppose that Alice prepares a state in $d$ possible orthogonal states, then clearly Bob can always access this information completely, and therefore $I_{\text{acc}}=\log_2 d$. When Alice prepares non-orthogonal states, this value will be smaller. Now, the Holevo bound is an upper bound to the accessible information:
$$
S(X:Y)\leq I_{\text{acc}} \leq S(\rho)-\sum_x p_x S(\rho^x)\leq N
$$
with $\rho=\sum_x p_x \rho^x$, and $N=\log_2 \dim {\cal H}$. In particular it follows that $N$ qubits can be used for information transmission of at most $N$ bits. Now we generalise this in the scenario that Charlie prepares a bipartite state and gives a half both to Alice and Bob, who can only perform LOCC \cite{BHSS03,GJKKR04,HOSS04}. Note that the local accessible information is bounded by the accessible information $I^{\text{LOCC}}_{\text{acc}}\leq I_{\text{acc}}$. A Holevo-like bound can be obtained in this way (we present a strongly simplified version):
\begin{theorem}
The local accessible information of a quantum state $\rho \in {\cal L}({{\cal H}_A} \otimes {{\cal H}_B})$ is bounded by
$$
I^{\text{LOCC}}_{\text{acc}}(\rho) \leq N- E_F(\rho)- E_F(\tilde \rho),
$$
with $N=\log_2 \dim {\cal H}$. Here $\tilde\rho$ is the residual state, the state left-over by Alice and Bob after a particular measurement process. The amount of entanglement $E_F(\tilde\rho)$ that may be extracted during the process satisfies $E_{\text{distilled}}\leq E_F(\tilde \rho)$.
\end{theorem}

Thus to some extent, the entanglement (both initial and final) and local accessibility are complementary quantities. In our example above, the local accessible information $I^{\text{LOCC}}_{\text{acc}}(\rho_{2m})=2$, while $N=4m$ and $E_F(\rho_{2m})=2m$, so that $E_{\text{distilled}}\leq E_F(\tilde \rho)\leq 2m-2$ in the process of information extraction. The bound is here saturated; as in the process of information extraction, we distill $2m-2$ ebits and the extraction of locally accessible information is in this case the best distillation procedure (see above). The upper bound of local accessible information hence gives an upper bound on the entanglement of distillation that can be obtained through local distinguishing (see also \cite{SSL05}). We will encounter distillation protocols based on local distinguishing in the next section.

\section{Some important distillation protocols}
\label{distprot}
In this section we review some important distillation protocols. We analyse them in some detail for bipartite qubit systems, and only point to the relevant literature for higher level systems as already in the bipartite qubit scenario the main principles of distillation can be captured. From a practical point of view, it is also probably the most interesting. Note however that we discuss distillation protocols from a theoretical point of view, with disregard for practical limitations and operational simplicity. The foundations of all current distillation protocols have been set out in two brilliant papers Ref.~\cite{BDSW96, BBPSSW96} which we will follow closely. It is somehow remarkable how relatively little progress has been made since these were originally published.

\subsection{Recursive protocols}
\subsubsection{The recurrence protocol (IBM protocol)}
Being the first protocol \cite{BDSW96, BBPSSW96}, the recurrence protocol is also the simplest. Alice and Bob start with a large pool of identical qubit states with fidelity $Y=\Tr(\rho P_+)\geq 1/2$. In every round of the protocol two pairs $\rho^{\otimes 2}(Y)$ get consumed, and with some probability, a single pair $\rho'(Y')$ is produced with $Y'>Y$. Doing this a large number of times, the original pool of states $\rho(Y)$ will be transformed into a smaller pool of states $\rho'(Y')$. Repeating the protocol, this time on the new pool of states, they can increase the fidelity further, and so on (hence the name recurrence protocol).

So let Alice and Bob take two identical pairs with fidelity $Y$. In the first step of the protocol, they will bring
both pairs in the standard isotropic form (see also Chapter~\ref{chapsmidt} and Section~\ref{ssulo})
$$
\rho'=\frac{1}{4+\beta}(\ido + \beta P_+).
$$
States of this form are invariant under local bilateral unitaries $\rho'= (U\otimes U^*)\rho' (U\otimes U^*)^\dagger$ and it follows that we can bring any state into the isotropic form by performing a random local unitary (this is the so-called twirling operation) without changing its fidelity.

The unitary XOR (or CNOT) \cite{NC00} defined by
$$
U_{\text{\tiny XOR}}=|00\ra\la 00|+|01\ra\la 01|+|11\ra\la 10|+|10\ra\la 11|,
$$
behaves just like a classical XOR, flipping the second bit, conditionally on the first. Now the second step of the recurrence protocol consists of both parties implementing the local unitary XOR on their two halves. For the sequel we will call the first pair the source pair and the second pair the target pair. This local unitary operation called BXOR (bilateral XOR) entangles both Alice's and Bob's local halves, while it leaves the global entanglement between Alice and Bob unaltered. 

\begin{figure}
\begin{center}
\psfrag{f}{$Y$}
\psfrag{fp}{$Y'$}
\includegraphics[width=13.5cm,height=10cm]{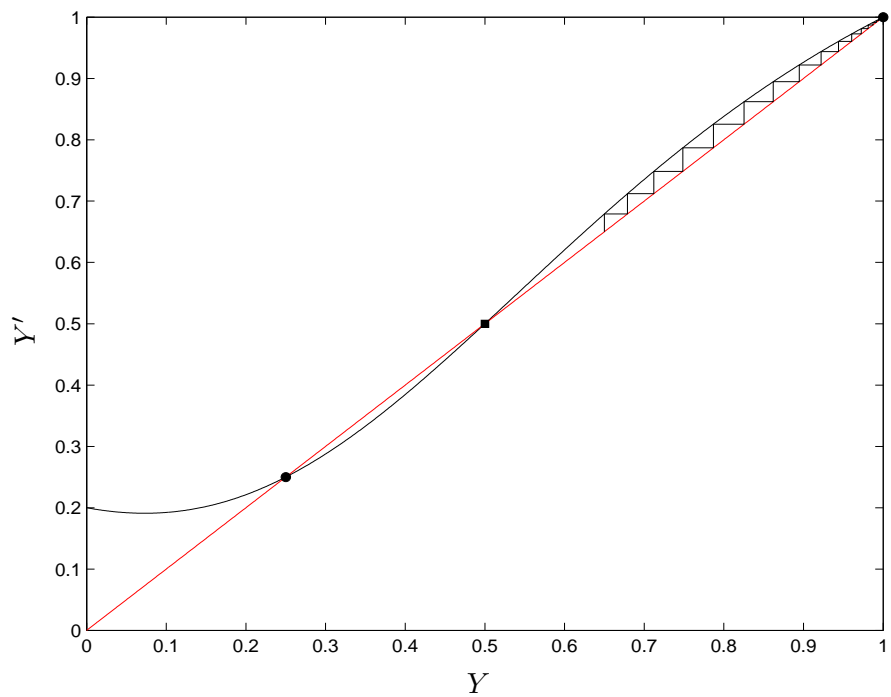}
\caption[One iteration in the recurrence protocol]{One iteration in the recurrence protocol. The curved line is $Y'(Y)$, while the staircase curve illustrates the evolution of the fidelity through repeated recurrence rounds, starting with $Y=0.65$. The black dots denote attractive fix-points, while the black square denotes a repulsive fix-point. This figure is an adaption of a similar one published in Ref.~\cite{AB02}.}
\label{reccu}
\end{center}
\end{figure}

Finally, Alice and Bob measure the second pair in the basis $\{|0\ra,|1\ra\}$. If the outcomes of their measurement coincide, it is not so hard to show that the fidelity of the first pair will change as
$$
Y'=\frac{9Y^2 + (1-Y)^2}{9Y^2+6Y(1-Y)+5(1-Y)^2}.
$$
Figure~\ref{reccu} illustrates this relation, and it can be seen that the fidelity strictly increases when $Y>1/2$. The first pair will no longer be isotropic, so that for further rounds in the protocol, the first step (twirling) will have to be repeated. When the measurement outcomes do not coincide, the first pair has to be disregarded, and the protocol was unsuccessful. It is important to note at this point that the exchange of the measurement outcomes requires two-way communication. 

As remarked at the start of this section, this round of the protocol will have to be repeated many times, until a sufficiently large pool of states with fidelity $Y'$ has been obtained. Then a new round can be started, using the new pool of states. In this way the recurrence protocol is able to transforms states with fidelity $Y>1/2$ into states with fidelity arbitrary close to one. The probability of doing so however tends to zero as $Y\rightarrow 1$, as in every round at least half of the states are disregarded.

\subsubsection{The QPA protocol (Oxford protocol)}
The QPA protocol \cite{DEJMPS96, Macchiavello98} was introduced in the context of quantum cryptography (Quantum Privacy Amplification) and is an improved version of the recurrence protocol. It achieves a certain fidelity with higher probability and does not require twirling. In fact it works in an identical fashion as the recurrence protocol, except that the twirling operation is replaced by a rotation. For completeness, the protocol consists of the following steps:
\begin{enumerate}
\item Denote $|\phi\ra=\frac{1}{\sqrt 2}(|0\ra - i|1\ra)$ and define a unitary operator $R=|\phi\ra\la 0| -i |\phi^*\ra\la 1|$. Then Alice implements $R$ on each of her qubits and Bob implements $R^\dagger$ on his qubits. It can be verified that this amounts to ordering the coefficients $p_{ij}$ (see below), such that $p_{00}\geq p_{01}\geq p_{10}\geq p_{11}$ \cite{DNMV02} .
\item The parties implement the unitary BXOR on their two halves.
\item Alice and Bob measure the second pair in the basis $\{|0\ra,|1\ra\}$ and keep the first pair if and only if their measurement outcomes coincide. 
\end{enumerate}

To see that this is a distillation protocol, let us write an arbitrary state in the Bell basis
$$
\rho=\sum_{i,j=0}^1 p_{ij} |B_{ij}\ra\la B_{ij}| + \sum_{i,j\neq k,l}^1 p_{ij,kl} |B_{ij}\ra\la B_{kl}|,
$$
with
\begin{align*}
|B_{00}\ra=\frac{1}{\sqrt 2} (|00\ra + |11\ra), \quad & \quad |B_{10}\ra=\frac{1}{\sqrt 2} (|00\ra - |11\ra),\\
|B_{01}\ra=\frac{1}{\sqrt 2} (|01\ra + |10\ra), \quad & \quad |B_{11}\ra=\frac{1}{\sqrt 2} (|01\ra - |10\ra).
\end{align*}
Here the first index will be called the phase index, and the second the shift index. 

In this way, we can describe one step in the protocol as the mapping $\{p_{ij}, p_{ij,kl}\} \rightarrow \{\tilde p_{ij}, \tilde p_{ij,kl}\}$. Now by going through the algebra it can verified that the coefficients $\tilde p_{ij}$ only depend on $p_{ij}$, more precisely:
\begin{align*}
\tilde p_{00}=\frac{p^2_{00}+p^2_{11}}{P}, \qquad & \qquad \tilde p_{10}=\frac{2p_{01}p_{10}}{P},\\
\tilde p_{01}=\frac{p^2_{01}+p^2_{10}}{P}, \qquad & \qquad \tilde p_{11}=\frac{2p_{00}p_{11}}{P},
\end{align*}
with $P=(p_{00}+p_{11})^2+(p_{01}+p_{10})^2$. It is not completely straightforward, but it can be proven \cite{Macchiavello98} that when $p_{00}=\Tr(\rho P_+)> 1/2$ the protocol converges to $P_+$ (that is $p_{00}\rightarrow 1$).

\subsubsection{Local permutations}
\label{ssslp}
Several improvements of the early recursive protocols have been proposed \cite{Metwally01, HLC03, MS00, OK99, MO06}. It has also been realised that the recurrence and the QPA protocol can be interpreted as coming from a quantum stabilizer code, and how to convert an arbitrary quantum stabilizer code to a distillation protocol \cite{Matsumoto03, WMU05, AG04}. As it turns out, all these different schemes can to a large extend be captured in the so-called local permutations schemes \cite{DNMV02, HDD04}.

The main idea is to start with $n$ pairs every round, and at the end measure $m$ pairs. Such a protocol will be called an $[n,m]$-protocol, the QPA and recurrence protocols are thus $[2,1]$ protocols. Naturally, for $n>2$ the XOR operation will have to be replaced by some multipartite unitary entangling operator. One way of doing this stems from analysing the BXOR operation in a bit more detail. Note that the XOR operation can be written as
$$
U_{\text{\tiny XOR}}|i\ra |j\ra=|i\ra |i\oplus j\ra,
$$
with $\oplus$ addition modulus 2. From this follows that the BXOR operation acts on two Bell states as (note that as usual the two Bell states are shared by both parties) \cite{VW02b}
$$
U_{\text{\tiny BXOR}}|B_{ij}\ra |B_{kl}\ra=|B_{i \oplus k,j}\ra |B_{k ,j\oplus l}\ra.
$$
Thus although BXOR is an entangling operation, it happens to be non-entangling on Bell states\footnote{To see that BXOR, or thus XOR is entangling it is sufficient to consider an example such as $U_{\text{\tiny XOR}} (|00\ra +|10\ra)= |00\ra+|11\ra$.}. In view of the above we write $ijkl \overset{\text{\tiny BXOR}}{\longrightarrow} i'k'j'l'$ and interpreting $ijkl$ as a bitstring, we can attach a number from 1 to $4^2$ to it. It is then not so hard to show that in this sense the BXOR corresponds to a permutation of $4^2$ elements. Generalising to $n$ pairs, one can consider permutations of $4^n$ elements, such that the corresponding unitary is local. The aim of working with this restricted set of unitaries, rather than say, all local unitaries or even all permutations is to make optimisation easier. Furthermore, an elegant closed expression for these special permutations is known \cite{DNMV02}. The problem of finding an optimal $[n,m]$-protocol is highly non-trivial, as a trade off has to be found between the increase in fidelity and the probability at which this happens. Also, in general, the choice of the permutation will be state dependent. Once a suitable permutation has been found, the distillation protocol is identical to the QPA protocol, except that in the second step the BXOR need to be replaced with the corresponding local unitary and in the third step, $m$ pairs need to be measured. 

Note that recurrence protocols can be readily generalised to higher level states $\rho$ with $\Tr(\rho P_+)>1/d$ by replacing the XOR by some multilevel entangling unitary \cite{ADGJ01, HH97, MDN03, BMD05}.

\subsection{Hashing and breeding}
\label{hashbreed}
The hashing protocol \cite{BDSW96, SS96}, unlike the recurrence schemes has a strictly positive yield for Bell diagonal states with $S(\rho)\leq 1$. Somewhat contradictorily, the protocol does only need 1-way communication. In this section we present a simplified version, the breeding protocol \cite{BBPSSW96}, which assumes that Alice and Bob share initially $S(\rho)$ pure singlets (we call them here pre-distilled singlets) per pair of $\rho$. However as it turns out, both protocols give rise to the same yield.

In the breeding protocol Alice and Bob share an arbitrarily large amount $n$ of identical Bell diagonal states:
$$
\rho^{\otimes n}=\sum_{k_1,l_1\cdots k_n,l_n} p_{k_1,l_1\cdots k_n,l_n} P_{k_1l_1} \otimes \cdots \otimes P_{k_nl_n},
$$
with $P_{a b}=|B_{ab}\ra\la B_{ab}|$. Note that any state can be made Bell diagonal by a certain discrete twirl operation (see Section~\ref{ssulo}). The main idea of the breeding/hashing protocols is now to interpret the above as Alice and Bob sharing $P_{k_1l_1} \otimes \cdots \otimes P_{k_nl_n}$ with probability $p_{k_1,l_1\cdots k_n,l_n}$. This way of reasoning is perfectly valid as the density operator contains all physical observable information, independently of the preparation procedure, and the state $\rho^{\otimes n}$ can be thought of as prepared in this way. The goal of Alice and Bob is then to identify the specific sequence $(\vec k, \vec l)=(k_1, \cdots, k_n, l_1, \cdots l_n)$, after which Bob can simply apply local unitary transformations to obtain $P^{\otimes n}_{00}$.

The breeding protocol uses pre-distilled singlets as target states for BXOR operations to gain information on the source states characterised by the sequence $(\vec k, \vec l)$. If Alice and Bob implement a BXOR with a pure target $|B_{0m}\ra$:
$$
U_{\text{\tiny BXOR}}|B_{kl}\ra |B_{0m}\ra=|B_{k,l}\ra |B_{0,l\oplus m}\ra,
$$
then the shift index of the target pair toggles when the source pair has shift index $l=1$, while the source pair remains unchanged. So when Alice and Bob perform BXOR operations on each pair of a subset of $n'$ pairs with a pure singlet as a fixed target state, they can identify the number of source pairs with shift index 1, by measuring (locally) the shift index of the target state. Thus, this procedure allows them to identify the parity of the shift indices in the subset. Similarly, by replacing BXOR by a different local unitary, they can determine the parity of the phase indices in the subset. 

The question is now how many parity checks are needed to gain complete information of the sequence $(\vec k, \vec l)$. This is important, as each parity check consumes 1 pair of their initially pre-distilled singlets. A total random sequence of $n$ pairs needs $2n$ parity checks to uniquely identify, so in doing so we would lose more entanglement than we gain. However, for $n\rightarrow \infty$ the sequence will belong to the class of $2^{nS(\rho)}$ typical sequences with probability going to one. It can be shown that the probability that two different sequences have the same parities after $r$ checks on random subsets is smaller than $\frac{1}{2^r}$. So typically, after every parity check the list of typically sequences reduces by a factor $\frac{1}{2}$ and thus $r=nS(\rho)$ parity checks will be sufficient to identity the sequence in the asymptotic regime (see Ref.~\cite{BBPSSW96} for a rigorous analysis). The yield of the breeding protocol is therefore
$$
 \lim_{n\rightarrow \infty} \frac{n-nS(\rho)}{n}=1-S(\rho),
$$
which agrees with the hashing inequality~(\ref{hashineq}). Both hashing and breeding can be generalised to multilevel systems \cite{VW02b}. As a side remark, note that there is a close correspondence between 1-way distillation protocols and quantum error correcting codes. The most important implication of this correspondence is the quantum noisy coding theorem, which follows from the hashing inequality (recall that this inequality gives a lower bound on the 1-way distillable entanglement) \cite{BDSW96, SS96, HHH00b, DW03c}. 

\begin{figure}
\begin{center}
\psfrag{f}{$\Tr(\rho F)$}
\psfrag{Best}{$E_{\tilde D}$}
\psfrag{EF}{$E_F$}
\psfrag{RE}{$E_R$}
\psfrag{Hash}{$H$}
\includegraphics[width=13.5cm]{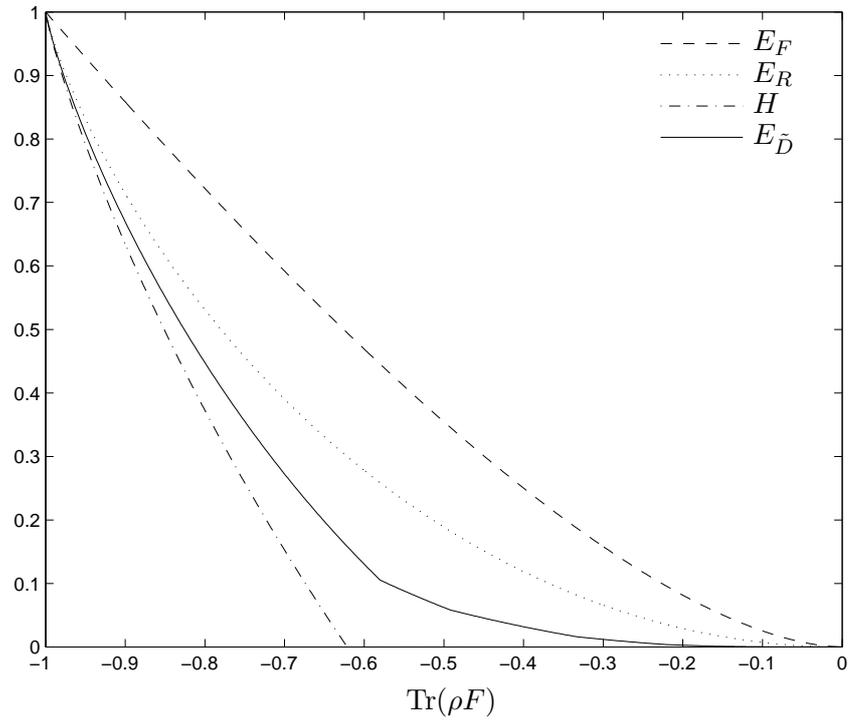}
\caption[Entanglement measures and the two-qubit Werner states]{Entanglement measures $E_F$ and $E_R=E^{PPT}_R=E_{\text{Rains}}=E^{PPT\infty}_R$ as a function of the parameter $\Tr(\rho F)$ for the Werner states. $H$ and $E_{\tilde D}$ are the distillation rates of the hashing protocol and the protocol of Ref.~\cite{HDD06} respectively (the data was obtained courtesy of E.~Hostens).}
\label{measurestwo}
\end{center}
\end{figure}

In Figure~\ref{measurestwo}, the value of distillable entanglement by the hashing protocol (denoted $H$) has been plotted for the Werner states. As can be seen from the figure, the hashing protocol works only for $\Tr(\rho F)<-0.62$. This can be bypassed by preceding the hashing protocol by a few recurrence steps, increasing the fidelity above this threshold. While the recurrence protocol exploits two-way communication, only the hashing protocol is a proper asymptotic protocol. By introducing two-way communication on the asymptotic level, small improvements can be made \cite{VV04, HDD06}. The currently best known distillation rate for the Werner states $E_{\tilde D}$ has been plotted in Figure~\ref{measurestwo}.

\chapter{Distillability, activation and {NPT} bound entanglement}
\label{chapnptbound}
Having encountered PPT bound entangled states in the previous chapter, this chapter deals with the more general question of distillability of general quantum states. We call a state distillable if and only if the distillable entanglement is strictly larger than zero. In three seminal papers \cite{HH97, HHH97, HHH98} the Horodeccy derived the fundamentals of the theory of distillability. Namely they showed that (i) any two-qubit entangled state can be distilled; (ii) any general $d$-level state $\rho$ is distillable if and only if there exists a two-qubit subspace on which $\rho^{\otimes n}$ is entangled for some $n$; (iii) any state violating the reduction criterion can be distilled. The second statement has a number of nontransparent reformulations. Here we present a unified presentation of these and other distillability theorems, using tools borrowed from the separability problem. This allows us to systematically construct positive maps, with the same properties as in (iii) of the reduction criterion \cite{Clarisse04}. We give several other applications, including a numerical method for checking distillability which we apply to estimate the volume of distillable states. (Sections~\ref{ssulo}-\ref{utodt}). 

In the next section we address the question of NPT bound entanglement in detail. After reviewing previous results \cite{DSSTT00, Watrous03, DCLB99}, we show that this question is equivalent to a specific Schmidt number problem and to a separability problem. In particular, we show it is equivalent to the existence of a certain class of PPT entangled states. We end this section with a set of one parameter states which we conjecture to exhibit all forms of distillability (the rainbow states) \cite{Clarisse05}.
 
As bound entangled states are not directly useful for typical quantum information tasks such as teleportation, the question arises whether they have any non-classical use at all. We will see that bound entanglement can be activated. That is, bound entanglement can be released with the help of other (possible bound) entangled states. A first activation-like effect was presented in Ref.~\cite{HHH98b}, but the main results were obtained in Ref.~\cite{EVWW01, Masanes05}. This is the topic of Section~\ref{actbent}.

\section{Classic results}
\label{ssulo}
\subsection{States symmetric under local unitaries}
In previous chapters we have encountered the isotropic states and the Werner states. These states are invariant under certain local unitaries, and because of their symmetry they have been exhaustively used in the study of distillability. In this section, we review briefly the general theory of states symmetric under local unitaries, as developed in Ref.~\cite{VW01}.

Let $G_0$ be a subgroup of the unitaries, or possibly the whole group of the unitaries. Then we can define the group $G$ of the unitaries consisting of all pairs of the form $U \otimes U'$ acting on a Hilbert space ${\cal H}={\cal H}_A\otimes {\cal H}_B$, where $U\in G_0$ and $U'$ is some given function of $U$. The set of bipartite states left invariant by the group $G$ will also be denoted as $UU'$-invariant instead of $G$-invariant. An arbitrary state can be projected onto an $UU'$-invariant state by twirling it
$$
{\cal T}_G(\rho) = \int_G dU ( U\otimes U') \rho (U\otimes U')^\dagger.
$$
Here the integral is performed according to the Haar measure on $G_0$. In particular, for a finite group, the integral can be replaced by the sum
$$
{\cal T}_G(\rho) = \frac{1}{|G|}\sum_{U\in G} U\otimes U' \rho (U\otimes U')^\dagger.
$$
The commutant $G'$ of the group $G$ is defined as the set of all operators invariant under $G$-twirling, or equivalently the set of all operators $A$ such that $[U,A]=0$ for all $U\in G$. It follows that the set of bipartite states left invariant by $G$ is just the intersection of the state space with the commutant $G'$ of the group $G$. 
Note that there always exist a finite basis of operators $B(G')$ spanning the commutant $G'$. Now for any operator $C$ it is straightforward to check that
$$
\Tr ( {\cal T}_G(\rho) C)=\Tr ( \rho {\cal T}_G(C) ).
$$
This relation has a number of important consequences:
\begin{enumerate}
\item A $G$-invariant state $\rho$ is uniquely determined by the expectation values $\Tr ( \rho A )$ for all $A\in G'$. In particular, if $B(G')$ is finite, then the set of $G$-invariant states can be characterised by a finite set of parameters $g_A=\Tr(\rho A)$, for all $A\in B(G')$.
\item For an arbitrary state $\rho$, the parameters $g_A=\Tr(\rho A)$, for all $A\in B(G')$, are invariant under twirling. From the previous point it follows that in order to calculate the image of $\rho$ under twirling it is sufficient to calculate these parameters.
\item Suppose we need to calculate an extreme for some $G$-invariant state $\rho$ such as $\min_C \Tr ( \rho C)$, where the operators $C$ belong to some set $\cal C$. It is then sufficient to minimise over the (smaller) set) ${\cal T}_G({\cal C})$.
\end{enumerate}

For the Werner states \cite{Werner89}, which are $UU$-invariant, the commutant is spanned by $\ido$ and $F$. The isotropic states \cite{HH97, Rains98} are $UU^*$-invariant and a basis is given by $\ido$ and $P_+$. Note that for qubits, the Werner and isotropic states are related via a local unitary, while this does not hold for higher dimensions.
As we have seen in the previous section, some distillation protocols make extensive use of twirling. As remarked in Section~\ref{gsr} for the Werner and isotropic states, it can be shown that the twirl can be written as a finite sum (\cite{DCLB99, BDSW96}). In practice however, twirling can be done simply by performing a single random $UU$ or $UU^*$ transformation and forgetting which specific unitary was used. 

The Bell diagonal states encountered in the previous chapter are invariant under the group
$$
G=\{\ido \otimes \ido, -\sigma_x\otimes \sigma_x, -\sigma_y\otimes \sigma_y, -\sigma_z\otimes \sigma_z\},
$$
with as usual $\sigma_x,\sigma_y,\sigma_z$ the Pauli matrices. A generalisation of Bell diagonal states and their symmetry group for arbitrary dimensions is presented in Ref.~\cite{VW02b}.

From basic symmetry groups one can generate others using tensor products. Consider the case where we have two symmetry groups $G=\{U\otimes U'\}$ and $K=\{V\otimes V'\}$ acting respectively on ${\cal H}_1={\cal H}_{A}\otimes {\cal H}_{B}$ and ${\cal H}_2={\cal H'}_{A}\otimes {\cal H'}_{B}$. Let $B_G$ and $B_K$ be a basis for the $UU'$- and the $VV'$-invariant states respectively. 
Then a basis for the $UU'VV'$-invariant states acting on ${\cal H}_1\otimes {\cal H}_2$ will be given by $B_G \otimes B_K$. In what follows, we will number the systems belonging to party $A$ with odd numbers and party $B$ with even numbers. As an example, a basis for the $UUVV$-invariant states is given by the operators $\{\ido_{12}\otimes \ido_{34}, F_{12} \otimes \ido_{34}, \ido_{12}\otimes F_{34}, F_{12}\otimes F_{34} \}$.
Imposing the extra condition that the coefficients of $F_{12} \otimes \ido_{34}$ and $\ido_{12}\otimes F_{34}$ should be the same, we end up with the so-called $UUVVF$-invariant states as introduced in Ref.~\cite{VW01}, where they were used as a counterexample to the additivity conjecture for the relative entropy of entanglement. We return to this set of states in Section~\ref{nptbound}.

\subsection{Fundamental theorems}
In this section we present the three basic theorems (Theorems~\ref{thhhh}, \ref{distilth} and \ref{ftthree}) on distillability by the Horodeccy \cite{HH97, HHH97, HHH98}, as mentioned in the introduction. We also present a first result \cite{HH97} on NPT bound entanglement. Because of their importance we present full proofs.

\begin{theorem}
All entangled two-qubit states $\rho$ are distillable ($E_D(\rho)>0$).
\label{thhhh}
\end{theorem}
\begin{proof}
The proof consists of locally transforming $\rho$ into a state $\rho''$, which has a singlet fraction larger than a half. In Section~\ref{hashbreed} we have seen that for such states the distillable rate is strictly positive.

Since $\rho$ is an entangled two-qubit state it has an negative partial transposition $\rho^{T_B}\not \geq 0$. We use the same technique as in (\ref{ewfiltering}) to transform the state such that it is detectable by a suitable entanglement witness. Let us denote the eigenvector corresponding to the negative eigenvalue of $\rho^{T_B}$ with $|\psi\ra$, so that $\la\psi| \rho^{T_B}|\psi\ra<0$. Now any vector $|\psi\ra$ can be written as $|\psi\ra= A\otimes \ido |\psi_+\ra$ (see (\ref{filtering})). With this we get
$$
\Tr[(A^\dagger \otimes \ido) \rho( A \otimes \ido) P^{T_B}_+]<0.
$$
Now let Alice perform a filtering operation corresponding to the Kraus operators $A$. Then the state becomes 
$$
\rho'=\frac{(A^\dagger \otimes \ido) \rho( A \otimes \ido)}{\Tr[(A^\dagger \otimes \ido) \rho( A \otimes \ido)]}.
$$
In this way we have transformed the state such that $\Tr \rho'F <0$, as $P^{T_B}_+=F$. Now in ${\cal L}(\mathbb C^2\otimes \mathbb C^ 2)$ the swap operator can be written as $F=\ido-2 P_-$, with $P_-$ the projector upon the state $|\psi_-\ra=(|01\ra-|10\ra)/\sqrt{2}$. So that we have
$$
\Tr \rho' P_- > \frac{1}{2},
$$
a simple local rotation can then map $\rho'$ onto a state $\rho''$ such that $\Tr \rho'' P_+ > \frac{1}{2}$. 
\end{proof}

The next theorem is the fundamental theorem of distillation, we present a slightly different proof as compared to Ref.~\cite{HHH98}.
\begin{theorem}
\label{distilth}
A state $\rho$ on ${\cal H}={\cal H}_A \otimes {\cal H}_B$ is distillable if and only if there exist two dimensional projectors $P:{\cal H}^{\otimes n}_A \rightarrow \mathbb C^2 $ and $Q:{\cal H}^{\otimes n}_B \rightarrow \mathbb C^2 $ such that for some $n\geq 1$ the state 
\begin{align}
\label{pqpq}
\rho''=(P\otimes Q)\rho^{\otimes n} (P\otimes Q)^{\dagger}
\end{align}
is entangled. Since the resulting state acts on $ \mathbb C^2\otimes \mathbb C^2$, this is equivalent to $\rho''$ having a negative partial transposition.
\end{theorem}
\begin{proof}
The condition is clearly sufficient, as when such projectors can be found, many copies of $\rho''$ can be produced, and from the previous theorem it follows that these can be distilled. Suppose now that the state $\rho$ is distillable, and thus $E_D(\rho)=r>0$. Let us fix arbitrarily small $\epsilon'>0$. From the definition of distillable entanglement (\ref{defentdis}), it follows that for any $\epsilon>0$ there exists a number $n$ and a trace preserving LOCC $\Lambda$ such that
$$
\|\Lambda(\rho^{\otimes n})-{P^{\otimes rn}_+}\|<\epsilon,
$$
with $P_+$ the maximally entangled two-qubit singlet. This implies that, by choosing $\epsilon$ small enough, and disregarding all but one approximate copy of $P_+$, we can find an LOCC protocol $\Lambda'$ such that 
$$
\|\Lambda'(\rho^{\otimes n})-{P_+}\|<\epsilon',
$$
We write this LOCC protocol in the separable form (with $N$ a normalisation factor)
\begin{align}
\label{tiasum}
\rho'=\Lambda'(\rho^{\otimes n})=\frac{1}{N}\sum_i (A_i\otimes B_i) \rho^{\otimes n} (A_i\otimes B_i)^\dagger.
\end{align}
Since this is approximately $P_+$ we know that $\rho'$ is entangled and we can assume that all $(A_i\otimes B_i)$ project onto the same two-qubit space. Thus every $A_i$ and $B_i$ can be written as
$$
A_i=|0\ra\la\psi_{Ai}| +|1\ra\la\phi_{Ai}|, \qquad B_i=|0\ra\la\psi_{Bi}| +|1\ra\la\phi_{Bi}|.
$$
Clearly, since $\rho'$ is entangled, at least one term in this sum (\ref{tiasum}) must be entangled, for instance for $i=1$. We denote the projector on the space spanned by $\{\phi_{A1},\psi_{A1}\}$ as $P$ and the projector on $\{\phi_{B1},\psi_{B1}\}$ as $Q$. We have thus shown that the following two-qubit state
$$
\rho''=(A_1\otimes B_1) \rho^{\otimes n} (A_1\otimes B_1)^\dagger= (A_1\otimes B_1)(P\otimes Q) \rho^{\otimes n}(P\otimes Q)^\dagger (A_1\otimes B_1)^\dagger
$$
is entangled. It follows that also the two-qubit state $(P\otimes Q) \rho^{\otimes n}(P\otimes Q)^\dagger$ must be entangled.
\end{proof}

If the condition of the theorem is satisfied for a particular number $n$, then we call the state \emph{pseudo-$n$-copy distillable} or in short \emph{$n$-distillable} \cite{DCLB99, DSSTT00}. This result has proven to be extremely useful and is often used as a definition of distillability. For completeness, we state the following corollary (see Ref.~\cite{HH97, DSSTT00}):

\begin{corollary}
\label{cpmdist}
A state $\rho$ on ${\cal H}={\cal H}_A \otimes {\cal H}_B$ is $n$-distillable if and only if there exist some local operators $A$ and $B$ mapping onto $\mathbb C^2$ such that 
$$
\Tr(F (A\otimes B) \rho^{\otimes n} (A \otimes B)^\dagger)<0
$$ 
or 
$$
\frac{\Tr(P_+ (A\otimes B) \rho^{\otimes n} (A \otimes B)^\dagger)}{\Tr((A\otimes B) \rho^{\otimes n} (A \otimes B)^\dagger)}>1/2,
$$ 
with $P_+$ and $F$ acting on $\mathbb C^2\otimes \mathbb C^2$.
\end{corollary}
\begin{proof}
This corollary follows directly from the previous theorem and the fact that any entangled two-qubit state violates both the partial transpose and reduction criterion. Via local filtering, see equation (\ref{ewfiltering}) and the subsequent discussion (see also the proof of Theorem~\ref{thhhh}), one can then transform these states such that they are detected by the witnesses $F$ and $\ido -2P_+$.
\end{proof}

\begin{theorem}
All states violating the reduction criterion are distillable.
\label{ftthree}
\end{theorem}
\begin{proof}
In a similar fashion as in the proof of the previous corollary, a state $\rho$ violating the reduction criterion can be locally transformed into a state $\rho'$, which can be detected by the witness $\ido -dP_+$, or
$$
\Tr(\rho' P_+)> \frac{1}{d}.
$$
As remarked in Section~\ref{ssslp}, for such states a recurrence protocol is available. To get a finite yield, one can then use a multilevel generalisation of the hashing protocol, see Section~\ref{hashbreed}. 

Another way \cite{Alb01} of showing this is to $UU^*$-twirl the state $\rho'$, after which it will become isotropic. The fidelity remains invariant under $UU^*$-twirling so that the state is still entangled. Now both parties apply on their half the operation $ P=|0\ra\la 0| +|1\ra\la 1|$. This projects the state $\rho'$ onto a two-qubit isotropic state $\rho''$ with $\Tr(\rho'' P_+)> \frac{1}{2}$, and such states are distillable, as we have seen in the discussion of distillation protocols (Section~\ref{distprot}).
\end{proof}

The second proof implies that all states violating the reduction criterion are distillable if and only if all entangled isotropic states are. A dual result can be obtained for the partial transpose criterion and the Werner states:

\begin{theorem}
\label{wernernptbound}
All entangled NPT states can be distilled if and only if all entangled Werner states can be distilled.
\end{theorem}
\begin{proof}
Since entangled Werner states have a negative partial transposition, one direction is trivial. To prove the other direction, note that as before, any NPT state $\rho$ can be locally transformed into a state $\rho'$ such that 
$$
\Tr(\rho' F)<0.
$$
Now twirling this state, we obtain an entangled Werner state $\rho''$ (because $\Tr(\rho' F)=\Tr(\rho'' F)<0$). If we are able to distill all Werner states, then also the original state is distillable.
\end{proof}
In Section~\ref{nptbound} we will present strong evidence that some of the entangled Werner states cannot be distilled, and hence for the existence of NPT bound entanglement.

We end this section on the fundamental theorems with a remark on finite copy distillation. We have seen that any distillable state is (pseudo)-$n$-copy distillable. The prefix $pseudo$ reflects that if we project upon a two dimensional subspace, we are only half way through our distillation process. Indeed, in the next step we would like to repeat this procedure $m$ times on batches of $n$ copies of $\rho$, giving us $m$ copies of the qubit pair $\rho'$. 
Finally we can use existing protocols to extract maximally entangled singlets from $\rho'^{\otimes m}$ for $m\rightarrow \infty$. Thus in general, even when a state is $n$-distillable we might need an infinite number of copies to get a finite yield of distillation.

It is then natural to ask whether there are any states which can be distilled using a finite number of copies, and in particular if there are any states which are single copy distillable. Clearly, pure states are single copy distillable
and so are the quasi-pure states encountered in Section~\ref{quasipure}. For generic qubit mixed states, it has been proven \cite{LMP98, Kent98} that this is not possible. For qubits the best strategy to increase the entanglement of formation is transforming the state into a Bell diagonal state \cite{KLM99}. 

The explicit form of this transformation was found in Ref.~\cite{VDD01b, CWYA02}. See Ref.~\cite{HD05} for bounds on single copy distillation of arbitrary level systems. Despite that in general single-copy distillation is not possible, the Horodeccy showed the existence of an effect which they called quasi-distillation \cite{HHH99b} (see also \cite{VDD01b}). In this scenario, it is possible to extract a pure maximally entangled singlet from some (specially constructed) single mixed state, but with probability going to zero, as the fidelity of the output state approaches this maximally entangled singlet. 

The problem of finite copy distillation for more than one copy of the state has been solved for qubits in Ref.~\cite{Jane02}. It was shown that $n\geq 2$ copies of a two-qubit mixed state $\rho$ can be distilled into a pure entangled state if and only if the range of $\rho$ cannot be spanned by product states. Surprisingly, this result is independent of $n$. This result is important as it shows that for entangled qubits, which are all 1-distillable, we need typically an infinite number of copies to distill pure entangled states. The reason is that although we can get a state with very high fidelity given a large number of copies (for instance by means of the recurrence protocol), in general this final state will still be mixed entangled. Only in the asymptotic limit of the number of copies, it is then possible to obtain a pure state.

\section{Uniform treatment of distillability}
\label{utodt}
This section contains an original contribution first published in Ref.~\cite{Clarisse04}. The motivation behind this work was an attempt to answer the following questions:
\begin{quote}
What is the special role the reduction criterion plays in the story of distillation? Are there any other positive maps with the same property? Can we find a decomposable map which is stronger than the reduction criterion, but so that all states detected by it are still distillable?
\end{quote}

\subsection{Borrowing tools from the separability problem}
We have seen in Chapter~\ref{entsepchap} that separable states form a convex set, from which follows that there exists an entanglement witness for each entangled state. Using the Jamio{\l}kowski isomorphism between operators and maps, this can be translated in terms of positive maps. The idea is to apply a similar reasoning to the set of distillable states. The crux is that 1-undistillable states also form a convex set. Indeed, from Theorem~\ref{distilth} and the linearity of the partial transpose, it follows that mixing 1-undistillable states can never yield a distillable state. 

The question is now to determine the corresponding witnesses. Notice that from Theorem~\ref{distilth} it follows \cite{DCLB99} that distillability is equivalent to the existence of a Schmidt rank two state $|\psi\ra=c_1|a_1,b_1\ra+c_2|a_2,b_2\ra$, with $\{|a_1\ra, |a_2\ra\}$ two orthonormal vectors in ${\cal H}^{\otimes n}_A$ and $\{|b_1\ra, |b_2\ra\}$ two orthonormal vectors in ${\cal H}^{\otimes n}_B$, and some $n$ such that 
\begin{align}
\la \psi | (\rho^{\otimes n})^{T_B}) |\psi\ra=\la \psi | (\rho^{T_B})^{\otimes n}) |\psi\ra<0.
\end{align}
Thus $D=|\psi\ra\la\psi|^{T_B}$, with $|\psi\ra$ Schmidt rank two, can be interpreted as a distillability witness. The following observation generalises this.
\begin{obs}
\label{firstobservation}
A state $\rho$ is 1-undistillable if and only if 
$$
\Tr(D\rho)\geq 0
$$
for all operators $D=\sigma^{T_B}$ with $\sigma$ a Schmidt number two state. 
\end{obs}
This observation is compatible with the fact that PPT states cannot be distilled as $D$ is decomposable by construction.
Note that to study the phenomenon of distillation it is sufficient to characterise the 1-distillable states; $n$-distillable states $\rho$ can be characterised by looking at one copy of $\rho^{\otimes n}$. 

Now in a similar fashion as the separability problem and Schmidt number problem, we have to look at map $\Lambda$ associated with the operator $D$ as given by the Jamio{\l}kowski isomorphism
$$
D=d(\ido\otimes \Lambda) P_+.
$$
Let us start with the simple case where $D=|\psi\ra\la\psi|^{T_B}$. As $D$ is a decomposable operator, the map $\Lambda$ is decomposable and of the form
$$
\Lambda= T \circ \Lambda^{CP}, \qquad \text{with} \qquad |\psi\ra\la\psi|=d(\ido\otimes \Lambda^{CP}) P_+.
$$
What more can we say about this completely positive map $\Lambda^{CP}$? Recall that the general form of such a map is given by (see Theorem~\ref{choipillis})
\begin{align}
\label{krauss}
\Lambda^{CP}(A)=\sum_i{V_i A V^\dagger_i},
\end{align}
where $V_i$ are arbitrary operators. Now let $|\psi\ra=\sum_{i=1}^k c_i |a_i b_i\ra$ be an arbitrary Schmidt rank $k$ state, we obtain after some algebra that the associated map\footnote{This fact was independently obtained in Ref.~\cite{AP03}.} is given by $\Lambda^{CP}(A)=V A V^\dagger$, with 
$$
V=\sum_{i=1}^k c_i |b_i\ra\la a^*_i|.
$$
In our case $k=2$, so that $V$ will have rank 2. 

Let us now move to the case where $D=\sigma^{T_B}$, with $\sigma$ an arbitrary Schmidt number two state. A similar line of reasoning shows that the associated map is given by $\Lambda=T \circ \Lambda^{CP}$, with $\Lambda^{CP}$ a CP map with rank two Kraus operators $V_i$. We call maps $\Lambda$ defined in this way \emph{two-decomposable}\footnote{This term has different meaning in other contexts, such as in Ref.~\cite{LMM03}.}. Note that the adjoint map of a two-decomposable map is two-decomposable. It is now an easy exercise to derive the necessary and sufficient condition of one-undistillability in terms of two-decomposable maps. For completeness, we present the derivation.

\begin{theorem}
\label{maintheorem}
A state $\rho$ is 1-undistillable if and only if
$$
(\openone\otimes \Lambda)(\rho)\geq 0
$$
for all two-decomposable maps $\Lambda$.
\end{theorem}
\begin{proof}
Suppose $\rho$ is 1-distillable, so that there exists a $D=|\psi\ra\la\psi|^{T_B}$, with $|\psi\ra$ a Schmidt rank-2 vector such that $\Tr(D\rho)<0$. Using the associated map this can be written as
$$
\Tr[(\openone \otimes\Lambda)(P_+)\rho]=\Tr[(\openone \otimes\Lambda^\dagger)(\rho)P_+]<0,
$$
and since $P_+$ is positive it follows that $\openone \otimes\Lambda^\dagger(\rho)\not\geq 0$, i.e.\ we have found a two-decomposable map that detects the state. To prove the converse, let $\Lambda=\Lambda^{CP}\circ T$ be a two-decomposable map such that $\openone \otimes\Lambda(\rho)$ has a negative eigenvalue. Denoting by $|\phi\ra$ the corresponding eigenvector, we get
\begin{align}
\label{keyeq}
\la\phi|\openone\otimes\Lambda^{CP}(\rho^{T_B}) |\phi\ra=\Tr[(\openone \otimes\Lambda^{\dagger})(|\phi\ra\la\phi|)\rho^{T_B}] <0.
\end{align}
It is sufficient to consider $\Lambda^{CP}(A)=V A V^\dagger$, with $V$ a rank-2 operator, therefore $\openone \otimes \Lambda^\dagger(|\phi\ra\la\phi|)= |\phi'\ra\la\phi'|$, with $|\phi'\ra$ Schmidt rank 2. Thus $\Tr( \rho^{T_B}|\phi'\ra\la\phi'|)<0$, and $\rho$ is 1-distillable.
\end{proof}

When $D=|\psi\ra\la\psi|^{T_B}$ the associated map is of the form
$$
\rho \rightarrow (\ido\otimes V) \rho^{T_B} (\ido\otimes V)^\dagger.
$$
It is instructive to compare this to Theorem~\ref{distilth}. The above mapping is basically a projection of the partial transposed state on a ${\mathbb C}^n \otimes {\mathbb C}^2$ (this special case was also derived in Ref.~\cite{DSSTT00}). Theorem~\ref{distilth} characterizes distillability in an analogous manner, but instead projects onto ${\mathbb C}^2 \otimes {\mathbb C}^2$. Thus regarded in this way, the above theorem is both a generalisation and a simplification of the main theorem of distillability.

Note that given a two-decomposable map $\Lambda$, the above result implies that undistillable states must satisfy both $(\openone\otimes \Lambda)(\rho)\geq 0$ and the dual criterion ($\Lambda\otimes \openone )(\rho)\geq 0$. As the reader might have guessed, the reduction criterion is just an example of a two-decomposable map, which we will show below. We also give numerous other applications of our Theorem~\ref{maintheorem} and Observation~\ref{firstobservation}, starting with an elementary derivation of some alternative formulations of distillability.

\subsection{Application 1: Derivation of other distillability theorems}
In this section we rederive two distillability theorems. The original derivations were quite lengthy, and as we will show, both theorems follow straightforwardly from our Observation~\ref{firstobservation} and some basic theorems on Schmidt numbers from Chapter~\ref{chapsmidt}. The following theorem was first derived in Ref.~\cite{DSSTT00}, see also the related characterisation in Ref.~\cite{SBL00}.
\begin{theorem}
\label{divi}
Let $\rho$ be a state and $S$ be the completely positive map defined by
$$
\rho=(\openone \otimes S)P_+. 
$$
Let $T$ denote the transpose map. Then $\rho$ is 1-undistillable if and only if $\Lambda=T\circ S$ is two-positive, and more generally $n$-undistillable if and only if $\Lambda^{\otimes n}$ is two-positive.
\end{theorem}
\begin{proof}
For undistillable states $\rho$ we have that 
$$
\Tr{(\rho D)}=\Tr{(\rho^{T_B} D^{T_B})}>0,
$$
for all $D^{T_B}$ having Schmidt number two. From the isomorphism between states positive on Schmidt number $k$ states and $k$-positive maps (Theorem~\ref{lemk1}), we deduce that the map associated with $\rho^{T_B}$ is two-positive for undistillable states. This map is just given by $\Lambda$ as defined in the statement of the theorem. Thus $\rho$ is 1-undistillable if and only if $\Lambda$ is two-positive. The second part follows trivially by applying the first part on $\rho^{\otimes n}$ and some elementary algebra.
\end{proof}

The second theorem is a slightly simplified version of the one presented in Ref.~\cite{KLC01}.
\begin{theorem}
\label{klctheorem}
Let $P_2$ be the projector onto a maximally entangled state acting on ${\cal H}_1={\cal H}_{A_1} \otimes {\cal H}_{B_1}={\mathbb C}^2 \otimes {\mathbb C}^2$. Then for an arbitrary operator $X$ acting on ${\cal H}_2={\cal H}_{A_2} \otimes {\cal H}_{B_2}$ we can define
\begin{align}
\label{waws}
W_X=P_2 \otimes X^{T_B}.
\end{align}
A state $\rho$ acting on ${\cal H}_2$ is $n$-undistillable if and only if $W_{\rho^{\otimes n}}$ is an entanglement witness.
\end{theorem}
\begin{proof}
From Observation~\ref{firstobservation} we have that $\rho$ is 1-undistillable if and only if $\rho^{T_B}$ is positive on Schmidt number two states. The theorem then follows directly from Lemma~\ref{combinedlemma}, and from the fact that $\rho$ is $n$-undistillable if and only if $\rho^{\otimes n}$ is 1-undistillable.
\end{proof}

\subsection{Application 2: A plethora of reduction-like maps}
\label{apptwo}
In this section we show that the map associated with the reduction criterion is two-decomposable, which immediately proves that all states violating the reduction criterion can be distilled. This greatly simplifies the original proof which relied on a series of protocols: filtering, twirling and distillation of isotropic states. Let us construct the operator $\sigma$ from the Schmidt rank-2 vectors $|\psi_{ij}\ra=|ij\ra-|ji\ra$, $i\neq j$. Adding the corresponding projectors we get
$$
\sigma=\sum_{i<j} |\psi_{ij}\ra\la\psi_{ij}|=\sum_{i\neq j}{(|ij\ra\la ij|-|ij\ra\la ji|)}=\openone-F.
$$
From this follows that $D=\sigma^{T_B}=\openone-dP_+$ is positive on undistillable states. The associated map is given by $\Lambda_1(A)=\Tr(A)\openone-A$. This is just the map used by the reduction criterion for entanglement. Thus if a state $\rho$ satisfies $\openone\otimes \rho_B-\rho \not \geq 0$ or $\rho_A\otimes\openone -\rho\not \geq 0$ then it is 1-distillable. In this way we have also proven that the map is decomposable \cite{HH97}, and using our results one obtains readily the explicit Choi-Kraus form of the map. Next, we construct some other elementary examples by combining different Schmidt rank-2 vectors:

$\diamondsuit$ Taking our vectors $|\psi_{ij}\ra=|ij\ra+|ji\ra$ we obtain
$$
\sigma=\sum_{i\neq j}{(|ij\ra\la ij|+|ij\ra\la ji|)}=\openone+F-2Z,
$$
with $Z=\sum_i|ii\ra\la ii|$. Therefore $D=\sigma^{T_B}=\ido+dP_+ - 2Z$ and the associated map is given by $\Lambda_2(A)=\Tr A\ido+A-2\diag{A}$. The map $diag$ maps all off-diagonal elements to zero and leaves the diagonal itself invariant.

$\diamondsuit$ For Schmidt rank-2 vectors of the form $|\psi_{ij}\ra=|ii\ra+|jj\ra$, we get:
$$
\sigma=\sum_{i\neq j}{(|ii\ra\la ii|+|ii\ra\la jj|)}=dP_+ +(d-2)Z,
$$
and thus $D=\sigma^{T_B}=F+(d-2)Z$; the associated map is given by 
$\Lambda_3(A)=A^T+(d-2)\diag{A}$. 

$\diamondsuit$ If we take Schmidt rank-2 vectors of the form $|\psi_{ij}\ra=|ii\ra-|jj\ra$, we get:
$$
\sigma=\sum_{i\neq j}{(|ii\ra\la ii|-|ii\ra\la jj|)}=-dP_+ +dZ,
$$
and thus $D=\sigma^{T_B}=-F+ dZ$; the associated map is given by 
$\Lambda_4(A)=-A^T+d\diag{A}$.

$\diamondsuit$ In fact, \emph{every} operator with Schmidt number 2 gives us a strong distillability witness. In Chapter~\ref{chapsmidt} we have seen that the special isotropic state
$$
\sigma=(d-2)\openone+(2d-1)dP_+
$$
allows a Schmidt rank-2 decomposition. We find $D=\sigma^{T_B}=(d-2)\openone+(2d-1)F$ and for the corresponding map $\Lambda_5(A)=(d-2)\Tr A \openone + (2d-1)A^T$.

It is worth noting that although these maps are just like the reduction map two-decomposable, they lack any physical interpretation. The reduction criterion on the other hand, has several intuitive properties, linking with other notions in quantum information (see for instance \cite{PVP00,CAG97}). As an example, note that the hash inequality only gives a non-trivial lower bound on the entanglement of distillation for states violating the reduction criterion \cite{HHH00b}. We believe therefore that the presented maps deserve further investigation.

There are however a few general statements we can make about the introduced maps. As we have seen for the separability problem, given an entanglement witness $D$, the dual positive map $\Lambda$ will detect much more states (see the discussion after Theorem~\ref{corhor2}). In a similar fashion as is shown there, the map $\Lambda$ corresponds to the class of witnesses $(A\otimes \openone) D (A^{\dagger}\otimes \openone)$, for arbitrary $A$. It also implies that the criteria $\openone \otimes \Lambda(\rho) \geq 0$ and $\Lambda \otimes \openone (\rho) \geq 0$ are insensitive to local transformations by one of the parties (thus to filtering operations). Indeed, suppose Alice performs a general measurement, with measurement operators $A_1$ and $A_2$ satisfying $A^\dagger_1A_1 +A^\dagger_2A_2=\openone$. Then the state $\rho$ will be transformed into $\rho_i=(A_i\otimes \openone) \rho (A^{\dagger}_i\otimes \openone)/p_i$. It follows that, if the original state did not violate the criteria, then the transformed state does not either. The map corresponds to the operator witness, together with all possible local filtering operations by one of the parties.

We have shown how to construct maps that detect distillability, applicable in arbitrary dimensions, which can be easily evaluated on states. There is however a catch. As can be seen from the reconstruction of the reduction criterion, the witness from which the map is derived, is a convex combination of \emph{many} witnesses. So it is possible that one of those witnesses (and the associated map) detects a state, while the sum does not. In other words, the map could be weaker than expected at first sight. The sum of the negative eigenvalues of $D$, or the entanglement of $D^{T_B}$, can be seen as a measure for the strength of a certain map.

The basic requirement for a distillability witness $D$ to detect a two-distillable state $\rho$ which is not 1-distillable is that $D^{T_B}$ must be entangled with respect to the first and second pair. Indeed suppose $D=|\psi\ra\la\psi|^{T_B}$ with $|\psi\ra=|\psi_1\ra\otimes |\psi_2\ra$ separable, then 
$$
\Tr(D^{T_B}\rho^{\otimes 2T_B})=\Tr(|\psi_1\ra\la\psi_1| \rho)\Tr(|\psi_2\ra\la\psi_2| \rho)>0,
$$
 since the vectors $|\psi_i\ra$ have at most Schmidt rank 2. Consider the witness for the reduction criterion on two pairs: 
$$
D^{T_B}=\openone -F= \openone_1 \otimes \openone_2 - F_1 \otimes F_2.
$$
But from Ref.~\cite{Alb01} we can write $F=P_S-P_A$ and $\openone=P_S+P_A$ with $P_S$ and $P_A$ projection operators onto the symmetric and antisymmetric subspaces respectively. Substitution yields 
$$
D^{T_B}= P_{S1} \otimes P_{A2} + P_{A1}\otimes P_{S2},
$$
 so that $D^{T_B}$ is separable with respect to the different pairs. Note that this property is not adhered by the map since $(A\otimes \openone) D^{T_B} (A^{\dagger}\otimes \openone)$ could be entangled with respect to the two pairs, even if $D$ itself is not. So it is not at all obvious that collective application of the reduction criterion (for instance on $\rho ^{\otimes 2}$) does not yield a stronger criterion.

It was proven in Ref.~\cite{HH97} that it is not the case, namely, if $\openone \otimes \Lambda_1 (\rho^{\otimes 2})\not\geq 0$ then $\openone \otimes \Lambda_1 (\rho)\not\geq 0$. It is readily verified that all introduced maps share this property. In the case of the reduction criterion, the reverse is also true. This can be proven as follows. From Ref.~\cite{HH97} we have
$$
\openone_A \otimes \Lambda_B (\rho_1 \otimes \rho_2)= [\openone_{A1} \otimes \rho_{B1}] \otimes [\openone_{A2} \otimes \Lambda_{B2}] (\rho_2) + [\openone_{A1} \otimes \Lambda_{B1}] ( \rho_1) \otimes \rho_2.
$$
Now suppose $\openone \otimes \Lambda (\rho_i)$ is not a positive operator, so that there exists a vector $|\psi\ra$ with negative expectation value. From the above expression follows that $\la\Psi|\openone_A \otimes \Lambda_B (\rho_1 \otimes \rho_2)|\Psi\ra<0$, with $|\Psi\ra=|\psi_1\ra \otimes |\psi_2\ra$. So applying the reduction criterion to one or more pairs is completely equivalent. A similar statement and proof applies for $\Lambda_4$, but for the other maps, this reverse statement is not true.

\subsection{Application 3: Random Schmidt-2 robustness and distillability}
The following proposition, first presented in Ref.~\cite{Clarisse05b} uses upper bounds to the random Schmidt-2 robustness as defined in Section~\ref{rsrsct} to construct a distillability criterion:

\begin{prop}
Let $\rho$ be an arbitrary bipartite state, such that $\rho^{T_B}$ has negative eigenvalues. Let $|\psi\ra$ be the eigenvector corresponding to a negative eigenvalue $\lambda$ and let $ \tilde{R}_{r2}(\psi)$ be an upper bound to its random Schmidt-2 robustness. Then $\rho$ is 1-distillable if 
$$
\lambda < - \tilde{R}_{r2}(\psi).
$$
\end{prop}
\begin{proof}
From the definition, we have that
$$
W=|\psi\ra\la\psi|+\tilde{R}_{r2}(\psi)\ido
$$
has Schmidt number two. Now from Observation~\ref{firstobservation} follows that $\rho$ is 1-distillable if $\Tr (W \rho^{T_B})<0$. This can we rewritten as
$$
\la\psi|\rho^{T_B}|\psi\ra + \tilde{R}_{r2}(\psi)= \lambda + \tilde{R}_{r2}(\psi) <0.
$$
\end{proof}

This proposition provides an important reason to find an exact analytical formula for $R_{r2}(\psi)$. Note that this distillability criterion only depends on the minimum eigenvalue and the corresponding eigenvector. It is easy to see that it is independent of the reduction criterion as we show by means of an example. Consider the state
\begin{align}
\rho=\frac{1}{16}\left[ \begin{array}{rrrrrrrrr}
 1 & 0 & 0 & 0 & 0 & 0 & 0 & 0 & 0 \\
 0 & 2 & 0 & -1 & 0 & 0 & 0 & 0 & 0 \\
 0 & 0 & 2 & 0 & 0 & 0 & 2 & 0 & 0 \\
 0 & -1 & 0 & 2 & 0 & 0 & 0 & 0 & 0 \\
 0 & 0 & 0 & 0 & 1 & 0 & 0 & 0 & 0 \\
 0 & 0 & 0 & 0 & 0 & 2 & 0 & 2 & 0 \\
 0 & 0 & 2 & 0 & 0 & 0 & 2 & 0 & 0 \\
 0 & 0 & 0 & 0 & 0 & 2 & 0 & 2 & 0 \\
 0 & 0 & 0 & 0 & 0 & 0 & 0 & 0 & 2
\end{array} \right].
\end{align}
It is easy to check that the reduction criterion is not useful here. The partial transposition $\rho^{T_B}$
has an eigenvector $|\psi\ra=\frac{1}{\sqrt{3}}( |00\ra+|11\ra-|22\ra)$ and corresponding eigenvalue $-\frac{1}{8}$. Since $-{R}_{r2}(\psi)=-\frac{1}{15}>-\frac{1}{8}$, $\rho$ is distillable. 

An unconditional distillability criterion, which does only depend on the minimum eigenvalue of $\rho^{T_B}$ would follow from Conjecture~\ref{mainconject}. Pending on the proof of this conjecture, we have that $\rho$ is distillable whenever $\rho^{T_B}$ has an eigenvalue $\lambda \leq -1/14$. 

\subsection{Application 4: Volume of 1-distillable states}
\label{peasantsec}
In this section we will outline a powerful algorithm for the detection of distillable states (first published in Ref.~\cite{Clarisse05}). As before, it is clear that we can confine ourselves to the study of 1-distillability. 
A first numerical method for this problem was devised by D\"ur et~al.\ \cite{DCLB99} starting from the basic minimisation over Schmidt rank two vectors $|\psi\ra$:
\begin{align}
\label{bprob}
\min_{\psi \in SR2} \la \psi | \rho^{T_B}|\psi \ra \overset{?}{<} 0.
\end{align}
They convert this problem into a minimisation of the minimum eigenvalue of certain matrices. Their method involves also calculating an inverse of a matrix in every step. 

In view of our discussion above, we will perform this minimisation not over the distillability witnesses, but over the much stronger associated two-decomposable maps. Thus for a given state $\rho$, we can reduce the problem (see also again Ref.~\cite{DSSTT00}) to checking whether there exists an operator $P=|0\ra\la a| + |1\ra\la b|$, with $\la b| a \ra=0$ such that
$$
\label{todoeq}
(\ido \otimes P) \rho^{T_B} (\ido \otimes P)^\dagger \overset{?}{\not \geq} 0.
$$
One way of doing this is to parametrize a countable subset of vectors which is dense within all vectors, such as the one introduced in Ref.~\cite{HB04}. Explicitly their set takes the form $G=\{\sum \lambda_i |i\ra | (\lambda_1, \lambda_2,\cdots, \lambda_d ) \in \tilde G_{\mathbb{N}} \}$ with 
$$ 
\tilde G_N=\left\{\left(\frac{p_1}{q_1}e^{2\pi i \frac{r_1}{s_1}}, ,\frac{p_2}{q_2}e^{2\pi i \frac{r_2}{s_2}},\cdots, \sqrt{1-\sum_l \frac{p^2_l}{q^2_l}}e^{2\pi i \frac{r_d}{s_d}}\right)\right \}
$$
 for $ 0 < p_i\leq q_i \leq N; 0< r_i\leq s_i \leq N $.
Thus for every $N$, we can construct sets of pairs $|a\ra$ and $|b\ra$, and taking $N$ increasingly large we will detect all 1-distillable states, except those arbitrary close to the boundary of the convex set of 1-undistillable states.

In practice two improvements can be made which greatly enhance the performance. First note that the countable subset above will yield vectors $|a\ra$ and $|b\ra$ not necessarily orthogonal. Furthermore, clearly such a countable subset will in general not pick vectors uniformly distributed according to the Haar measure. One way of overcoming this is to take for $|a\ra$ and $|b\ra$ two columns of a random unitary. Here orthogonality and uniformity are automatically guaranteed. The associated algorithm works well for low dimensional density matrices. For higher dimensions, a local optimisation of the $|a\ra$ and $|b\ra$ yielding the minimum after a certain cut-off number of tests turns out to work well. The explicit algorithm has been reproduced in Appendix~\ref{dtmatlab}.

\begin{figure}
\begin{center}
\includegraphics[height=8cm]{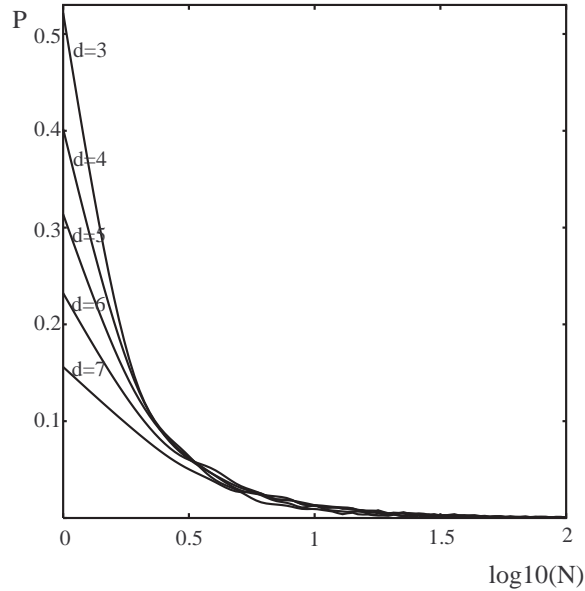}
\caption[The probability of detecting 1-distillability in $N$ tests]{The probability of detecting 1-distillability as a function of the test number for random states drawn from ${\cal D} \times {\cal U}$. For visual purposes only the first 100 tests are depicted and the points have been smoothened out to curves.} 
\label{probtest}
\end{center}
\end{figure}

Let us apply this to give a numerical estimate of the volume of 1-distillable states for low dimensional quantum states. A similar numerical estimate has been carried out for entangled states \cite{ZHSL98, Zyczkowski99}. When talking about volumes on the set of density operators, it is clear that the results will depend on the measure. We choose the measure applied in Ref.~\cite{ZHSL98} as it seems very natural.

A general bipartite quantum state $\rho$ acting on ${\cal H}_A \otimes {\cal H}_B$ can be expanded by virtue of the spectral decomposition as
$$
\rho=UDU^\dagger,
$$
with $D=(d_{ii})$ diagonal and $U$ unitary. The measure we are going to use is the product measure ${\cal D} \times {\cal U}$. Here ${\cal D}$ represents the uniform distribution of the points on the manifold given by $\sum_i d_{ii}=1$. A simple method for generating such a distribution from independent uniformly distributed random numbers chosen in the interval $(0,1)$ is outlined in Appendix A of Ref.~\cite{ZHSL98}. Similarly, ${\cal U}$ is chosen to be the uniform measure on unitary matrices (the Haar measure). To generate random unitaries according to this measure one can use the algorithm from Ref.~\cite{ZK94, PZK98} which relies on a decomposition of a general unitary in two-dimensional unitary transformations. A simpler method for generating Haar unitary matrices is as follows \cite{Stewart80, ER05}. Take a random matrix $A$, whose entries are complex numbers which are independently normally distributed with mean zero. The polar decomposition $A=PU$, with $P$ positive, then yields $U$ distributed according to ${\cal U}$.

\begin{figure}
\begin{center}
\includegraphics[height=9cm]{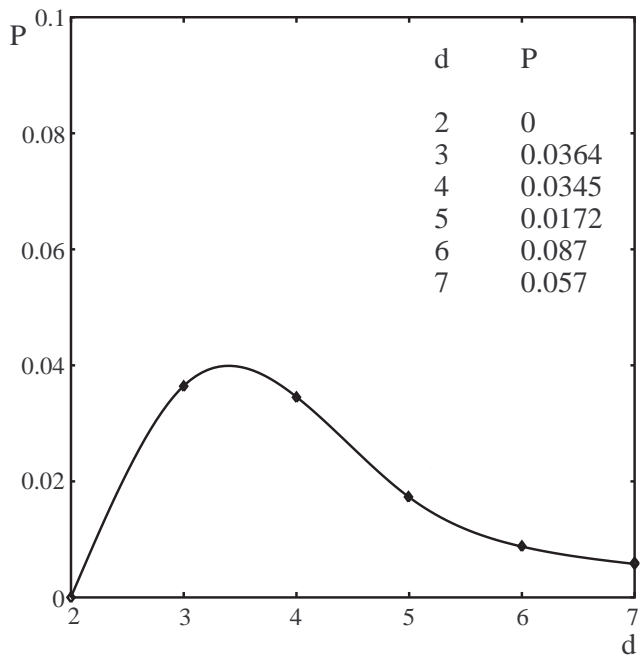}
\caption[Volume of NPT distillable states]{Probability of a random NPT state drawn from ${\cal D} \times {\cal U}$ to be 1-undistillable. The curve has been drawn to guide the eye.} 
\label{volnpt}
\end{center}
\end{figure}

We have tested the our method on $10^5$ density matrices acting on $\mathbb{C}^d \otimes \mathbb{C}^d$ for $d=3,4,5,6,7$. The estimated probability for proving the distillability of a state as a function of the test number is displayed in Figure~\ref{probtest}. A striking feature is that the vast majority of the distillable states were detected in the first few tests. This is especially true for low dimensional states. For example for $d=3$ about half of the NPT states are found to be distillable in the first test.  For $d=7$ the probability of finding a distillable state in the first test is about $1/6$. 

\begin{figure}
\begin{center}
\includegraphics[height=9cm]{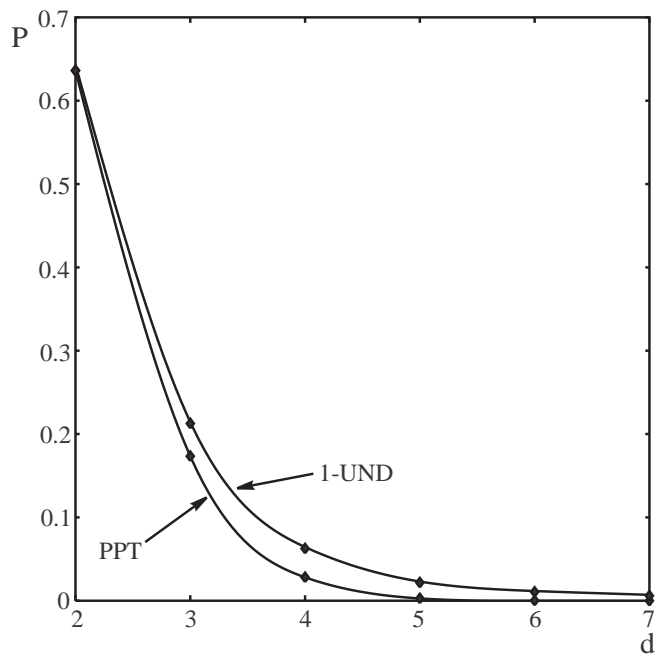}
\caption[Volume of distillable states]{Probability of a random state drawn from ${\cal D} \times {\cal U}$ to be 1-undistillable.} 
\label{volall}
\end{center}
\end{figure}

In Figure~\ref{volnpt} the probability of an NPT state being 1-undistillable is plotted versus the dimension $d$. To obtain sufficient precision we carried out $10^5$ random tests per state, and in addition $10^4d$ optimisation steps seeking for a local minimum. Of course, this method does not guarantee to detect every 1-distillable state, but we obtain an upper bound of the number of undistillable states. The figure suggests that the volume of 1-undistillable states drops to zero for high dimensions. In Figure~\ref{volall} the same graph is drawn, but now PPT states are included. 

It is tempting to conclude from this numerical evidence that bound entangled states are primarily a phenomenon present in low dimensional quantum systems.
In high dimensional systems most undistillable states are therefore situated in the immediate neighbourhood of the set of separable states. Our results are consistent with the fact that bound entanglement for continuous variables is a rare phenomenon \cite{HCL01}. In particular it was shown that the subset of undistillable states is nowhere dense in the set of bipartite continuous variable states. From this it follows that the set of undistillable continuous variables states does not contain any open ball, an argument which was made explicit for separable states in Ref.~\cite{ESP01}. Given an infinite dimensional separable state one can construct sequences of closer and closer states all of which are entangled. Following the same methods one can explicitly construct a distillable state in any $\epsilon$-neighbourhood in the trace norm of any state \cite{Eisert}. However, note that in Ref.~\cite{HLW05} a parametrized family of measures on states was introduced, which in some region yields states primarily 1-undistillable.

\section{The question of NPT bound entanglement}
\label{nptbound}
\subsection{Early results}
As we have seen in Theorem~\ref{wernernptbound}, the question of the existence of NPT bound entanglement is equivalent to the existence of undistillable entangled Werner states. Numerical evidence seems to suggest that some Werner states are indeed undistillable. However, the problem is still open, and constitutes one of the major challenges in quantum information theory. What has been shown \cite{DSSTT00,DCLB99} is that there are entangled Werner states that are $n$-undistillable for every $n$, however the range of states for which one is able to prove this, vanishes as $n$ goes to infinity. We will prove this partial result following the method presented in Ref.~\cite{DCLB99}. We start with the following key lemma.

\begin{lemma}
\label{ciracineq}
Let $P$ be the projector onto the maximally entangled $d$-level state and $Q=\ido-P$, then for any Schmidt rank two vector $|\psi\ra$ we have:
\begin{enumerate}
\item $\la \psi | \ido^{\otimes N-k}\otimes P^{\otimes k} | \psi \ra \leq \frac{2}{d^k}$
\item $\la \psi | Q^{\otimes N-k} \otimes P^{\otimes k} | \psi \ra \leq \frac{2}{d^k}$
\item $\la \psi | Q^{\otimes N} | \psi \ra \geq (1-\frac{2}{d})^N$
\end{enumerate}
\end{lemma}
\begin{proof}
(1) For $k=1$, this is clearly the case, as the overlap of the maximally entangled state and a Schmidt rank two vector is at most $2/d$ (see Lemma~\ref{fefl}). For $k>1$ this follows from the case $k=1$ and $P_d^{\otimes k}=P_{d^k}$.

(2) This follows from $Q\leq \ido$.

(3) For $N=1$, this follows directly from (1) and the definition $Q=\ido-P$. We prove it for arbitrary $N$ by induction, so let it be true for a certain $N\geq 1$. Now 
$$
\la \psi | Q^{\otimes (N + 1)} | \psi \ra =\Tr[ |\psi\ra\la\psi| Q \otimes Q^{\otimes N}].
$$
Writing this trace as $\Tr(\cdot)=\Tr(\Tr_1(\cdot))$, where we first trace out over the first pair, we get
$$
\Tr[ |\psi\ra\la\psi| Q \otimes Q^{\otimes N}]=\Tr [ \Tr_1 (|\psi\ra\la\psi| Q )Q^{\otimes N} ].
$$
Now since $Q$ is separable (see (\ref{isotropic})), it is easy to verify that $\Tr_1 (|\psi\ra\la\psi| Q )$ can be written as $\sum_i c_i |\psi_i\ra\psi_i |$ with $\psi_i$ Schmidt rank two. Also from the case $N=1$ follows that $\Tr[\sum_i c_i |\psi_i\ra\la\psi_i |]\geq 1-2/d$.
\end{proof}

This lemma is practically all we need to present the main result of this section.
\begin{theorem}
The Werner states 
$$
\rho = \frac{1}{d^2+\beta d}(\ido + \beta F)
$$
with $-1\leq \beta \leq 1$ are entangled for $\beta<-1/d$ or $-1\leq \Tr(\rho F)<0$. They are 1-distillable if and only if $\beta<-1/2$ and $n$-undistillable for $\beta>-1/d+\epsilon_n$, for some $\epsilon_n>0$.
\end{theorem}
\begin{proof}
The first part follows from the fact that positivity of the partial transposition is a necessary and sufficient condition for separability of the Werner states \cite{VW01}. The Werner states are 1-distillable for $\beta<-1/2$, as follows from the overlap of the partial transposition with a Schmidt rank two vector $|\psi\ra$
$$
\la \psi | \rho^{T_B}|\psi\ra=1 + \beta \la \psi | d P_+ |\psi\ra \geq 1 + 2 \beta ,
$$
where we have used Lemma~\ref{fefl}. For $n$ copies let us put $\alpha=1+\beta d$ and $\ido=P+Q$ so that 
$$
\rho^{T_B \otimes n} = Q^{\otimes n} + \alpha R.
$$
Here $R$ consists of all the other terms in the tensor product expansion. Note that all these terms contain at least one $P$. From the previous lemma follows then that for any Schmidt rank two vector $|\psi\ra$
$$
\la \psi | \rho^{T_B \otimes n}|\psi\ra\geq (1-\frac{2}{d})^N + \alpha \la \psi | R|\psi\ra.
$$
Applying again the previous lemma, we also have that we can calculate a lower bound for $\alpha \la \psi | R|\psi\ra$. Thus we can always choose $\alpha$ sufficiently small so that $\la \psi | \rho^{T_B \otimes n}|\psi\ra\geq 0$ for which $\rho$ will be $n$-undistillable. We refer to the original reference and our next section for qualitative results on the value of $\epsilon_n$.
\end{proof}

From this theorem it was conjectured that all Werner states with $\beta>-1/2$ are undistillable, although $\epsilon_n \rightarrow 0$ for $n\rightarrow \infty$. Apart from some numerical evidence, the main reason was that at the time these results appeared, not a single example of a state was known which is 2-distillable but 1-undistillable.
In the meantime, several such states have appeared \cite{SST01,EVWW01,VW02}. These first examples arose in the context of activation (see Section~\ref{actbent}). Important progress was made by Watrous \cite{Watrous03} who constructed a one-parameter set of distillable states which are $n$-undistillable in some range. We will discuss these states in the next section. Watrous's result does not seem too unlikely, as the notion of $n$-distillability, as we explained before, is really as much a mathematical as a physical notion. A distillation process will typically require a very large number of copies, even if the state is $1$-distillable.

\subsection{The UUVVF-invariant states}
In this section we will discuss the distillability properties of a certain class of highly symmetric two-parameter states, which includes the Watrous states and two copies of the Werner states, and hence is ideal for the study of distillability. We will see that over a wide range of the parameters, the states are probably undistillable. Other examples of conjectured NPT bound entangled states can be found in Ref.~\cite{BR03}. 

The set of states we are going to study are the $UUVVF$-invariant states introduced in Section~\ref{ssulo}. A convenient parametrisation of the set of $UUVVF$-invariant states is given by
\begin{align*}
\rho=\ido_{12}\otimes\ido_{34}+\frac{\epsilon d-1}{d}(\ido_{12}\otimes F_{34} + 
 F_{12} \otimes \ido_{34}) + \frac{1-2\epsilon d+\delta d^2}{d^2}F_{12}\otimes F_{34}.
\end{align*}
The set of density operators is restricted by the following inequalities
\begin{align*}
(d-1)^2+2\epsilon d(d-1)+\delta d^2 & \geq 0, \\
d^2-1+2\epsilon d-\delta d^2 & \geq 0,\\
(d+1)^2-2\epsilon d (d+1)+\delta d^2 & \geq 0.
\end{align*}
This set of states includes the Werner \cite{Werner89} states (2 pairs in $d\otimes d$ for $\delta =\epsilon^2$ and 1 pair in $d^2\otimes d^2$ for $\epsilon=1/d$) and the Watrous \cite{Watrous03} states ($1-2\epsilon d+\delta d^2=d^2$). The PPT states are just the separable states (see Ref.~\cite{VW01}). 
The states are entangled for $\epsilon<0$ or $\delta<0$ which follows from 
\begin{align}
\label{ptbrho}
\rho^{T_B}=Q_{12}\otimes Q_{34} + d\epsilon(P_{12}\otimes Q_{34} + Q_{12}\otimes P_{34}) + \delta d^2 P_{12}\otimes P_{34},
\end{align}
with as usual $Q=\ido-P$. 

\begin{figure}
\begin{center}
\includegraphics[width=14.5cm]{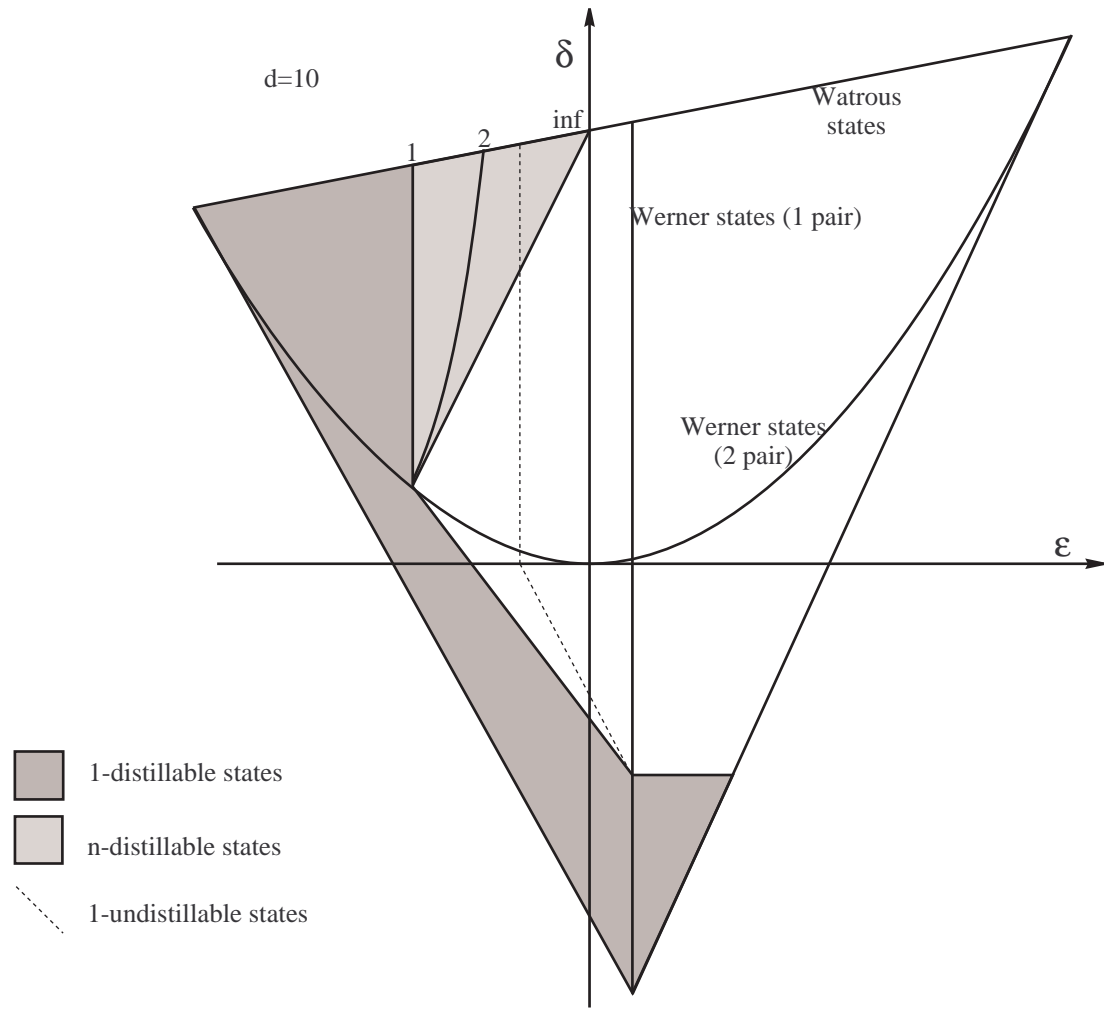}
\caption[The $UUVVF$-invariant states]{$UUVVF$-invariant states. All states satisfying $\epsilon\geq 0$ and $\delta\geq 0$ are separable. Shaded areas mark out distillable states ($1$,$2$ and $\infty$- copy distillable). The states lying above and to the right of the dotted line are states for which we were able to show 1-undistillability.} 
\label{fig1}
\end{center}
\end{figure}

Let us now investigate the distillation properties of these states. To prove distillability, all we need to do is to find Schmidt rank two vectors $\psi$ such that $\la \psi | \rho^{T_B}|\psi\ra<0$. We present three such vectors, which we conjecture to detect all 1-distillable states:
\begin{align}
|\psi\ra_A&=|00\ra_{12} \otimes (|00\ra+|11\ra)_{34}, \nonumber\\
|\psi\ra_B&=|01\ra_{12} \otimes (|00\ra+|11\ra)_{34}, \label{ABC}\\
|\psi\ra_C&=\sum_{ij}|ii\ra_A|jj\ra_B+|i(i+1)\ra_A|j(j+1)\ra_B\nonumber.
\end{align}
which detect distillability for the states respectively satisfying
\begin{gather*}
d^2+3d(\epsilon d-1)+2(1-2\epsilon d +\delta d^2)<0,\\
\epsilon <\frac{1}{d}-\frac{1}{2}, \\
\delta <\frac{1}{d^2}-\frac{1}{2}.
\end{gather*}
This set of states is shaded dark grey in Figure~\ref{fig1}.

Now let us derive results in the other direction, namely which states are undistillable? First let us show that for arbitrary $n$ we can find NPT states which are $n$-undistillable. We will use the inequalities from Lemma~\ref{ciracineq}.
From (\ref{ptbrho}) it follows that all potentially negative terms contain at least one $P$ term. But we will also have a term $\la \Psi | Q^{\otimes 2N} | \Psi \ra \geq (1-\frac{2}{d})^{2N}$, and this term can always dominate when we choose $\epsilon$ and $\delta$ small enough. Thus for each $n$, as long as we choose $\epsilon$ \emph{or} $\delta$ small enough (but one or both negative), we obtain an $n$-undistillable NPT state. 

Next, let us derive states that are definitely 1-undistillable. From (\ref{ptbrho}) and the given inequalities it follows straightforwardly that the states satisfying
$$
(1-2/d)^2+\min(4\epsilon,0)+\min(2\delta,0)\geq 0,
$$
are 1-undistillable. Note that this set does not touch the set of 1-distillable states. The reason for this is that although the inequalities are sharp, the sum of such inequalities are not. In some regions we can get a better bound by rewriting (\ref{ptbrho}) as
\begin{align*}
\rho^{T_B}=\ido_{12}\otimes \ido_{34} + (\epsilon d-1)(P_{12}\otimes Q_{34} + Q_{12}\otimes P_{34}) + (\delta d^2-1) P_{12}\otimes P_{34}, 
\end{align*}
and we find that all states such that
$$
d^2+\min(4d(\epsilon d-1),0)+\min(2(\delta d^2 -1),0) \geq 0
$$
are 1-undistillable. This second set of states is depicted in Figure~\ref{fig1}.

To conclude our discussion of the distillation properties we will discuss the Watrous states \cite{Watrous03} of which the general form is given by
\begin{align*}
\rho=\ido_{1,2}\otimes\ido_{3,4}+\frac{\epsilon d-1}{d}(\ido_{1,2}\otimes F_{3,4} + 
 F_{1,2} \otimes \ido_{3,4}) + F_{1,2}\otimes F_{3,4},
\end{align*}
with $1+1/d>\epsilon >1/d-1$. The states are entangled if and only if $\epsilon<0$ and definitely 1-distillable if $\epsilon <1/d-1/2$. We will show that all entangled Watrous states are distillable. Suppose now we have two pairs of entangled Watrous states, the second pair having indices $5,6$ and $7,8$. Using the identities
\begin{align*}
\Tr((P_{1,5}\otimes P_{2,6})(\ido_{1,2}\otimes F_{5,6}))&=1/d, \\
\Tr((P_{1,5}\otimes P_{2,6})(F_{1,2}\otimes \ido_{5,6}))&=1/d, \\ 
\Tr((P_{1,5}\otimes P_{2,6})(F_{1,2}\otimes F_{5,6}))&=1,
\end{align*}
it can be verified that projecting upon $P_{1,5}\otimes P_{2,6}$ will yield a new Watrous state
$\rho'$ with parameter
$$
\epsilon'=\epsilon\Bigl(\frac{2(\epsilon d+d^2-1)}{d^2 \epsilon^2+d^2-1}\Bigr).
$$
Since $\epsilon'<\epsilon$ for $0>\epsilon>1/d-1$, the state will be more entangled than the state we started from for each $\epsilon<0$. One can repeat this protocol on many pairs until finally $\epsilon<1/d-1/2$, at which point we obtain a 1-distillable state. More generally the protocol can be applied to the whole set of states; a straightforward but tedious calculation leads to
\begin{align*}
\epsilon'&=\frac{\epsilon(d^2\delta + d^2-1)}{d^2 \epsilon^2 +d^2-1}, \\
\delta'&=\frac{\epsilon^2(d^2-1)+d^2\delta^2}{d^2 \epsilon^2 +d^2-1}.
\end{align*}
The states which are 2-distillable with this protocol are depicted in Figure~\ref{fig1}. Repeating the protocol recursively, it is not hard to show that all states satisfying
$$
\delta>\frac{3d^2+4d-8}{2d(d-2)}\epsilon+1-\frac{1}{d^2},
$$
are distillable (see again Figure~\ref{fig1}). In Section~\ref{apptwo} we have seen that the basic requirement for a distillability witness $W$, to detect 2-distillability of a state $\rho$ (or thus 1-distillability of $\rho^{\otimes 2}$) which is not 1-distillable, is that $W^{T_B}$ be entangled with respect to the two pairs. The Watrous states allow us to illustrate this. Indeed from the above protocol we can deduce that $W_2$ on $\rho^{\otimes 2}$ will detect 2-distillable states that are not $1$-distillable, with 
$$
W_2=(P_{15}\otimes P^T_{26}\otimes P_{B,3478})^{T_{2468}}.
$$
Here $T$ is the normal transposition, $T_{2468}$ is the transposition with respect to the subsystems $2,4,6$ and $8$ and $P_{B}$ is the projector onto $|\psi\ra_B$ defined in (\ref{ABC}). The projectors $P_{15}\otimes P_{26}$ are responsible for the entanglement between the first and second pair.

Using the numerically method outlined in Section~\ref{peasantsec} we have checked numerically the distillability of the $UUVVF$-invariant states over the complete range of parameters for 1 and 2 copies for $d=3$. We easily recovered the proposed boundaries for 1-distillability. For two copies, the states act on a $\mathbb{C}^{d^4} \otimes \mathbb{C}^{d^4}$ Hilbert space and numerical matrix manipulations in a space of this magnitude seem very hard. 
Fortunately, the states are sparse and our method only requires minimisation of the minimum eigenvalue of a $2d^2\times 2d^2$ matrix. We were able to detect distillability for the Watrous states in $\epsilon>1/d-1/2$ readily and exhaustive testing suggest that also the proposed boundary for 2-distillability is correct. This for the first time provides strong evidence that the Werner states are 4-undistillable for $\beta>-1/2$.

\subsection{As a Schmidt number problem}
As we have seen, the NPT bound entangled conjecture is equivalent to the statement that there exists \emph{no} Schmidt rank two vector $|\psi\ra$ such that
\begin{align}
\label{origfor}
\la \psi | {\rho^{T_B {\otimes n}}} |\psi\ra < 0,
\end{align}
for all $n\geq 1$ and $\rho^{T_B} = \ido -\frac{d}{2} P$. In other words, it seems that affirmation of the conjecture would have to be in the form of an impossibility proof as opposed to a constructive proof.
 We will reformulate this conjecture in a more tractable form, namely as a special instance of the separability problem, for which a large number of tools are present. As a steppingstone we first show how to translate the distillability problem into the problem of detecting Schmidt number 3. In the next subsection we then reformulate it as a separability problem. We will do so for one, two and $n$ copies of the Werner states. 

Let us start with one copy of $\rho$, for which the answer is known. As $\rho$ belongs to the set of $UU$-invariant states, $\rho^{T_B}$ will belong to the set of $UU^*$-invariant states (the isotropic states) and hence will be invariant under the $UU^*$-twirl: ${\cal T}_{UU^*}(\rho^{T_B})=\rho^{T_B}$. From this follows that 
$$
\la \psi |\rho^{T_B} |\psi\ra = \Tr(P_\psi\rho^{T_B})=\Tr( {\cal T}_{UU^*}(P_\psi)\rho^{T_B}),
$$
here ${\cal T}_{UU^*}(P_\psi)$ is the operator $P_{\psi}=|\psi\ra\la\psi|$ after application of the $UU^*$-twirl. Thus we do not need to check over the whole set of Schmidt rank two vectors, but instead over the restricted set of isotropic states with Schmidt number 2. Recall (see (\ref{isotropic}) and Ref.~\cite{TH00, SBL00}) that if we parametrise the isotropic states as $\rho_\alpha=\ido+\alpha P$ then $\rho_\alpha$ has Schmidt number $k$ when
$$
\alpha \leq \frac{d(kd-1)}{d-k}.
$$
From this follows that $\rho= \ido +\beta F$ is 1-distillable if and only if $\Tr(\rho^{T_B} \rho_\alpha)<0$, with $\alpha=\frac{d(2d-1)}{d-2}$. Going through the algebra we recover that all Werner states $\rho_\beta$ with $\beta<-1/2$ are 1-distillable.

\begin{figure}
\begin{center}
\includegraphics[height=9cm]{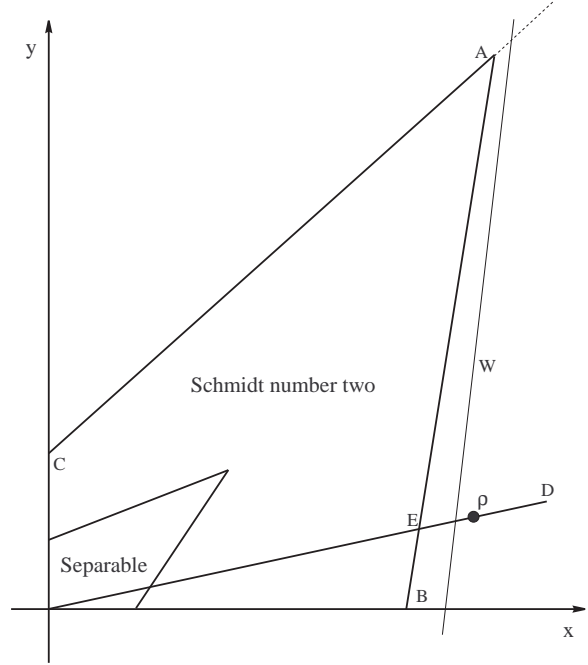}
\caption{The $UU^*VV^*F$-invariant states.} 
\label{uuvvs}
\end{center}
\end{figure}

Let us now look at two copies of the Werner states $\rho^{\otimes 2}$; as pointed out before, these states belong to the larger class of the $UUVVF$-invariant states. Thus the relevant dual set is the set of the $UU^*VV^*F$-invariant states. For convenience we write the operators in the order of the indices 1,2,3,4 and omit these indices. With this in mind we parametrise the $UU^*VV^*F$-invariant states as
$$
\rho=Q\otimes Q +x (Q\otimes P + P\otimes Q) +y P \otimes P,
$$
with $x,y>0$. These states are separable for $d^2-2d(x-1)+(1-2x+y)\leq 0$ and $d^2-(1-2x+y)\leq 0$ as depicted in Figure~\ref{uuvvs}. The set of Schmidt number 2 states contains at least the convex hull of the points $(x,y)$
\begin{align*}
A&=\biggl(\frac{(3d-4)(d+1)}{2d-4},\frac{2(d+1)^2(d-1)}{d-2} \biggr) \\
B&=\biggl(\frac{d^2-1}{d-2},0\biggr)\\
C&=\biggl(0,\frac{2(d^2-1)^2}{d^2-2}\biggr)
\end{align*}
obtained by twirling the Schmidt rank two vectors from equation~(\ref{ABC}). All the states lying above the $CA$-line have a Schmidt number larger than two. This follows from the fact that the operator $\ido \otimes \ido - \frac{d^2}{2} P\otimes P$ is positive on Schmidt number 2 states. If the Werner states are 2-undistillable then also all the states lying to the right of the $AB$-line have Schmidt number at least three. This follows easily by evaluating the expectation value of $\rho$ at $(\ido -\frac{d}{2}P)^{\otimes 2}$. 
Note that it is sufficient to show that all the states $\rho$ on for instance the $ED$-segment (see Figure~\ref{uuvvs}) have Schmidt number 3.
This can be seen as follows: let $\rho$ approach $E$, and suppose we can show that each $\rho$ has Schmidt number 3. This would imply the existence of a hyperplane $W$ separating $\rho$ from the set of Schmidt number 2 states. For $\rho$ arbitrarily close to $E$ this hyperplane would be parallel to the $AB$-segment, otherwise cutting it. Therefore this would show that all the states lying to the right of the $AB$-hyperplane have Schmidt number at least three.

For the general case we need to consider the set of $U_1U^*_1\cdots U_nU^*_nF$-invariant states. Here the subindex in $U_i$ refers to the subsystem the unitary operator is acting and the $F$ denotes any permutation of the subsystems. In what follows we will call these states $U_iU^*_iF$-invariant states. We parametrise them as
\begin{align*}
\rho={\tilde Q}^{\otimes n}+x_1(P\otimes {\tilde Q}^{\otimes n-1} + {\tilde Q}\otimes P\otimes {\tilde Q}^{\otimes n-2}+\cdots)+ \\
 x_2(P^{\otimes 2}\otimes {\tilde Q}^{\otimes n-2} + P\otimes {\tilde Q}\otimes P \otimes {\tilde Q}^{\otimes n-2}+\cdots) + \\ 
\cdots + x_n P^{\otimes n},
\end{align*}
where we have found it convenient to now use the normalised ${\tilde Q}=Q/(d^2-1)$. The relevant hyperplane is given by 
$$
\Tr\Bigl[\Bigl(Q+\Bigl(1-\frac{d}{2}\Bigr)P\Bigr)^{\otimes n} \rho\Bigr]=1+\sum_{i=1}^n \binom{n}{i} \Bigl ( 1-\frac{d}{2}\Bigr)^i x_i=0.
$$
We can also write this as $1+\sum_{i=1}^n \binom{n}{i} (-1)^i {\tilde x}_i=0$ with ${\tilde x}_i= \bigl(\frac{d}{2}-1\bigr)^i x_i$. From now on we will continue to work in these normalised variables. Next we will generalise the idea developed for two copies. First we need to show that the hyperplane is spanned by Schmidt number 2 states. Then in order to check distillability, it will be enough to find the boundary between two and three Schmidt number along a line from the origin to an interior point of the points spanning the hyperplane.

An independent set of Schmidt number 2 states spanning the hyperplane is easily obtained as follows (compare to the case for two copies).
Let $|\psi\ra_1=\frac{1}{\sqrt{2}} |01\ra^{\otimes n-1} \otimes (|00\ra+|11\ra)$. Twirling this state will yield an $U_iU^*_iF$-invariant state with coordinates $(\frac{1}{n},0,\cdots,0)$. Similarly, twirling $|\psi\ra_2=\frac{1}{\sqrt{2}} |00\ra\otimes |01\ra^{\otimes n-2} \otimes (|00\ra+|11\ra)$ will yield an $U_iU^*_iF$-invariant state with coordinates $({\tilde x}_1\neq 0,{\tilde x}_2\neq 0,0,\cdots,0)$. 
In general 
$$
|\psi\ra_k=\frac{1}{\sqrt{2}} |00\ra^{\otimes k} \otimes |01\ra^{\otimes n-k-1} \otimes (|00\ra+|11\ra)
$$
will yield a state with coordinates ${\tilde x}_i \neq 0$ for $i\leq k+1$ ${\tilde x}_i = 0$ for $i> k+1$. It is evident that all points will lie on the hyperplane and that they form an independent set, spanning the hyperplane. An interior point can for instance be obtained from the first point, as $(\frac{1}{n}+\epsilon(1-\frac{1}{n}),\epsilon,\cdots,\epsilon)$ for sufficiently small $\epsilon$. It can readily be verified that this point belongs to the hyperplane by using the identity $\sum_{i=2}^n \binom{n}{i}(-1)^i=n-1$.

Thus $n$-undistillability of the Werner states beyond the 1-distillability boundary is equivalent to the statement that the $U_iU^*_iF$-invariant state with coordinates $(\frac{1}{n}+\epsilon(1-\frac{1}{n})+\delta,\epsilon,\cdots,\epsilon)$ has Schmidt number 3 for $\epsilon>0$ small enough and all $\delta >0$.

\subsection{As a separability problem}
In this section we cast the distillability problem as a special instance of the separability problem. Theorem~\ref{klctheorem} provides the key ingredient to this, which applied to the conjecture implies that
\begin{align}
W_X=P_2 \otimes X^{T_B},
\end{align}
is an entanglement witness for all $n\geq 1$ with $X=\rho^{\otimes n}$ and $\rho=(\ido-F/2)$. We will again use the local symmetry, and present a dual positive formulation of the conjecture. The analysis will be analogous to the reformulation as a Schmidt number problem and a continuous comparison of this section with the previous one is very instructive.

\begin{figure}
\begin{center}
\includegraphics[height=9cm]{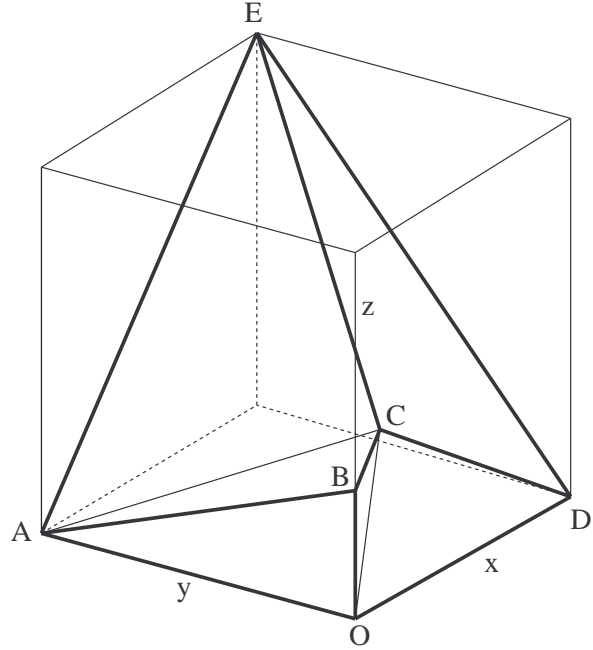}
\caption[The $UU^*VV^*$-invariant states]{The state space of the $UU^*VV^*$-invariant states. The point C lies in the $xz$-plane such that the points A, B, C and E lie in the same plane.} 
\label{threed}
\end{center}
\end{figure}

For one pair, we need to prove that $P_2\otimes (\ido-\frac{d}{2}P)$ is an entanglement witness. As before, it will be sufficient to characterize the subset of the separable states of the general $UU^*VV^*$ invariant states. Here $U$ acts on a two-dimensional Hilbert space. We parametrize the $UU^*VV^*$-invariant states as
$$
\rho=\tilde Q_2\otimes\tilde Q+xP_2\otimes \tilde Q+y\tilde Q_2\otimes P+zP_2\otimes P,
$$
with $x,y,z\geq 0$. The separable states are a subset of the states with positive partial transpose, which satisfy the inequalities
\begin{align*}
3z\leq\frac{1+3x}{d-1}-y & \qquad & \text{$ABCE$-plane}\\
z\leq\frac{1-x}{d+1}+y & \qquad & \text{$ECD$-plane} \\
z\geq \frac{x-1}{d-1}+y & \qquad & \text{$EAD$-plane}.
\end{align*}
Thus the states with positive partial transposition are contained in the polyhedron spanned by the points OABCDE as in Fig~\ref{threed}. By twirling the pure separable states $|0101\ra$, $|0100\ra$, $|0001\ra$ and $\sum_{i,j=0}^1|ijij\ra$ one obtains the states represented respectively by the points $O,A,C,D$ and $E$. Note that $B$ is not in this list. Indeed, as 
we know that $P_2\otimes (\ido-\frac{d}{2}P)$ is an entanglement witness, all states satisfying $z>2x/(d-2)$ are entangled. In particular the states in the tetrahedron spanned by the points $ABCO$ are PPT entangled. Conversely, knowing that the polyhedron $ABCO$ is PPT entangled immediately proves that $P_2\otimes (\ido-\frac{d}{2}P)$ is an entanglement witness. 

For two copies, we need to consider the $UU^*(V_1V_1^*V_2V_2^*F)$-invariant states which we parametrise as
\begin{align*}
\rho= \tilde{Q}_2 \otimes(\tilde{Q}\otimes \tilde{Q} + y_1 (P \otimes \tilde{Q}+ \tilde{Q} \otimes P) +y_2 P\otimes P)+ \\
P_2\otimes (x_0 \tilde{Q} \otimes \tilde{Q} + x_1 (P \otimes \tilde{Q}+ \tilde{Q} \otimes P) +x_2 P\otimes P).
\end{align*}
We will not attempt to completely classify the separable subset, instead it is enough to look at what happens in the neighbourhood of the hyperplane 
\begin{align*}
\Tr\Bigl[P_2\otimes \Bigl(Q+\Bigl(1-\frac{d}{2}\Bigr)P\Bigr)^{\otimes 2} \rho\Bigr]= 
x_0-(d-2)x_1+\Bigl(\frac{d-2}{2}\Bigr)^2 x_2=0.
\end{align*}
We now show that this hyperplane is spanned by separable states. Consider the following pure separable states and their coordinates $(y_1,y_2,x_0,x_1,x_2)$ after action of the $UU^*(V_1V_1^*V_2V_2^*F)$ twirl:
\begin{align*}
|01\ra|01\ra|01\ra: & \quad (0,0,0,0,0) \\
|01\ra|00\ra|01\ra: & \quad \Biggl(\frac{1}{2(d-1)},0,0,0,0 \Biggr) \\
|01\ra|00\ra|00\ra: & \quad \Biggl(\frac{d-1}{d^2-d-1},\frac{1}{d^2-d-1},0,0,0 \Biggr) \\
\sum_{i,j=0}^1 |ij\ra|ij\ra|00\ra:&\quad \Biggl(\frac{1}{2(d-1)},0,\frac{d-2}{3d},\frac{3d-4}{6d(d-1)},\frac{2}{3d(d-1)}\Biggr) \\
\sum_{i,j=0}^1 |ij\ra|ij\ra|01\ra: &\quad \Biggl(0,0,\frac{d-2}{3d},\frac{1}{3d},0\Biggr).
\end{align*}
These coordinates can be verified after a tedious but straightforward calculation.

 A point in the interior of the convex hull of these points can be obtained by averaging these coordinates. In this case one obtains the state with coordinates $x_0=2(d-2)/(15d)$, $x_1=(5d-6)/(30d(d-1))$, $x_2=2/(15 d(d-1))$, $y_1=d(2d-3)/(5(d-1)(d^2-d-1))$ and $y_2=1/(5(d^2-d-1))$. From this follows that the Werner states are 2-undistillable if and only if the $UU^*(V_1V_1^*V_2V_2^*F)$-invariant states with coordinates $(y_1,y_2,x_0,x_1+\epsilon,x_2)$ are entangled for all $\epsilon>0$.

\newpage
Let us now move to $n$ copies. The relevant set of states is the set of $UU^*(V_iV^*_iF)$-invariant states
\begin{align*}
\rho={\tilde Q}_2 \otimes {\tilde Q}^{\otimes n} + 
{\tilde Q}_2 \otimes [ y_1(P\otimes {\tilde Q}^{\otimes n-1} + {\tilde Q}\otimes P\otimes {\tilde Q}^{\otimes n-2}+\cdots) +\\ y_2(P^{\otimes 2}\otimes {\tilde Q}^{\otimes n-2} + \cdots)+\cdots + y_n P^{\otimes n-1}]+ 
P_2 \otimes [ x_0 {\tilde Q}^{\otimes n} + \\ x_1(P\otimes {\tilde Q}^{\otimes n-1} + {\tilde Q}\otimes P\otimes {\tilde Q}^{\otimes n-2}+\cdots)+ \cdots + x_n P^{\otimes n}], 
\end{align*}
with $x_i,y_i\geq 0$. The relevant hyperplane is given by
\begin{align*}
\Tr\Bigl[P_2\otimes \Bigl(Q+\Bigl(1-\frac{d}{2}\Bigr)P\Bigr)^{\otimes n} \rho\Bigr]= 
\sum_{i=1}^n \binom{n}{i} \Bigl ( 1-\frac{d}{2}\Bigr)^i x_i=0.
\end{align*}
Renormalising $x_i$, this can be rewritten as $\sum_{i=1}^n \binom{n}{i} (-1)^i {\tilde x}_i=0$ with ${\tilde x}_i= \bigl(\frac{d}{2}-1\bigr)^i x_i$. 
Next we will show that this hyperplane touches the set of separable states by constructing a set of $2n+1$ separable states spanning the hyperplane. The first $n+1$ states are obtained by twirling 
$$
|\psi\ra_k=|01\ra |00\ra^{\otimes k} |01\ra^{\otimes n-k},
$$
 for $k=0,\cdots, n$. The twirled state will be $UU^*(V_iV^*_iF)$-invariant and will satisfy ${\tilde x}_i = 0$ for all $i$ and ${\tilde y}_j \neq 0$ for $j\leq k$ and ${\tilde y}_j = 0$ for $j>k$. 
The last $n$ states are obtained by twirling 
$$
|\psi\ra_k=\sum_{i,j=0}^1|ij\ra|ij\ra |00\ra^{\otimes n-k-1} |01\ra^{\otimes k},
$$
for $k=0,\cdots, n-1$. The twirled state will be $UU^*(V_iV^*_iF)$-invariant and the $x$ coordinates will satisfy $x_j=0$ for $j> n-k$ and $x_j\neq 0$ for $j\leq n-k$. Therefore the coordinates of the $2n+1$ states are linearly independent and the choice of $|\psi\ra_k$ guarantees that the states will lie in the hyperplane.

An interior point in the convex hull of these $2n+1$ points can be obtained by choosing $\tilde x_0=\epsilon$ and $\tilde y_j=\tilde x_j=\epsilon$, for $\epsilon$ sufficiently small. One verifies that this point belongs to the hyperplane using the identity $\sum_{i=0}^{n-1}\binom{n}{i} (-1)^i =-1$.
From this follows that the conjecture is equivalent to the statement that the $UU^*(V_iV^*_iF)$-invariant states with coordinates $\tilde y_i=\epsilon$ , $\tilde x_i=\epsilon$ and $\tilde x_1=\epsilon+\delta$, where $j=1,\cdots,n$ and $i=0,2,\cdots,n$, are entangled for $\epsilon>0$ small enough and all $\delta >0$.

\subsubsection{Discussion}
The general separability problem has been proven to be NP-hard \cite{Gurvits03}. However, above we showed that the distillability problem can be reformulated as the question of entanglement of a \emph{particular} set of one parameter states. The first and by far still the most elegant tool for detecting entanglement is the partial transposition criterion \cite{Peres93}. However, it is not useful to solve the dual entanglement problem, as it was proven in Ref.~\cite{KLC01} that $W_{\rho^{\otimes n}_W}$ is a decomposable entanglement witness if and only if $\rho$ has positive partial transposition. Hence $W_{\rho^{\otimes n}_W}$ will be either a non-decomposable witness or no entanglement witness at all depending on whether $\rho^{\otimes n}_W$ is 1-undistillable or 1-distillable. Thus in order to solve the dual problem, we will either need some powerful tool for detecting PPT-entanglement or a tool for proving separability. However, in the latter case, the original formulation (see (\ref{origfor})) seems easier, and for this purpose we presented an efficient algorithm for detecting distillability in Section~\ref{peasantsec}.

Probably the most powerful method for detecting entanglement is the complete family of separability criteria introduced by Doherty et~al.\ \cite{DPS01, DPS03}, as reviewed in Section~\ref{numap}. Recall that their method yields an analytical provable entanglement witness from the output of the algorithm. In principle therefore, the distillability problem can be solved for any number of copies, using our dual formulation together with the algorithm associated with the complete family of separability criteria. The numerical method introduced by Eisert et al.~\cite{EHGC04} can also be employed for this goal. However, note that the reformulation of the distillability problem as a separability problem is unnecessary here, as their method allows for a simple way of checking whether or not an operator is an entanglement witness. Using analogous techniques, it can be extended to testing whether or not an operator is a Schmidt number 3 witness \cite{Eisert}. It should be stressed that the family of criteria introduced by Doherty et~al.\ has the major advantage that an \emph{analytical} canonical entanglement witness can be extracted.
Unfortunately, we were unable to test either algorithm, as both methods seem only practical for low dimensional systems. The interesting case, two copies of the Werner states for $d=3$, could hence not be tested. It would be worth investigating whether these numerical methods could be simplified for the highly symmetrical states we are interested in.

Numerical solutions are one option, another possible approach would be one of a more indirect nature. A powerful method for proving that a certain state is entangled is to show that, when shared by two parties, the state can enhance typical quantum operations such as teleportation or distillation. We will come back to this topic in Section~\ref{actbent}. One of the main results is \cite{Masanes05} that \emph{every} entangled state can enhance the so-called conclusive teleportation fidelity of some other state. This characterisation of entangled states is promising as a way of proving that a certain state is PPT entangled. In Ref.~\cite{Ishizaka04} a class of PPT states was constructed which was shown to provide overall convertibility of pure entangled states. In particular it was shown to be able to increase the Schmidt number of a pure state. Now the PPT entangled states we obtained for one copy of the Werner states (tetrahedron ABCO in Figure~\ref{threed}) are of a similar form, and although not necessarily in an optimal way, they too can help increase the Schmidt number of a pure state. Similar activation effects can be expected from the conjectured PPT states derived from two copies of the Werner states.

\subsection{The rainbow states}
In this section, as a curiosity, we present a set of states which we conjecture to exhibit all the different forms of distillability. The set of states was inspired by the results of Ref.~\cite{Ishizaka04}.

The states are $UUVV$-invariant, but with subsystems of unequal dimension (we omit the indices $\{1,2\}$ and $\{3,4\}$)
\begin{align*}
\rho=\ido_m\otimes\ido_d+\frac{d\epsilon -1}{d}\ido_m\otimes F_d + \frac{m\epsilon -1}{m}F_m \otimes \ido_d + \frac{1-(m+d)\epsilon+ d m \delta}{dm}F_m\otimes F_d.
\end{align*}
In what follows we will assume that $3\leq m<d$. The set of density operators is restricted by the following inequalities
\begin{align*}
1+\delta+2\epsilon+\frac{1}{md}-(\epsilon+1)\frac{m+d}{md}& \geq 0\\
1-\frac{1}{md}+\epsilon\frac{m+d}{md}-\delta +\frac{1}{m}-\frac{1}{d}&\geq 0\\
1+\delta-2\epsilon+\frac{1}{md}+(1-\epsilon)\frac{m+d}{md}& \geq 0
\end{align*}
The partial transpose is given by
\begin{align*}
\rho^{T_B}=Q_m\otimes Q_d+m\epsilon P_m\otimes Q_d+ d\epsilon Q_m\otimes P_d+md\delta P_m\otimes P_d,
\end{align*}
and it is easy to see that the states are NPT iff $\epsilon<0$ or $\delta<0$. Now we will show that these states also include some PPT entangled ones. It is easy to see from equation (\ref{fefl}) in Section~\ref{iswsec}, that $\ido_d-dP_d/m$ is positive on Schmidt number $m$ states; it then follows from Lemma~\ref{combinedlemma} that
\begin{align*}
mP_m\otimes (\ido_d-\frac{d}{m}P_d)
\end{align*}
is an entanglement witness. The partial transpose of an entanglement witness is again an entanglement witness:
\begin{align*}
F_m\otimes (\ido_d-\frac{1}{m}F_d).
\end{align*}
Applying this entanglement witness on $\rho$ we find that the states satisfying
\begin{align*}
\epsilon m^2 (d^2-1)+dm\delta(m-d)<0
\end{align*}
are (PPT) entangled.

\begin{figure}
\begin{center}
\includegraphics[width=14cm]{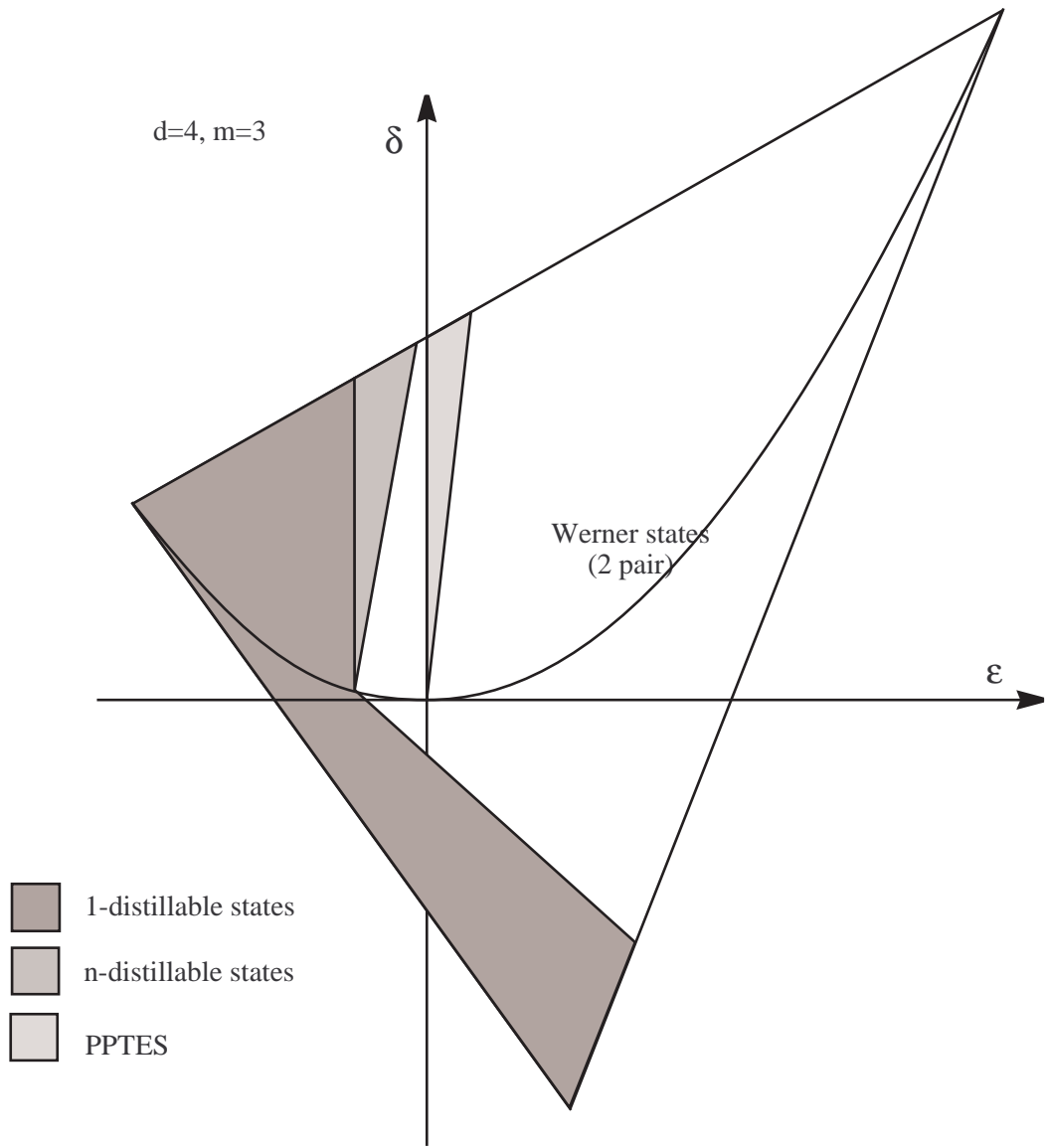}
\caption[The rainbow states]{The rainbow states: the top line of the triangle covers the whole spectrum of different types of distillability.}
\label{fig2}
\end{center}
\end{figure}

Let us now look at the distillation properties. Similar arguments to the ones we used for the $UUVVF$-invariant states apply here. For $\epsilon$ or $\delta$ sufficiently small, one can find $n$-undistillable states. The following vectors provide apparently the optimal boundaries for the 1-distillable states (we will only be interested in the states having $\delta>0$).

1. $|\psi\ra=|00\ra_A|00\ra_B+|10\ra_A|10\ra_B$ gives rise to
\begin{align*}
2+2(d\epsilon-1)/d+4(m\epsilon-1)/m+ 4(1-(m+d)\epsilon+dm\delta)/(md)<0
\end{align*}

2. $|\psi\ra=|00\ra_A|01\ra_B+|10\ra_A|11\ra_B$ gives rise to
\begin{align}
\epsilon <\frac{1}{m}-\frac{1}{2}.
\end{align}

Now let us take two pairs
\begin{align*}
\rho_1=\ido^{34}_m\otimes\ido^{12}_d+\frac{d\epsilon -1}{d}\ido^{34}_m\otimes F^{12}_d + \frac{m\epsilon -1}{m}F^{34}_m \otimes \ido^{12}_d + \nonumber\\ +\frac{1-(m+d)\epsilon+ d m \delta}{dm}F^{34}_m\otimes F^{12}_d,
\end{align*}
and
\begin{align*}
\rho_2=\ido^{78}_m\otimes\ido^{56}_d+\frac{d\epsilon -1}{d}\ido^{78}_m\otimes F^{56}_d + \frac{m\epsilon -1}{m}F^{78}_m \otimes \ido^{56}_d + \nonumber\\ +\frac{1-(m+d)\epsilon+ d m \delta}{dm}F^{78}_m\otimes F^{56}_d .
\end{align*}
Taking both pairs together, and projecting upon $P^{1,5}_d\otimes P^{2,6}_d$, we end up with a $UUVVF$-invariant state in $m^2\otimes m^2$ with 
\begin{align*}
\epsilon'&=\frac{\epsilon(d^2\delta + d^2-1)}{d^2 \epsilon^2 +d^2-1} \\
\delta'&=\frac{\epsilon^2(d^2-1)+d^2\delta^2}{d^2 \epsilon^2 +d^2-1}.
\end{align*}
Now we know when these states can be distilled. The states distillable with this protocol are shown in Figure~\ref{fig2}. One verifies that projecting upon $P^{1,5}_m\otimes P^{2,6}_m$ performs worse.
The set $1-\frac{1}{md}+\epsilon\frac{m+d}{md}-\delta +\frac{1}{m}-\frac{1}{d}=0$ contains states of all kinds: 1-distillable, $n$-undistillable but $n+1$-distillable, NPT undistillable (conjectured), PPT bound entangled and separable. 

\section{Activation of bound entanglement}
\label{actbent}
As bound entanglement cannot be distilled, it cannot be directly used for typical quantum information tasks, such as teleportation. Activation of bound entanglement is any use of bound entanglement that allows for operations which are not possible by means of LOCC only.

The first of these effects was discovered by the Horodeccy \cite{HHH98b}. They showed that a specific state can be quasi-distilled with the aid of an unlimited supply of particular PPT bound entangled states, while this is not possible without them. Although this is a relatively weak result, it was the first indication that bound entanglement can be activated. After this first result, a series of negative results were obtained. Linden and Popescu \cite{LP99b} showed rigorously that some PPT bound entangled state cannot teleport states with better than classical fidelity. Vedral \cite{Vedral99} subsequently demonstrated that the PPT distillable entanglement (that is the distillable entanglement of a state in the presence of an unlimited supply of bound entangled states) is still bounded by the relative entropy of entanglement (see also Ref.~\cite{DMSST99}).

Major progress on activation was made by Shor et~al.\ \cite{SST01}, and their results were later generalised by Eggeling et~al.\ \cite{EVWW01}. We discuss this last paper in some detail and sketch the main proofs. Their work is based on the following theorem by Rains \cite{Rains00}, which gives a characterisation of the maximal PPT fidelity of distillation $F_d$. This quantity is defined as the maximum overlap a state can have with the maximally entangled state $P_+$ after application of a PPT and trace preserving operation.

\begin{theorem}
The maximal PPT fidelity of distillation of a $d$-level bipartite state $\rho$ is given by
$$
F_d(\rho)=\max_\Lambda\Tr(P_+\Lambda(\rho))=\max_A\Tr(\rho A ),
$$
where the maximum is taken over all PPT and trace preserving maps $\Lambda$ or all Hermitian operators $A$ such that
$0\leq A\leq \ido$ and $-\ido\leq dA^{T_B}\leq \ido$.
\end{theorem}
\begin{proof} (sketch)
Since $P_+$ commutes with all unitaries of the form $U\otimes U^*$ we may assume without loss of generality that $\Lambda(\rho)$ is an isotropic state
$$
\Lambda(\rho)=\Tr(\rho B)(\ido-P_+) +\Tr(\rho A) P_+,
$$
where it is easy to see that the coefficients can always be written as linear functionals on $\rho$. By the definition of $\Lambda$, the operator $\Lambda(\rho)$ has to be a proper quantum state, so we have to require that $0\leq A,B\leq \ido$ and $(d^2-1)B+A=\ido$. All that is left to do is account for the PPT property of $\Lambda$, that is if $\rho>0$ then
$$
\Lambda(\rho^{T_B})^{T_B}=\Tr(\rho B^{T_B})(\ido-\frac{F}{d}) +\Tr(\rho A^{T_B}) \frac{F}{d}\geq 0.
$$
It is easy to see that this is equivalent to $-\ido\leq dA^{T_B}\leq \ido$.
\end{proof}

We are now in the position of proving the main result of Eggeling et~al. Ref.~\cite{EVWW01}.

\begin{theorem}
\label{egtheor}
With the aid of a bound entangled state any bipartite NPT state is 1-distillable.
\end{theorem}
\begin{proof}
Denoting $P_{\text{n}}$ the projector onto the negative eigenspace of $\rho^{T_B}$ we define 
$$
A=\frac{1}{d}(\ido-\epsilon P^{T_B}_{\text{n}}), 
$$
 with $0<\epsilon\leq \min(2,\| P^{T_B}_{\text{n}}\|^{-1}_\infty )$ and $\| \cdot \|_\infty$ the operator norm (the maximum singular value). It is easy to check that the operator defined in this way meets the conditions of Rains' Theorem. So that 
$$
\Tr(\rho A)=\frac{1+\epsilon {\cal N}(\rho)}{d},
$$
where ${\cal N}(\rho)>0$ is the negativity. Therefore Rains' theorem implies that there exists a PPT and trace preserving operation $\Lambda$ such that 
$$
\Tr(\Lambda(\rho) P_+)=\Tr(\rho' P_+)>\frac{1}{d}.
$$
and such states are 1-distillable (see Theorem~\ref{ftthree}). The proof ends by noting that any PPT preserving map can be implemented locally using a shared PPT bound entangled state (see Section~\ref{entopssec}).
\end{proof}
An alternative proof using entanglement witnesses can be found in Ref.~\cite{KLC01}. It was later also shown that any NPT state can be 1-distilled using only an infinitesimal amount of bound entanglement, as this is possible for the Werner states \cite{VW02} .

The implications of this result are non-trivial. As pointed out in Ref.~\cite{SST01}, it probably implies the superadditivity of the distillable entanglement. Indeed, assuming the existence of NPT bound entanglement we have that 
there exist states $\rho_1$ (NPT) and $\rho_2$ (PPT) such that $E_D(\rho_1)=E_D(\rho_2)=0$, but
$$
E_D(\rho_1\otimes\rho_2)>0,
$$
where the inequality follows from the previous theorem. This effect is called superactivation, and has been proven to exist in a multipartite setting \cite{SST00} (see also Ref.~\cite{DC00, Smolin00}). Equally it follows from the existence of NPT bound entanglement that the distillable entanglement is nonconvex, as convexity would imply
\begin{align*}
0&=E_D(\rho_1 \otimes |00\ra\la 00|)+E_D(\rho_2 \otimes |11\ra\la 11|) \\
&\geq E_D(\rho_1 \otimes |00\ra\la 00|+ \rho_2 \otimes |11\ra\la 11|)=E_D(\rho)>0.
\end{align*}
The last inequality follows from the fact that the states $\rho_1 \otimes |00\ra\la 00|$ and $\rho_2 \otimes |11\ra\la 11|$ can be locally discriminated. So that when Alice and Bob are given a large number of states $\rho$ they can recover a large supply of $\rho_1$ and $\rho_2$, which they can then distill by tensoring them together. In any case, such states $\rho$ are examples of 2-distillable states which are not 1-distillable. In Ref.~\cite{Horodecki01} Micha{\l} Horodecki brilliantly argues on why convexity of the distillable entanglement is not to be expected. The argument is based on the fact that the distillable entanglement is a measure characterising the entanglement of source rather than that of individual states (see also \cite{Vidal02}).

Another activation effect was demonstrated by Ishizaka \cite{Ishizaka04}, who showed that with the aid of specific bound entanglement, pure entanglement is completely interconvertible: all pure entangled states can be converted into each other with some probability, regardless of the Schmidt rank (this was already implicit in Ref.~\cite{APE03}). Theorem~\ref{egtheor} implies that \emph{all} NPT bound entanglement can be activated, as well as some \emph{particular} PPT entangled states. The question remains if every PPT bound entangled state can be activated. Recently, this question was answered affirmatively by Masanes \cite{Masanes05}. He showed that any entangled state can enhance the teleportation fidelity of some other state. A special case can be formulated in the context of distillation:
\begin{theorem}
For any entangled state $\rho$ there exists a 1-undistillable state $\sigma$ such that $\rho\otimes \sigma$ is 1-distillable.
\end{theorem}
Although this result is of tremendous importance, the proof works by contradiction, and reveals nothing about the state $\sigma$. Not surprisingly, as if this was the case, it would solve the separability problem! A generalisation of this result to multipartite systems was presented in Ref.~\cite{Masanes05b}. Related results were obtained in \cite{VV03, Brandao05b}, where it was shown that the generalised robustness (which is nonzero for entangled states) has an operational meaning as the maximum percentual increase an entangled state can provide in the fidelity of teleportation of another state.

\appendix
\chapter{Two Matlab programs}
\section{General permutation}
The $m$-file below is a simple implementation of an arbitrary permutation of a density matrix, as defined in (\ref{perdeff}). It is listed here, as some complicated implementations of for instance the partial transpose criterion are circulating. The procedure below is transparent and general. It can also be used to exchange subsystems within a density matrix.

\label{gpmatlab}
\begin{verbatim}
function rho2=permcrit(rho,p)
%permcrit(rho,[1 2 4 3]) calculates the partial transposition 
%permcrit(rho,[2 4 1 3]) calculates the realigned matrix
%permcrit(rho,[3 4 1 2]) exchanges the first with the second subsystem
%permcrit(rho,[3 5 4 2 6 1]) random permutation for tripartite system
[d1,d2]=size(rho); [s,t]=size(p); n=t/2; d=round(d1^(1/n)); q(p)=1:t;
if s~=1 || floor(t/2)~=ceil(t/2) || ~isequal(q(p),1:t) || length(q)~=t 
error('Not a valid permutation'); end;
if d^n~=d1 || d1~=d2, error('Incompatible dimensions'); end;
rho2=zeros(d1,d1); vec=zeros(t,1); dnw=power(d,n-1:-1:0);
for i=0:d1-1, for j=0:d1-1 
 r=1; c=1; i2=i; j2=j; 
 for w=1:n, vec(w*2-1)=floor(i2/dnw(w)); vec(w*2)=floor(j2/dnw(w));
 i2=i2-vec(w*2-1)*dnw(w); j2=j2-vec(w*2)*dnw(w); end; 
 vec=vec(p); 
 for w=1:n, r=r+vec(w*2-1)*dnw(w); c=c+vec(w*2)*dnw(w); end;
 rho2(r,c)=rho(i+1,j+1); 
end; end;
\end{verbatim}

\section{Distillability test}
\label{dtmatlab}
The $m$-file below checks whether a state is 1-distillable with the algorithm outlined in Section~\ref{peasantsec}. 
Input parameters are the partial transposed state, the dimension of the Hilbert spaces of the parties and the number of tests. As the tests involve randomisation, it is recommended to set the number of tests at least to $10^4$, and to repeat the test a number of times. If the output is negative, however, the state is guaranteed to be distillable.

\begin{verbatim}
function mineig=distest(rhotb, d, nrtests)
idxs=zeros(1); mineig=1; precision=-10^(-8);
for yy=1:d
 idxs(2*yy-1)=(yy-1)*d+1; idxs(2*yy)=(yy-1)*d+2; 
end;
if issparse(rhotb), EE=sparse(d*d,d*d);
else, EE=zeros(d*d,d*d); end;
for aa=1:nrtests 
 [UU,R]=qr(randn(d)+i*randn(d)); 
 UU=UU*diag(exp(2*pi*i*rand(d,1))); UU(3:d,:)=0;
 for k=1:d:d*d, EE(k:k+d-1,k:k+d-1)=UU; end;
 B=EE*rhotb*EE'; mreig=min(real(eig(full(B(idxs,idxs)))));
 if mreig<mineig, mineig=mreig; UUb=UU; end;
 if mineig<precision, return; end;
end;
for loop=1:nrtests/d
for counter=1:2
 te1=UUb(counter,:); 
 for mn=1:d
 UUb(counter,mn)=randn(1,1)+i*randn(1,1);
 UUb(counter,:)=UUb(counter,:)/norm(UUb(counter,:));
 for k=1:d:d*d, EE(k:k+d-1,k:k+d-1)=UUb; end;
 B=EE*rhotb*EE'; mreig=min(real(eig(full(B(idxs,idxs)))));
 if mreig<mineig, mineig=mreig; te1=UUb(counter,:);
 else, UUb(counter,:)=te1; end; 
 end; 
 if mineig<precision, return; end;
end; end;
mineig=1;
\end{verbatim}

\newpage
\thispagestyle{plain}
\addcontentsline{toc}{chapter}{Bibliography}
\bibliographystyle{plainnatl}
\bibliography{ent}

\begin{thebibliography}{342}
\providecommand{\natexlab}[1]{#1}
\providecommand{\url}[1]{\texttt{#1}}
\expandafter\ifx\csname urlstyle\endcsname\relax
  \providecommand{\doi}[1]{doi: #1}\else
  \providecommand{\doi}{doi: \begingroup \urlstyle{rm}\Url}\fi
\providecommand{\eprint}[2][]{\url{#2}}

\bibitem[Ac{\'i}n et~al.(2001)Ac{\'i}n, Bru{\ss}, Lewenstein, and
  Sanpera]{ABLS01}
A.~Ac{\'i}n, D.~Bru{\ss}, M.~Lewenstein, and A.~Sanpera.
\newblock Classification of mixed three-qubit states.
\newblock \emph{Physical Review Letters}, 87:\penalty0 40401, 2001,
  \eprint{quant-ph/0103025}.

\bibitem[Alber et~al.(2001)Alber, Delgado, Gisin, and Jex]{ADGJ01}
G.~Alber, A.~Delgado, N.~Gisin, and I.~Jex.
\newblock Efficient bipartite quantum state purification in arbitrary
  dimensional {H}ilbert spaces.
\newblock \emph{Journal of Physics A: Mathematical and General}, 34:\penalty0
  8821--8833, 2001, \eprint{quant-ph/0102035}.

\bibitem[Ambainis and Gottesman(2004)]{AG04}
A.~Ambainis and D.~Gottesman.
\newblock The minimum distance problem for two-way entanglement purification.
\newblock \emph{IEEE Transactions on Information Theory}, 2004,
  \eprint{quant-ph/0310097}.

\bibitem[Arrighi and Patricot(2004)]{AP03}
P.~Arrighi and C.~Patricot.
\newblock On quantum operations as quantum states.
\newblock \emph{Annals of Physics}, 311:\penalty0 26--52, 2004,
  \eprint{quant-ph/0307024}.

\bibitem[Aschauer and Briegel(2003)]{AB02}
H.~Aschauer and H.~J. Briegel.
\newblock Quantum communication and decoherence.
\newblock In \emph{Coherent Evolution in Noisy Environment}. Springer Verlag,
  2003.

\bibitem[Aspect et~al.(1982)Aspect, Dalibard, and Roger]{ADR82}
A.~Aspect, J.~Dalibard, and G.~Roger.
\newblock Experimental test of {B}ell's inequalities using time-varying
  analyzers.
\newblock \emph{Physical Review Letters}, 49:\penalty0 1804--1807, 1982.

\bibitem[Audenaert et~al.(2001)Audenaert, Eisert, Jan{\'{e}}, Plenio, Virmani,
  and Moor]{AEJPVM01}
K.~Audenaert, J.~Eisert, E.~Jan{\'{e}}, M.~B. Plenio, S.~Virmani, and B.~De
  Moor.
\newblock The asymptotic relative entropy of entanglement.
\newblock \emph{Physical Review Letters}, 87:\penalty0 217902, 2001,
  \eprint{quant-ph/0103096}.

\bibitem[Audenaert et~al.(2002)Audenaert, Moor, Vollbrecht, and Werner]{AMVW02}
K.~Audenaert, B.~De Moor, K.~G.~H. Vollbrecht, and R.~F. Werner.
\newblock Asymptotic relative entropy of entanglement for orthogonally
  invariant states.
\newblock \emph{Physical Review A}, 66:\penalty0 032310, 2002,
  \eprint{quant-ph/0204143}.

\bibitem[Audenaert et~al.(2003)Audenaert, Plenio, and Eisert]{APE03}
K.~Audenaert, M.~B. Plenio, and J.~Eisert.
\newblock Entanglement cost under positive-partial-transpose-preserving
  operations.
\newblock \emph{Physical Review Letters}, 90:\penalty0 027901, 2003,
  \eprint{quant-ph/0207146}.

\bibitem[Audenaert and Braunstein(2003)]{AB03}
K.~M.~R. Audenaert and S.~L. Braunstein.
\newblock On strong superadditivity of the entanglement of formation.
\newblock \emph{Communications in Mathematical Physics}, 246:\penalty0
  443--452, 2003, \eprint{quant-ph/0303045}.

\bibitem[Augusiak and Horodecki(2004)]{AH04}
R.~Augusiak and P.~Horodecki.
\newblock Bound entanglement can maximally violate {B}ell inequalities: quantum
  entanglement is not equivalent to quantum security.
\newblock 2004, \eprint{quant-ph/0405187}.

\bibitem[Badzi{\c{a}}g et~al.(2003)Badzi{\c{a}}g, Horodecki, Sen(De), and
  Sen]{BHSS03}
P.~Badzi{\c{a}}g, M.~Horodecki, A.~Sen(De), and U.~Sen.
\newblock Locally accessible information: How much can the parties gain by
  cooperating.
\newblock \emph{Physical Review Letters}, 91:\penalty0 117901, 2003,
  \eprint{quant-ph/0304040}.

\bibitem[Bandyopadhyay(2000)]{Bandyopadhyay00}
S.~Bandyopadhyay.
\newblock Qubit and entanglement assisted optimal entanglement concentration.
\newblock \emph{Physical Review A}, 62:\penalty0 032308, 2000,
  \eprint{quant-ph/9911013}.

\bibitem[Bandyopadhyay and Roychowdhury(2004)]{BR03}
S.~Bandyopadhyay and V.~Roychowdhury.
\newblock Maximally-disordered distillable quantum states.
\newblock \emph{Physical Review A}, 69:\penalty0 040302(R), 2004,
  \eprint{quant-ph/0302043}.

\bibitem[Bandyopadhyay et~al.(2005)Bandyopadhyay, Ghosh, and
  Roychowdhury]{BGR04}
S.~Bandyopadhyay, S.~Ghosh, and V.~Roychowdhury.
\newblock Non-full rank bound entangled states satisfying the range criterion.
\newblock \emph{Physical Review A}, 71:\penalty0 012316, 2005,
  \eprint{quant-ph/0406023}.

\bibitem[Bell(1964)]{Bell64}
J.~S. Bell.
\newblock On the {E}instein {P}odolsky {R}osen paradox.
\newblock \emph{Physics}, 1:\penalty0 195--200, 1964.

\bibitem[Bell(1966)]{Bell66}
J.~S. Bell.
\newblock On the problem of hidden variables in quantum mechanics.
\newblock \emph{Reviews of Modern Physics}, 38:\penalty0 447--452, 1966.

\bibitem[Benatti et~al.(2004)Benatti, Floreanini, and Piani]{BFP04}
F.~Benatti, R.~Floreanini, and M.~Piani.
\newblock Non-decomposable quantum dynamical semigroups and bound entangled
  states.
\newblock \emph{Open Systems and Information Dynamics}, 11:\penalty0 325--338,
  2004, \eprint{quant-ph/0411095}.

\bibitem[Bennett et~al.(1996{\natexlab{a}})Bennett, Bernstein, Popescu, and
  Schumacher]{BBPS96}
C.~H. Bennett, H.~J. Bernstein, S.~Popescu, and B.~Schumacher.
\newblock Concentrating partial entanglement by local operations.
\newblock \emph{Physical Review A}, 53:\penalty0 2046--2052,
  1996{\natexlab{a}}, \eprint{quant-ph/9511030}.

\bibitem[Bennett et~al.(1996{\natexlab{b}})Bennett, Brassard, Popescu,
  Schumacher, Smolin, and Wootters]{BBPSSW96}
C.~H. Bennett, G.~Brassard, S.~Popescu, B.~Schumacher, J.~A. Smolin, and W.~K.
  Wootters.
\newblock Purification of noisy entanglement and faithful teleportation via
  noisy channels.
\newblock \emph{Physical Review Letters}, 76:\penalty0 722--725,
  1996{\natexlab{b}}, \eprint{quant-ph/9511027}.

\bibitem[Bennett et~al.(1996{\natexlab{c}})Bennett, Di{V}incenzo, Smolin, and
  Wootters]{BDSW96}
C.~H. Bennett, D.~P. Di{V}incenzo, J.~A. Smolin, and W.~K. Wootters.
\newblock Mixed-state entanglement and quantum error correction.
\newblock \emph{Physical Review A}, 54:\penalty0 3824--3851,
  1996{\natexlab{c}}, \eprint{quant-ph/9604024}.

\bibitem[Bennett et~al.(1999{\natexlab{a}})Bennett, Di{V}incenzo, Fuchs, Mor,
  Rains, Shor, Smolin, and Wootters]{BDFMRSSW98}
C.~H. Bennett, D.~P. Di{V}incenzo, C.~A. Fuchs, T.~Mor, E.~Rains, P.~W. Shor,
  J.~A. Smolin, and W.~K. Wootters.
\newblock Quantum nonlocality without entanglement.
\newblock \emph{Physical Review A}, 59:\penalty0 1070--1091,
  1999{\natexlab{a}}, \eprint{quant-ph/9804053}.

\bibitem[Bennett et~al.(1999{\natexlab{b}})Bennett, Di{V}incenzo, Mor, Shor,
  Smolin, and Terhal]{BDMSST99}
C.~H. Bennett, D.~P. Di{V}incenzo, T.~Mor, P.~W. Shor, J.~A. Smolin, and B.~M.
  Terhal.
\newblock Unextendible product bases and bound entanglement.
\newblock \emph{Physical Review Letters}, 82:\penalty0 5385,
  1999{\natexlab{b}}, \eprint{quant-ph/9808030}.

\bibitem[Bertlmann et~al.(2002)Bertlmann, Narnhofer, and Thirring]{BNT02}
R.~A. Bertlmann, H.~Narnhofer, and W.~Thirring.
\newblock Geometric picture of entanglement and {B}ell inequalities.
\newblock \emph{Physical Review A}, 66:\penalty0 032319, 2002,
  \eprint{quant-ph/0111116}.

\bibitem[Bhatia(1997)]{Bhatia97}
R.~Bhatia.
\newblock \emph{Matrix Analysis}.
\newblock Springer Verlag, New York, 1997.

\bibitem[Bohm(1951)]{Bohm51}
D.~Bohm.
\newblock \emph{Quantum theory}.
\newblock Prentice-Hall, New York, 1951.

\bibitem[Bohr(1935)]{Bohr35}
N.~Bohr.
\newblock Can quantum-mechanical description of physical reality be considered
  complete?
\newblock \emph{Physical Review}, 48:\penalty0 696--702, 1935.

\bibitem[Bombin and Martin-Delgado(2005)]{BMD05}
H.~Bombin and M.~A. Martin-Delgado.
\newblock Entanglement distillation protocols and number theory.
\newblock \emph{Physical Review A}, 72:\penalty0 032313, 2005,
  \eprint{quant-ph/0503013}.

\bibitem[Bose et~al.(1998)Bose, Vedral, and Knight]{BVK98}
S.~Bose, V.~Vedral, and P.~L. Knight.
\newblock Multiparticle generalization of entanglement swapping.
\newblock \emph{Physical Review A}, 57:\penalty0 822--829, 1998,
  \eprint{quant-ph/9708004}.

\bibitem[Bose et~al.(1999)Bose, Vedral, and Knight]{BVK99}
S.~Bose, V.~Vedral, and P.~L. Knight.
\newblock Purification via entanglement swapping and conserved entanglement.
\newblock \emph{Physical Review A}, 60:\penalty0 194--197, 1999,
  \eprint{quant-ph/9812013}.

\bibitem[Bourennane et~al.(2004)Bourennane, Eibl, Kurtsiefer, Weinfurter,
  G{\"u}hne, Hyllus, Bru{\ss}, Lewenstein, and Sanpera]{BEKWGHBLS03}
M.~Bourennane, M.~Eibl, C.~Kurtsiefer, H.~Weinfurter, O.~G{\"u}hne, P.~Hyllus,
  D.~Bru{\ss}, M.~Lewenstein, and A.~Sanpera.
\newblock Witnessing multipartite entanglement.
\newblock \emph{Physical Review Letters}, 92:\penalty0 087902, 2004,
  \eprint{quant-ph/0309043}.

\bibitem[Boyd and Vandenberghe(2004)]{BV04}
S.~Boyd and L.~Vandenberghe.
\newblock \emph{Convex Optimization}.
\newblock Cambridge University Press, 2004.

\bibitem[Brand{\~a}o(2005{\natexlab{a}})]{Brandao05}
F.~G. S.~L. Brand{\~a}o.
\newblock Quantifying entanglement with witness operators.
\newblock \emph{Physical Review A}, 72:\penalty0 022310, 2005{\natexlab{a}},
  \eprint{quant-ph/0503152}.

\bibitem[Brand{\~a}o(2005{\natexlab{b}})]{Brandao05b}
F.~G. S.~L. Brand{\~a}o.
\newblock Quantifying the activation power of bipartite entangled states.
\newblock 2005{\natexlab{b}}, \eprint{quant-ph/0510078}.

\bibitem[Brand{\~a}o and Vianna(2006)]{BV04b}
F.~G. S.~L. Brand{\~a}o and R.~O. Vianna.
\newblock Witnessed entanglement.
\newblock \emph{International Journal of Quantum Information}, 4:\penalty0
  331--340, 2006, \eprint{quant-ph/0405096}.

\bibitem[Breuer(2006)]{Breuer06}
H.-P. Breuer.
\newblock An optimal entanglement criterion for mixed quantum states.
\newblock \emph{Physical Review Letters}, 97:\penalty0 080501, 2006,
  \eprint{quant-ph/0605036}.

\bibitem[Bru{\ss}(2002)]{Bruss02}
D.~Bru{\ss}.
\newblock Characterizing entanglement.
\newblock \emph{Journal of Mathematical Physics}, 43:\penalty0 4237--4251,
  2002, \eprint{quant-ph/011078}.

\bibitem[Bru{\ss} and Peres(2000)]{BP99}
D.~Bru{\ss} and A.~Peres.
\newblock Construction of quantum states with bound entanglement.
\newblock \emph{Physical Review A}, 61:\penalty0 30301, 2000,
  \eprint{quant-ph/9911056}.

\bibitem[Bru{\ss} et~al.(2002)Bru{\ss}, Cirac, Horodecki, Hulpke, Kraus,
  Lewenstein, and Sanpera]{BCHHKLS01}
D.~Bru{\ss}, J.~I. Cirac, P.~Horodecki, F.~Hulpke, B.~Kraus, M.~Lewenstein, and
  A.~Sanpera.
\newblock Reflections upon separability and distillability.
\newblock \emph{Journal of Modern Optics}, 49:\penalty0 1399--1418, 2002,
  \eprint{quant-ph/0110081}.

\bibitem[Carteret et~al.(2000)Carteret, Higuchi, and Sudbery]{CHS00}
H.~A. Carteret, A.~Higuchi, and A.~Sudbery.
\newblock Multipartite generalisation of the schmidt decomposition.
\newblock \emph{Journal of Mathematical Physics}, 41:\penalty0 7932--7939,
  2000, \eprint{quant-ph/0006125}.

\bibitem[Cavalcanti(2006)]{Cavalcanti06}
D.~Cavalcanti.
\newblock Connecting the generalized robustness and the geometric measure of
  entanglement.
\newblock \emph{Physical Review A}, 73:\penalty0 044302, 2006,
  \eprint{quant-ph/0602031}.

\bibitem[Cen et~al.(2002)Cen, Wu, Yang, and An]{CWYA02}
L.X. Cen, N.-J. Wu, F.-H. Yang, and J.-H. An.
\newblock Local transformation of mixed states of two qubits to {B}ell diagonal
  states.
\newblock \emph{Physical Review A}, 65:\penalty0 052318, 2002,
  \eprint{quant-ph/0203092}.

\bibitem[Cerf et~al.(1997)Cerf, Adami, and Gingrich]{CAG97}
N.~J. Cerf, C.~Adami, and R.~M. Gingrich.
\newblock Reduction criterion for separability.
\newblock \emph{Physical Review A}, 60:\penalty0 898--909, 1997,
  \eprint{quant-ph/9710001}.

\bibitem[Chen and Wu(2003)]{CW02}
K.~Chen and L.~A. Wu.
\newblock A matrix realignment method for recognizing entanglement.
\newblock \emph{Quantum Information and Computation}, 3:\penalty0 193--202,
  2003, \eprint{quant-ph/0205017}.

\bibitem[Chen and Wu(2002)]{CW02b}
K.~Chen and L.~A. Wu.
\newblock The generalized partial transposition criterion for separability of
  multipartite quantum states.
\newblock \emph{Physics Letters A}, 306:\penalty0 14--20, 2002,
  \eprint{quant-ph/0208058}.

\bibitem[Chen et~al.(2005)Chen, Albeverio, and Fei]{CAF05}
K.~Chen, S.~Albeverio, and S.-M. Fei.
\newblock Entanglement of formation of bipartite quantum states.
\newblock \emph{Physical Review Letters}, 95:\penalty0 210501, 2005,
  \eprint{quant-ph/0511155}.

\bibitem[Chen and Yang(2002)]{CY02b}
Y.~X. Chen and D.~Yang.
\newblock The relative entropy of entanglement of {S}chmidt correlated states
  and distillation.
\newblock 2002, \eprint{quant-ph/0204152}.

\bibitem[Chen et~al.(2003)Chen, Jin, and Yang]{CJY03}
Y.-X. Chen, J.-S Jin, and D.~Yang.
\newblock Distillation of multiple copies of {B}ell states.
\newblock \emph{Physical Review A}, 67:\penalty0 014302, 2003,
  \eprint{quant-ph/0204004}.

\bibitem[Choi(1975{\natexlab{a}})]{Choi75}
M.-D. Choi.
\newblock Completely positive linear maps on complex matrices.
\newblock \emph{Linear Algebra and its Applications}, 10:\penalty0 285--290,
  1975{\natexlab{a}}.

\bibitem[Choi(1975{\natexlab{b}})]{Choi75b}
M.-D. Choi.
\newblock Positive semidefinite biquadratic forms.
\newblock \emph{Linear Algebra and its Applications}, 12:\penalty0 95--100,
  1975{\natexlab{b}}.

\bibitem[Choi(1977)]{Choi77}
M.-D. Choi.
\newblock Extremal positive semidefinite forms.
\newblock \emph{Mathematische Annalen}, pages 1--18, 1977.

\bibitem[Choi(1980)]{Choi80}
M.-D. Choi.
\newblock Some assorted inequalities for positive linear maps on
  {$C^*$}-algebras.
\newblock \emph{Journal of Operator Theory}, 4:\penalty0 271--285, 1980.

\bibitem[Choi(1982)]{Choi82}
M.-D. Choi.
\newblock Positive linear maps.
\newblock \emph{Proceedings of Symposia in Pure Mathematics}, 38:\penalty0
  583--589, 1982.

\bibitem[Christandl(2006)]{Christandl06}
M.~Christandl.
\newblock \emph{The structure of bipartite quantum states: {I}nsights from
  group theory and cryptography}.
\newblock PhD thesis, University of Cambridge, 2006.

\bibitem[Christandl and Winter(2004)]{CW03}
M.~Christandl and A.~Winter.
\newblock Squached entanglement {--} {A}n additive entanglement measure.
\newblock \emph{Journal of Mathematical Physics}, 45:\penalty0 829--840, 2004,
  \eprint{quant-ph/0308088}.

\bibitem[Chru{\'s}ci{\'n}ski and Kossakowski(2006)]{CK06b}
D.~Chru{\'s}ci{\'n}ski and A.~Kossakowski.
\newblock On the structure of entanglement witnesses and new class of positive
  indecomposable maps.
\newblock 2006, \eprint{quant-ph/0606211}.

\bibitem[Cirac(2004)]{cirac}
J.~I. Cirac.
\newblock Private communication, 2004.

\bibitem[Cirac et~al.(2001)Cirac, D{\"u}r, Kraus, and Lewenstein]{CDKL01}
J.~I. Cirac, W.~D{\"u}r, B.~Kraus, and M.~Lewenstein.
\newblock Entangling operations and their implementation using a small amount
  of entanglement.
\newblock \emph{Physical Review Letters}, 86:\penalty0 544--547, 2001,
  \eprint{quant-ph/0007057}.

\bibitem[Clarisse(2005)]{Clarisse04}
L.~Clarisse.
\newblock Characterization of distillability of entanglement in terms of
  positive maps.
\newblock \emph{Physical Review A}, 71:\penalty0 032332, 2005,
  \eprint{quant-ph/0403073}.

\bibitem[Clarisse(2006{\natexlab{a}})]{Clarisse05}
L.~Clarisse.
\newblock The distillability problem revisited.
\newblock \emph{Quantum Information and Computation}, 6:\penalty0 539--560,
  2006{\natexlab{a}}, \eprint{quant-ph/0510035}.

\bibitem[Clarisse(2006{\natexlab{b}})]{Clarisse05b}
L.~Clarisse.
\newblock On the {S}chmidt robustness of pure states.
\newblock \emph{Journal of Physics A: Mathematical and General}, 39:\penalty0
  4239--4249, 2006{\natexlab{b}}, \eprint{quant-ph/0512012}.

\bibitem[Clarisse(2006{\natexlab{c}})]{Clarisse06}
L.~Clarisse.
\newblock Construction of bound entangled edge states with special ranks.
\newblock \emph{Physics Letters A}, 2006{\natexlab{c}},
  \eprint{quant-ph/0603283}.

\bibitem[Clarisse and Wocjan(2005)]{CW05}
L.~Clarisse and P.~Wocjan.
\newblock On independent permutation separability criteria.
\newblock \emph{Quantum Information and Computation}, 6:\penalty0 277--288,
  2005, \eprint{quant-ph/0504160}.

\bibitem[Clarisse et~al.(2005)Clarisse, Ghosh, Severini, and Sudbery]{CGSS05}
L.~Clarisse, S.~Ghosh, S.~Severini, and A.~Sudbery.
\newblock Entangling power of permutations.
\newblock \emph{Physical Review A}, 72:\penalty0 012314, 2005,
  \eprint{quant-ph/0502040}.

\bibitem[Clauser and Shimony(1978)]{CS78}
J.~F. Clauser and A.~Shimony.
\newblock Bell's theorem: experimental tests and implications.
\newblock \emph{Reports of Progress in Physics}, 41:\penalty0 1881--1927, 1978.

\bibitem[Clauser et~al.(1969)Clauser, Horne, Shimony, and Holt]{CHSH69}
J.~F. Clauser, A.~M. Horne, A.~Shimony, and R.A. Holt.
\newblock Proposed experiment to test local hidden-variable theories.
\newblock \emph{Physical Review Letters}, 23:\penalty0 880--884, 1969.

\bibitem[{d}e Pillis(1967)]{Pillis67}
J.~{d}e Pillis.
\newblock Linear transformations which preserve hermitian and positive
  semidefinite operators.
\newblock \emph{Pacific Journal of Mathematics}, 23:\penalty0 129--137, 1967.

\bibitem[Dehaene et~al.(2003)Dehaene, den Nest, Moor, and Verstraete]{DNMV02}
J.~Dehaene, M.~Van den Nest, B.~De Moor, and F.~Verstraete.
\newblock Local permutations of products of {B}ell states and entanglement
  distillation.
\newblock \emph{Physical Review A}, 67:\penalty0 022310, 2003,
  \eprint{quant-ph/0207154}.

\bibitem[Deutsch et~al.(1996)Deutsch, Ekert, Jozsa, Macchiavello, Popescu, and
  Sanpera]{DEJMPS96}
D.~Deutsch, A.~Ekert, R.~Jozsa, C.~Macchiavello, S.~Popescu, and A.~Sanpera.
\newblock Quantum privacy amplification and the security of quantum
  cryptography over noisy channels.
\newblock \emph{Physical Review Letters}, 77:\penalty0 2818--2821, 1996,
  \eprint{quant-ph/9604039}.

\bibitem[Devetak and Winter(2005)]{DW03c}
I.~Devetak and A.~Winter.
\newblock Distillation of secret key and entanglement from quantum states.
\newblock \emph{Proceedings of the Royal Society London A}, 461:\penalty0
  207--235, 2005, \eprint{quant-ph/0306078}.

\bibitem[Di{V}incenzo and Terhal(2002)]{DT00}
D.~P. Di{V}incenzo and B.~M. Terhal.
\newblock Product bases in quantum information theory.
\newblock In A.~Fokas, A.~Grigoryan, T.~Kibble, and B.~Zegarlinski, editors,
  \emph{13th International Congress of Mathematical Physics}, pages 399--407.
  International Press, Boston, 2002.

\bibitem[Di{V}incenzo et~al.(2000)Di{V}incenzo, Shor, Smolin, Terhal, and
  Thapliyal]{DSSTT00}
D.~P. Di{V}incenzo, P.~W. Shor, J.~A. Smolin, B.~M. Terhal, and A.~V.
  Thapliyal.
\newblock Evidence for bound entangled states with negative partial transpose.
\newblock \emph{Physical Review A}, 61:\penalty0 062312, 2000,
  \eprint{quant-ph/9910026}.

\bibitem[Di{V}incenzo et~al.(2003)Di{V}incenzo, Mor, Shor, Smolin, and
  Terhal]{DMSST99}
D.~P. Di{V}incenzo, T.~Mor, P.~W. Shor, J.~A. Smolin, and B.~M. Terhal.
\newblock Unextendible product bases, uncompletable product bases and bound
  entanglement.
\newblock \emph{Communications in Mathematical Physics}, 238:\penalty0
  379--410, 2003, \eprint{quant-ph/9908070}.

\bibitem[Doherty et~al.(2002)Doherty, Parrilo, and Spedalieri]{DPS01}
A.~C. Doherty, P.~A. Parrilo, and F.~M. Spedalieri.
\newblock Distinguishing separable and entangled states.
\newblock \emph{Physical Review Letters}, 88:\penalty0 187904, 2002,
  \eprint{quant-ph/0112007}.

\bibitem[Doherty et~al.(2004)Doherty, Parrilo, and Spedalieri]{DPS03}
A.~C. Doherty, P.~A. Parrilo, and F.~M. Spedalieri.
\newblock Complete family of separability criteria.
\newblock \emph{Physical Review A}, 69:\penalty0 022308, 2004,
  \eprint{quant-ph/0308032}.

\bibitem[Doherty et~al.(2005)Doherty, Parrilo, and Spedalieri]{DPS04}
A.~C. Doherty, P.~A. Parrilo, and F.~M. Spedalieri.
\newblock Detecting multipartite entanglement.
\newblock \emph{Physical Review A}, 71:\penalty0 032333, 2005,
  \eprint{quant-ph/0407143}.

\bibitem[Donald et~al.(2001)Donald, Horodecki, and Rudolph]{DHR01b}
M.~J. Donald, M.~Horodecki, and O.~Rudolph.
\newblock The uniqueness theorem for entanglement measures.
\newblock 2001, \eprint{quant-ph/0105017v1}.

\bibitem[Donald et~al.(2002)Donald, Horodecki, and Rudolph]{DHR01}
M.~J. Donald, M.~Horodecki, and O.~Rudolph.
\newblock The uniqueness theorem for entanglement measures.
\newblock \emph{Journal of Mathematical Physics}, 43:\penalty0 4252--4272,
  2002, \eprint{quant-ph/0105017}.

\bibitem[D{\"u}r(2001)]{Dur01}
W.~D{\"u}r.
\newblock Multipartite bound entangled states that violate {B}ell's inequality.
\newblock \emph{Physical Review Letters}, 87:\penalty0 230402, 2001,
  \eprint{quant-ph/0107050}.

\bibitem[D{\"u}r and Cirac(2000{\natexlab{a}})]{DC00}
W.~D{\"u}r and J.~I. Cirac.
\newblock Activating bound entanglement in multi{--}particle systems.
\newblock \emph{Physical Review A}, 62:\penalty0 022302, 2000{\natexlab{a}},
  \eprint{quant-ph/0002028}.

\bibitem[D{\"u}r and Cirac(2001)]{DC01}
W.~D{\"u}r and J.~I. Cirac.
\newblock Nonlocal operations: purification, storage, compression, tomography,
  and probabilistic implementation.
\newblock \emph{Physical Review A}, 64:\penalty0 012317, 2001,
  \eprint{quant-ph/0012148}.

\bibitem[D{\"u}r and Cirac(2000{\natexlab{b}})]{DC99}
W.~D{\"u}r and J.~I. Cirac.
\newblock Classification of multi{--}qubit mixed states: separability and
  distillability properties.
\newblock \emph{Physical Review A}, 61:\penalty0 042314, 2000{\natexlab{b}},
  \eprint{quant-ph/9911044}.

\bibitem[D{\"u}r et~al.(2000)D{\"u}r, Cirac, Lewenstein, and Bru{\ss}]{DCLB99}
W.~D{\"u}r, J.~I. Cirac, M.~Lewenstein, and D.~Bru{\ss}.
\newblock Distillability and partial transposition in bipartite systems.
\newblock \emph{Physical Review A}, 61:\penalty0 062313, 2000,
  \eprint{quant-ph/9910022}.

\bibitem[Eckert and Kreutzmann(2002)]{EK02}
K.~Eckert and H.~Kreutzmann.
\newblock Quantum-information-theory wintersemester 2000/2001.
\newblock http://lodda.iqo.uni-hannover.de/, 2002.

\bibitem[Eckert et~al.(2005)Eckert, G{\"u}hne, Hulpke, Hyllus, Korbicz,
  Mompart, Bru{\ss}, Lewenstein, and Sanpera]{EGHHKMBLS02}
K.~Eckert, O.~G{\"u}hne, F.~Hulpke, P.~Hyllus, J.~Korbicz, J.~Mompart,
  D.~Bru{\ss}, M.~Lewenstein, and A.~Sanpera.
\newblock Entanglement properties of composite quantum systems.
\newblock In T.~Beth and G.~Leuchs, editors, \emph{Quantum Information
  Processing}, chapter~7, pages 83--99. Wiley, 2005.

\bibitem[Edelman and Rao(2005)]{ER05}
A.~Edelman and N.~R. Rao.
\newblock Random matrix theory.
\newblock \emph{Acta Numerica}, 14:\penalty0 1--65, 2005.

\bibitem[Eggeling et~al.(2001)Eggeling, Vollbrecht, Werner, and Wolf]{EVWW01}
T.~Eggeling, K.~G.~H. Vollbrecht, R.~F. Werner, and M.~M. Wolf.
\newblock Distillability via protocols respecting the positivity of partial
  transpose.
\newblock \emph{Physical Review Letters}, 87:\penalty0 257902, 2001,
  \eprint{quant-ph/0104095}.

\bibitem[Einstein et~al.(1935)Einstein, Podolsky, and Rosen]{EPR35}
A.~Einstein, B.~Podolsky, and N.~Rosen.
\newblock Can quantum-mechanical description of physical reality be considered
  complete?
\newblock \emph{Physical Review}, 47:\penalty0 777--780, 1935.

\bibitem[Eisert(2005-2006)]{Eisert}
J.~Eisert.
\newblock Private communication, 2005-2006.

\bibitem[Eisert and Briegel(2001)]{EB01}
J.~Eisert and H.~J. Briegel.
\newblock The {S}chmidt measure as a tool for quantifying multi-particle
  entanglement.
\newblock \emph{Physical Review A}, 64:\penalty0 022306, 2001,
  \eprint{quant-ph/0007081}.

\bibitem[Eisert et~al.(2000)Eisert, Felbinger, Papadopoulos, Plenio, and
  Wilkens]{EFPPW99}
J.~Eisert, T.~Felbinger, P.~Papadopoulos, M.~B. Plenio, and M.~Wilkens.
\newblock Classical information and distillable entanglement.
\newblock \emph{Physical Review Letters}, 84:\penalty0 1611--1614, 2000,
  \eprint{quant-ph/9907021}.

\bibitem[Eisert et~al.(2002)Eisert, Simon, and Plenio]{ESP01}
J.~Eisert, C.~Simon, and M.~B. Plenio.
\newblock On the quantification of entanglement in infinite dimensional quantum
  systems.
\newblock \emph{Journal of Physics A}, 35:\penalty0 3911--3923, 2002,
  \eprint{quant-ph/0112064}.

\bibitem[Eisert et~al.(2005)Eisert, Hyllus, G{\"u}hne, and Curty]{EHGC04}
J.~Eisert, P.~Hyllus, O.~G{\"u}hne, and M.~Curty.
\newblock Complete hierarchies of efficient approximations to problems in
  entanglement theory.
\newblock \emph{Physical Review A}, 70:\penalty0 062317, 2005,
  \eprint{quant-ph/0407135}.

\bibitem[Ekert and Knight(1995)]{EK95}
A.~Ekert and P.~L. Knight.
\newblock Entangled quantum systems and the {S}chmidt decomposition.
\newblock \emph{American Journal of Physics}, 63:\penalty0 415--423, 1995.

\bibitem[Eom and Kye(2000)]{EK00}
M.-H. Eom and S.-H. Kye.
\newblock Duality for positive linear maps in matrix algebras.
\newblock \emph{Mathematica Scandinavia}, 80:\penalty0 130--142, 2000.

\bibitem[Fan(2002)]{Fan02}
H.~Fan.
\newblock A note on separability criteria for multipartite state.
\newblock 2002, \eprint{quant-ph/0210168}.

\bibitem[Fei et~al.(2006)Fei, Li-Jost, and Sun]{FLJS06}
S.-M. Fei, X.~Li-Jost, and B.-Z. Sun.
\newblock A class of bound entangled states.
\newblock \emph{Physics Letters A}, 352:\penalty0 321--325, 2006,
  \eprint{quant-ph/0603105}.

\bibitem[Franson(1989)]{Franson89}
J.~D. Franson.
\newblock Bell inequality for position and time.
\newblock \emph{Physical Review Letters}, 62:\penalty0 2205--2208, 1989.

\bibitem[Freedman and Clauser(1972)]{FC72}
S.~J. Freedman and J.~F. Clauser.
\newblock Experimental test of local hidden-variable theories.
\newblock \emph{Physical Review Letters}, 28:\penalty0 938--941, 1972.

\bibitem[Fry and Walther(2002)]{FW02}
E.~S. Fry and T.~Walther.
\newblock Atom based tests of the {B}ell inequalities: the legacy of {J}ohn
  {B}ell continues.
\newblock In R.A. Bertlmann and A.~Zeilinger, editors, \emph{Quantum
  [Un]speakables}, pages 103--117. Springer, Berlin-Heidelberg-New York, 2002.

\bibitem[Ghosh et~al.(2001)Ghosh, Kar, Roy, Sen({D}e), and Sen]{GKRSS01}
S.~Ghosh, G.~Kar, A.~Roy, A.~Sen({D}e), and U.~Sen.
\newblock Distinguishability of the {B}ell states.
\newblock \emph{Physical Review Letters}, 87:\penalty0 277902, 2001,
  \eprint{quant-ph/0106148}.

\bibitem[Ghosh et~al.(2004)Ghosh, Kar, and Roy]{GKR03}
S.~Ghosh, G.~Kar, and A.~Roy.
\newblock Local cloning of {B}ell states and distillable entanglement.
\newblock \emph{Physical Review A}, 69:\penalty0 052312, 2004,
  \eprint{quant-ph/0311062}.

\bibitem[Ghosh et~al.(2005)Ghosh, Joag, Kar, Kunkri, and Roy]{GJKKR04}
S.~Ghosh, P.~Joag, G.~Kar, S.~Kunkri, and A.~Roy.
\newblock Locally accessible information and distillation of entanglement.
\newblock \emph{Physical Review A}, 71:\penalty0 012321, 2005,
  \eprint{quant-ph/0403134}.

\bibitem[Gisin(1991)]{Gisin91}
N.~Gisin.
\newblock Bell inequality holds for all non-product states.
\newblock \emph{Physics Letters A}, 154:\penalty0 201--202, 1991.

\bibitem[Gisin(1996)]{Gisin96}
N.~Gisin.
\newblock Hidden quantum nonlocality revealed by local filters.
\newblock \emph{Physics Letters A}, 210:\penalty0 151--156, 1996.

\bibitem[Gleason(1957)]{Gleason57}
A.~M. Gleason.
\newblock Measures on the closed subspaces of a {H}ilbert space.
\newblock \emph{Journal of Mathematics and Mechanics}, 6\penalty0 (6):\penalty0
  885--893, 1957.

\bibitem[G{\"u}hne et~al.(2002{\natexlab{a}})G{\"u}hne, Hyllus, Bru{\ss},
  Ekert, Lewenstein, Macchiavello, and Sanpera]{GHBELMS02}
O.~G{\"u}hne, P.~Hyllus, D.~Bru{\ss}, A.~Ekert, M.~Lewenstein, C.~Macchiavello,
  and A.~Sanpera.
\newblock Detection of entanglement with few local measurements.
\newblock \emph{Physical Review A}, 66:\penalty0 062305, 2002{\natexlab{a}},
  \eprint{quant-ph/0205089}.

\bibitem[G{\"u}hne et~al.(2002{\natexlab{b}})G{\"u}hne, Hyllus, Bru{\ss},
  Ekert, Lewenstein, Macchiavello, and Sanpera]{GHBELMS02b}
O.~G{\"u}hne, P.~Hyllus, D.~Bru{\ss}, A.~Ekert, M.~Lewenstein, C.~Macchiavello,
  and A.~Sanpera.
\newblock Experimental detection of entanglement via witness operators and
  local measurements.
\newblock \emph{Physical Review A}, 66:\penalty0 62305, 2002{\natexlab{b}},
  \eprint{quant-ph/0210134}.

\bibitem[G{\"u}hne et~al.(2006)G{\"u}hne, Mechler, T{\'o}th, and Adam]{GMTA06}
O.~G{\"u}hne, M.~Mechler, G.~T{\'o}th, and P.~Adam.
\newblock Entanglement criteria based on local uncertainty relations are
  strictly stronger than the computable cross norm criterion.
\newblock \emph{Physical Review A}, 74:\penalty0 010301(R), 2006,
  \eprint{quant-ph/0604050}.

\bibitem[Gurvits(2003)]{Gurvits03}
L.~Gurvits.
\newblock Classical deterministic complexity of {E}dmonds' problem and quantum
  entanglement.
\newblock In \emph{Proceedings of the 35th ACM Symposium on Theory of
  Computing}, pages 10--19, New York, 2003.

\bibitem[Ha(2003)]{Ha03}
K.-C. Ha.
\newblock A class of atomic positive linear maps in matrix algebras.
\newblock \emph{Linear Algebra and its Applications}, 359:\penalty0 277--290,
  2003.

\bibitem[Ha and Kye(2004)]{HK03}
K.-C. Ha and S.-H. Kye.
\newblock Construction of entangled states with positive partial transposes
  based on indecomposable positive linear maps.
\newblock \emph{Physics Letters A}, 325:\penalty0 315--324, 2004,
  \eprint{quant-ph/0310109}.

\bibitem[Ha and Kye(2005)]{HK05}
K.-C. Ha and S.-H. Kye.
\newblock Construction of {$3\otimes 3$} entangled edge states with positive
  partial transposes.
\newblock \emph{Journal of Physics A: Mathematical and General}, 38:\penalty0
  9039--9050, 2005, \eprint{quant-ph/0509079}.

\bibitem[Ha et~al.(2003)Ha, Kye, and Park]{HKP03}
K.-C. Ha, S.-H. Kye, and Y.~S. Park.
\newblock Entangled states with positive partial transposes arising from
  indecomposable positive linear maps.
\newblock \emph{Physics Letters A}, 313:\penalty0 163--174, 2003,
  \eprint{quant-ph/0305005}.

\bibitem[Hall(2006)]{Hall06}
W.~Hall.
\newblock Constructions of indecomposable positive maps based on a new
  criterion of indecomposability.
\newblock 2006, \eprint{quant-ph/0607035}.

\bibitem[Hardy(1999)]{Hardy99}
L.~Hardy.
\newblock A method of areas for manipulating the entanglement properties of one
  copy of a two-particle pure state.
\newblock \emph{Physical Review A}, 60:\penalty0 1912--1923, 1999,
  \eprint{quant-ph/9903001}.

\bibitem[Harrow and Nielsen(2003)]{HN03}
A.~W. Harrow and M.~A. Nielsen.
\newblock How robust is a quantum gate in the presence of noise.
\newblock \emph{Physical Review A}, 68:\penalty0 012308, 2003,
  \eprint{quant-ph/0301108}.

\bibitem[Hayashi(2006)]{Hayashi02}
M.~Hayashi.
\newblock General formulas for fixed-length quantum entanglement concentration.
\newblock \emph{IEEE Transactions on Information Theory}, 52:\penalty0
  1904--1921, 2006, \eprint{quant-ph/0206187}.

\bibitem[Hayashi and Matsumoto(2001)]{HM01}
M.~Hayashi and K.~Matsumoto.
\newblock Variable length universal entanglement concentration by local
  operations and its application to teleportation and dense coding.
\newblock 2001, \eprint{quant-ph/0109028}.

\bibitem[Hayashi et~al.(2003)Hayashi, Koashi, Matsumoto, Morikoshi, and
  Winter]{HKMMW02}
M.~Hayashi, M.~Koashi, K.~Matsumoto, F.~Morikoshi, and A.~Winter.
\newblock Error exponents for entanglement concentration.
\newblock \emph{Journal of Physics A: Mathematical and General}, 36:\penalty0
  36 527--553, 2003, \eprint{quant-ph/0206097}.

\bibitem[Hayden et~al.(2006)Hayden, Leung, and Winter]{HLW05}
P.~Hayden, D.~W. Leung, and A.~Winter.
\newblock Aspects of generic entanglement.
\newblock \emph{Communications in Mathematical Physics}, 265:\penalty0 95--117,
  2006, \eprint{quant-ph/0407049}.

\bibitem[Hayden et~al.(2001)Hayden, Horodecki, and Terhal]{HHT00}
P.~M. Hayden, M.~Horodecki, and B.~M. Terhal.
\newblock The asymptotic entanglement cost of preparing a quantum state.
\newblock \emph{Journal of Physics A: Mathematical and General}, 34:\penalty0
  6891--6898, 2001, \eprint{quant-ph/0008134}.

\bibitem[Henderson and Vedral(2000)]{HV00}
L.~Henderson and V.~Vedral.
\newblock Information, relative entropy of entanglement and irreversibility.
\newblock \emph{Physical Review Letters}, 84:\penalty0 2263--2267, 2000,
  \eprint{quant-ph/9909011}.

\bibitem[Hill(1973)]{Hill73}
R.~D. Hill.
\newblock Linear transformations which preserve {H}ermitian matrices.
\newblock \emph{Linear Algebra and its Applications}, 6:\penalty0 257--262,
  1973.

\bibitem[Hiroshima and Hayashi(2004)]{HH04}
T.~Hiroshima and M.~Hayashi.
\newblock Finding a maximally correlated state - {S}imultaneous {S}chmidt
  decomposition of bipartite pure states.
\newblock \emph{Physical Review A}, 70:\penalty0 030302(R), 2004,
  \eprint{quant-ph/0405107}.

\bibitem[Horn and Johnson(1985)]{HJ85}
R.~A. Horn and C.~R. Johnson.
\newblock \emph{Matrix analysis}.
\newblock Cambridge University Press, Cambridge, United Kingdom, 1985.

\bibitem[Horn and Johnson(1991)]{HJ91}
R.~A. Horn and C.~R. Johnson.
\newblock \emph{Topics in Matrix Analysis}.
\newblock Cambridge University Press, Cambridge, 1991.

\bibitem[Horodecki et~al.(2005{\natexlab{a}})Horodecki, Horodecki, Horodecki,
  and Oppenheim]{HHHO03}
K.~Horodecki, M.~Horodecki, P.~Horodecki, and J.~Oppenheim.
\newblock Secure key from bound entanglement.
\newblock \emph{Physical Review Letters}, 94:\penalty0 160502,
  2005{\natexlab{a}}, \eprint{quant-ph/0309110}.

\bibitem[Horodecki et~al.(2005{\natexlab{b}})Horodecki, Pankowski, Horodecki,
  and Horodecki]{HPHH05}
K.~Horodecki, L.~Pankowski, M.~Horodecki, and P.~Horodecki.
\newblock Low dimensional bound entanglement with one-way distillable
  cryptographic key.
\newblock 2005{\natexlab{b}}, \eprint{quant-ph/0506203}.

\bibitem[Horodecki(2001)]{Horodecki01}
M.~Horodecki.
\newblock Entanglement measures.
\newblock \emph{Quantum Information and Computation}, 1:\penalty0 3--26, 2001.

\bibitem[Horodecki and Horodecki(1999)]{HH97}
M.~Horodecki and P.~Horodecki.
\newblock Reduction criterion of separability and limits for a class of
  distillation protocols.
\newblock \emph{Physical Review A}, 59:\penalty0 4206--4216, 1999,
  \eprint{quant-ph/9708015}.

\bibitem[Horodecki et~al.(1996{\natexlab{a}})Horodecki, Horodecki, and
  Horodecki]{HHH96}
M.~Horodecki, P.~Horodecki, and R.~Horodecki.
\newblock Separability of mixed states: necessary and sufficient conditions.
\newblock \emph{Physics Letters A}, 223:\penalty0 1--8, 1996{\natexlab{a}},
  \eprint{quant-ph/9605038}.

\bibitem[Horodecki et~al.(1997)Horodecki, Horodecki, and Horodecki]{HHH97}
M.~Horodecki, P.~Horodecki, and R.~Horodecki.
\newblock Inseparable two spin-{$\frac{1}{2}$} density matrices can be
  distilled to a singlet form.
\newblock \emph{Physical Review Letters}, 78:\penalty0 574--577, 1997,
  \eprint{quant-ph/9607009}.

\bibitem[Horodecki et~al.(1998{\natexlab{a}})Horodecki, Horodecki, and
  Horodecki]{HHH98}
M.~Horodecki, P.~Horodecki, and R.~Horodecki.
\newblock Mixed-state entanglement and distillation: is there a {`}bound{'}
  entanglement in nature?
\newblock \emph{Physical Review Letters}, 80:\penalty0 5239--5242,
  1998{\natexlab{a}}, \eprint{quant-ph/9801069}.

\bibitem[Horodecki et~al.(1999{\natexlab{a}})Horodecki, Horodecki, and
  Horodecki]{HHH99}
M.~Horodecki, P.~Horodecki, and R.~Horodecki.
\newblock Limits for entanglement measures.
\newblock \emph{Physical Review Letters}, 84:\penalty0 2014--2017,
  1999{\natexlab{a}}, \eprint{quant-ph/9908065}.

\bibitem[Horodecki et~al.(1999{\natexlab{b}})Horodecki, Horodecki, and
  Horodecki]{HHH99b}
M.~Horodecki, P.~Horodecki, and R.~Horodecki.
\newblock General teleportation channel, singlet fraction and
  quasi-distillation.
\newblock \emph{Physical Review A}, 60:\penalty0 1888--1898,
  1999{\natexlab{b}}, \eprint{quant-ph/9807091}.

\bibitem[Horodecki et~al.(2000{\natexlab{a}})Horodecki, Horodecki, and
  Horodecki]{HHH00b}
M.~Horodecki, P.~Horodecki, and R.~Horodecki.
\newblock Unified approach to quantum capacities: towards quantum noisy coding
  theorem.
\newblock \emph{Physical Review Letters}, 85:\penalty0 433--436,
  2000{\natexlab{a}}, \eprint{quant-ph/0003040}.

\bibitem[Horodecki et~al.(2001{\natexlab{a}})Horodecki, Horodecki, and
  Horodecki]{Alb01}
M.~Horodecki, P.~Horodecki, and R.~Horodecki.
\newblock Mixed-state entanglement and quantum communication.
\newblock In \emph{Quantum Information: An Introduction to Basic Theoretical
  Concepts and Experiments}. Springer, 2001{\natexlab{a}}.

\bibitem[Horodecki et~al.(2001{\natexlab{b}})Horodecki, Horodecki, and
  Horodecki]{HHH00}
M.~Horodecki, P.~Horodecki, and R.~Horodecki.
\newblock Separability of $n$-particle mixed states: necessary and sufficient
  conditions in terms of linear maps.
\newblock \emph{Physics Letters A}, 283:\penalty0 1--7, 2001{\natexlab{b}},
  \eprint{quant-ph/0006071}.

\bibitem[Horodecki et~al.(2003{\natexlab{a}})Horodecki, Sen(De), and
  Sen]{HSS02b}
M.~Horodecki, A.~Sen(De), and U.~Sen.
\newblock Rates of asymptotic entanglement transformations for bipartite mixed
  states: {M}aximally entangled states are not special.
\newblock \emph{Physical Review A}, 67:\penalty0 062314, 2003{\natexlab{a}},
  \eprint{quant-ph/0207031}.

\bibitem[Horodecki et~al.(2003{\natexlab{b}})Horodecki, Sen(De), Sen, and
  Horodecki]{HSSH03}
M.~Horodecki, A.~Sen(De), U.~Sen, and K.~Horodecki.
\newblock Local indistinguishability: more nonlocality with less entanglement.
\newblock \emph{Physical Review Letters}, 90:\penalty0 047902,
  2003{\natexlab{b}}, \eprint{quant-ph/0301106}.

\bibitem[Horodecki et~al.(2003{\natexlab{c}})Horodecki, Shor, and
  Ruskai]{HSR03}
M.~Horodecki, P.~W. Shor, and M.~B. Ruskai.
\newblock General entanglement breaking channels.
\newblock \emph{Reviews of Mathematical Physics}, 15:\penalty0 629--641,
  2003{\natexlab{c}}, \eprint{quant-ph/0302031}.

\bibitem[Horodecki et~al.(2004)Horodecki, Oppenheim, Sen(De), and Sen]{HOSS04}
M.~Horodecki, J.~Oppenheim, A.~Sen(De), and U.~Sen.
\newblock Distillation protocols: output entanglement and local mutual
  information.
\newblock \emph{Physical Review Letters}, 93:\penalty0 170503, 2004,
  \eprint{quant-ph/0405185}.

\bibitem[Horodecki et~al.(2006)Horodecki, Horodecki, and Horodecki]{HHH02}
M.~Horodecki, P.~Horodecki, and R.~Horodecki.
\newblock Separability of mixed quantum states: {L}inear contractions and
  permutation criteria.
\newblock \emph{Open Systems and Information Dynamics}, 13:\penalty0 103--111,
  2006, \eprint{quant-ph/0206008}.

\bibitem[Horodecki(2003)]{Horodecki03}
P.~Horodecki.
\newblock Direct estimation of elements of quantum states algebra and
  entanglement detection via linear contractions.
\newblock \emph{Physics Letters A}, 319:\penalty0 1--7, 2003.

\bibitem[Horodecki(1997)]{Horodecki97}
P.~Horodecki.
\newblock Separability criterion and inseparable mixed states with positive
  partial transposition.
\newblock \emph{Physics Letters A}, 232:\penalty0 333--339, 1997,
  \eprint{quant-ph/9703004}.

\bibitem[Horodecki and Demianowicz(2005)]{HD05}
P.~Horodecki and M.~Demianowicz.
\newblock Unconditional fidelity thresholds in single copy distillation and
  some aspects of quantum error correction.
\newblock 2005, \eprint{quant-ph/0501105}.

\bibitem[Horodecki and Lewenstein(2000)]{HL00}
P.~Horodecki and M.~Lewenstein.
\newblock Bound entanglement and continuous variables.
\newblock \emph{Physical Review Letters}, 85:\penalty0 2657--2660, 2000,
  \eprint{quant-ph/0001035}.

\bibitem[Horodecki et~al.(1998{\natexlab{b}})Horodecki, Horodecki, and
  Horodecki]{HHH98b}
P.~Horodecki, M.~Horodecki, and R.~Horodecki.
\newblock Bound entanglement can be activated.
\newblock \emph{Physical Review Letters}, 82:\penalty0 1056--1059,
  1998{\natexlab{b}}, \eprint{quant-ph/9806058}.

\bibitem[Horodecki et~al.(1998{\natexlab{c}})Horodecki, Horodecki, and
  Horodecki]{HHH98c}
P.~Horodecki, R.~Horodecki, and M.~Horodecki.
\newblock Entanglement and thermodynamical analogies.
\newblock \emph{Acta Physica Slovacia}, 48:\penalty0 141--156,
  1998{\natexlab{c}}, \eprint{quant-ph/9805072}.

\bibitem[Horodecki et~al.(2000{\natexlab{b}})Horodecki, Lewenstein, Vidal, and
  Cirac]{HLVC00}
P.~Horodecki, M.~Lewenstein, G.~Vidal, and I.~Cirac.
\newblock Operational criterion and constructive checks for the separability of
  low rank density matrices.
\newblock \emph{Physical Review A}, 62:\penalty0 032310, 2000{\natexlab{b}},
  \eprint{quant-ph/0002089}.

\bibitem[Horodecki et~al.(2003{\natexlab{d}})Horodecki, Cirac, and
  Lewenstein]{HCL01}
P.~Horodecki, J.~I. Cirac, and M.~Lewenstein.
\newblock Bound entanglement for continuous variables is a rare phenomenon.
\newblock In S.~L. Braunstein and A.~K. Pati, editors, \emph{Quantum
  Information with Continuous Variables}, pages 211--230. Springer,
  2003{\natexlab{d}}.

\bibitem[Horodecki et~al.(2003{\natexlab{e}})Horodecki, Smolin, Terhal, and
  Thapliyal]{HSTT99}
P.~Horodecki, J.~A. Smolin, B.~M. Terhal, and A.~V. Thapliyal.
\newblock Rank two bipartite bound entangled states do not exist.
\newblock \emph{Theoretical Computer Science}, 293:\penalty0 589--596,
  2003{\natexlab{e}}, \eprint{quant-ph/9910122}.

\bibitem[Horodecki et~al.(1996{\natexlab{b}})Horodecki, Horodecki, and
  Horodecki]{HHH96c}
R.~Horodecki, M.~Horodecki, and P.~Horodecki.
\newblock Teleportation, {B}ell's inequalities and inseparability.
\newblock \emph{Physics Letters A}, 222:\penalty0 21--25, 1996{\natexlab{b}},
  \eprint{quant-ph/9606027}.

\bibitem[Hostens et~al.(2004)Hostens, Dehaene, and Moor]{HDD04}
E.~Hostens, J.~Dehaene, and B.~De Moor.
\newblock The equivalence of two approaches to the design of entanglement
  distillation protocols.
\newblock 2004, \eprint{quant-ph/0406017}.

\bibitem[Hostens et~al.(2006)Hostens, Dehaene, and Moor]{HDD06}
E.~Hostens, J.~Dehaene, and B.~De Moor.
\newblock Asymptotic adaptive bipartite entanglement distillation protocol.
\newblock \emph{Physical Review A}, 73:\penalty0 062337, 2006,
  \eprint{quant-ph/0602205}.

\bibitem[Hsieh et~al.(2004)Hsieh, Li, and Chuu]{HLC03}
J.-Y. Hsieh, C.-M. Li, and D.-S. Chuu.
\newblock A simplification of entanglement purification.
\newblock \emph{Physics Letters A}, 328:\penalty0 94--101, 2004,
  \eprint{quant-ph/0304069}.

\bibitem[Hsu(2002)]{Hsu02}
L.-Y. Hsu.
\newblock Optimal entanglement purification via entanglement swapping with
  least classical communication.
\newblock \emph{Physics Letters A}, 297:\penalty0 126--128, 2002.

\bibitem[Hulpke and Bru{\ss}(2005)]{HB04}
F.~Hulpke and D.~Bru{\ss}.
\newblock A two-way algorithm for the entanglement problem.
\newblock \emph{Journal of Physics A: Mathematical and General}, 38:\penalty0
  5573--5579, 2005, \eprint{quant-ph/0407179}.

\bibitem[Hulpke et~al.(2004)Hulpke, Bru{\ss}, Lewenstein, and Sanpera]{HBLS04}
F.~Hulpke, D.~Bru{\ss}, M.~Lewenstein, and A.~Sanpera.
\newblock Simplifying {S}chmidt number witnesses via higher-dimensional
  embeddings.
\newblock \emph{Quantum Information and Computation}, 4:\penalty0 207--221,
  2004, \eprint{quant-ph/0401118}.

\bibitem[Hyllus et~al.(2005)Hyllus, G{\"u}hne, Bru{\ss}, and
  Lewenstein]{HGBL05}
P.~Hyllus, O.~G{\"u}hne, D.~Bru{\ss}, and M.~Lewenstein.
\newblock Relations between entanglement witnesses and {B}ell inequalities.
\newblock \emph{Physical Review A}, 72:\penalty0 012321, 2005,
  \eprint{quant-ph/0504079}.

\bibitem[Ioannou(2006)]{Ioannou06}
L.~M. Ioannou.
\newblock Deterministic computational complexity of the quantum separability
  problem.
\newblock 2006, \eprint{quant-ph/0603199}.

\bibitem[Isham(1995)]{Isham95}
C.~J. Isham.
\newblock \emph{Lectures on Quantum Theory: Mathematical and Structural
  Foundations}.
\newblock Imperial College Press, London, 1995.

\bibitem[Ishizaka(2004{\natexlab{a}})]{Ishizaka03}
S.~Ishizaka.
\newblock Binegativity and geometry of entangled states in two qubits.
\newblock \emph{Physical Review A}, 69:\penalty0 020301(R), 2004{\natexlab{a}},
  \eprint{quant-ph/0308056}.

\bibitem[Ishizaka(2004{\natexlab{b}})]{Ishizaka04}
S.~Ishizaka.
\newblock Bound entangled states provide overall convertibility of pure
  entangled states.
\newblock \emph{Physical Review Letters}, 93:\penalty0 190501,
  2004{\natexlab{b}}, \eprint{quant-ph/0403016}.

\bibitem[Jamio{\l}kowski(1972)]{Jamiolkowski72}
A.~Jamio{\l}kowski.
\newblock Linear transformations which preserve trace and positive
  semidefiteness of operators.
\newblock \emph{Reports on Mathematical Physics}, 3\penalty0 (4):\penalty0
  275--278, 1972.

\bibitem[Jammer(1974)]{Jammer74}
M.~Jammer.
\newblock \emph{The conceptual development of quantum mechanics. The
  interpretation of quantum mechanics in historical perspective.}
\newblock John Wiley, 1974.

\bibitem[Jan{\'{e}}(2002)]{Jane02}
E.~Jan{\'{e}}.
\newblock Purification of two-qubit mixed states.
\newblock \emph{Quantum Information and Computation}, 2:\penalty0 348--354,
  2002, \eprint{quant-ph/0205107}.

\bibitem[Jonathan and Plenio(1999{\natexlab{a}})]{JP99}
D.~Jonathan and M.~B. Plenio.
\newblock Minimal conditions for local pure-state entanglement manipulation.
\newblock \emph{Physical Review Letters}, 83:\penalty0 1455--1458,
  1999{\natexlab{a}}, \eprint{quant-ph/9903054}.

\bibitem[Jonathan and Plenio(1999{\natexlab{b}})]{JP99b}
D.~Jonathan and M.~B. Plenio.
\newblock Entanglement-assisted local manipulation of pure quantum states.
\newblock \emph{Physical Review Letters}, 83:\penalty0 3566--3569,
  1999{\natexlab{b}}, \eprint{quant-ph/9905071}.

\bibitem[Karnas and Lewenstein(2001)]{KL00}
S.~Karnas and M.~Lewenstein.
\newblock Separable approximations of density matrices of composite quantum
  systems.
\newblock \emph{Journal of Physics A: Mathematical and General}, 34:\penalty0
  6919--6937, 2001, \eprint{quant-ph/0011066}.

\bibitem[Kent(1998)]{Kent98}
A.~Kent.
\newblock Entangled mixed states and local purification.
\newblock \emph{Physical Review Letters}, 81:\penalty0 2839--2841, 1998,
  \eprint{quant-ph/9805088}.

\bibitem[Kent et~al.(1999)Kent, Linden, and Massar]{KLM99}
A.~Kent, N.~Linden, and S.~Massar.
\newblock Optimal entanglement enhancement for mixed states.
\newblock \emph{Physical Review Letters}, 83:\penalty0 2656--2659, 1999,
  \eprint{quant-ph/9902022}.

\bibitem[Keyl(2002)]{Keyl02}
M.~Keyl.
\newblock Fundamentals of quantum information theory.
\newblock \emph{Physics Reports}, 369:\penalty0 431--548, 2002,
  \eprint{quant-ph/0202122}.

\bibitem[Kochen and Specker(1967)]{KS67}
S.~Kochen and E.~P. Specker.
\newblock The problem of hidden variables in quantum mechanics.
\newblock \emph{Journal of Mathematics and Mechanics}, 17\penalty0
  (1):\penalty0 59--87, 1967.

\bibitem[Kraus et~al.(2000)Kraus, Cirac, Karnas, and Lewenstein]{KCKL00}
B.~Kraus, J.~I. Cirac, S.~Karnas, and M.~Lewenstein.
\newblock Separability in {$2\otimes N$} composite quantum systems.
\newblock \emph{Physical Review A}, 61:\penalty0 062302, 2000,
  \eprint{quant-ph/9912010}.

\bibitem[Kraus et~al.(2002)Kraus, Lewenstein, and Cirac]{KLC01}
B.~Kraus, M.~Lewenstein, and J.~I. Cirac.
\newblock Characterization of distillable and activable states using
  entanglement witnesses.
\newblock \emph{Physical Review A}, 65:\penalty0 042327, 2002,
  \eprint{quant-ph/0110174}.

\bibitem[Kraus(1983)]{Kraus83}
K.~Kraus.
\newblock \emph{States, Effects and Operations}.
\newblock Lecture Notes in Physics. Springer, Berlin, Heidelberg, 1983.

\bibitem[Labuschagne et~al.(2003)Labuschagne, Majewski, and Marciniak]{LMM03}
L.~Labuschagne, W.~Majewski, and M.~Marciniak.
\newblock On {$k$}-decomposability of positive maps.
\newblock 2003, \eprint{math-ph/0306017}.

\bibitem[Lamehi-Rachti and Mittig(1976)]{LRM76}
M.~Lamehi-Rachti and W.~Mittig.
\newblock Quantum mechanics and hidden variables: {A} test of {B}ell's
  inequality by the measurement of the spin correlation in low-energy
  proton-proton scattering.
\newblock \emph{Physical Review D}, 14:\penalty0 2543--2555, 1976.

\bibitem[Lax(2002)]{Lax02}
P.~D. Lax.
\newblock \emph{Functional Analysis}.
\newblock Wiley, 2002.

\bibitem[Lewenstein and Sanpera(1998)]{LS97}
M.~Lewenstein and A.~Sanpera.
\newblock Separability and entanglement of composite quantum systems.
\newblock \emph{Physical Review Letters}, 80:\penalty0 2261--2264, 1998,
  \eprint{quant-ph/9707043}.

\bibitem[Lewenstein et~al.(1999)Lewenstein, Cirac, and Karnas]{LCK99}
M.~Lewenstein, J.~I. Cirac, and S.~Karnas.
\newblock Separability and entanglement in {$2\otimes N$} composite quantum
  systems.
\newblock 1999, \eprint{quant-ph/9903012}.

\bibitem[Lewenstein et~al.(2000{\natexlab{a}})Lewenstein, Bru{\ss}, Cirac,
  Kraus, Ku{\'s}, Samsonowicz, Sanpera, and Tarrach]{LBCKKSST00}
M.~Lewenstein, D.~Bru{\ss}, J.~I. Cirac, B.~Kraus, M.~Ku{\'s}, J.~Samsonowicz,
  A.~Sanpera, and R.~Tarrach.
\newblock Separability and distillability in composite quantum systems -- a
  primer --.
\newblock \emph{Journal of Modern Optics}, 47:\penalty0 2481--2499,
  2000{\natexlab{a}}, \eprint{quant-ph/0006064}.

\bibitem[Lewenstein et~al.(2000{\natexlab{b}})Lewenstein, Kraus, Cirac, and
  Horodecki]{LKCH00}
M.~Lewenstein, B.~Kraus, J.~I. Cirac, and P.~Horodecki.
\newblock Optimization of entanglement witness.
\newblock \emph{Physical Review A}, 62:\penalty0 052310, 2000{\natexlab{b}},
  \eprint{quant-ph/0005014}.

\bibitem[Lewenstein et~al.(2000{\natexlab{c}})Lewenstein, Kraus, Horodecki, and
  Cirac]{LKHC00}
M.~Lewenstein, B.~Kraus, P.~Horodecki, and J.~I. Cirac.
\newblock Characterization of separable states and entanglement witness.
\newblock \emph{Physical Review A}, 63:\penalty0 044304, 2000{\natexlab{c}},
  \eprint{quant-ph/0005112}.

\bibitem[Li and Woerdeman(1997)]{LW97}
C.-K. Li and H.~J. Woerdeman.
\newblock Special classes of positive and completely positive maps.
\newblock \emph{Linear Algebra and its Applications}, 255:\penalty0 247--258,
  1997.

\bibitem[Linden and Popescu(1999)]{LP99b}
N.~Linden and S.~Popescu.
\newblock Bound entanglement and teleportation.
\newblock \emph{Physical Review A}, 59:\penalty0 137--140, 1999,
  \eprint{quant-ph/9807069}.

\bibitem[Linden et~al.(1998)Linden, Massar, and Popescu]{LMP98}
N.~Linden, S.~Massar, and S.~Popescu.
\newblock Purifying noisy entanglement requires collective measurements.
\newblock \emph{Physical Review Letters}, 81:\penalty0 3279--3282, 1998,
  \eprint{quant-ph/9805001}.

\bibitem[Lo and Popescu(1997)]{LP97}
H.-K. Lo and S.~Popescu.
\newblock Concentrating entanglement by local actions{---}beyond mean values.
\newblock \emph{Physical Review A}, 63:\penalty0 022301, 1997,
  \eprint{quant-ph/9707038}.

\bibitem[Lo and Popescu(1999)]{LP99}
H.-K. Lo and S.~Popescu.
\newblock Classical communication cost of entanglement manipulation: is
  entanglement an interconvertible resource?
\newblock \emph{Physical Review Letters}, 83:\penalty0 1459--1462, 1999,
  \eprint{quant-ph/9902045}.

\bibitem[Lockhart and Steiner(2002)]{LS00}
R.~B. Lockhart and M.~J. Steiner.
\newblock Preserving entanglement under decoherence and sandwiching all
  separable states.
\newblock \emph{Physical Review A}, 65:\penalty0 022107, 2002,
  \eprint{quant-ph/0009090}.

\bibitem[Lockhart et~al.(2002)Lockhart, Steiner, and Gerlach]{LSG00}
R.~B. Lockhart, M.~J. Steiner, and K.~Gerlach.
\newblock Geometry and product states.
\newblock \emph{Quantum Information and Computation}, 2:\penalty0 333--347,
  2002, \eprint{quant-ph/0010013}.

\bibitem[Macchiavello(1998)]{Macchiavello98}
C.~Macchiavello.
\newblock On the analytical convergence of the qpa procedure.
\newblock \emph{Physics Letters A}, 246:\penalty0 385--388, 1998,
  \eprint{quant-ph/9807074}.

\bibitem[Mair et~al.(2001)Mair, Vaziri, Weihs, and Zeilinger]{MVWZ01}
A.~Mair, A.~Vaziri, G.~Weihs, and A.~Zeilinger.
\newblock Entanglement of orbital angular momentum states of photons.
\newblock \emph{Nature}, 412:\penalty0 313--316, 2001,
  \eprint{quant-ph/0104070}.

\bibitem[Majewski(2000)]{Majewski00}
W.~A. Majewski.
\newblock Some remarks on separability of states.
\newblock 2000, \eprint{quant-ph/0003007}.

\bibitem[Majewski(2004{\natexlab{a}})]{Majewski04}
W.~A. Majewski.
\newblock Positive maps, states, entanglement and all that; some old and new
  problems.
\newblock 2004{\natexlab{a}}, \eprint{quant-ph/0411043}.

\bibitem[Majewski(2004{\natexlab{b}})]{Majewski04b}
W.~A. Majewski.
\newblock On positive maps, entanglement and quantization.
\newblock \emph{Open Systems and Information Dynamics}, 11:\penalty0 43--52,
  2004{\natexlab{b}}, \eprint{quant-ph/0404030}.

\bibitem[Majewski and Marciniak(2001)]{MM01}
W.~A. Majewski and M.~Marciniak.
\newblock On a characterization of positive maps.
\newblock \emph{Journal of Physics A: Mathematical and General}, 34:\penalty0
  5863--5874, 2001.

\bibitem[Majewski and Marciniak(2004)]{MM04}
W.~A. Majewski and M.~Marciniak.
\newblock {$k$}-decomposability of positive maps.
\newblock 2004, \eprint{quant-ph/0411035}.

\bibitem[Maneva and Smolin(2000)]{MS00}
E.~N. Maneva and J.~A. Smolin.
\newblock Improved two-party and multi-party purification protocols.
\newblock In S.~J. Lomonaco and H.~E. Brandt, editors, \emph{AMS Contemporary
  Mathematics Series: Quantum Computation and Quantum Information}, volume 305,
  pages 203--212. American Mathematical Society, 2000.

\bibitem[Mart{\'i}n-Delgado and Navascu{\'e}s(2003)]{MDN03}
M.~A. Mart{\'i}n-Delgado and M.~Navascu{\'e}s.
\newblock Distillation protocols for mixed states of multilevel qubits and the
  quantum renormalisation group.
\newblock \emph{European Physics Journal D}, 27:\penalty0 169--180, 2003,
  \eprint{quant-ph/0301099}.

\bibitem[Masanes(2006)]{Masanes05}
L.~Masanes.
\newblock All entangled states are useful for information processing.
\newblock \emph{Physical Review Letters}, 96:\penalty0 150501, 2006,
  \eprint{quant-ph/0508071}.

\bibitem[Masanes(2005)]{Masanes05b}
L.~Masanes.
\newblock Useful entanglement can be extracted from all nonseparable states.
\newblock 2005, \eprint{quant-ph/0510188}.

\bibitem[Matsumoto and Hayashi(2005)]{MH05}
K.~Matsumoto and M.~Hayashi.
\newblock Universal distortion-free entanglement concentration.
\newblock 2005, \eprint{quant-ph/0509140}.

\bibitem[Matsumoto and Yura(2004)]{MY03}
K.~Matsumoto and F.~Yura.
\newblock Entanglement cost of antisymmetric states and additivity of capacity
  of some quantum channels.
\newblock \emph{Journal of Physics A: Mathematical and General}, 37:\penalty0
  L167--L171, 2004, \eprint{quant-ph/0306009}.

\bibitem[Matsumoto et~al.(2004)Matsumoto, Shimono, and Winter]{MSW02}
K.~Matsumoto, T.~Shimono, and A.~Winter.
\newblock Remarks on additivity of the {H}olevo channel capacity and of the
  entanglement of formation.
\newblock \emph{Communications in Mathematical Physics}, 246:\penalty0
  427--442, 2004, \eprint{quant-ph/0206148}.

\bibitem[Matsumoto(2003)]{Matsumoto03}
R.~Matsumoto.
\newblock Conversion of a general quantum stabilizer code to an entanglement
  distillation protocol.
\newblock \emph{Journal of Physics A: Mathematical and General}, 36:\penalty0
  8113--8127, 2003, \eprint{quant-ph/0209091}.

\bibitem[Metwally(2001)]{Metwally01}
N.~Metwally.
\newblock A more efficient entanglement purification.
\newblock \emph{Physical Review A}, 66:\penalty0 054302, 2001,
  \eprint{quant-ph/0109051}.

\bibitem[Metwally and Obada(2006)]{MO06}
N.~Metwally and A.-S. Obada.
\newblock More efficient purifying sheme via controlled-controlled {NOT} gate.
\newblock \emph{Physics Letters A}, 352:\penalty0 45--48, 2006.

\bibitem[Morikoshi(2000)]{Morikoshi99}
F.~Morikoshi.
\newblock Recovery of entanglement lost in entanglement manipulation.
\newblock \emph{Physical Review Letters}, 84:\penalty0 3189, 2000,
  \eprint{quant-ph/9911019}.

\bibitem[Morikoshi and Koashi(2001)]{MK01}
F.~Morikoshi and M.~Koashi.
\newblock Deterministic entanglement concentration.
\newblock \emph{Physical Review A}, 64:\penalty0 022319, 2001,
  \eprint{quant-ph/0107120}.

\bibitem[Nielsen(1999)]{Nielsen98}
M.~A. Nielsen.
\newblock Conditions for a class of entanglement transformations.
\newblock \emph{Physical Review Letters}, 83:\penalty0 436--439, 1999,
  \eprint{quant-ph/9811053}.

\bibitem[Nielsen and Chuang(2000)]{NC00}
M.~A. Nielsen and I.~L. Chuang.
\newblock \emph{Quantum Computation and Quantum Information}.
\newblock Cambridge University Press, 2000.

\bibitem[Nielsen et~al.(1998)Nielsen, Caves, Schumacher, and Barnum]{NCSB97}
M.~A. Nielsen, C.~M. Caves, B.~Schumacher, and H.~Barnum.
\newblock Information-theoretic approach to quantum error correction and
  reversible measurement.
\newblock \emph{Proceedings of the Royal Society London A}, 454:\penalty0
  277--304, 1998, \eprint{quant-ph/9706064}.

\bibitem[Opatrn{\'y} and Kurizki(1999)]{OK99}
T.~Opatrn{\'y} and G.~Kurizki.
\newblock Optimization approach to entanglement distillation.
\newblock \emph{Physical Review A}, 60\penalty0 (1):\penalty0 167--172, 1999,
  \eprint{quant-ph/9811090}.

\bibitem[Osaka(1991)]{Osaka91}
H.~Osaka.
\newblock Indecomposable positive maps in low dimensional matrix algebras.
\newblock \emph{Linear Algebra and its Applications}, 153:\penalty0 73--83,
  1991.

\bibitem[Oxenrider and Hill(1985)]{OH85}
C.~J. Oxenrider and R.~D. Hill.
\newblock On the matrix reorderings {$\Gamma$} and {$\Psi$}.
\newblock \emph{Linear Algebra and its Applications}, 69:\penalty0 205--212,
  1985.

\bibitem[Ozawa(2000)]{Ozawa00}
M.~Ozawa.
\newblock Entanglement measures and the {Hilbert}-{Schmidt} distance.
\newblock \emph{Physics Letters A}, 268:\penalty0 158--160, 2000,
  \eprint{quant-ph/0002036}.

\bibitem[Pan et~al.(1998)Pan, Bouwmeester, Weinfurter, and Zeilinger]{PBWZ98}
J.-W. Pan, D.~Bouwmeester, H.~Weinfurter, and A.~Zeilinger.
\newblock Experimental entanglement swapping: entangling photons that never
  interacted.
\newblock \emph{Physical Review Letters}, 80:\penalty0 3891--3894, 1998.

\bibitem[Paunkovi{\'c} et~al.(2002)Paunkovi{\'c}, Omar, Bose, and
  Vedral]{POBV02}
N.~Paunkovi{\'c}, Y.~Omar, S.~Bose, and V.~Vedral.
\newblock Entanglement concentration using quantum statistics.
\newblock \emph{Physical Review Letters}, 88:\penalty0 187903, 2002,
  \eprint{quant-ph/0112004}.

\bibitem[Peres(1993)]{Peres93}
A.~Peres.
\newblock \emph{Quantum theory: concepts and methods}.
\newblock Kluwer Academic Publishers, Dordrecht, 1993.

\bibitem[Peres(1996{\natexlab{a}})]{Peres96}
A.~Peres.
\newblock Separability criterion for density matrices.
\newblock \emph{Physical Review Letters}, 77:\penalty0 1413--1415,
  1996{\natexlab{a}}, \eprint{quant-ph/9604005}.

\bibitem[Peres(1996{\natexlab{b}})]{Peres96b}
A.~Peres.
\newblock Collective tests for quantum nonlocality.
\newblock \emph{Physical Review A}, 54:\penalty0 2685, 1996{\natexlab{b}},
  \eprint{quant-ph/9603023, quant-ph/9609016}.

\bibitem[Peres(1999)]{Peres99}
A.~Peres.
\newblock All the {B}ell inequalities.
\newblock \emph{Foundation of Physics}, 29:\penalty0 589--614, 1999,
  \eprint{quant-ph/9807017}.

\bibitem[Piani(2006)]{Piani04}
M.~Piani.
\newblock A class of {$2^N \times 2^N$} bound entangled states revealed by
  non-decomposable maps.
\newblock \emph{Physical Review A}, 73:\penalty0 012345, 2006,
  \eprint{quant-ph/0411098}.

\bibitem[Piani and Mora(2006)]{PM06}
M.~Piani and C.~E. Mora.
\newblock Class of {PPT} bound entangled states associated to almost any set of
  pure entangled states.
\newblock 2006, \eprint{quant-ph/0607061}.

\bibitem[Pittenger(2003)]{Pittenger02}
A.~O. Pittenger.
\newblock Unextendible product bases and the construction of inseparable
  states.
\newblock \emph{Linear Algebra and its Applications}, 359:\penalty0 235--248,
  2003, \eprint{quant-ph/0208028}.

\bibitem[Plenio(2005)]{Plenio05}
M.~B. Plenio.
\newblock The logarithmic negativity: A full entanglement monotone that is not
  convex.
\newblock \emph{Physical Review Letters}, 95:\penalty0 090503, 2005,
  \eprint{quant-ph/0505071}.

\bibitem[Plenio and Virmani(2005)]{PV05}
M.~B. Plenio and S.~Virmani.
\newblock An introduction to entanglement measures.
\newblock 2005, \eprint{quant-ph/0504163}.

\bibitem[Plenio et~al.(2000)Plenio, Virmani, and Papadopoulos]{PVP00}
M.~B. Plenio, S.~Virmani, and P.~Papadopoulos.
\newblock Operator monotones, the reduction criterion and the relative entropy.
\newblock \emph{Journal of Physics A: Mathematical and General}, 33:\penalty0
  L193--L197, 2000, \eprint{quant-ph/0002075}.

\bibitem[Poluikis and Hill(1981)]{PH81}
J.~A. Poluikis and R.~D. Hill.
\newblock Completely positive and {H}ermitian preserving linear
  transformations.
\newblock \emph{Linear Algebra and its Applications}, 35:\penalty0 1--10, 1981.

\bibitem[Popescu(1994)]{Popescu94}
S.~Popescu.
\newblock Bell's inequalities versus teleportation: what is nonlocality?
\newblock \emph{Physical Review Letters}, 72\penalty0 (6):\penalty0 797--799,
  1994.

\bibitem[Popescu(1995)]{Popescu95}
S.~Popescu.
\newblock Bell's inequalities and density matrices: revealing {`}hidden{'}
  nonlocality.
\newblock \emph{Physical Review Letters}, 74\penalty0 (14):\penalty0
  2619--2622, 1995, \eprint{quant-ph/9502005}.

\bibitem[Popescu and Rohrlich(1992)]{PR92}
S.~Popescu and D.~Rohrlich.
\newblock Generic quantum nonlocality.
\newblock \emph{Physics Letters A}, 166:\penalty0 293--297, 1992.

\bibitem[Popescu and Rohrlich(1997)]{PR97}
S.~Popescu and D.~Rohrlich.
\newblock Thermodynamics and the measure of entanglement.
\newblock \emph{Physical Review A}, 56:\penalty0 3319--3321(R), 1997,
  \eprint{quant-ph/9610044}.

\bibitem[Po{\'z}niak et~al.(1998)Po{\'z}niak, {\.{Z}}yczkowski, and
  Ku{\'s}]{PZK98}
M.~Po{\'z}niak, K.~{\.{Z}}yczkowski, and M.~Ku{\'s}.
\newblock Composed ensembles of random unitary matrices.
\newblock \emph{Journal of Physics A}, 31:\penalty0 1059--1071, 1998,
  \eprint{chao-dyn/9707006}.

\bibitem[Rains(2000{\natexlab{a}})]{Rains00}
E.~M. Rains.
\newblock A semidefinite program for distillable entanglement.
\newblock \emph{IEEE Transactions on Information Theory}, 47:\penalty0
  2921--2933, 2000{\natexlab{a}}, \eprint{quant-ph/0008047}.

\bibitem[Rains(2000{\natexlab{b}})]{Rains00b}
E.~M. Rains.
\newblock Erratum: Bound on distillable entanglement.
\newblock \emph{Physical Review A}, 63:\penalty0 019902(E), 2000{\natexlab{b}},
  \eprint{quant-ph/9809082v2}.

\bibitem[Rains(1997)]{Rains97}
E.~M. Rains.
\newblock Entanglement purification via separable superoperators.
\newblock 1997, \eprint{quant-ph/9707002}.

\bibitem[Rains(1999{\natexlab{a}})]{Rains98}
E.~M. Rains.
\newblock Bound on distillable entanglement.
\newblock \emph{Physical Review A}, 60:\penalty0 179--184, 1999{\natexlab{a}},
  \eprint{quant-ph/9809082v1}.

\bibitem[Rains(1999{\natexlab{b}})]{Rains99}
E.~M. Rains.
\newblock Rigorous treatment of distillable entanglement.
\newblock \emph{Physical Review A}, 60\penalty0 (1):\penalty0 173--178,
  1999{\natexlab{b}}, \eprint{quant-ph/9809078}.

\bibitem[Rarity and Tapster(1990)]{RT90}
J.~Rarity and P.~Tapster.
\newblock Experimental violation of {B}ell's inequality based on phase and
  momentum.
\newblock \emph{Physical Review Letters}, 64:\penalty0 2495--2498, 1990.

\bibitem[Redhead(1986)]{Redhead87}
M.~Redhead.
\newblock \emph{Incompletness, Nonlocality and Realism: A Prolegomenon to the
  Philosophy of Quantum Mechanics}.
\newblock Clarendon Press, Oxford, 1986.

\bibitem[Reznick(2000)]{Reznick00}
B.~Reznick.
\newblock Some concrete aspects of {H}ilbert's 17th problem.
\newblock \emph{Contemporary Mathematics}, 253:\penalty0 251--272, 2000.

\bibitem[Robertson(1985)]{Robertson85}
A.~G. Robertson.
\newblock Positive projections on {$C^*$}-algebra and an extremal positive map.
\newblock \emph{Journal of the London Mathematical Society}, 32:\penalty0
  133--140, 1985.

\bibitem[Rockafellar(1970)]{Rockafellar70}
R.~T. Rockafellar.
\newblock \emph{Convex Analysis}.
\newblock Princeton University Press, 1970.

\bibitem[Rowe et~al.(2001)Rowe, Kielpinski, Meyer, Sackett, Itano, Monroe, and
  Wineland]{RKMSIMW01}
M.~Rowe, D.~Kielpinski, V.~Meyer, C.~Sackett, W.~Itano, C.~Monroe, and
  D.~Wineland.
\newblock Experimental violation of a {B}ell's inequality with efficient
  detection.
\newblock \emph{Nature}, 409:\penalty0 791--794, 2001.

\bibitem[Rudolph(2000)]{Rudolph00}
O.~Rudolph.
\newblock A separability criterion for density operators.
\newblock \emph{Journal of Physics A: Mathematical and General}, 33:\penalty0
  3951--3955, 2000, \eprint{quant-ph/0002026}.

\bibitem[Rudolph(2001)]{Rudolph00b}
O.~Rudolph.
\newblock A new class of entanglement measures.
\newblock \emph{Journal of Mathematical Physics}, 42:\penalty0 5306--5314,
  2001, \eprint{quant-ph/0005011}.

\bibitem[Rudolph(2002{\natexlab{a}})]{Rudolph02}
O.~Rudolph.
\newblock Further results on the cross norm criterion for separability.
\newblock 2002{\natexlab{a}}, \eprint{quant-ph/0202121}.

\bibitem[Rudolph(2002{\natexlab{b}})]{Rudolph02b}
O.~Rudolph.
\newblock A note on "a matrix realignment method for recognizing entanglement"
  quant-ph/0205017 v1.
\newblock 2002{\natexlab{b}}, \eprint{quant-ph/0205054}.

\bibitem[Rudolph(2003)]{Rudolph02c}
O.~Rudolph.
\newblock Some properties of the computable cross norm criterion for
  separability.
\newblock \emph{Physical Review A}, 67:\penalty0 032312, 2003,
  \eprint{quant-ph/0212047}.

\bibitem[Rudolph(2004)]{Rudolph04}
O.~Rudolph.
\newblock Computable cross-norm criterion for separability.
\newblock \emph{Letters in Mathematical Physics}, 70:\penalty0 57--64, 2004.

\bibitem[Ruskai(2002)]{Ruskai02}
M.~B. Ruskai.
\newblock Inequalities for quantum entropy: a review with conditions for
  equality.
\newblock \emph{Journal of Mathematical Physics}, 43:\penalty0 4358--4375,
  2002, \eprint{quant-ph/0205064}.

\bibitem[Ruskai(2003)]{Ruskai03}
M.~B. Ruskai.
\newblock Qubit entanglement breaking channels.
\newblock \emph{Reviews of Mathematical Physics}, 15:\penalty0 643--662, 2003,
  \eprint{quant-ph/0302032}.

\bibitem[Sanpera et~al.(2001)Sanpera, Bru{\ss}, and Lewenstein]{SBL00}
A.~Sanpera, D.~Bru{\ss}, and M.~Lewenstein.
\newblock Schmidt number witnesses and bound entanglement.
\newblock \emph{Physical Review A}, 63:\penalty0 050301(R), 2001,
  \eprint{quant-ph/0009109}.

\bibitem[Schatten(1970)]{Schatten70}
R.~Schatten.
\newblock \emph{Norm ideals of completely continuous operators}.
\newblock Springer, Berlin, Heidelberg, 2nd edition, 1970.

\bibitem[Schmidt(1907)]{Schmidt06}
E.~Schmidt.
\newblock Zur {T}heorie der linearen und nichtlinearen {I}ntegralgleichungen.
\newblock \emph{Mathematische Annalen}, 63:\penalty0 433--476, 1907.

\bibitem[Schr{\"o}dinger(1935)]{Schrodinger35}
E.~Schr{\"o}dinger.
\newblock Discussion of probability distributions between separated systems.
\newblock \emph{Proceedings of the Cambridge Philosophical Society},
  31:\penalty0 555--563, 1935.

\bibitem[Schumacher and Westmoreland(2000)]{SW00}
B.~Schumacher and M.~D. Westmoreland.
\newblock Relative entropy in quantum information theory.
\newblock In S.~J. Lomonaco and H.~E. Brandt, editors, \emph{AMS Contemporary
  Mathematics Series: Quantum Computation and Quantum Information}, volume 305,
  pages 265--290. American Mathematical Society, 2000.

\bibitem[Sen(De) et~al.(2005{\natexlab{a}})Sen(De), Sen, and Lewenstein]{SSL05}
A.~Sen(De), U.~Sen, and M.~Lewenstein.
\newblock Distillation protocols that involve local distinguishing: composing
  upper and lower bounds on locally accessible information.
\newblock 2005{\natexlab{a}}, \eprint{quant-ph/0505137}.

\bibitem[Sen(De) et~al.(2005{\natexlab{b}})Sen(De), Sen, Lewenstein, and
  Sanpera]{SSLS05}
A.~Sen(De), U.~Sen, M.~Lewenstein, and A.~Sanpera.
\newblock Lectures on quantum information. {C}hapter 1: {T}he separability
  problem.
\newblock 2005{\natexlab{b}}, \eprint{quant-ph/0508032}.

\bibitem[Shi et~al.(2000)Shi, Jiang, and Guo]{SJG00}
B.-S. Shi, Y.-K. Jiang, and G.-C. Guo.
\newblock Optimal entanglement purification via entanglement swapping.
\newblock \emph{Physical Review A}, 62:\penalty0 054301, 2000,
  \eprint{quant-ph/0005125}.

\bibitem[Shimono(2002)]{Shimono02}
T.~Shimono.
\newblock Lower bound for entanglement cost of antisymmetric states.
\newblock 2002, \eprint{quant-ph/0203039}.

\bibitem[Shimony(1995)]{Shimony95}
A.~Shimony.
\newblock Degree of entanglement.
\newblock \emph{Annals New York Academy of Sciences}, 755:\penalty0 675--678,
  1995.

\bibitem[Shor(2002)]{Shor02}
P.~W. Shor.
\newblock Additivity of the classical capacity of entanglement-breaking quantum
  channels.
\newblock \emph{Journal of Mathematical Physics}, 43:\penalty0 4334--4340,
  2002, \eprint{quant-ph/0201149}.

\bibitem[Shor(2004)]{Shor03}
P.~W. Shor.
\newblock Equivalence of additivity questions in quantum information theory.
\newblock \emph{Communications in Mathematical Physics}, 246:\penalty0
  453--472, 2004, \eprint{quant-ph/0305035}.

\bibitem[Shor and Smolin(1996)]{SS96}
P.~W. Shor and J.~A. Smolin.
\newblock Quantum error-correcting codes need not completely reveal the error
  syndrome.
\newblock 1996, \eprint{quant-ph/9604006}.

\bibitem[Shor et~al.(2001)Shor, Smolin, and Terhal]{SST01}
P.~W. Shor, J.~A. Smolin, and B.~M. Terhal.
\newblock Nonadditivity of bipartite distillable entanglement follows from
  conjecture on bound entangled {W}erner states.
\newblock \emph{Physical Review Letters}, 86:\penalty0 2681--2684, 2001,
  \eprint{quant-ph/0010054}.

\bibitem[Shor et~al.(2003)Shor, Smolin, and Thapliyal]{SST00}
P.~W. Shor, J.~A. Smolin, and A.~V. Thapliyal.
\newblock Superactivation of bound entanglement.
\newblock \emph{Physical Review Letters}, 90:\penalty0 107901, 2003,
  \eprint{quant-ph/0005117}.

\bibitem[Sloane()]{sloan}
N.~J.~A. Sloane.
\newblock The on-line encyclopedia of integer published electronically
  at:{\newline}
  {\href{http://www.research.att.com/~njas/sequences/index.html}{http://www.re%
search.att.com/\symbol{126}njas/sequences/}.}

\bibitem[Smolin(2000)]{Smolin00}
J.~A. Smolin.
\newblock A four-party unlockable bound-entangled state.
\newblock \emph{Physical Review A}, 63:\penalty0 032306, 2000,
  \eprint{quant-ph/0001001}.

\bibitem[Steiner(2003)]{Steiner03}
M.~Steiner.
\newblock Generalised robustness of entanglement.
\newblock \emph{Physical Review A}, 67:\penalty0 054305, 2003,
  \eprint{quant-ph/0304009}.

\bibitem[Stewart(1980)]{Stewart80}
G.~W. Stewart.
\newblock The efficient generation of random orthogonal matrices with an
  application to condition estimators.
\newblock \emph{SIAM Journal of Numerical Analysis}, 17:\penalty0 403--409,
  1980.

\bibitem[Stinespring(1955)]{Stinespring55}
W.~F. Stinespring.
\newblock Positive functions on {$C^*$}-algebras.
\newblock \emph{Proceedings of the American Mathematical Society}, 6:\penalty0
  211--216, 1955.

\bibitem[St{\o}rmer(1963)]{Stormer63}
E.~St{\o}rmer.
\newblock Positive linear maps of operator algebras.
\newblock \emph{Acta Mathematica}, 110:\penalty0 233--278, 1963.

\bibitem[St{\o}rmer(1982)]{Stormer82}
E.~St{\o}rmer.
\newblock Decomposable positive maps on {$C^*$}-algebras.
\newblock \emph{Proceedings of the American Mathematical Society}, 86:\penalty0
  402--404, 1982.

\bibitem[Sudbery(2005)]{Sudbery05}
A.~Sudbery.
\newblock The power of entanglement.
\newblock Plenary lecture at the Fourth International Symposium on Quantum
  Theory and Symmetries, Varna, Bulgaria, 16 August 2005, 2005.

\bibitem[Sudbery(2006)]{Tony06}
A.~Sudbery.
\newblock Private communication, 2006.

\bibitem[Takasaki and Tomiyama(1983)]{TT83}
T.~Takasaki and J.~Tomiyama.
\newblock On the geometry of positive maps in matrix algebras.
\newblock \emph{Mathematische Zeitschrift}, 184:\penalty0 101--108, 1983.

\bibitem[Tang(1986)]{Tang86}
W.S. Tang.
\newblock On positive linear maps between matrix algebras.
\newblock \emph{Linear Algebra and its Applications}, 79:\penalty0 33--44,
  1986.

\bibitem[Terhal(2000)]{Terhal00}
B.~M. Terhal.
\newblock Bell inequalities and the separability criterion.
\newblock \emph{Physics Letters A}, 217:\penalty0 319--326, 2000,
  \eprint{quant-ph/9911057}.

\bibitem[Terhal(2001{\natexlab{a}})]{Terhal01}
B.~M. Terhal.
\newblock Detecting quantum entanglement.
\newblock \emph{Theoretical Computer Science}, 287:\penalty0 313--335,
  2001{\natexlab{a}}, \eprint{quant-ph/0101032}.

\bibitem[Terhal(2001{\natexlab{b}})]{Terhal01b}
B.~M. Terhal.
\newblock A family of indecomposable positive linear maps based on entangled
  quantum states.
\newblock \emph{Linear Algebra and its Applications}, 323:\penalty0 61--73,
  2001{\natexlab{b}}, \eprint{quant-ph/9810091}.

\bibitem[Terhal and Horodecki(2000)]{TH00}
B.~M. Terhal and P.~Horodecki.
\newblock A {S}chmidt number for density matrices.
\newblock \emph{Physical Review A}, 61:\penalty0 040301(R), 2000,
  \eprint{quant-ph/9911117}.

\bibitem[Terhal and Vollbrecht(2000)]{TV00}
B.~M. Terhal and K.~G.~V. Vollbrecht.
\newblock The entanglement of formation for isotropic states.
\newblock \emph{Physical Review Letters}, 85:\penalty0 2625--2628, 2000,
  \eprint{quant-ph/0005062}.

\bibitem[Tittel et~al.(1998)Tittel, Brendel, Zbinden, and Gisin]{TBZG98}
W.~Tittel, J.~Brendel, H.~Zbinden, and N.~Gisin.
\newblock Violation of {B}ell inequalities by photons more than 10 km apart.
\newblock \emph{Physical Review Letters}, 81:\penalty0 3563--3566, 1998,
  \eprint{quant-ph/9806043}.

\bibitem[Tomiyama(1985)]{Tomiyama85}
J.~Tomiyama.
\newblock On the geometry of positive maps in matrix algebras. {II}.
\newblock \emph{Linear Algebra and its Applications}, 69:\penalty0 169--177,
  1985.

\bibitem[van Dam and Hayden(2003)]{DH02}
W.~van Dam and P.~Hayden.
\newblock Embezzling entangled quantum states.
\newblock \emph{Physical Review A}, 67:\penalty0 060302(R), 2003,
  \eprint{quant-ph/0201041}.

\bibitem[Vedral(2002)]{Vedral01}
V.~Vedral.
\newblock The role of relative entropy in quantum information theory.
\newblock \emph{Reviews of Modern Physics}, 74:\penalty0 197--234, 2002,
  \eprint{quant-ph/0102094}.

\bibitem[Vedral(1999)]{Vedral99}
V.~Vedral.
\newblock On bound entanglement assisted distillation.
\newblock \emph{Physics Letters A}, 262:\penalty0 121--124, 1999,
  \eprint{quant-ph/9908047}.

\bibitem[Vedral and Plenio(1998)]{VP98}
V.~Vedral and M.~B. Plenio.
\newblock Entanglement measures and purification procedures.
\newblock \emph{Physical Review A}, 57\penalty0 (3):\penalty0 1619--1633, 1998,
  \eprint{quant-ph/9707035}.

\bibitem[Vedral et~al.(1997)Vedral, Plenio, Rippin, and Knight]{VPRK97}
V.~Vedral, M.~B. Plenio, M.~A. Rippin, and P.~L. Knight.
\newblock Quantifying entanglement.
\newblock \emph{Physical Review Letters}, 78\penalty0 (12):\penalty0
  2275--2279, 1997, \eprint{quant-ph/9702027}.

\bibitem[Verstraete(2002)]{Verstraete02}
F.~Verstraete.
\newblock \emph{A study of entanglement in quantum information theory}.
\newblock PhD thesis, Katholieke Universiteit Leuven, 2002.

\bibitem[Verstraete and Verschelde(2002)]{VV02}
F.~Verstraete and H.~Verschelde.
\newblock On quantum channels.
\newblock 2002, \eprint{quant-ph/0202124}.

\bibitem[Verstraete and Verschelde(2003)]{VV03}
F.~Verstraete and H.~Verschelde.
\newblock Optimal teleportation with a mixed state of two qubits.
\newblock \emph{Physical Review Letters}, 90:\penalty0 097901, 2003,
  \eprint{quant-ph/0303007}.

\bibitem[Verstraete et~al.(2001)Verstraete, Dehaene, and Moor]{VDD01b}
F.~Verstraete, J.~Dehaene, and B.~De Moor.
\newblock Local filtering operations on two qubits.
\newblock \emph{Physical Review A}, 64:\penalty0 010101(R), 2001,
  \eprint{quant-ph/0011111}.

\bibitem[Vidal(2000)]{Vidal00}
G.~Vidal.
\newblock Entanglement monotones.
\newblock \emph{Journal of Modern Optics}, 47:\penalty0 355--376, 2000,
  \eprint{quant-ph/9807077}.

\bibitem[Vidal()]{Vidal02}
G.~Vidal.
\newblock On the continuity of asymptotic measures of entanglement.
\newblock \eprint{quant-ph/0203107}.

\bibitem[Vidal(1999)]{Vidal99}
G.~Vidal.
\newblock Entanglement of pure states for a single copy.
\newblock \emph{Physical Review Letters}, 83\penalty0 (5):\penalty0 1046--1049,
  1999, \eprint{quant-ph/9902033}.

\bibitem[Vidal and Cirac(2001)]{VC01}
G.~Vidal and J.~I. Cirac.
\newblock Irreversibility in asymptotic manipulations of entanglement.
\newblock \emph{Physical Review Letters}, 86:\penalty0 5803--5806, 2001,
  \eprint{quant-ph/0102036}.

\bibitem[Vidal and Cirac(2002)]{VC01b}
G.~Vidal and J.~I. Cirac.
\newblock When only two thirds of the entanglement can be distilled.
\newblock \emph{Physical Review A}, 65:\penalty0 012323, 2002,
  \eprint{quant-ph/0107051}.

\bibitem[Vidal and Tarrach(1999)]{VT99}
G.~Vidal and R.~Tarrach.
\newblock Robustness of entanglement.
\newblock \emph{Physical Review A}, 59:\penalty0 141--155, 1999,
  \eprint{quant-ph/9806094}.

\bibitem[Vidal and Werner(2002)]{VW01b}
G.~Vidal and R.~F. Werner.
\newblock A computable measure of entanglement.
\newblock \emph{Physical Review A}, 65:\penalty0 032314, 2002,
  \eprint{quant-ph/0102117}.

\bibitem[Vidal et~al.(2000)Vidal, Jonathan, and Nielsen]{VJN99}
G.~Vidal, D.~Jonathan, and M.~A. Nielsen.
\newblock Approximate transformations and robust manipulation of bipartite pure
  state entanglement.
\newblock \emph{Physical Review A}, 62:\penalty0 012304, 2000,
  \eprint{quant-ph/9910099}.

\bibitem[Vidal et~al.(2002)Vidal, D{\"u}r, and Cirac]{VDC01}
G.~Vidal, W.~D{\"u}r, and J.~I. Cirac.
\newblock Entanglement cost of mixed states.
\newblock \emph{Physical Review Letters}, 89:\penalty0 027901, 2002,
  \eprint{quant-ph/0112131}.

\bibitem[Vollbrecht and Verstraete(2005)]{VV04}
K.~G.~H. Vollbrecht and F.~Verstraete.
\newblock Interpolation of recurrence and hashing entanglement distillation
  protocols.
\newblock \emph{Physical Review A}, 71:\penalty0 062325, 2005,
  \eprint{quant-ph/0404111}.

\bibitem[Vollbrecht and Werner(2001)]{VW01}
K.~G.~H. Vollbrecht and R.~F. Werner.
\newblock Entanglement measures under symmetry.
\newblock \emph{Physical Review A}, 64:\penalty0 062307, 2001,
  \eprint{quant-ph/0010095}.

\bibitem[Vollbrecht and Wolf(2002)]{VW02}
K.~G.~H. Vollbrecht and M.~M. Wolf.
\newblock Activating {NPT} distillation with an infinitesimal amount of bound
  entanglement.
\newblock \emph{Physical Review Letters}, 88:\penalty0 247901, 2002,
  \eprint{quant-ph/0201103}.

\bibitem[Vollbrecht and Wolf(2003)]{VW02b}
K.~G.~H. Vollbrecht and M.~M. Wolf.
\newblock Efficient distillation beyond qubits.
\newblock \emph{Physical Review A}, 67:\penalty0 012303, 2003,
  \eprint{quant-ph/0208152}.

\bibitem[Vollbrecht et~al.(2004)Vollbrecht, Werner, and Wolf]{VWW03}
K.~G.~H. Vollbrecht, R.~F. Werner, and M.~M. Wolf.
\newblock On the irreversibility of entanglement distillation.
\newblock \emph{Physical Review A}, 69:\penalty0 062304, 2004,
  \eprint{quant-ph/0301072}.

\bibitem[Walgate et~al.(2000)Walgate, Short, Hardy, and Vedral]{WSHV00}
J.~Walgate, A.~J. Short, L.~Hardy, and V.~Vedral.
\newblock Local distinguishability of multipartite orthogonal quantum states.
\newblock \emph{Physical Review Letters}, 85:\penalty0 4972--4975, 2000,
  \eprint{quant-ph/0007098}.

\bibitem[Wang and Zanardi(2002)]{WZ02}
X.~Wang and P.~Zanardi.
\newblock Quantum entanglement of unitary operators on bi-partite systems.
\newblock \emph{Physical Review A}, 66:\penalty0 044303, 2002,
  \eprint{quant-ph/0207007}.

\bibitem[Watanabe et~al.(2006)Watanabe, Matsumoto, and Uyematsu]{WMU05}
S.~Watanabe, R.~Matsumoto, and T.~Uyematsu.
\newblock Improvement of stabilizer based entanglement distillation protocols
  by encoding operators.
\newblock \emph{Journal of Physics A: Mathematical and General}, 39:\penalty0
  4273--4290, 2006, \eprint{quant-ph/0506054}.

\bibitem[Watrous(2004)]{Watrous03}
J.~Watrous.
\newblock On the number of copies required for entanglement distillation.
\newblock \emph{Physical Review Letters}, 93:\penalty0 010502, 2004,
  \eprint{quant-ph/0312123}.

\bibitem[Wei and Goldbart(2003)]{WG03}
T.-C. Wei and P.~M. Goldbart.
\newblock Geometric measure of entanglement and applications to bipartite and
  multipartite quantum states.
\newblock \emph{Physical Review A}, 68:\penalty0 042307, 2003,
  \eprint{quant-ph/0307219}.

\bibitem[Wei et~al.(2004{\natexlab{a}})Wei, Altepeter, Goldbart, and
  Munro]{WAGM04}
T.-C. Wei, J.~B. Altepeter, P.~M. Goldbart, and W.~J. Munro.
\newblock Measures of entanglement in multipartite bound entangled states.
\newblock \emph{Physical Review A}, 70:\penalty0 022322, 2004{\natexlab{a}},
  \eprint{quant-ph/0308031}.

\bibitem[Wei et~al.(2004{\natexlab{b}})Wei, Ericsson, Goldbart, and
  Munro]{WEGM04}
T.-C. Wei, M.~Ericsson, P.~M. Goldbart, and W.~J. Munro.
\newblock Connections between relative entropy of entanglement and geometric
  measure of entanglement.
\newblock \emph{Quantum Information and Computation}, 4:\penalty0 252--272,
  2004{\natexlab{b}}, \eprint{quant-ph/0405002}.

\bibitem[Wellens and Ku{\'s}(2001)]{WK01}
T.~Wellens and M.~Ku{\'s}.
\newblock Separable approximation for mixed state of composite quantum systems.
\newblock \emph{Physical Review A}, 65:\penalty0 052302, 2001,
  \eprint{quant-ph/0104098}.

\bibitem[Werner(1989)]{Werner89}
R.~F. Werner.
\newblock Quantum states with {E}instein-{P}odolsky-{R}osen correlations
  admitting a hidden-variable model.
\newblock \emph{Physical Review A}, 40:\penalty0 4277--4281, 1989.

\bibitem[Werner and Wolf(2001)]{WW01b}
R.~F. Werner and M.~M. Wolf.
\newblock Bound entangled {G}aussian states.
\newblock \emph{Physical Review Letters}, 86:\penalty0 3658--3661, 2001,
  \eprint{quant-ph/0009118}.

\bibitem[Witte and Trucks(1999)]{WT98}
C.~Witte and M.~Trucks.
\newblock A new entanglement measure induced by the {H}ilbert-{S}chmidt norm.
\newblock \emph{Physics Letters A}, 257:\penalty0 14--20, 1999,
  \eprint{quant-ph/9811027}.

\bibitem[Wocjan and Horodecki(2005)]{WH05b}
P.~Wocjan and M.~Horodecki.
\newblock Characterization of combinatorically independent permutation
  separability criteria.
\newblock \emph{Open Systems and Information Dynamics}, 12:\penalty0 331--345,
  2005, \eprint{quant-ph/0503129}.

\bibitem[Woerdeman(2003)]{Woerdeman03}
H.~J. Woerdeman.
\newblock Checking {$2\times M$} quantum separability via semidefinite
  programming.
\newblock \emph{Physical Review A}, 67:\penalty0 010303(R), 2003,
  \eprint{quant-ph/0301058}.

\bibitem[Wootters(2001)]{Wootters01}
W.~K. Wootters.
\newblock Entanglement of formation and concurrence.
\newblock \emph{Quantum Information and Computation}, 1:\penalty0 27--44, 2001.

\bibitem[Wootters(1997)]{Wootters97}
W.~K. Wootters.
\newblock Entanglement of formation of an arbitrary state of two qubits.
\newblock \emph{Physical Review Letters}, 80:\penalty0 2245--2248, 1997,
  \eprint{quant-ph/9709029}.

\bibitem[Woronowicz(1976)]{Woronowicz76}
S.~L. Woronowicz.
\newblock Positive maps of low dimensional matrix algebras.
\newblock \emph{Reports on Mathematical Physics}, 10:\penalty0 165--183, 1976.

\bibitem[Wu and Zhang(2000)]{WZ00}
S.~Wu and Y.~Zhang.
\newblock Calculating the relative entropy of entanglement.
\newblock 2000, \eprint{quant-ph/0004018}.

\bibitem[Yang and Chen(2004)]{YC03}
D.~Yang and Y.-X. Chen.
\newblock Mixture of multiple copies of maximally entangled states is
  quasi-pure.
\newblock \emph{Physical Review A}, 69:\penalty0 024302, 2004,
  \eprint{quant-ph/0304194}.

\bibitem[Yang et~al.(2005)Yang, Horodecki, Horodecki, and Synak-Radtke]{YHHS05}
D.~Yang, M.~Horodecki, R.~Horodecki, and B.~Synak-Radtke.
\newblock Irreversibility for all bound entangled states.
\newblock \emph{Physical Review Letters}, 95:\penalty0 190501, 2005,
  \eprint{quant-ph/0506138}.

\bibitem[Yu and Liu(2005)]{YL04}
S.~Yu and N.l. Liu.
\newblock Entanglement detection by local orthogonal observables.
\newblock \emph{Physical Review Letters}, 95:\penalty0 150504, 2005,
  \eprint{quant-ph/0412220}.

\bibitem[Yura(2003)]{Yura03}
F.~Yura.
\newblock Entanglement cost of three-level antisymmetric states.
\newblock \emph{Journal of Physics A: Mathematical and General}, 36:\penalty0
  L237--L242, 2003, \eprint{quant-ph/0302163}.

\bibitem[Yurke and Stoler(1992{\natexlab{a}})]{YS92}
B.~Yurke and D.~Stoler.
\newblock Bell's-inequality experiments using independent-particle sources.
\newblock \emph{Physical Review A}, 46:\penalty0 2229--2234,
  1992{\natexlab{a}}.

\bibitem[Yurke and Stoler(1992{\natexlab{b}})]{YS92b}
B.~Yurke and D.~Stoler.
\newblock Einstein-{P}odolsky-{R}osen effects from independent particle
  sources.
\newblock \emph{Physical Review Letters}, 68:\penalty0 1251--1254,
  1992{\natexlab{b}}.

\bibitem[Zanardi(2001)]{Zanardi00}
P.~Zanardi.
\newblock Entanglement of quantum evolutions.
\newblock \emph{Physical Review A}, 63:\penalty0 040304(R), 2001,
  \eprint{quant-ph/0010074}.

\bibitem[Zanardi et~al.(2000)Zanardi, Zalka, and Faoro]{ZZF00}
P.~Zanardi, C.~Zalka, and L.~Faoro.
\newblock On the entangling power of quantum evolutions.
\newblock \emph{Physical Review A}, 62:\penalty0 030301(R), 2000,
  \eprint{quant-ph/0005031}.

\bibitem[{\.{Z}}ukowski et~al.(1993){\.{Z}}ukowski, Zeilinger, Horne, and
  Ekert]{ZZHE93}
M.~{\.{Z}}ukowski, A.~Zeilinger, M.~A. Horne, and A.~K. Ekert.
\newblock {E}vent-{R}eady-{D}etectors: {B}ell experiment via entanglement
  swapping.
\newblock \emph{Physical Review Letters}, 71:\penalty0 4287--4290, 1993.

\bibitem[{\.{Z}}yczkowski(1999)]{Zyczkowski99}
K.~{\.{Z}}yczkowski.
\newblock Volume of the set of mixed entangled states {II}.
\newblock \emph{Physical Review A}, 60:\penalty0 3496--3507, 1999,
  \eprint{quant-ph/9902050}.

\bibitem[{\.{Z}}yczkowski and Bengtsson(2004)]{ZB04}
K.~{\.{Z}}yczkowski and I.~Bengtsson.
\newblock On duality between quantum maps and quantum states.
\newblock \emph{Open Systems and Information Dynamics}, 11:\penalty0 3--42,
  2004, \eprint{quant-ph/0401119}.

\bibitem[{\.{Z}}yczkowski and Ku{\'s}(1994)]{ZK94}
K.~{\.{Z}}yczkowski and M.~Ku{\'s}.
\newblock Random unitary matrices.
\newblock \emph{Journal of Physics A: Mathematical and General}, 27:\penalty0
  4235--4245, 1994.

\bibitem[{\.{Z}}yczkowski et~al.(1998){\.{Z}}yczkowski, Horodecki, Sanpera, and
  Lewenstein]{ZHSL98}
K.~{\.{Z}}yczkowski, P.~Horodecki, A.~Sanpera, and M.~Lewenstein.
\newblock On the volume of the set of mixed entangled states.
\newblock \emph{Physical Review A}, 58:\penalty0 883--892, 1998,
  \eprint{quant-ph/9804024}.

\end{thebibliography}
\end{document}